\documentclass[PhD]{uclathes}

\usepackage{caption}
\usepackage{subcaption}
\usepackage{graphicx}
\usepackage{cite}
\usepackage{url}
\usepackage[cmex10]{amsmath}
\usepackage{amssymb}
\usepackage{algorithm}
\usepackage{algpseudocode}
\usepackage{cases}
\usepackage{color}
\usepackage{soul}
\usepackage{xcolor}
\usepackage{framed}
\usepackage{cite}
\usepackage{epstopdf}
\usepackage{calc}
\usepackage{array}
\usepackage{amsmath}
\usepackage{comment}
\usepackage{dsfont}
\usepackage{amsthm}
\usepackage{algorithm}
\usepackage{algorithmicx}
\usepackage{algpseudocode}
\usepackage{bbm}

\DeclareGraphicsExtensions{.pdf,.eps,.png,.jpg,.mps}
\DeclareMathOperator{\diag}{diag}

\newtheorem{theorem}{Theorem}
\newtheorem{corollary}{Corollary}

\title          {Interference Mitigation and Resource Allocation in Underlay Cognitive Radio Networks}
\author         {Shailesh Digambar Chaudhari}
\department     {{Electrical and Computer Engineering}}
\degreeyear     {2018}

\chair          {Danijela Cabric}
\member         {Babak Daneshrad}
\member         {Lieven Vandenberghe}
\member         {Mario Gerla}

\vitaitem   {2006--2010}{\hspace{4mm} B. Tech. in Electronics and Telecommunication, College of Engineering, \\ \hspace*{4mm} Pune, India.}
                
\vitaitem   {2010--2011}
{\hspace{4mm} Software Engineer, Cisco Systems India Pvt. Ltd.}

\vitaitem   {2011--2013}
				{\hspace{4mm} M.S. in Electrical Engineering, University of California, Los Angeles.}

\vitaitem   {2013--2017}
				{\hspace{4mm} Graduate Student Reseearcher and Teaching Assistant, UCLA.}

\vitaitem   {Winter 2013}
				{\hspace{4mm} Intern, Silvus Technologies, Inc.}
				
\vitaitem   {Summer 2016}
				{\hspace{4mm} Intern, Qualcomm Technologies, Inc.}

\vitaitem   {2017--2018}
{\hspace{4mm} Graduate Student Reseearcher and Teaching Fellow, UCLA.}

\publication    {\cite{chaudhari2019_tccn}{S. Chaudhari}, D. Cabric, ``{Power Control and Frequency Band Selection Policies for Underlay MIMO Cognitive Radio}", \textit{in IEEE Trans. on Cognitive Comm. and Networks, 2019.}\\ \\	
	\cite{chaudhari2018_tccn} {S. Chaudhari}, D. Cabric, ``{QoS Aware Power Allocation and Secondary User Selection in Massive MIMO Cognitive Radio Networks}", \textit{in IEEE Trans. on Cognitive Comm. and Networks, 2018.}\\ \\
	\cite{chaudhari2016_tccn}{S. Chaudhari}, D. Cabric, ``{Cyclic Weighted Centroid Algorithm for Primary User Localization in the Presence of Interference}", \textit{in IEEE Trans. on Cognitive Comm. and Networks, 2016.}\\ \\ 	
	\cite{chaudhari2017}{S. Chaudhari}, D. Cabric, ``{Unsupervised Frequency Clustering Algorithm for Null Space Estimation in Wideband Spectrum Sharing Networks}", \textit{in IEEE GlobalSIP, 2017.}\\ \\
	\cite{chaudhari2017_kuiper}{S. Chaudhari}, D. Cabric, ``{Kuiper Test based Modulation Level Classification under Unknown Frequency Selective Channels}", \textit{in IEEE GlobalSIP, 2017.}\\ \\
	\cite{yan2017_camsap} H.Yan, {S. Chaudhari}, D. Cabric, ``Wideband Channel Tracking for mmWave MIMO System with Hybrid Beamforming Architecture", \textit{(invited paper) in IEEE CAMSAP, 2017.}\\ \\
	\cite{wang2017_asilomar}X Wang, {S. Chaudhari}, M. Laghate, and D. Cabric, ``Wideband Spectrum Sensing Measurement Results using Tunable Front-End and FPGA Implementation", \textit{in Asilomar conf. on Signals, Systems, and Computers, 2017.}\\ \\
	\cite{laghate2017_dyspan}M. Laghate, {S. Chaudhari}, and D. Cabric, ``USRP N210 Demonstration of Wideband Sensing and Blind Hierarchical Modulation Classification", \textit{in IEEE DySPAN Workshop: Battle of the ModRecs, 2017.}\\ \\
	\cite{chaudhari2017a}{S. Chaudhari}, D. Cabric, ``{Feasibility of Serving K Secondary Users in Underlay Cognitive Radio networks using Massive MIMO}", \textit{in ITG conf. on Systems, Comm., Coding (SCC), 2017.}\\ \\
	\cite{chauhari2015_asilomar}{S. Chaudhari}, D. Cabric, ``{Downlink Transceiver Beamforming and Admission Control for Massive MIMO Cognitive Radio Networks}", \textit{in Asilomar conf. on Signals, Systems and Computers, 2015.}\\ \\
	\cite{chaudhari2014_globecom}{S. Chaudhari}, D. Cabric, ``{Cyclic Weighted Centroid Localization for Spectrally Overlapped Sources in Cognitive Radio Networks}", \textit{IEEE Globecom, 2014.}	
}

\abstract {Due to ever increasing usage of wireless devices and data hungry applications, it has become necessary to improve the spectral efficiency of existing wireless networks. One way of improving spectral efficiency is to share the spectrum amongst different coexisting networks and serve multiple devices simultaneously. Spectrum sharing mechanisms for coexistence of a licensed network, such as LTE, with an unlicensed network, such as Wi-Fi, are being considered in the recent literature and standardizations. In order to enable the coexistence between licensed and unlicensed users, it is necessary to include interference mitigation techniques to protect the licensed primary users (PUs) from harmful interference. Typical interference mitigation mechanisms are based on spectrum sensing and cognitive radio (CR), wherein unlicensed secondary users (SUs) observe the spectrum and utilize it when licensed PUs are inactive. Thus, the SUs utilize empty time-slots in the shared spectrum to avoid the interference. The spectral efficiency can be further improved if the SUs are allowed to transmit concurrently with PUs by exploiting the spatial dimension provided by multiple antenna techniques.

The underlay CR paradigm allows such coexistence where SUs transmit its signal in the same time-slots as PUs by exploiting the spatial and frequency resources in the network. In order to exploit the spatial dimension, SUs can utilize the location coordinates of PUs to steer its signal away from PUs to mitigate the interference. The SU transmitter can also employ multiple antenna techniques to serve a large number of devices. Further, the SUs can utilize frequency bands occupied by PUs by dynamically selecting the frequency band that provides the highest rate. In this work, we develop techniques for PU location estimation, spatial resource allocation and frequency band selection for SUs in underlay CR networks.

We begin by considering the problem of estimation of PU location coordinates in a network of SUs in the presence of spectrally overlapped interference. A localization algorithm based on cyclostationary properties of the PU signal is proposed in order to mitigate the impact of the interference. The proposed scheme identifies and eliminates the SUs in the vicinity of the interferer thereby improving the localization accuracy.

Next, we propose a low-complexity algorithm to solve a resource allocation and interference control problem in a network where secondary BS (SBS) is equipped with a large antenna array. The proposed algorithm selects maximum number of SUs and allocates power for downlink transmission from the SBS, while keeping the interference to PUs below a specified limit. It has been shown that the proposed low-complexity algorithm provides optimum solution if the number of antennas at the SBS is order of magnitude larger than the number of SUs and PUs in the network.

Finally, we analyze power control and frequency band selection policies for a SU transmitter-receiver pair that can select one out of multiple available frequency band in each time-slot to maximize its achievable rate. We derive an expression for transmit power in a frequency band as a function of interference constraints, PU traffic statistics in the frequency band, and temporal correlation of channels. We show that instead of hopping to a different frequency band in each time-slot, the SU can stay on one frequency band in order to maximize its own rate while keeping the interference toward PUs to a predetermined level.

}

\begin {document}
\makeintropages

%
%

\chapter{Introduction}
\label{chapter_introduction}
\section{Underlay cognitive radio for spectrum sharing}
In recent years, wireless networks are required serve a large number of data-hungry devices and applications as the annual mobile data traffic is projected to exceed 30.6 exabytes by the end of this decade \cite{CISCO:datatraffic}. 
In order to serve a multitude of devices, the next generation 5G networks need to utilize spectral resources, namely, time and frequency, more efficiently. Multiple spectrum sharing technologies are being considered for upcoming wireless networks to enable co-existence of licensed network, such as LTE, with unlicensed network, such as Wi-Fi. The examples of such spectrum sharing technologies include:
\begin{enumerate}
	\item LTE-Licensed Assisted Access (LAA): Coexistence of LTE carriers with WiFi devices in 5GHz unlicensed band \cite{3gpp_36_889}.	
	\item MulteFire: Standalone LTE-like system for spectrum sharing with Wi-Fi in unlicensed 5GHz band \cite{MulteFire_1_0}.
	\item Citizen Broadcast Radio Service (CBRS): Newly opened spectrum sharing band in the United States at frequencies 3550-3700MHz \cite{fcc_15_47A1}.
\end{enumerate}

The above technologies borrow from the concept of Cognitive Radio (CR) networks that enable efficient utilization of spectral resources by allowing resource sharing between licensed primary users (PUs) and unlicensed secondary users (SUs). The main challenge in implementing such spectrum sharing network is to mitigate interference caused by SUs to PUs. Cognitive radio networks  mitigate the interference to PUs in two ways. 

In the first paradigm, called interweave CR, the SUs and PUs are served in the orthogonal time-frequency block. Therefore, the SUs in interweave CR network need to either look for frequency bands unused by PUs or wait for PU transmission to stop before SUs start using the spectrum. Such mechanism has been used in Listen-Before-Talk (LBT) techniques used in LAA and MulteFire technologies that enable coexistence of its users with Wi-Fi users. In the LBT technique, the LAA/MulteFire user first senses the wireless medium to check if the frequency band is unused. If the band is found to be empty, then LAA/MulteFire user can utilize it to transmit its own signal. In such a network, the throughput of SUs, e.g.. LAA/MulteFire user, is limited by the activity of existing Wi-Fi users and also the ability of SUs to quickly find the unoccupied time-frequency block. This limits the efficiency of the shared spectrum.

In order to further improve the spectral efficiency, the second paradigm called underlay CR can be used, which is the focus of this dissertation. In the underlay CR network, SUs are allowed to transmit in the same time-frequency block as PUs if the interference to PUs is kept below a specified limit \cite{biglieri2012a}. It is the responsibility of the SUs to contain the interference to PUs. Therefore, the SUs in an underlay CR network operate to achieve two objectives: maximize its own rate and keep the interference to PUs below a specified limit. In order to achieve these two objective, SUs may use multiple antenna beamforming techniques to transmit its signal in orthogonal signal space of the PUs. It can also use location coordinates of PUs for smart routing and avoid interference to PUs.

Such underlay CR methods can potentially improve the spectrum efficiency of LAA and MulteFire systems where  LAA/MulteFire users utilize advanced spectrum sharing techniques that make use of multiple antennas \cite{qcom_5GNRSS}. The underlay CR also finds applications in upcoming CBRS band where 150MHz wide spectrum between 3550-3700MHz is shared between users belonging to three layers of priorities: 1) Incumbents, 2) Priority Access License (PAL), and 3) General Authorized Access (GAA). Incumbents users, that are associated with the US military, are protected from interference from PAL and GAA users using geographical protection zones. PAL users are Primary Users who may purchase the license to provide services to its users, while GAA users are SUs who can utilize the spectrum as long as the interference to PAL users is below a specified limit. Therefore, GAA users need to utilize advanced interference mitigation and multiple antenna techniques to efficiently utilize the shared spectrum. In this dissertation, we develop three techniques for SUs to achieve the objectives of interference mitigation and efficient spectrum utilization.

Firstly, SUs in the underlay CR network can mitigate the interference to PU by steering their signal away from the PU. The location coordinates of the PU are useful in this regard \cite{Wang2011}. In the first part of this dissertation, we develop an algorithm to estimate the location of PU using the signal received on SUs in the network. The first algorithm was published in \cite{chaudhari2014_globecom}. An improved version was then published in \cite{chaudhari2016_tccn}.

Secondly, in order to limit the interference to PUs, the SUs can employ beamforming techniques using multiple antennas to steer null towards PUs. The SU transmitter, in this case, utilizes the channel state information (CSI) between itself and the PU receiver in order to transmit its signal in the null space of channels to PUs. Spectral efficiency of an underlay CR network is increased if the SU transmitter, e.g. secondary base station (SBS), serves multiple SUs in the same time-frequency block while still satisfying the interference constraints to PUs. This is possible if the SBS is equipped with a large antenna array. The SBS can then use massive MIMO beamforming techniques to achieve dual goal of efficient spectrum utilization and limiting the interference to PUs. In the second part of the dissertation, we develop an optimization framework and propose an algorithm to simultaneously serve maximum number of SUs from the SBS, while limiting the interference to PUs below the specified limit. We published first work on this topic in \cite{chauhari2015_asilomar}, which was then extended in \cite{chaudhari2018_tccn}.

In the underlay CR paradigm, the SU equipped with multiple antennas can transmit in the same time-frequency block as PU. In the third part of the dissertation, we extend this system to a wideband scenario where the SU can select one out of multiple frequency bands to transmit its signal. In the third part of this dissertation, we study frequency band selection and power control policies to achieve the dual goal of maximizing the rate of the SU while limiting the interference to PUs in those bands. This work was published in \cite{chaudhari2019_tccn}.

We specifically address the following challenges in the above three tasks.


\begin{figure*}
	\centering
	\includegraphics[width=0.9 \columnwidth]{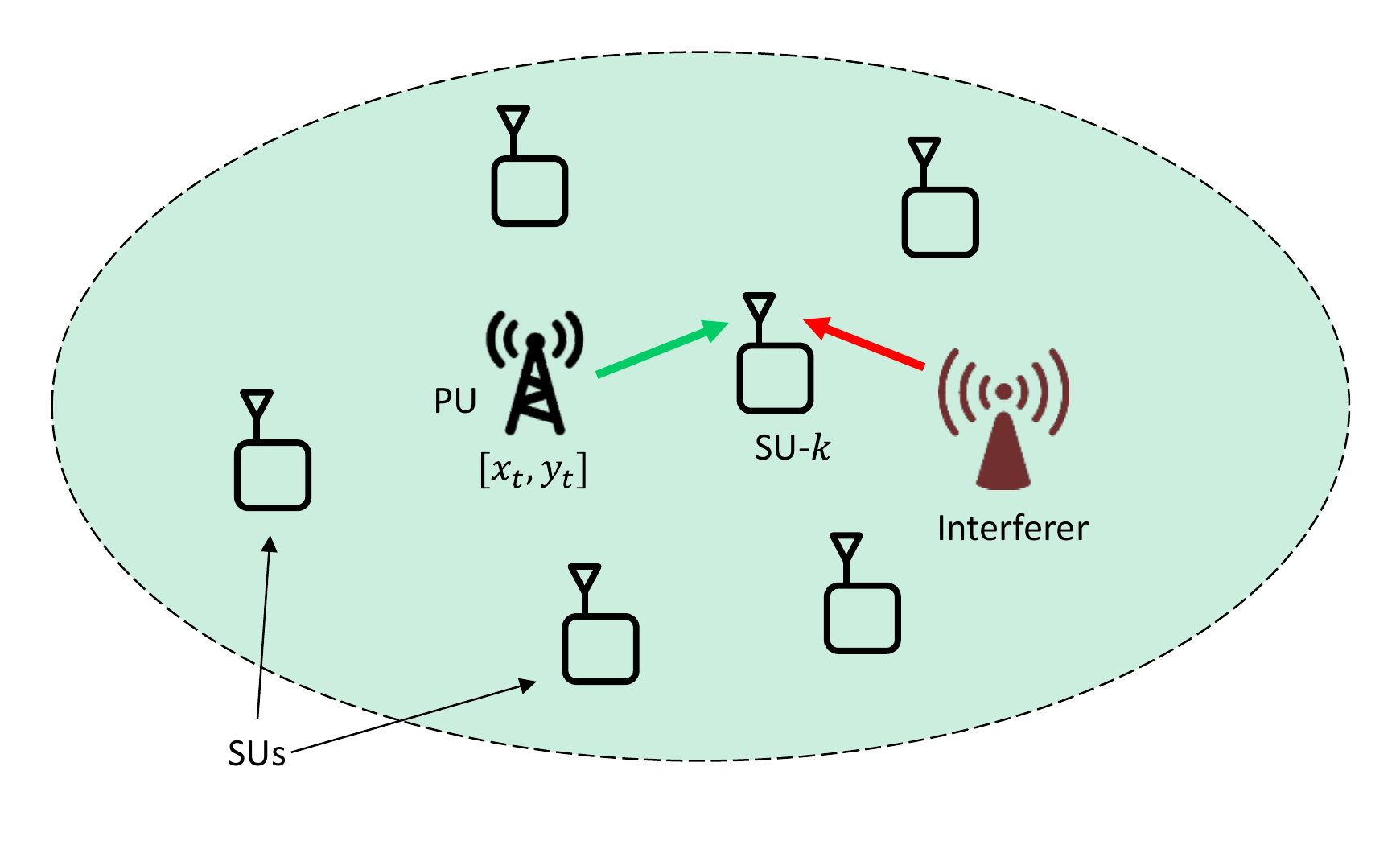}
	\caption{\small (Question 1) How to estimate PU location coordinates $[x_t, y_t]$ in the presence of spectrally overlapped interference?}
	\label{fig:task1}
\end{figure*}

\section{Challenges and objectives}
The location coordinates of the PU are typically estimated using the received signal on SUs in the CR network as shown in Fig. \ref{fig:task1}. However, if the PU signal is corrupted by a spectrally overlapped interferer, the error in location estimate increases as the interferer power increases. Our first objective is to develop a technique to estimate the location of a PU transmitter in the presence of interference using only the received signal on the SUs.

\begin{figure*}
	\centering
	\includegraphics[width=0.9 \columnwidth]{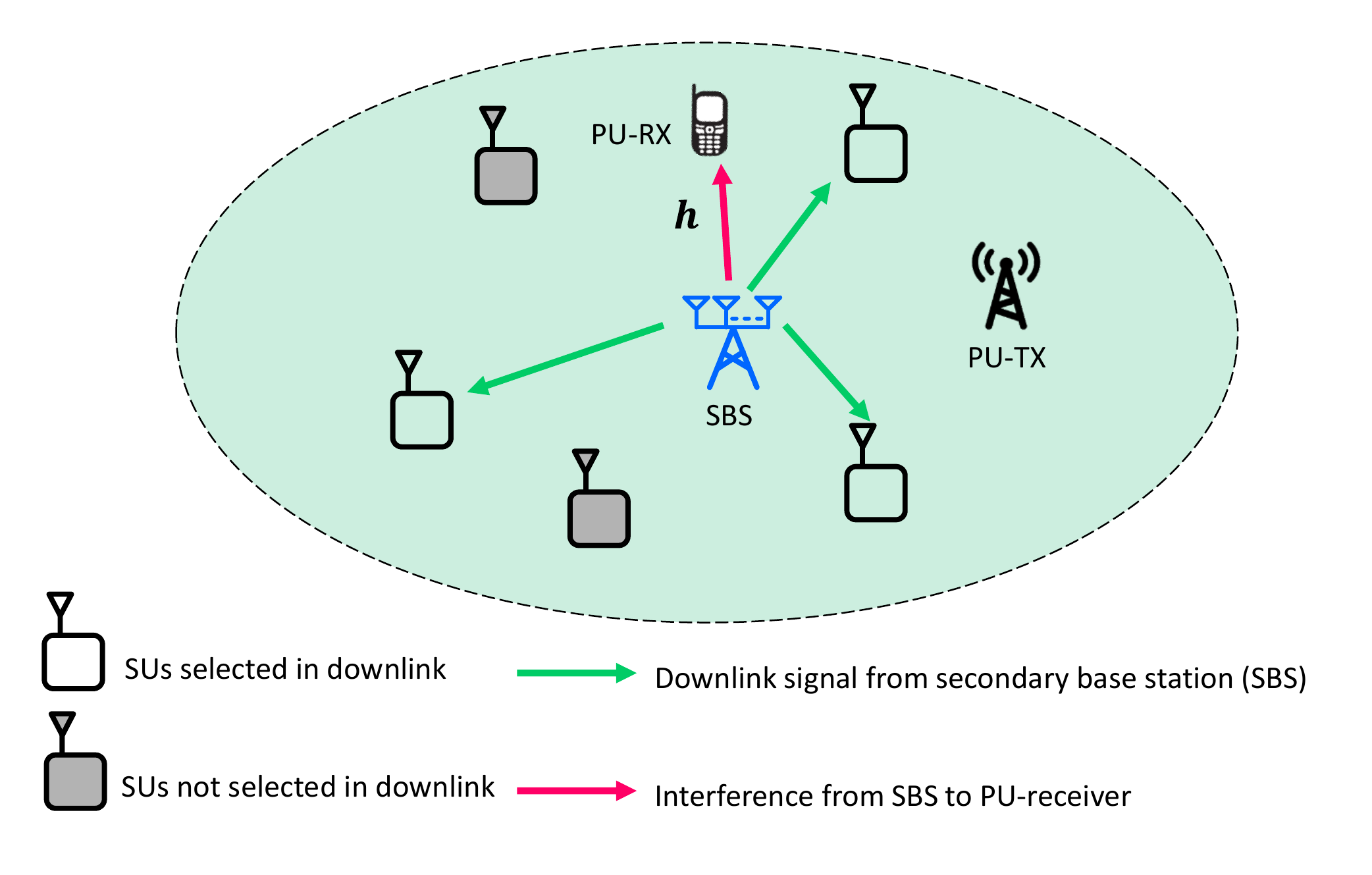}
	\caption{\small (Question 2) How to select maximum number of SUs and control interference to PU-receiver with imperfect knowledge of $\mathbf{h}$?}
	\label{fig:task2}
\end{figure*}

In order to mitigate the interference to PUs using multiple antenna techniques, the SU transmitter, e.g. Secondary BS, requires knowledge of the CSI between itself and the PU receiver as shown in Fig. \ref{fig:task2}. The SBS, equipped with a large antenna array, can form a transmit beamforming vectors using the CSI so that the interference to PU is zero. However, if only imperfect CSI is available at the SBS, then the PU receives non-zero interference. The interference at the PU depends on which SUs are selected in the downlink and how much power is allocated to them by the SBS. A judicious user selection and power allocation is required from the SBS to control the interference and simultaneously serve maximum number of SUs. Our second objective is to develop such mechanism for the network shown in Fig. \ref{fig:task2}.

\begin{figure*}
	\centering
	\includegraphics[width=0.9 \columnwidth]{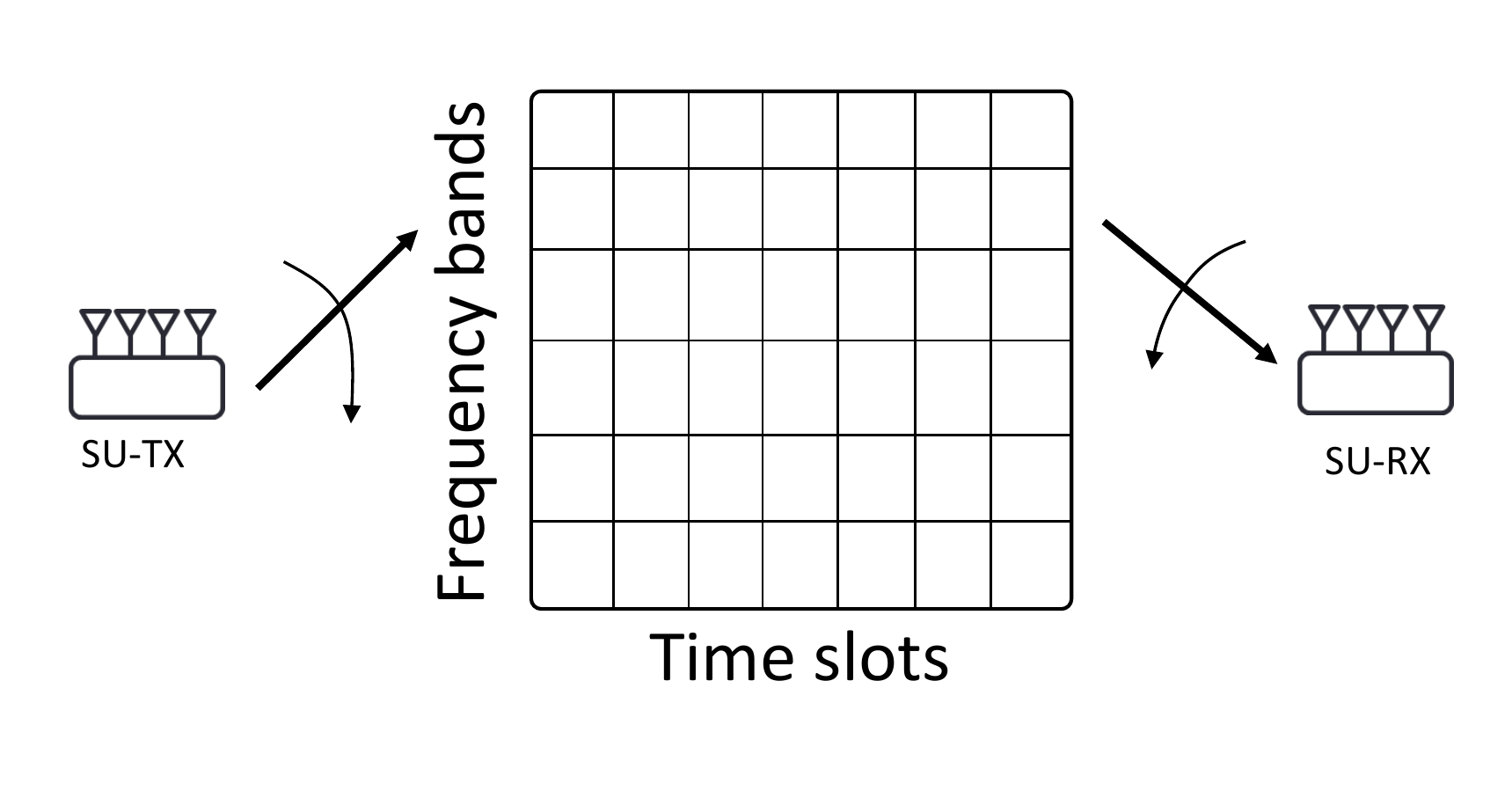}
	\caption{\small (Question 3) For a SU transmitter-receiver pair equipped with multiple antennas, what is the best frequency hopping and power allocation policy in each time-slot?}
	\label{fig:task3}
\end{figure*}

In underlay CR network, where SUs are equipped with multiple antennas, the rate of the SU transmitter-receiver link depends on the transmit power. When there are multiple frequency bands available for the SU as shown in Fig. \ref{fig:task3}, the best frequency band in each slot is the one where the SU can transmit maximum power without causing harmful interference to the PU in that band. The maximum transmit power in each band depends on the traffic pattern of the PUs in that band as well as temporal correlation of physical channels, which determines how fast the channels between SUs and PUs change from one time slot to the next. Our third objective is to determine the power control and frequency band selection policy for SUs in such underlay MIMO CR network.

\section{Contributions of dissertation}
The contributions of this dissertation are as follows:

\subsection{Primary transmitter localization in the presence of spectrally overlapped interference}
We start by considering a location estimation problem in a network of SUs with a non-cooperative target PU. This is a non-cooperative localization problem in the sense that the PU does not assist SUs in order to estimate its location. The SUs receive signal from the PU as well as from a spectrally overlapped interference in the same time-frequency block. Since the interference signal cannot be separated from the PU signal in time-frequency domains, we utilize distinct cyclostationary properties of the PU signal in order to estimate its location. We propose Cyclic Weighted Centroid Algorithm (Cyclic WCL) that estimates the location as a weighted centroid of location of SUs and the weights are computed using the cyclic autocorrelation of signal received at each SU. Further, an improved Cyclic WCL is developed that identifies the SUs in the vicinity of the interferer and eliminate them from the localization process. This elimination significantly reduces the location estimation error. We propose a statistic called \textit{feature variation coefficient} in order to identify and eliminate the SUs in vicinity of the interferer. We show that if only received signal strength is used for localization of the PU as in case of traditional WCL algorithm, then the estimation error increases with increased interference power, while the proposed Cyclic WCL is shown to be robust to it.

\subsection{User selection and power allocation in massive MIMO CR network}
Next, we consider a secondary user selection and power allocation problem in a CR network with a massive MIMO base station equipped with a large antenna array. The massive MIMO base station serves multiple SUs while limiting the interference to existing PUs. A new optimization framework is proposed in order to select the maximum number of SUs and compute power allocations in order to satisfy instantaneous rate requirements of SUs. The optimization framework also aims to restrict the interference to PUs below a predefined threshold using available imperfect CSI at the base station. In order to obtain a feasible solution for power allocation and user selection, we propose a low-complexity algorithm called Delete-su-with-Maximum-Power-allocation (DMP). Theoretical analysis is provided to compute the interference to PUs and the number of SUs exceeding the required rate. The analysis and simulations show that the proposed DMP algorithm outperforms the state-of-the art selection algorithm in terms of serving more users with given rate constraints, and it approaches the optimal solution if the number of antennas is an order of magnitude greater than the number of users.

\subsection{Power control and frequency band selection for MIMO cognitive radio}
In the third part, we study power control and frequency band selection policies for underlay MIMO cognitive radio with the objective of maximizing the rate of SU while keeping the interference to primary users (PUs) below a specified threshold. We assume that the SU employs beamforming to transmit its signal in the null space of channels to PUs. Time-varying MIMO channels are considered to study the performance of the following policies: fixed band fixed power (FBFP), fixed
band dynamic power (FBDP), dynamic band fixed power (DBFP). The expression for transmit power in a frequency band is
derived as a function of traffic of PU transmitter-receiver link in that band. Theoretical analysis is presented to show that the FBDP policy provides higher rate than FBFP policy. The analysis also shows that dynamic band selection policies such as round robin, random, and polices based on multi-armed bandit framework result in higher interference to PUs as compared to fixed band policies. We also provide an expression for the gap between the rate achieved by an optimal genie-aided clairvoyant policy and the FBFP policy. It is observed that this gap reduces under slow-varying channels and as the number of SU antennas is increased. This implies that instead of hopping to a different frequency band in each time-slot, the SU can stay on one band in order to maximize its rate while protecting PUs from harmful interference.

\section{Organization of dissertation}
The rest of the dissertation is organized as follows. The PU localization problem is tackled in Chapter 2 where we present the proposed Cyclic WCL algorithm. The problem formulation for secondary user selection and power allocation is provided in Chapter 3 along with the proposed DMP algorithm. Power control and frequency band selection policies for MIMO CR are studied in Chapter 4. Finally, the concluding remarks are provided in Chapter 5.

\chapter{Localization of Primary User in the Presence of Spectrally Overlapped Interference}
\label{chapter_localization}

\section{Introduction}
\label{sec:Introduction}
The knowledge of PU location is required in  cognitive radio networks for advanced spectrum sharing techniques including location-aware smart routing and location-based interference management \cite{Zhu2012}. In this chapter, we address the problem of estimating the location coordinates of a non-cooperative target PU in the presence of a spectrally overlapped interference in a cognitive radio network. Under spectrally overlapped interference, the traditional Weighted Centroid Localization (WCL) algorithm \cite{Wang2011} results in higher localization errors, since it uses only the received signal power for localization. In this work, we utilize distinct cyclostationary features of the overlapping signals that arise from different symbol rates, in order to estimate the location of the target transmitter. We propose Cyclic WCL and improved Cyclic WCL algorithms to estimate the target location in the presence of spectrally overlapped interference.

	
Spectrally overlapped interference arises in the following scenario. Consider transmitter localization problem in a CR network in the presence of a jammer \cite{Camilo2012, Fragkiadakis2013, Hlavacek2014} that transmits energy in the same band as Tx-1. In this case, the traditional WCL algorithm \cite{Wang2011} results in higher localization error because it uses only the received signal power at each CR. On the other hand, the proposed Cyclic WCL and improved Cyclic WCL provide robustness against such interference by using distinct cyclostationary properties of the Tx-1 signal.

Non-cooperative localization is commonly used in CR networks where the target transmitter does not cooperate with CRs in the localization process. In this scenario, localization techniques based on Time of Arrival (ToA) or Time-Delay of Arrival (TDoA) are not applicable. In this chapter, we consider techniques that do not require ToA or TDoA information and estimates the target location using the received signal at the CRs in the network. We use the terms CR and SU interchangeably in this chapter. 

\subsection{Related work}
Several non-cooperative localization techniques have been proposed in the literature to estimate the location coordinates of the target transmitter. These techniques can be broadly classified as range-based \cite{Langendoen2003, Xiao2007, Chen2010} and range-free 
\cite{Bulusu2000, Blumenthal2007,  Laurendeau2010,  Ma2010,  Wang2011, Mariani2012, Xu2011, Zhang2012, Nanda2012, Kong2013, Feng2013, Zhao2014}. Range-based techniques  provide better estimate of the target location than range-free techniques, but require accurate knowledge of  wireless propagation properties such as path-loss exponent. Accurate information of the path-loss exponent is difficult to obtain and it may not be available. In such a case, range-free techniques such as centroid localization \cite{Bulusu2000, Blumenthal2007, Laurendeau2010} are used to perform coarse-grained target localization without any knowledge of the path-loss exponent. 

In the Weighted Centroid Localization (WCL) algorithm \cite{Blumenthal2007, Laurendeau2010}, the target location is approximated as the weighted average of all CR locations in the network. The CR locations are known at the central processing node where the WCL algorithm is implemented.  There are different variations of the WCL algorithm proposed in the literature to improve the localization performance \cite{Wang2011, Mariani2012, Xu2011, Zhang2012, Nanda2012, Kong2013, Feng2013, Zhao2014}. Theoretical analysis of the WCL algorithm and its distributed implementation are presented in \cite{Wang2011}. Different weighting strategies and CR selection techniques are presented in \cite{Mariani2012} to reduce the adverse impact of the border effect and to improve root mean square error (RMSE) performance of the algorithm. The technique presented in \cite{Xu2011} selects reliable CRs that are located closer to the target and uses them in the localization process. Papers \cite{Zhang2012} and \cite{Nanda2012} propose direction vector hop (DV-hop) and received signal power based fuzzy logic interference model, respectively, to assign WCL weights. The WCL algorithm that takes into account self-localization error of CRs is presented in \cite{Zhao2014}.

In the algorithms presented in \cite{Blumenthal2007, Laurendeau2010, Wang2011, Mariani2012, Xu2011, Zhang2012, Nanda2012, Kong2013, Feng2013, Zhao2014}, the weights for each CR location are computed based on the received signal power from the target. However, in the presence of a spectrally overlapped interference in the network, the received signal power at each CR is the summation of powers received from the target and the interferer. Therefore, localization errors in existing algorithms increase significantly due to the spectrally overlapped interference. Hence, there is a need to modify the WCL algorithm when a spectrally overlapped interferer is present in the network. 

\subsection{Summary of contributions and outline}
In this chapter, we propose Cyclic WCL algorithm to estimate the location coordinates of the target transmitter (Tx-1) in a CR network in the presence of a spectrally overlapped interference (Tx-2). The location coordinates of the CRs in the network are known at the central processing node, where the localization algorithm is implemented. The cyclostationary properties of the target signal are used to compute weights in the Cyclic WCL algorithm. The proposed localization algorithm does not require any knowledge of the path-loss model and the location of the interferer.

The contributions of this chapter are as follows:
\begin{enumerate}
	\item
	A cyclostationarity-based localization algorithm referred to as Cyclic WCL is proposed in order to estimate the target location in the presence of a spectrally overlapped interference. In the proposed algorithm, the central processing node estimates the target location as a weighted sum of the CR locations, where weights for CR locations are computed based on the cyclic autocorrelation (CAC) of the received signal at that CR.
	
	\item
	Theoretical analysis of the proposed algorithm is presented. The RMSE in the location estimate is computed as a function of the target and the interferer locations, their transmitted powers, and the CR locations.
	
	\item
	The performance of Cyclic WCL is further improved by eliminating CRs in the vicinity of the interferer from the localization process. The ratio of the variance and mean of the square of absolute value of the CAC of the received signal at each CR is used to identify and eliminate CRs in the vicinity of the interferer. Theoretical analysis for the improved Cyclic WCL is also presented in order to compute the RMSE.
\end{enumerate}

This chapter is organized as follows. Section \ref{sec:Model} describes the system model, Cyclic WCL and improved Cyclic WCL algorithms. Section \ref{sec:Theoretical} presents theoretical analysis to compute the RMSE. Simulation results in various scenarios are provided in Section \ref{sec:Results}. Finally, Section \ref{sec:summary_c1} provides summary of this work. 

In this chapter, we denote vectors by bold, lowercase letters, e.g., $\mathbf{a}$ or $\boldsymbol{\theta}$. Matrices are denoted by bold, uppercase letters, e.g., $\mathbf{A}$. Scalars are denoted by non-bold letters, e.g., $x_k$ or $R_{r_k}$. Transpose, conjugate, trace and determinant of matrices are denoted by $(.)^T$, $(.)^*$, $\operatorname{Tr}(.)$, and $\operatorname{det}(.)$, respectively. Finally, the terms Tx-1 and target are used interchangeably. 

\section{System model and algorithm}
\label{sec:Model}
\subsection{System model}
\begin{figure*}
	\centering
	\includegraphics[width=0.7 \columnwidth]{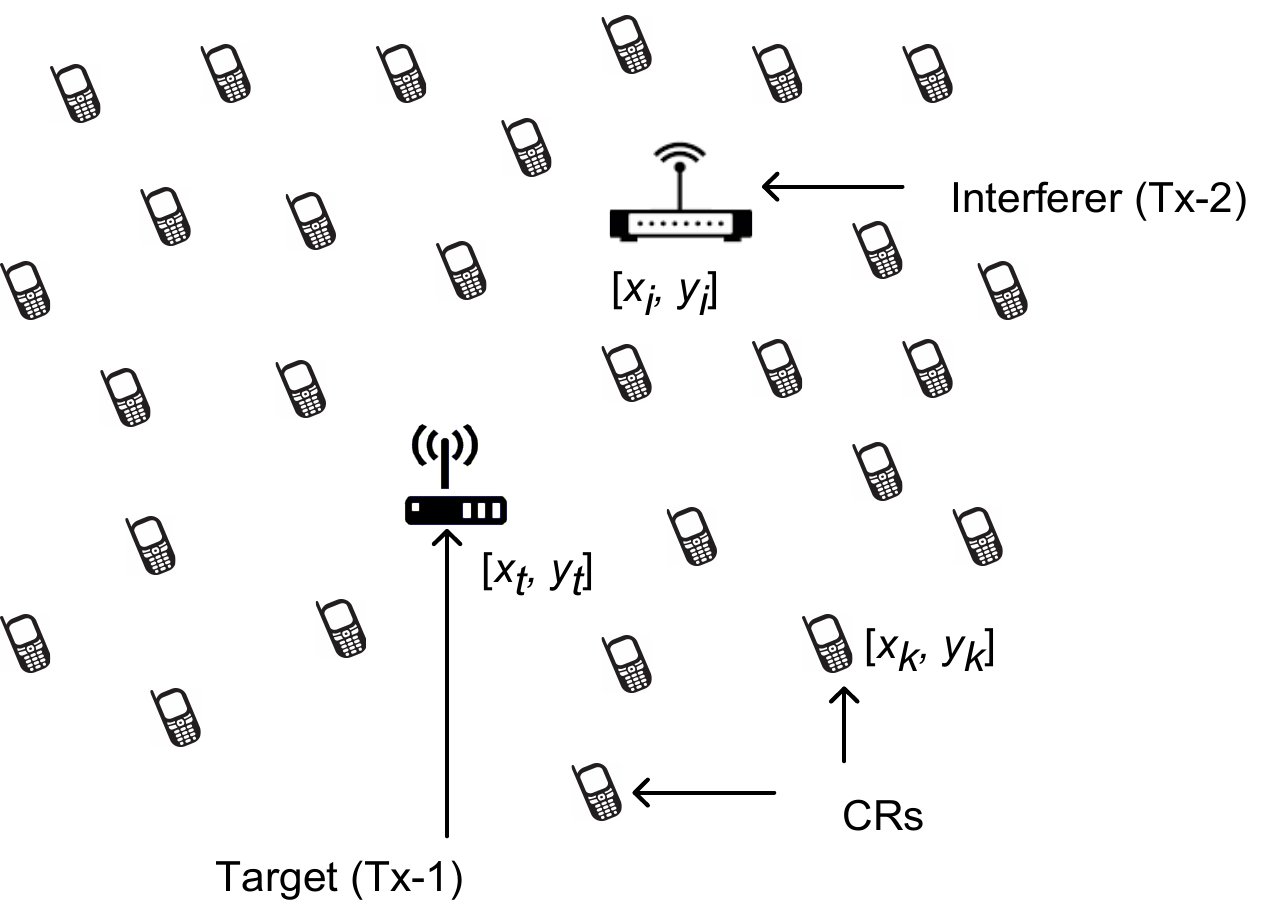}
	\caption{\small System schematic: CR/SU network with a target (Tx-1) and a spectrally overlapped interferer (Tx-2). The location coordinates of the target and the interferer are [$x_t, y_t$] and [$x_i, y_i$] respectively. The CRs are located at [$x_k, y_k$], $k = 1,2,\cdots,K$.}
	\label{fig:system_schematic}
\end{figure*}
We consider a target-centric CR network, where the target is located at the origin and its coordinates are given by $\mathbf{L_{t}}=[x_{t},y_{t}]^{T}=[0,0]^{T}$. There are $K$ CRs in the network, distributed in a square of side $a$ meters around the origin. The locations of the CRs are known at the central processing node and are denoted by $\mathbf{L_{k}}=[x_{k},y_{k}]^{T}, 1\leq k\leq K$. The interferer is located at $\mathbf{L_{i}}=[x_{i},y_{i}]^{T}$. The knowledge of the interferer's location is not required while estimating the target's location coordinates. We assume that the target and the interferer locations are different, i.e., $\mathbf{L_t} \neq \mathbf{L_i} $. The schematic of the system is shown in Fig.~\ref{fig:system_schematic}.

Let $p_t$ and $p_i$ denote transmitted powers from the target and the interferer, respectively. The signals transmitted from the target and the interferer are $\sqrt{p_t}s_t(t)$ and $\sqrt{p_i}s_i(t)$, respectively. Both $s_t$ and $s_i$ are unit power signals. The signals $s_t$ and $s_i$ partially or completely overlap in the frequency domain. 
The algorithm presented in this chapter is applicable to single carrier as well as multi-carrier OFDM signals, since both signal types exhibit cyclostationary properties. The single carrier and multi-carrier signal models are described below.
\subsubsection{Single carrier signal model}
\label{sec:single_carrier_model}
Single carrier target and interference signals can be expressed as
\begin{align}
\nonumber
s_{t}(n)=s_{t}(nT_{s})&=\sum_{l=-\infty}^{\infty}a_{l}g(nT_{s}-lT_{g})e^{j2\pi f_tnT_s}\\
&=\sum_{l=-\infty}^{\infty}a_{l}g_{n,l}e^{j2\pi f_tnT_s} \text{ and}
\label{eq:single_carrier_signal_model}
\end{align}
\begin{align}
\nonumber
s_{i}(n)=s_{i}(nT_{s})&=\sum_{l=-\infty}^{\infty}b_{l}h(nT_{s}-lT_{h})e^{j2\pi f_inT_s}\\
&=\sum_{l=-\infty}^{\infty}b_{l}h_{n,l}e^{j2\pi f_inT_s},
\label{eq:stsi}
\end{align}
where $T_{s}$ is the sampling period, $a_{l}$ and $b_{l}$ are transmitted
data symbols, $g$ and $h$ are pulse shaping filters, and $T_{g}$ and
$T_{h}$ are symbol periods of the signals $s_t$ and $s_i$, respectively. The carrier frequencies of $s_t$ and $s_i$ are $f_t$ and $f_i$, respectively. The data symbols $a_{l}$ and $b_{l}$ are assumed to be i.i.d. and zero mean. For simplicity of notation, we write $g_{n,l} = g(nT_s -lT_g)$ and $h_{n,l} = h(nT_s -lT_h)$. 

\subsubsection{Multi-carrier signal model}
\label{sec:multi_carrier_model}
An OFDM target signal with $N_{c,t}$ sub-carriers and sub-carrier spacing $\Delta f_{t}$ can be expressed as \cite{Sutton2008}
\begin{align}
\nonumber s_t(n) &= \sum_{l=-\infty}^{\infty}\sum_{\kappa=-\frac{N_{c,t}}{2}}^{\frac{N_{c,t}}{2}-1}c_{\kappa,l}e^{j2\pi (f_t + \kappa\Delta f_{t} )n T_s}g(nT_s - l T_g)\\
& = \sum_{l=-\infty}^{\infty}\sum_{\kappa=-\frac{N_{c,t}}{2}}^{\frac{N_{c,t}}{2}-1}c_{\kappa,l} g_{n,l} e^{j2\pi \kappa \Delta f_{t} n T_s} e^{j2\pi f_t n T_s},
\label{eq:ofdm_signal_model}
\end{align}
where $c_{\kappa,l}$ are data symbols on $\kappa^{th}$ sub-carrier, $g_{n,l}=g(nT_s -lT_g)$ is the window function. The duration of OFDM symbol is  $T_g = 1/\Delta f_t + T_{cp}$, where $T_{cp}$ is the duration of cyclic prefix. The data symbols $c_{\kappa,l}$ are assumed to be i.i.d. and zero mean. 
Similarly, a multi-carrier interfering signal can be expressed as
\begin{align}
\nonumber s_i(n) &= \sum_{l=-\infty}^{\infty}\sum_{\kappa=-\frac{N_{c,i}}{2}}^{\frac{N_{c,i}}{2}-1}d_{\kappa,l}e^{j2\pi (f_i + \kappa\Delta f_{i} )n T_s}h(nT_s - l T_h)\\
& = \sum_{l=-\infty}^{\infty}\sum_{\kappa=-\frac{N_{c,i}}{2}}^{\frac{N_{c,i}}{2}-1}d_{\kappa,l}h_{n,l} e^{j2\pi \kappa \Delta f_{i} n T_s} e^{j2\pi f_i n T_s},
\end{align}
where $N_{c,i}, d_{\kappa,l}$, and $h_{n,l}$ denote the number of sub-carriers, symbols on $\kappa^{th}$ sub-carrier, and the window function, respectively.

Let $\alpha_t$ and $\alpha_i$ be the cyclic frequencies of the target and the interferer signal, respectively. For single carrier signals, the values of the cyclic frequencies of a signal depend on its modulation type and symbol rate \cite{Rebeiz2013}. The cyclic frequencies are at integer multiples of the symbol rate. It should be noted that OFDM signals also have cyclic frequencies at integer multiples of symbol rate due to the existence of cyclic prefix\cite{Sutton2008}. 

In this chapter, we assume that the cyclic frequencies of the target and interferer signals are different, i.e., $\alpha_i \neq \alpha_t$. The cyclic frequency of the target signal is known at the CRs in the network. To simplify the analysis, we consider sampling frequency $f_s=1/T_s > \text{max}(\alpha_t, \alpha_i)$. 

Let us denote the powers received at the $k^{th}$ CR from the target and the interferer by $p_{t,k}$ and $p_{i,k}$, respectively ($p_{t,k}^{\text{dB}}$ and $p_{i,k}^{\text{dB}}$ in logarithmic scale). In order to compute  the received powers, we take into account path-loss and shadowing effects. We assume that the signals $s_t$ and $s_i$ undergo independent shadowing due to the location difference in the target and the interferer. Therefore, $p_{t,k}^{\text{dB}}$ and  $p_{i,k}^{\text{dB}}$ are given by
\begin{align}
\nonumber  {p_{t,k}^{\text{dB}}} &= {p_t^{\text{dB}}} - 10\gamma \log \left(\frac{{||\mathbf{{L_k} - {L_t}}||}}
{{{d_0}}}\right ) - {q_{t,k}} \text{ and}\\
{p_{i,k}^{\text{dB}}} &= {p_i^{\text{dB}}} - 10\gamma \log \left(\frac{{||\mathbf{{L_k} - {L_i}}||}}
{{{d_0}}} \right) - {q_{i,k}},
\label{eq:p_tk_p_ik}
\end{align}
where $p_t^\text{dB}$ is the power transmitted (in dB) by the target and $p_i^\text{dB}$ is the power transmitted by the interferer, $\gamma$ is the path-loss exponent and $d_0$ is reference distance. The variables $q_{t,k}$ and $q_{i,k}$ are used to model the shadowing effect on the target and the interferer powers. As mentioned above, $q_{t,k}$ and $q_{i,k}$ are independent variables and are not identical due to different locations of the target and the interferer. 

Let us consider a vector $\mathbf{q_t} = [q_{t,1},q_{t,2},...q_{t,K}]^T$ consisting of shadowing variables for the power received from the target at $K$ CRs. We consider uncorrelated shadowing at $K$ CRs in the network. For log-normal shadowing effect, $\mathbf{q_t}$ is modeled as a Gaussian random vector with zero mean and covariance matrix  $\sigma_q^2\mathbf{I_K}$, where $\mathbf{I_K}$ is a $K$x$K$ identity matrix ($\mathbf{q_t}\sim N(0,\sigma_q^2\mathbf{I_K})$). We assume that the shadowing statistics for the target and the interferer power are the same. Therefore, the vector corresponding to the interferer power is $\mathbf{q_i} =  [q_{i,1},q_{i,2},...q_{i,K}]^T \sim N(0,\sigma_q^2\mathbf{I_K})$. 

Further, the noise power at each CR is denoted by $\sigma_w^2$. The AWGN noise at $k^{th}$ CR is $w_k\sim CN(0,\sigma^2_w)$. Each CR samples the received signal at the sampling frequency $f_s=1/T_s$. The received signal at $k^{th}$ CR is given as
\begin{equation}
r_k(nT_s)=r_k(n)=\sqrt{p_{t,k}} s_t(n)+\sqrt{p_{i,k}} s_i(n)+ w_k(n).
\label{eq:rk}
\end{equation}

\subsection{Cyclic WCL}
\label{Cyclic WCL}
In Cyclic WCL, the CRs do not require the knowledge of transmitted powers ($p_{t}, p_{i}$) or received powers ($p_{t,k}, p_{i,k}$) in order to estimate the target location. Each CR in the network observes $N$ samples of the received signal $r_k$ and computes the non-asymptotic CAC of $r_k$ at the known cyclic frequency $\alpha_t=1/T_g$ of the target signal using
\begin{equation}
{\hat R_{{r_k}}^{\alpha_t}} = \frac{1}
{N}\mathop \sum \limits_{n = 0}^{N-1} |{r_k}(n){|^2}{e^{ - j2\pi {\alpha _t}n{T_s}}}.
\label{eq:cac_rk}
\end{equation}
In the above equation, (\^{}) indicates non-asymptotic estimate based on $N$ samples. In order to reduce the impact of the interferer signal at the cyclic frequency $\alpha_t$, the number of samples $N > 10 \lceil \frac{f_s}{\Delta\alpha}\rceil$, as shown in Appendix \ref{sec:lower_bound_N}, where $\Delta \alpha = |\alpha_t - \alpha_i|$ and $\lceil x\rceil$ denotes the smallest integer not less than $x$. In order to find $N$, the knowledge of $\Delta \alpha$ is required. In our system, $\alpha_t$ is known and $\alpha_i$ can be estimated using cyclostationary spectrum sensing \cite{Rebeiz2013}. From the estimate of $\alpha_i$, $\Delta\alpha$ and hence $N$ can be estimated. It should be noted that the knowledge of $\alpha_i$ is required only to obtain a lower bound on $N$ and the proposed algorithm does not depend on the accuracy of estimation of $\alpha_i$, as shown in results in Section \ref{sec:Results_Delta_Alpha}.

Using $\hat{R_{r_k}^{\alpha_t}}$ from (\ref{eq:cac_rk}), the central processing node estimates the location coordinates of the target as
\begin{equation}
\mathbf{{\hat L}_t} = \frac{{\sum\limits_{k = 1}^K {|{\hat R_{{r_k}}^{\alpha_t}}{|^2}} \mathbf{L_k}}}
{{\sum\limits_{k = 1}^K {|{\hat R_{{r_k}}^{\alpha_t}}{|^2}} }}.
\label{eq:lt}
\end{equation}
It should be noted that the locations of the CRs, the target and the interferer should remain constant until $N$ samples of the received signal are collected. Therefore, we assume that the locations do not change for time duration $NT_s$. We also assume that received powers $p_{t,k}$ and $p_{i,k}$ remain constant for this duration.

\begin{algorithm}[t]
	\caption{Cyclic WCL}
	\label{alg:cyclic_wcl}
	\begin{algorithmic}[1]
		\State Estimate $\Delta\alpha = |\alpha_t - \alpha_i|$ and select $N > 10 \lceil \frac{f_s}{\Delta\alpha}\rceil$.
		\State At each CR $k=1,2..K$, collect $N$ samples of the received signal $r_k(n), n=1,2..N$.
		\State At each CR, compute one CAC estimate: 
		\Statex ${\hat R_{{r_k}}^{\alpha_t}} = \frac{1}
		{N}\mathop \sum \limits_{n = 0}^{N-1} |{r_k}(n){|^2}{e^{ - j2\pi {\alpha _t}n{T_s}}}$
		\State Compute the target location estimate at the central processing node:
		$\mathbf{{\hat L}_t} = \frac{{\sum\limits_{k = 1}^K {|{\hat R_{{r_k}}^{\alpha_t}}{|^2}} \mathbf{L_k}}}
		{{\sum\limits_{k = 1}^K {|{\hat R_{{r_k}}^{\alpha_t}}{|^2}} }}$
	\end{algorithmic}
\end{algorithm}

\subsection{Improved Cyclic WCL}
\label{sec:Improved Cyclic WCL}
From the expression of the Cyclic WCL estimates in (\ref{eq:lt}), it is observed that the target location estimates are the weighted average of the CR locations and the weights are computed using the strength (square of absolute value) of the CAC of the received signal at each CR. Therefore, the target location estimate $\mathbf{{\hat L}_t}$ is closer to the CR which has the highest value of $\hat R_{r_k}$. From (\ref{eq:rk}) and (\ref{eq:lt}), intuitively it is expected that the location estimates are closer to the target's actual location when the interferer power is low. As the interferer power increases, the value of $\hat{R}_{r_k}$ at CRs in the vicinity of the interferer increases. Therefore, the location estimates move away from the target and towards the interferer's location. However, by eliminating the CRs in the vicinity of the interferer from the localization process, the impact of the interferer can be reduced. Based on this argument, we propose the improved Cyclic WCL algorithm that reduces the location estimation error due to the interferer by eliminating the CRs in the vicinity of the interferer.

First, let us define the transmit power ratio ($\rho$) and the received power ratio at the $k^{th}$ CR ($\rho_k$) as
\begin{align}
\rho = \frac{p_t}{p_i} \text{, and }
\rho_k = \frac{p_{t,k}}{p_{i,k}}. 
\end{align}
In the above equation, it is observed that if the $k^{th}$ CR is in the proximity of the interferer, we have $p_{i,k} > p_{t,k}$. On the other hand, for a CR in the proximity of the target, we have $p_{i,k} < p_{t,k}$. Therefore, the received power ratio ($\rho_k$) for a CR in the proximity of the interferer is smaller than for a CR in the proximity of the target. Further, we define Feature Variation Coefficient (FVC) at the $k^{th}$ CR as
\begin{equation}
\phi_k = \frac{\operatorname{var}(\hat R_{r_k}^{\alpha_t})}{\mathbb{E}[|\hat R_{r_k}^{\alpha_t}|^2]}.
\label{eq:phi_k}
\end{equation}
As shown in the Appendix \ref{sec:lower_bound_N} and \ref{sec:phi_k}, for a sufficiently large value of $N$ $\left(N > 10 \lceil \frac{f_s}{\Delta\alpha}\rceil \right)$, $\phi_k$ is a strictly monotonically decreasing function of $\rho_k$ and $0 \leq \phi_k \leq 1$ and it can be written as
\begin{align}
{\phi _k} &= \frac{{\operatorname{var} ({{\hat R}_{{r_k}}^{\alpha_t}})}}
{{\mathbb{E}[|{{\hat R}_{{r_k}}^{\alpha_t}}{|^2}]}} = \frac{\rho_k^2\operatorname{var} ({{\hat R}_{{s_t}}^{\alpha_t}}) + \mathbb{E}[|{{\hat R}_{{s_i}}^{\alpha_t}}{|^2}] + \rho_k \mathbb{E}[|{{\hat R}_{{s_t}{s_i}}^{\alpha_t}}{|^2}]}
{\rho_k^2\mathbb{E}[|{{\hat R}_{{s_t}}^{\alpha_t}}{|^2}] + \mathbb{E}[|{{\hat R}_{{s_i}}^{\alpha_t}}{|^2}] + \rho_k \mathbb{E}[|{{\hat R}_{{s_t}{s_i}}^{\alpha_t}}{|^2}]}.
\end{align}

\begin{figure}
	\centering
	\includegraphics[width= 0.7\columnwidth]{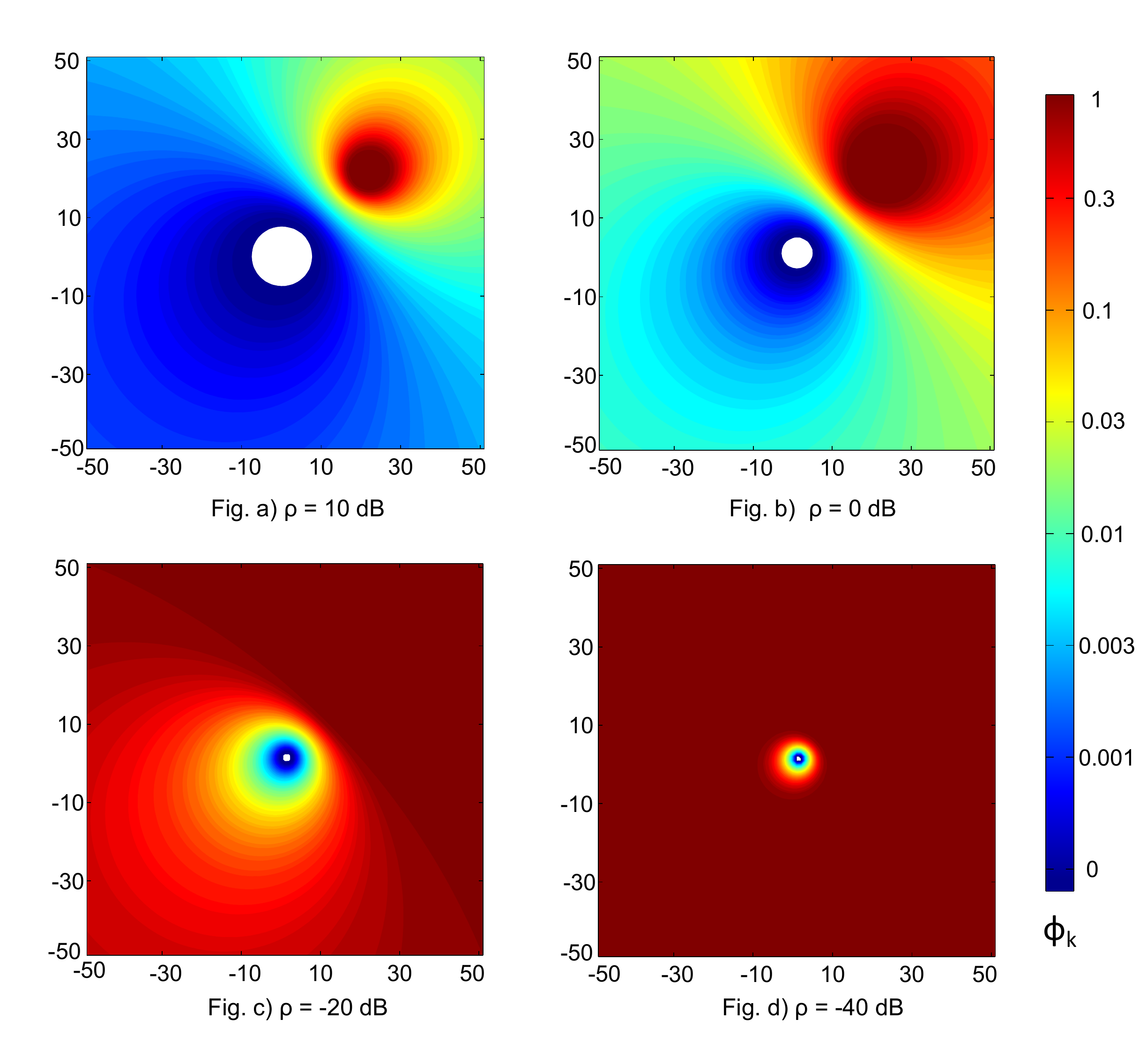}
	\caption{\small Contour of FVC ($\phi_k$) at different locations in the network for a particular transmit power ratio ($\rho$). Target is at [0, 0]. Interferer is at [20, 20]. The target signal $s_t$ is 4-QAM with BW = 20MHz and carrier frequency = 2.4GHz. The interferer signal $s_t$ is 4-QAM with BW = 25MHz and carrier frequency = 2.4GHz. Shadowing variance $\sigma_q = 0$dB.}
	\label{fig:fvc_contour}
\end{figure}

In order to illustrate how $\phi_k$ depends on the location of the $k^{th}$ CR, contour plots of $\phi_k$ at various locations in the network are shown in Fig. \ref{fig:fvc_contour}. It is observed that in the proximity of the target, $\phi_k \rightarrow 0$ . On the other hand, $\phi_k \rightarrow 1$ in the proximity of the interferer. In the four figures in Fig. \ref{fig:fvc_contour}, we can see how higher interferer power (lower $\rho$) changes the $\phi_k$ values at various locations in the network. 

From the above discussion, it is clear that if the central processing node has the knowledge of $\phi_k$ at each CR, it can set a threshold $\phi_0$ on the FVC value and eliminate CRs for which $\phi_k > \phi_0$, since they are closer to the interferer. Thus in the improved Cyclic WCL, each CR computes $\phi_k$ from the received signal according to (\ref{eq:phi_k}) and sends out this information to the central processing node. The CRs do not require the knowledge of $\rho$ and $\rho_k$ to compute $\phi_k$. 

It should be noted that in (\ref{eq:phi_k}), $\phi_k$ depends on ${\operatorname{var} ({{\hat R}_{{r_k}}^{\alpha_t}})}$ and ${\mathbb{E}[|{{\hat R}_{{r_k}}^{\alpha_t}}{|^2}]}$. In a practical system, the $k^{th}$ CR estimates the variance of ${{\hat R}_{{r_k}}^{\alpha_t}}$ and mean of $|{{\hat R}_{{r_k}}^{\alpha_t}}{|^2}$ using $M$ realizations of ${\hat R}_{{r_k}}^{\alpha_t}$. The FVC at $k^{th}$ CR is an estimate based on $M$ realizations of ${\hat R}_{{r_k}}^{\alpha_t}$ and is denoted by $\hat{\phi}_k^M$. Therefore, we have
\begin{align}
{\hat{\phi}_k^M} &= \frac{v_s}{e_s},
\label{eq:sample_fvc}
\end{align}
where $v_s$ is sample variance given by ${v_s} = \frac{1}{M-1}\sum_{i=1}^{M}|(\hat{R}_{r_k}^{\alpha_t})_i-m_s|^2$ with $m_s = \frac{1}{M}\sum_{i=1}^{M}(\hat{R}_{r_k}^{\alpha_t})_i$ and $e_s  = \frac{1}{M}\sum_{i=1}^{M}|(\hat{R}_{r_k}^{\alpha_t})_i|^2= \frac{M-1}{M}v_s + |m_s|^2$. Here $(\hat{R}_{r_k}^{\alpha_t})_i$ is the $i^{th}$ realization of $\hat{R}_{r_k}^{\alpha_t}$. The number of realizations $M$ required to estimate accurate value of $\phi_k$ is obtained using confidence interval of the estimate $\hat{\phi}_k^M$. The analysis to find sufficient number of realizations $M$ for satisfactory performance of the algorithm is presented in Section \ref{sec:phi_k_M}. 

From the knowledge of $\hat{\phi}_k^M$, the central processing node estimates the target location by including CRs for which $\hat{\phi}_k^M\leq\phi_0$:
\begin{equation}
\mathbf{{\hat L}_{t_{improved}}}(\phi_0) = \frac{{\sum\limits_{k = 1}^K {|({\hat R_{{r_k}}^{\alpha_t}})_M{|^2}} \mathbf{L_k}} \mathds{1}(\hat{\phi}_k^M \leq \phi_0)}
{{\sum\limits_{k = 1}^K {|({\hat R_{{r_k}}^{\alpha_t}})_M{|^2}} \mathds{1}(\hat{\phi}_k^M \leq \phi_0)}},
\label{eq:lt_modified}
\end{equation}
where $({\hat R_{{r_k}}^{\alpha_t}})_M$ is the $M^{th}$ realization of ${\hat R_{{r_k}}^{\alpha_t}}$ and $\mathds{1}(\hat{\phi}_k^M \leq \phi_0)$ is an indicator function that is 1 for $\hat{\phi}_k^M\leq\phi_0$ and 0 otherwise. It should be noted that the locations of the CRs, the target and the interferer should remain constant until $MN$ samples of the received signal are collected at each CR. We assume that the locations remain constant for the time duration $MNT_s$. We also assume that received powers $p_{t,k}$ and $p_{i,k}$ remain constant for this duration.

In (\ref{eq:lt_modified}),  $\phi_0$ is a design parameter of the improved Cyclic WCL. Two important points should be noted while selecting the value of $\phi_0$. First, as observed in  Fig. \ref{fig:fvc_contour}, if $\phi_0$ is reduced for a fixed transmit power ratio, then CRs in a smaller area are included in Cyclic WCL. It means that only the CRs confined in the area where $\hat{\phi}_k^M \leq \phi_0$ are included in the Cyclic WCL. Second, if $\phi_0$ is fixed and the interferer power is increased, again CRs in a smaller area are included in the algorithm. Computation of the optimum value $\phi_0^{opt}$ of the FVC threshold that minimizes the RMSE is presented in Section \ref{sec:optimum_fvc_threshold}. This computation requires the information of $p_{t,k}$, $p_{i,k}$, $s_t$ and $s_i$. Since the central processing node might not have this information, it computes a suboptimal threshold $\phi_0^{sub}$ as described below.

Without the loss of generality, consider $\hat{\phi}_1^M \leq \hat{\phi}_2^M ...\leq \hat{\phi}_K^M $. It can be deduced from the analysis presented in Section \ref{sec:optimum_fvc_threshold}, that the unique values of $||\mathbf{\hat{L}_{t_{improved}}}(\phi_0)||^2$ occur at $\phi_0 \in \{\hat{\phi}_1^M, \hat{\phi}_2^M,...\hat{\phi}_K^M\}$, where $||.||$ is 2-norm of the vector.  Further, the optimal value of $\phi_0$ that minimizes the RMSE is one out of $\{\hat{\phi}_1^M, \hat{\phi}_2^M,...\hat{\phi}_K^M\}$. It should be noted that as the value of $\phi_0$ approaches $\phi_0^{opt}$, $||\mathbf{\hat{L}_{t_{improved}}}(\phi_0)||^2$ approaches $||\mathbf{{L}_{t}}||^2$. On the other hand, when the value of $\phi_0$ is significantly different from $\phi_0^{opt}$, $||\mathbf{\hat{L}_{t_{improved}}}(\phi_0)||^2$ takes values that are significantly different from $||\mathbf{{L}_{t}}||^2$. Based on this notion, the central processing node clusters $||\mathbf{\hat{L}_{t_{improved}}}(\phi_0)||^2, \phi_0 \in \{\hat{\phi}_1^M, \hat{\phi}_2^M,...\hat{\phi}_K^M\}$ in two sets: $\mathbb{S}^{opt}$ and $\mathbb{S}^{non-opt}$, using k-means algorithm \cite{Hartigan1979}. If $||\mathbf{\hat{L}_{t_{improved}}}(\phi_0)||^2 \in \mathbb{S}^{opt}$, then the value of $||\mathbf{\hat{L}_{t_{improved}}}(\phi_0)||^2$ is closer to  $||\mathbf{{L}_{t}}||^2$. Similarly if $||\mathbf{\hat{L}_{t_{improved}}}(\phi_0)||^2 \in \mathbb{S}^{non-opt}$, then  the value of $||\mathbf{\hat{L}_{t_{improved}}}(\phi_0)||^2$ is significantly different from $||\mathbf{{L}_{t}}||^2$. The set containing $||\mathbf{\hat{L}_{t_{improved}}}(\hat{\phi}_K^M)||^2$ is identified as $\mathbb{S}^{non-opt}$. This is due to the fact that for $\phi_0 = \hat{\phi}_K^M$, all the CRs, including those in the vicinity of the interferer are included in the localization which makes the value of $||\mathbf{\hat{L}_{t_{improved}}}(\phi_0)||^2$ differ significantly from $||\mathbf{{L}_{t}}||^2$. The other set obtained by k-means algorithm is then identified as $\mathbb{S}^{opt}$. Further the suboptimal FVC threshold $\phi_0^{sub}$ is computed as the average of  $\{ \phi_0 : ||\mathbf{\hat{L}_{t_{improved}}}(\phi_0)||^2 \in \mathbb{S}^{opt}\}$. 
The algorithm steps of the improved Cyclic WCL are described in Algorithm \ref{alg:improved_cyclic_wcl}.
\begin{algorithm}[t]
	\caption{Improved Cyclic WCL}
	\label{alg:improved_cyclic_wcl}
	\begin{algorithmic}[1]
		\State Estimate $\Delta\alpha = |\alpha_t - \alpha_i|$ and select $N > 10 \lceil \frac{f_s}{\Delta\alpha}\rceil$.
		\For{i=1,2..M}
		\State At each CR $k=1,2..K$, collect $N$ samples of the received signal $r_k(n), n=1,2..N$.
		\State Compute CAC estimate: 
		\Statex $({\hat R_{{r_k}}^{\alpha_t}})_i = \frac{1}
		{N}\mathop \sum \limits_{n = 0}^{N-1} |{r_k}(n){|^2}{e^{ - j2\pi {\alpha _t}n{T_s}}}$
		\EndFor
		\State Using above $M$ realizations of ${\hat R_{{r_k}}^{\alpha_t}}$, compute $\hat{\phi}_k^M$ using (\ref{eq:sample_fvc}).
		\State Compute the suboptimal FVC threshold $\phi_0^{sub}$ and estimate the target location using (\ref{eq:lt_modified}).
	\end{algorithmic}
\end{algorithm}

\section{Theoretical analysis}
\label{sec:Theoretical}
\subsection{Analysis of Cyclic WCL}
\label{sec:Theoretical_Cyclic_WCL}
In order to analyze the Cyclic WCL algorithm, we first formulate the estimates of the x- and y-coordinates of the target in the form of ratios of quadratic forms of a Gaussian vector (RQGV). The RMSE in the target location estimates is computed using the RQGV form.

Let us define Cyclic Cross-Correlation (CCC) between signals any two signals $u(n)$ and $v(n)$ at cyclic frequency $\alpha_t$ as
\begin{equation}
\begin{gathered}
{{\hat R}_{uv}^{\alpha_t}} = \frac{1}
{N}\mathop \sum \limits_{n = 0}^{N-1} 2\operatorname{Re} \{ u(n)v{(n)^*}\} {e^{ - j2\pi {\alpha _t}n{T_s}}}. \hfill \\ 
\end{gathered} 
\label{eq:ccc_def}
\end{equation}
In the above equation, $(\hat{\text{       }})$ indicates the estimate based on $N$ samples and $(^*)$ indicates complex conjugate. Substituting (\ref{eq:rk}) in (\ref{eq:cac_rk}) and using the above definition, we can write the CAC of the received signal at $k^{th}$ CR as
\begin{align}
\nonumber
{{\hat R}_{{r_k}}^{\alpha_t}} ={\text{       }} &{p_{t,k}}{{\hat R}_{{s_t}}^{\alpha_t}} + {p_{i,k}}{{\hat R}_{{s_i}}^{\alpha_t}} + \sqrt {{p_{t,k}}{p_{i,k}}} {{\hat R}_{{s_t}{s_i}}^{\alpha_t}} \\&+ {{\hat R}_{{w_k}}^{\alpha_t}}  
{\text{       }} + \sqrt {{p_{t,k}}} {{\hat R}_{{s_t}{w_k}}^{\alpha_t}} + \sqrt {{p_{i,k}}} {{\hat R}_{{s_i}{w_k}}^{\alpha_t}}.   
\label{eq:cac_rk_expanded}
\end{align}
Now, let us define three vectors: $\boldsymbol{\hat \theta_r} , \boldsymbol{\hat \theta_i}$ and $\mathbf{p_k}$. The vectors $\boldsymbol{\hat \theta_r}$ and $\boldsymbol{\hat \theta_i}$ contain real and imaginary parts of CAC and CCC, and $\mathbf{p_k}$ contains powers received at the $k^{th}$ CR from the target and the interferer. Therefore, we have
\begin{align}
\nonumber
\boldsymbol{\hat \theta_r} = [ {\operatorname{Re} \{ {{\hat R}_{{s_t}}^{\alpha_t}}\}, {\text{ }}\operatorname{Re} \{ {{\hat R}_{{s_i}}^{\alpha_t}}\}, {\text{ }}\operatorname{Re} \{ {{\hat R}_{{s_t}{s_i}}^{\alpha_t}}\}}, \text{ }{\operatorname{Re} \{ {{\hat R}_{{w}}^{\alpha_t}}\} , 
	{\text{ }}\operatorname{Re} \{ {{\hat R}_{{s_tw}}^{\alpha_t}}\} ,{\text{ }}\operatorname{Re} \{ {{\hat R}_{s_iw}^{\alpha_t}}\}} ]^T,
\end{align}
\begin{align}
\nonumber
\boldsymbol{{\hat \theta_i }} = [\begin{matrix} {\operatorname{Im} \{ {{\hat R}_{{s_t}}^{\alpha_t}}\}, {\text{ }}\operatorname{Im} \{ {{\hat R}_{{s_i}}^{\alpha_t}}\}, {\text{ }}\operatorname{Im} \{ {{\hat R}_{{s_t}{s_i}}^{\alpha_t}}\}, \operatorname{Im} \{ {{\hat R}_{{w}}^{\alpha_t}}\}, {\text{ }}\operatorname{Im} \{ {{\hat R}_{{s_tw}}^{\alpha_t}}\} ,{\text{ }}\operatorname{Im} \{ {{\hat R}_{s_iw}^{\alpha_t}}\}} \end{matrix}]^T,
\end{align}
\begin{align}
\mathbf{p_k} &= {\left[ {{p_{t,k}},{\text{  }}{p_{i,k}},{\text{ }}\sqrt {{p_{t,k}}{p_{i,k}}}} ,\text{ } 1 ,\text{   }\sqrt {{p_{t,k}}},\text{  }\sqrt {{p_{i,k}}} \right]^T} .\hfill  
\label{eq:vector_def}
\end{align}
Using (\ref{eq:cac_rk_expanded}) and (\ref{eq:vector_def}),  we rewrite the estimate of x-coordinate of the target in terms of ratio of weighted sum of a vector norm:
\begin{equation}
{{\hat x}_t} = \frac{{\sum\limits_{k = 1}^K {|{{\hat R}_{{r_k}}^{\alpha_t}}{|^2}} {x_k}}}
{{\sum\limits_{k = 1}^K {|{{\hat R}_{{r_k}}^{\alpha_t}}{|^2}} }} = \frac{{{{\sum\limits_{k = 1}^K {\left\| {\left[ \begin{gathered}
						{\boldsymbol{\hat \theta_r }^T} \hfill \\
						{\boldsymbol{\hat \theta_i }^T} \hfill \\ 
						\end{gathered}  \right]\mathbf{p_k}} \right\|} }^2}{x_k}}}
{{\sum\limits_{k = 1}^K {{{\left\| {\left[ \begin{gathered}
						{\boldsymbol{\hat \theta_r }^T} \hfill \\
						{\boldsymbol{\hat \theta_i }^T} \hfill \\ 
						\end{gathered}  \right]\mathbf{p_k}} \right\|}^2}} }}.
\label{eq:xt}
\end{equation}
Further, we define power matrix $\mathbf{P} = [\mathbf{p_1, p_2,...p_K}]$ and position matrices $\mathbf{X} =$ \\$\diag(x_1, x_2,...,x_K)$ and $\mathbf{Y} = \diag(y_1, y_2,...y_K)$. From $\mathbf{P}$, $\mathbf{X}$ and $\mathbf{Y}$, symmetric matrices $\mathbf{A_x}$, $\mathbf{A_y}$, and $\mathbf{B}$ are obtained as shown below:
\begin{equation}
\begin{gathered}
\mathbf{A_x} = \diag(\mathbf{PXP}^T, \mathbf{PXP}^T), \\
\mathbf{A_y} = \diag(\mathbf{PYP}^T, \mathbf{PYP}^T),\\
\mathbf{B} = \diag(\mathbf{PP}^T, \mathbf{PP}^T). 
\end{gathered}
\label{eq:matrix_def}
\end{equation}
As shown in the Appendix \ref{app_rq}, $\hat x_t$ can be written in terms of $\mathbf{A_x}$, $\mathbf{B}$ and $\boldsymbol{\hat \theta}  = {\left[ {{\boldsymbol{\hat \theta_r }^T}{\text{ }}{\boldsymbol{\hat \theta_i }^T}} \right]^T}$ as:
\begin{equation}
{{\hat x}_t} = \frac{{{\boldsymbol{\hat \theta }^T}\mathbf{A_x}\boldsymbol{\hat \theta} }}
{{{\boldsymbol{\hat \theta }^T}\mathbf{B}\boldsymbol{\hat \theta} }}.
\label{eq:xt_2}
\end{equation}
Similarly, the estimate of the y-coordinate can be written as:
\begin{equation}
{{\hat y}_t} = \frac{{{\boldsymbol{\hat \theta }^T}\mathbf{A_y}\boldsymbol{\hat \theta} }}
{{{\boldsymbol{\hat \theta }^T}\mathbf{B}\boldsymbol{\hat \theta} }},
\label{eq:yt_2}
\end{equation}

The target location estimates in (\ref{eq:xt_2}) and (\ref{eq:yt_2}) are functions of the vector $\boldsymbol{\hat \theta} = {\left[ {{\boldsymbol{\hat \theta_r }^T}{\text{ }}{\boldsymbol{\hat \theta_i }^T}} \right]^T}$ that contains real and imaginary parts of estimates of CACs and CCCs as defined in (\ref{eq:vector_def}). The estimates of CACs and CCCs are computed using $N$ samples of corresponding signals and are modeled as Gaussian random variables for a sufficiently large value of $N$ \cite[Eqn. (20)]{Dandawate1994}.
It follows that the vector $\boldsymbol{\hat \theta}$ is a Gaussian vector and is a function of number of samples observed $N$. 

As shown in Appendix \ref{app_derivations}, CACs and CCCs are computed from moments of $s_t$ and $s_i$, which in turn are functions of data symbols, $a_l, b_l$ and pulse shapes $g_{n,l}, h_{n,l}$ for single carrier signals. For OFDM signals, the moments  are functions of subcarrier symbols $c_{\kappa,l}, d_{\kappa,l}$, subcarrier spacings $\Delta f_t, \Delta f_i$, and window functions $g_{n,l}, h_{n,l}$. Therefore, the mean $\mathbb{E}[\boldsymbol{\hat{\theta}}]$ and the covariance matrix $\boldsymbol{\Sigma_{\hat \theta}}$ of the Gaussian vector $\boldsymbol{\hat \theta}$ are derived in terms of these parameters as shown in Appendix \ref{app_derivations}.
The target location estimates in (\ref{eq:xt_2}) and (\ref{eq:yt_2}) also depend on the locations of CRs and the power received at the CRs through matrices $\mathbf{A_x}$, $\mathbf{A_y}$, and $\mathbf{B}$.

In order to compute the RMSE, we find the second moments of location estimates $\mathbb{E}[\hat{x}_t^2]$ and $\mathbb{E}[\hat{y}_t^2]$. It should be noted that $\hat{x}_t$ and $\hat{y}_t$ are in the RQGV form with vector $\boldsymbol{\hat \theta}$. The second moments of RQGV are given in  \cite[Thm. 6]{Magnus1986}. To utilize the result presented in \cite{Magnus1986}, the matrix $\mathbf{B}$ should be positive semidefinite. By definition, $\mathbf{B} = \diag(\mathbf{P}\mathbf{P^T}, \mathbf{P}\mathbf{P^T})$. Any matrix of the form $\mathbf{P}\mathbf{P^T}$ is positive semi-definite. Since $\mathbf{B}$ has  $\mathbf{P}\mathbf{P^T}$ on its diagonal, $\mathbf{B}$ is also a positive  semidefinite matrix. 

First, we compute Cholesky factorization of $\boldsymbol{\Sigma_{\hat \theta}}= \mathbf{CC^T}$. Then, the eigenvalue decomposition of $\mathbf{C^TBC}$ gives the orthogonal matrix $\mathbf{V}$ with eigenvectors on its columns and diagonal matrix $\boldsymbol{\Lambda}$ with eigenvalues on the diagonal. Define a matrix $\mathbf{A^*}=\mathbf{V^TC^TA_xCV}$ and a vector $\boldsymbol{\mu}=\mathbf{V^TC^{-1}\mathbb{E}[\boldsymbol{\hat{\theta}}]}$. Further,  $\mathbb{E}[\hat{x}_t^2]$ is computed, using \cite[Thm. 6]{Magnus1986}, as:
\begin{align}
\nonumber
&\mathbb{E}[{{\hat x}^2}] = E\left[ {{{\left( {\frac{{{\boldsymbol{\hat \theta }^T}\mathbf{A_x}\boldsymbol{\hat \theta} }}
				{{{\boldsymbol{\hat \theta }^T}\mathbf{B}\boldsymbol{\hat \theta }}}} \right)}^2}} \right]= {e^{ - \frac{1}
		{2}E{{[\boldsymbol{\hat \theta} ]}^T}\boldsymbol{\Sigma _{\hat \theta }}^{ - 1}\mathbb{E}[\boldsymbol{\hat \theta} ]}}\times \\
&\left\{ {2\int\limits_0^\infty  {t \text{ } \operatorname{det}(\boldsymbol{\Delta} ){e^{\frac{1}
				{2}\boldsymbol{\zeta ^T}\zeta }}} \left( {\operatorname{Tr}(\mathbf{R^2}) + 2\boldsymbol{\zeta ^T}\mathbf{R^2}\boldsymbol{\zeta} } \right)dt} \right. \hfill 
\left. { + \int\limits_0^\infty  {t \text{ } \operatorname{det}(\boldsymbol{\Delta} ){e^{\frac{1}
				{2}\boldsymbol{\zeta ^T}\boldsymbol{\zeta} }}} {{\left( {\operatorname{Tr}(\mathbf{R}) + \boldsymbol{\zeta ^T}\mathbf{R}\boldsymbol{\zeta} } \right)}^2}dt} \right\} \hfill
\label{eq:e_xt_2}
\end{align}
where, $\boldsymbol{\Delta}=(\mathbf{I_n}+2t\boldsymbol{\Lambda})^{-\frac{1}{2}}$, $\mathbf{R}=\mathbf{\Delta A^* \Delta}$, $\boldsymbol{\zeta} = \boldsymbol{\Delta\mu}$.

Similarly, $\mathbb{E}[\hat{y}_t^2]$ is obtained by replacing $\mathbf{A_x}$ with $\mathbf{A_y}$. From the second moments, we get the theoretical value of the RMSE ($\epsilon$) using
\begin{equation}
\epsilon= \sqrt{\mathbb{E}[\hat{x}_t^2]+\mathbb{E}[\hat{y}_t^2]}.
\label{eq:rmse}
\end{equation}
\subsection{Analysis of improved Cyclic WCL}
\label{sec:Theoretical_Improved_Cyclic_WCL}
\subsubsection{Estimation of feature variation coefficient $\phi_k$}
\label{sec:phi_k_M}
As mentioned in Section \ref{sec:Improved Cyclic WCL}, in the practical system, the $k^{th}$ CR estimates the value of $\phi_k$ using $M$ realizations. We denote the estimate of $\phi_k$ using $M$ realizations by $\hat{\phi}_k^M$ and it is given by $\hat{\phi}_k^M = v_s/e_s$, where $v_s$ and $e_s$ are sample variance of $\hat{R}_{r_k}^{\alpha_t}$ and sample mean of $|\hat{R}_{r_k}^{\alpha_t}|^2$, respectively.

The sample variance ${v_s}$ follows the Chi-square distribution, but it can be approximated by the Gaussian distribution for $M>50$ \cite[pp.118]{Box1978}. We consider $M>50$ for simplicity of analysis. The mean ($\mu_{v_{s}}$) and variance ($\sigma_{v_{s}}^2$) of $v_s$ are given by \cite{Garcia2008}
\begin{align}
\mu_{v_{s}} = \operatorname{var}(\hat{R}_{r_k}^{\alpha_t}) \text{\ and }
\sigma_{v_{s}}^2 = \frac{1}{M}\left[{\mu_4 - \frac{M-3}{M-1}\mu_{v_{s}}^2}\right],
\label{eq:mean_var_v_s}
\end{align}
where $\mu_4 = E\{|\hat{R}_{r_k}^{\alpha_t} - \mathbb{E}[\hat{R}_{r_k}^{\alpha_t}]|^4\}$.
Similarly, ${e_s}$ is a Gaussian random variable with mean $\mu_{e_{s}}$ and variance $\sigma_{e_{s}}^2$ as follows:
\begin{align}
\mu_{e_{s}} = \mathbb{E}[|\hat{R}_{r_k}^{\alpha_t}|^2]  , \text{\ and }
\sigma_{e_{s}}^2=\operatorname{var}(|\hat{R}_{r_k}^{\alpha_t}|^2)/M.
\label{eq:mean_var_e_s}
\end{align}
The analytical expressions of $\mu_{v_s}$, $\sigma_{v_s}$, $\mu_{e_s}$, $\sigma_{e_s}$ in terms of $\mathbb{E}[\boldsymbol{\hat{\theta}}]$, $\boldsymbol{\Sigma_{\hat{\theta}}}$ and $\mathbf{p_k}$ are presented in Appendix \ref{sec:phi_k_M}. It should be noted that $\sigma_{v_s}^2$ and $\sigma_{e_s}^2 \rightarrow 0$  as $M \rightarrow \infty$ since $v_s$ and $e_s$ are consistent estimators of $\operatorname{var}(\hat{R}_{r_k}^{\alpha_t})$ and $\mathbb{E}[|\hat{R}_{r_k}^{\alpha_t}|^2]$.


We compute the confidence interval for $\hat{\phi}_k^M$ using the fact that it is a ratio of two Gaussian variables $v_s$ and $e_s$. Let $\beta$ be the required confidence level, i.e., the probability that the true value of $\phi_k$ lies within the given confidence interval. Further, let $C$ be the center of the confidence interval corresponding to the confidence level $\beta$ and $S$ be the standard error in the computation of $\phi_k$. Then from \cite{Dunlap1986}, we get
\begin{gather}
\nonumber
C = \frac{\frac{v_s}{e_s} - z_\beta^2 \frac{\sigma_{v_se_s}}{e_s^2}}{Q} \text{,}  \\
S^2 = \frac{\sigma_{v_s}^2 - 2 \frac{v_s}{e_s}\sigma_{v_se_s}+\frac{v_s^2}{e_e^2}\sigma_{e_s}^2 - z_\beta^2 \frac{\sigma_{e_s}^2}{e_s^2}\left[\sigma_{v_s}^2 - \frac{\sigma_{v_se_s}^2}{\sigma_{e_s}^2}\right]}{e_s^2Q^2}
\label{eq:confidence_interval_1}
\end{gather}
and the confidence interval is $C.I. = C \pm z_\beta S$, where $Q = 1 - z_\beta^2 \sigma_{e_s}^2 / e_s^2$ and $z_\beta$ is Student's-t variable corresponding to the confidence level $\beta$ and the number of realizations $M$. The value of $z_\beta$ is obtained from standard tables such as \cite[Table 8.2]{Garcia2008}. 

It should be noted that $z_\beta S$ reduces with if the $M$ is increased. While computing $\hat{\phi}_k^M$, the number of realizations $M$ should be large enough to satisfy $z_\beta S < \delta$ for a small value of $\delta$. The value of $\delta$ is selected as the minimum difference between the FVC at any two CRs, i.e., $\delta = \min \limits_{i,j}({\phi_i}- {\phi_j}),$ for $i,j \in \{1,2,...K\}, i \neq j$. This value of $\delta$ ensures that only CRs for which $\phi_k \leq \phi_0$ are included in the algorithm with confidence level $\beta$. In other words, if $\phi_k \leq \phi_0$, then we have $\hat{\phi}_k^M\leq \phi_0$. Other other hand, if $\phi_k > \phi_0$, then we have $\hat{\phi}_k^M> \phi_0$ with probability $\beta$.



\subsubsection{RMSE in the improved Cyclic WCL Estimates}
\label{sec:Rmse_Improved_Cyclic_WCL}
As mentioned in the previous section, the improved Cyclic WCL includes only CRs with $\hat{\phi}_k^M \leq \phi_0$, where $\phi_0$ is the FVC threshold. In order to write the estimates of x- and y-coordinates as a function of $\phi_0$, we introduce a threshold based $K \times K$ diagonal matrix, called selection matrix $\mathbf{S_0}$. The $k^{th}$ diagonal element of $\mathbf{S_0}$ is 1 if $\hat{\phi}_k^M \leq \phi_0$ and 0 otherwise. In other words, the $k^{th}$ diagonal element of the matrix $\mathbf{S_0}$ is the indicator function $\mathds{1}(\hat{\phi}_k^M\leq\phi_0)$. The selection matrix is incorporated in previous definitions of symmetric matrices in (\ref{eq:matrix_def}) as 
\begin{align}
\nonumber  \mathbf{A_x'} &= \diag(\mathbf{PS_0XS_0}^T\mathbf{P}^T,\mathbf{PS_0XS_0}^T\mathbf{P}^T) \hfill, \\
\nonumber  \mathbf{A_y'} &= \diag(\mathbf{PS_0YS_0}^T\mathbf{P}^T,\mathbf{PS_0YS_0}^T\mathbf{P}^T) \hfill \text{, and} \\
\mathbf{B'} &= \diag(\mathbf{PS_0S_0}^T\mathbf{P}^T, \mathbf{PS_0S_0}^T\mathbf{P}^T). \hfill 
\label{eq:matrix_def_improved}
\end{align}
The location estimates using the improved Cyclic WCL are given as
\begin{equation}
{{\hat x}_{t}(\phi_0)} = \frac{{{\boldsymbol{\hat \theta }^T}\mathbf{A_x'}\boldsymbol{\hat \theta} }}
{{{\boldsymbol{\hat \theta }^T}\mathbf{B'}\boldsymbol{\hat \theta} }}{\text{ and }}{{\hat y}_{t}(\phi_0)} = \frac{{{\boldsymbol{\hat \theta }^T}\mathbf{A_y'}\boldsymbol{\hat \theta} }}
{{{\boldsymbol{\hat \theta }^T}\mathbf{B'}\boldsymbol{\hat \theta} }}.
\label{eq:xt_yt_improved}
\end{equation}

It should be noted that ${{\hat x}_{t}(\phi_0)}$ and ${{\hat y}_{t}(\phi_0)}$ are also in RQGV form in the Gaussian vector $\boldsymbol{\hat{\theta}}$. Further, in order to show that $\mathbf{B'}$ is positive semidefinite, we note that $\mathbf{PS_0S_0}^T\mathbf{P}^T$ is positive semi-definite, if there is at least one non-zero element on the diagonal of $\mathbf{S_0}$. The number of non-zero elements of the diagonal of $\mathbf{S_0}$ is the number of CRs satisfying $\hat{\phi}_k^M \leq \phi_0$. If $\phi_0$ is selected such that at least one CR satisfies $\hat{\phi}_k^M \leq \phi_0$, then $\mathbf{PS_0S_0}^T\mathbf{P}^T$ is positive semi-definite, which in turn makes $\mathbf{B'}$ also a positive semi-definite matrix. Therefore, for a fixed value of $\phi_0$, the RMSE for improved Cyclic WCL is computed in similar way as Cyclic WCL: 
\begin{equation}
\epsilon(\phi_0) = \mathbb{E}[\hat{x}_{t}^2 (\phi_0)] + \mathbb{E}[\hat{y}_{t}^2 (\phi_0)],
\label{eq:rmse_improved}
\end{equation}
where  $\mathbb{E}[\hat{x}_t^2 (\phi_0)]$  and $\mathbb{E}[\hat{y}_t^2 ( \phi_0)]$ are obtained by replacing $\mathbf{A_x}$, $\mathbf{A_y}$ and $\mathbf{B}$ with $\mathbf{A_x'}$ and $\mathbf{A_y'}$ and $\mathbf{B'}$, respectively in (\ref{eq:e_xt_2}).

\subsubsection{Optimum FVC threshold}
\label{sec:optimum_fvc_threshold}
The optimum value of the FVC threshold $\phi_0$ that minimizes the RMSE $\epsilon(\phi_0)$ is obtained with the knowledge of $\mathbf{A_x'}$, $\mathbf{A_y'}$, $\mathbf{B'}$, $\mathbb{E}[\boldsymbol{\hat{\theta}}]$ and $\boldsymbol{\Sigma_{\hat \theta}}$. Without loss of generality, let us consider  $\hat{\phi}_1^M \leq \hat{\phi}_2^M ...\leq \hat{\phi}_K^M$. For $\phi_0 < \hat{\phi}_1^M$, no CR will be used for localization and the target location estimates cannot be computed. Therefore, we must have $\phi_0 \geq \hat{\phi}_1^M$ for the improved Cyclic WCL algorithm to work. If the value of FVC threshold is $\phi_0 = \hat{\phi}_{k'}^M$ $(1\leq k' \leq K)$, then CRs $1,2,3..k'$  will be included in the localization process.

It should be noted that the RMSE for any value of $\phi_0$ in the range $[\hat{\phi}_{k'}^M, \hat{\phi}_{k'+1}^M)$ remains constant. This is because the same $k'$ CRs are used for localization if the value of $\phi_0$ is in the given range. Therefore, $\epsilon(\phi_0) = \epsilon(\hat{\phi}_{k'}^M) $ if $\phi_0 \in [\hat{\phi}_{k'}^M,\hat{\phi}_{k'+1}^M)$. It follows from the above argument that $\epsilon({\phi_0})$ has $K$ unique values in the domain of $\phi_0 \in [0,1]$ and the unique values are  $\{\epsilon(\hat{\phi}_{1}^M), \epsilon(\hat{\phi}_{2}^M),....\epsilon(\hat{\phi}_{K}^M)\}$. We obtain the unique values of $\epsilon({\phi_0})$ using (\ref{eq:rmse_improved}) at $\phi_0 \in \{\hat{\phi}_{1}^M, \hat{\phi}_{2}^M,...\hat{\phi}_{K}^M\}$. The optimum value of the FVC threshold is then given by $\{\phi_0^{opt}: \epsilon(\phi_0^{opt}) \leq \epsilon(\hat{\phi}_k^M), k=1,2..K \}$.


\subsection{{Complexity analysis}}
\label{sec:complexity}
In this section, we compare the computational complexities of the proposed algorithms, Cyclic WCL and improved Cyclic WCL, with traditional WCL algorithm. We assume that the number of operations (OPS) required for addition, subtraction and comparison is 1, and for multiplication and division is 10, as considered in \cite{Wang2011}. The number of OPS required for Cyclic WCL, using (\ref{eq:rk}) and (\ref{eq:lt}), is $21KN + 23K + 36$. Similarly, the number of OPS required for improved Cyclic WCL, using (\ref{eq:rk}), (\ref{eq:phi_k}), and (\ref{eq:lt_modified}), is $21MKN+24MK+19M+(71+\eta) K+17$. Here, $\eta K$ is the number of OPS required to obtain $\phi_0^{sub}$ using k-means clustering. The computational complexity of the k-means algorithm is $\mathcal{O}(K)$ \cite{Hamerly2015}. Therefore, we assume $\eta K$ is the number of OPS required for k-means clustering. 

Finally, the WCL algorithm in \cite{Wang2011} is a special case of Cyclic WCL with $\alpha_t=0$. The number of OPS required for WCL is $11KN+23K+26$. Therefore, the computational complexities of WCL and Cyclic WCL are $\mathcal{O}(KN)$, while that of improved Cyclic WCL is $\mathcal{O}(MKN)$.

\begin{figure}[t!]
	\centering
	\includegraphics[width=0.8\columnwidth]{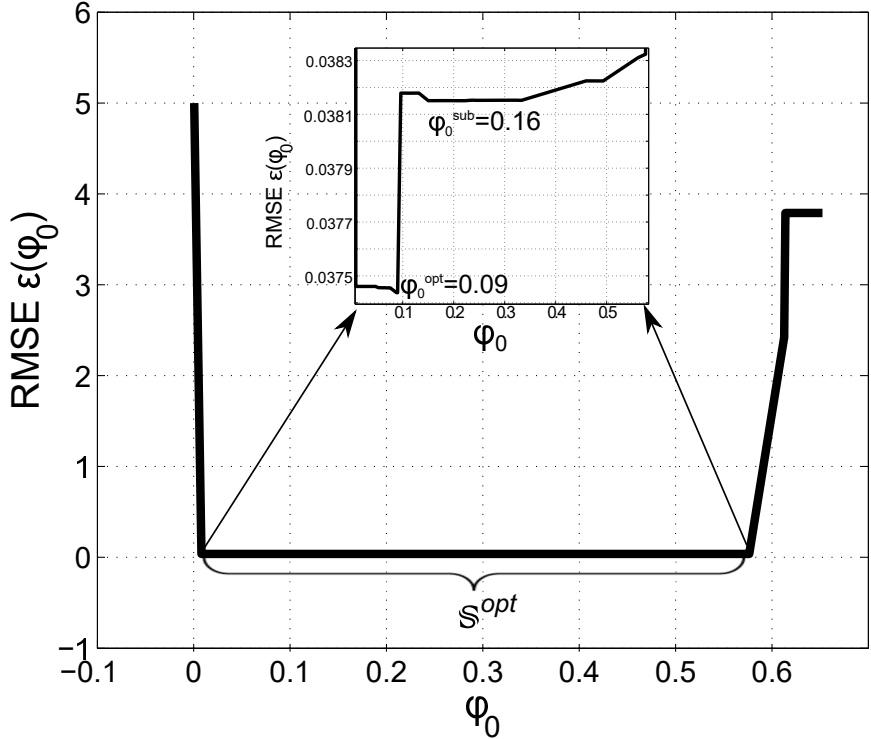}
	\caption{\small Improved Cyclic WCL: RMSE vs $\phi_0$. $\phi_0^{opt}=0.09, \epsilon(\phi_0^{opt})=0.03744 $. $\phi_0^{sub}=0.16, \epsilon(\phi_0^{sub})=0.03815$. $K = 50$. $\rho = -10$ dB. Locations of the CRs [$x_k, y_k$] and $x_k \in \{-40, -20, 0, 20, 40\}$ and $y_k \in \{-45, -35, -25, -15, -5, 5 , 15, 25, 35, 45\}$. Interferer location: $[20, 20]$.}
	\label{fig:rmse_vs_phi_0}
\end{figure}

\section{Simulation results and discussion}
\label{sec:Results}
We consider a CR network with $K$ CRs located in a square shaped area of size $100$m x $100$m. The path-loss exponent $\gamma = 3.8$ and noise PSD is $N_0 = -174$dBm/Hz. Initially, we show results with single carrier 4-QAM signals with carrier frequency $f_t=f_i = 2.4$GHz. The symbol rates of the target and  the interferer signals are $\alpha_t=20$MHz and $\alpha_i=25$MHz, respectively. The sampling frequency is $f_s = 200$MHz. The number of samples $N$ is selected such that $N>10 \lceil \frac{f_s}{\Delta\alpha}\rceil = 400$.

First, improved Cyclic WCL is studied in Section \ref{sec:Results_Improved_Cyclic_WCL}. Then, we compare the performances of improved Cyclic WCL and Cyclic WCL in Section \ref{sec:Results_Comparison_Cyclic_WCLs} in the absence of shadowing ($\sigma_q = 0$dB). The comparison between the traditional WCL, Cyclic WCL and the improved Cyclic WCL under the shadowing environment is presented in Section \ref{sec:Results_Comparison}. This section also shows results with OFDM signals. Finally, simulation results with multipath fading channels are shown in Section \ref{sec:Results_multipath}.
\subsection{Performance of the improved Cyclic WCL}
\label{sec:Results_Improved_Cyclic_WCL}
In  improved Cyclic WCL, the number of realizations $M$ are computed using confidence interval as described in Section \ref{sec:phi_k_M}. Further, following the analysis in Section \ref{sec:Rmse_Improved_Cyclic_WCL}, the RMSE in the localization estimates is obtained. Note that above analysis holds for given fixed locations of the CRs, the target and the interferer. Therefore, in order to evaluate the performance of the improved Cyclic WCL, first we show the simulation results for fixed locations of the CRs in Section \ref{sec:Results_Fixed_Locations}. In Section \ref{sec:Results_Uniform_Locations}, we consider uniformly distributed CRs in the network and compute the average RMSE over 1000 iterations.

\begin{figure}[t!]
	\centering
	\includegraphics[width=0.6\columnwidth]{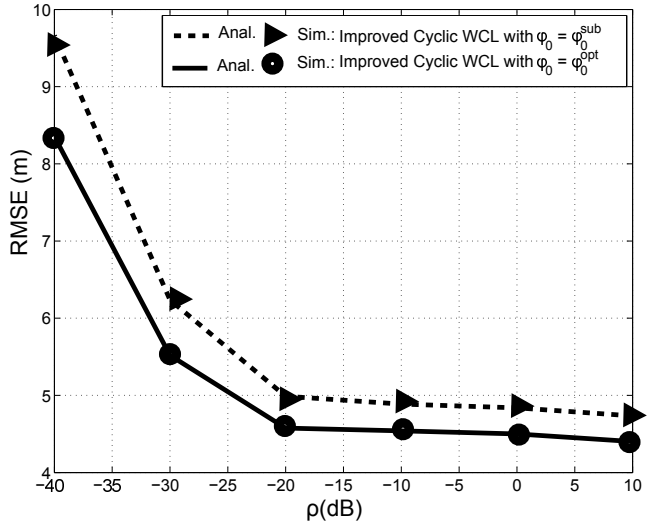}
	\caption{\small RMSE vs transmit power ratio ($\rho$) for improved Cyclic WCL with optimal and suboptimal FVC threshold. Number of CRs $K=50$. Number of Samples $N=500$. Interferer location: $[x_i,y_i]= [20, 20]$.}
	\label{fig:rmse_vs_rho_opt_subopt}
\end{figure}

\subsubsection{Performance of improved Cyclic WCL for fixed locations of CRs}
\label{sec:Results_Fixed_Locations}
The locations of $K=50$ CRs are fixed at $[x_k, y_k]$ such that $x_k \in \{-40, -20, 0, 20, 40\}$ and $y_k \in \{-45, -35, -25, -15, -5, 5 , 15, 25, 35, 45\}$. The target and the interferer are located at $[0, 0]$ and $[20, 20]$, respectively.

The number of realizations $M$ required to obtain $\hat{\phi}_k^M$ are computed with confidence level $\beta = 0.9$ and $\delta = 0.01$. 
For the given parameters, it was observed that $z_\beta S < 0.01$ for $M \geq 60$, therefore $M=60$ realizations are used to compute $\hat{\phi}_k^M$. Further, the optimum threshold $\phi_0$ was obtained as discussed in Section \ref{sec:optimum_fvc_threshold}. In the scenario considered here, $\phi_0^{opt} = 0.09$ and the corresponding RMSE is $\epsilon(\phi_0^{opt})=0.03744$. The impact of selecting different FVC thresholds on the error is shown in Fig. \ref{fig:rmse_vs_phi_0}. The y-axis in this figure also represents $||\boldsymbol{\hat{L}_{t_{improved}}}(\phi_0)||^2$ since the target is located at the origin. In this particular case, the k-means clustering results in $||\boldsymbol{\hat{L}_{t_{improved}}}(\phi_0)||^2 \in \mathbb{S}^{opt}$ for $0.08 \leq \phi_0 \leq 0.57$. Further, the suboptimal threshold $\phi_0^{sub} = 0.16$ and the corresponding RMSE is $\epsilon(\phi_0^{sub})=0.03815$. Hence, the localization error increased by only 0.0007m if suboptimal threshold is used in this setting.
\subsubsection{Performance of improved Cyclic WCL with uniformly distributed CRs}
\label{sec:Results_Uniform_Locations}
Now we consider that the CRs are uniformly distributed in the network. The average RMSE of 1000 realizations of CR locations is plotted against the transmit power ratio ($\rho$) in Fig. \ref{fig:rmse_vs_rho_opt_subopt}. The figure compares performance of the algorithm with optimal ($\phi_0^{opt}$) and suboptimal ($\phi_0^{sub}$) FVC threshold. As described in Section \ref{sec:Improved Cyclic WCL}, unlike $\phi_0^{opt}$,  the computation of $\phi_0^{sub}$ does not require the knowledge of $p_{t,k}, p_{i,k}, s_t$ and $s_i$. From Fig. \ref{fig:rmse_vs_rho_opt_subopt}, it can be observed that the performance of the improved Cyclic WCL with suboptimal FVC threshold is comparable to the performance with optimal threshold. The suboptimal threshold results in increased error of at most $1$m for transmit power ratio ranging from $10$ dB to $-40$dB. Therefore, the knowledge of the transmit powers of the target and the interferer and their signals is not necessary for satisfactory performance of the proposed coarse-grained localization algorithm.
\subsection{Comparison between Cyclic WCL and improved Cyclic WCL}
\label{sec:Results_Comparison_Cyclic_WCLs}
In this section, the impacts of the location of the interferer, CR density and the number of samples $N$ on Cyclic WCL and improved Cyclic WCL are studied. In each scenario, the algorithm is simulated for 1000 realizations with uniformly distributed CRs. The average RMSE over 1000 realizations is plotted. The improved Cyclic WCL is implemented with suboptimal threshold $\hat{\phi}_0^{sub}$.

\begin{figure}[t!]
	\centering
	\begin{subfigure}[c]{0.7\textwidth}
		\includegraphics[width=0.9\columnwidth]{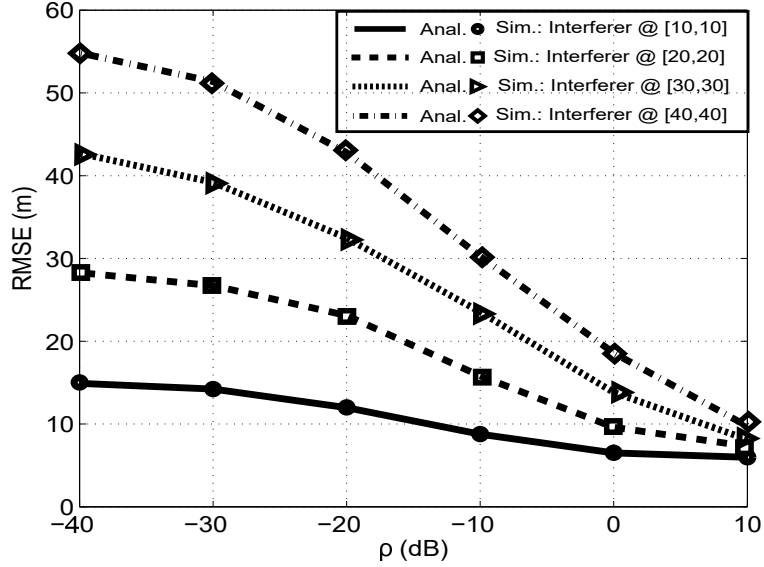}
		\caption{\small Cyclic WCL}
		\label{fig:diff_loc_cwcl}
	\end{subfigure}
	
	\begin{subfigure}[c]{0.8\textwidth}
		\includegraphics[width=0.9\columnwidth]{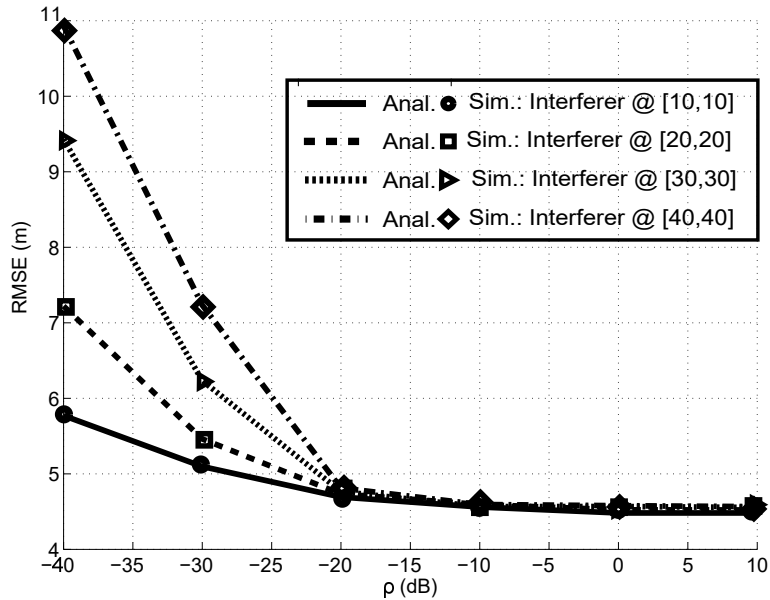}
		\caption{{\small Improved Cyclic WCL}}
		\label{fig:diff_loc_icwcl}
	\end{subfigure}
	\caption{\small Impact of interferer location on a) Cyclic WCL and b) Improved Cyclic WCL, with $p_t=10$dBm. Number of CRs $K= 50$. Number of samples $N=500$. Target location: $[0, 0]$.}
	\label{fig:rmse_vs_rho_diff_loc}
\end{figure}

\subsubsection{Impact of the location of the interferer}
The RMSE in the target location estimates for different locations of the interferer is shown in Fig. \ref{fig:rmse_vs_rho_diff_loc}. The results shown in the figures are counter-intuitive, since the interferer located further away from the target causes higher error as compared to the interferer located closer to the target, especially with increased interferer power. This is due to the fact that at high interferer power, the centroid of $|\hat{R}_{r_k}|^2$ is closer to the interferer. Further, if the interferer location is away from the target, the centroid and hence the target location estimates move away from the target and closer to the interferer. This phenomenon results in increased error as seen in the figure. For a fixed position of the interferer, it is observed that the RMSE increases with higher interferer power, since the impact of the interferer on the centroid of $|\hat{R}_{r_k}|^2$ becomes more prominent. {As shown in Fig.} \ref{fig:diff_loc_cwcl} {and} \ref{fig:diff_loc_icwcl}, {the RMSE in improved Cyclic WCL is significantly lower than RMSE in Cyclic WCL. Further, the RMSE for different locations converge at $\rho=10$ dB for Cyclic WCL, and at $\rho= -20$dB for improved Cyclic WCL. Therefore, improved Cyclic WCL provides more robustness against interferer's power and location.}

\begin{figure}
	\centering
	\begin{subfigure} [b]{0.6\textwidth}
		\includegraphics[width=\textwidth]{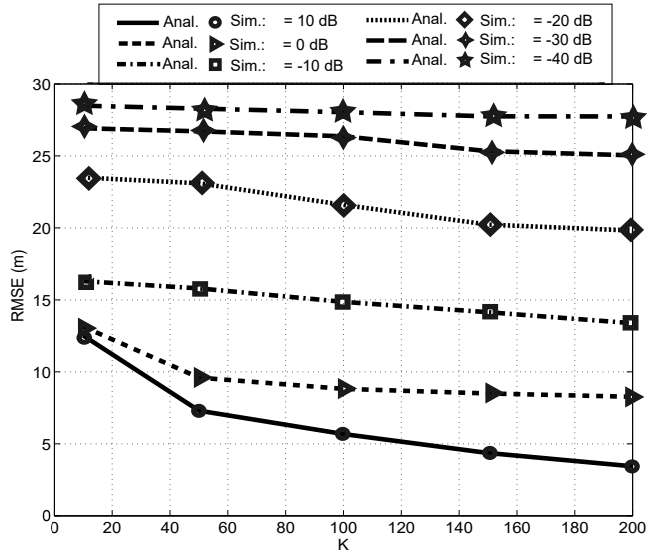}
		\caption{\small Cyclic WCL}
		\label{fig:diff_K_cwcl}
	\end{subfigure}
	\hspace{0.5em}
	\begin{subfigure}[b]{0.6\textwidth}
		\includegraphics[width=\textwidth]{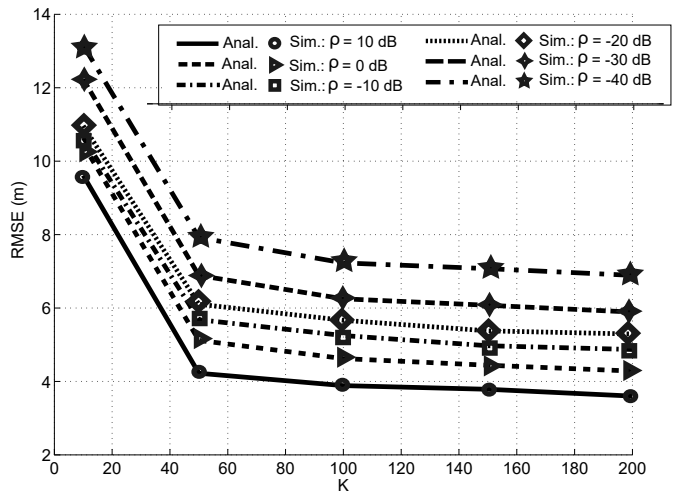}
		\caption{{\small Improved Cyclic WCL}}
		\label{fig:diff_K_icwcl}
	\end{subfigure}
	\caption{\small Impact of CR density on a) Cyclic WCL and b) Improved Cyclic WCL, with $p_t=10$dBm. Number of samples $N=500$. Target and interferer locations: $[0, 0]$, $[20, 20]$.}
	\label{fig:rmse_vs_rho_diff_K}
\end{figure}

\begin{figure}
	\centering
	\begin{subfigure}[c]{1\textwidth}
		\hspace*{25mm} \includegraphics[width=0.8\textwidth]{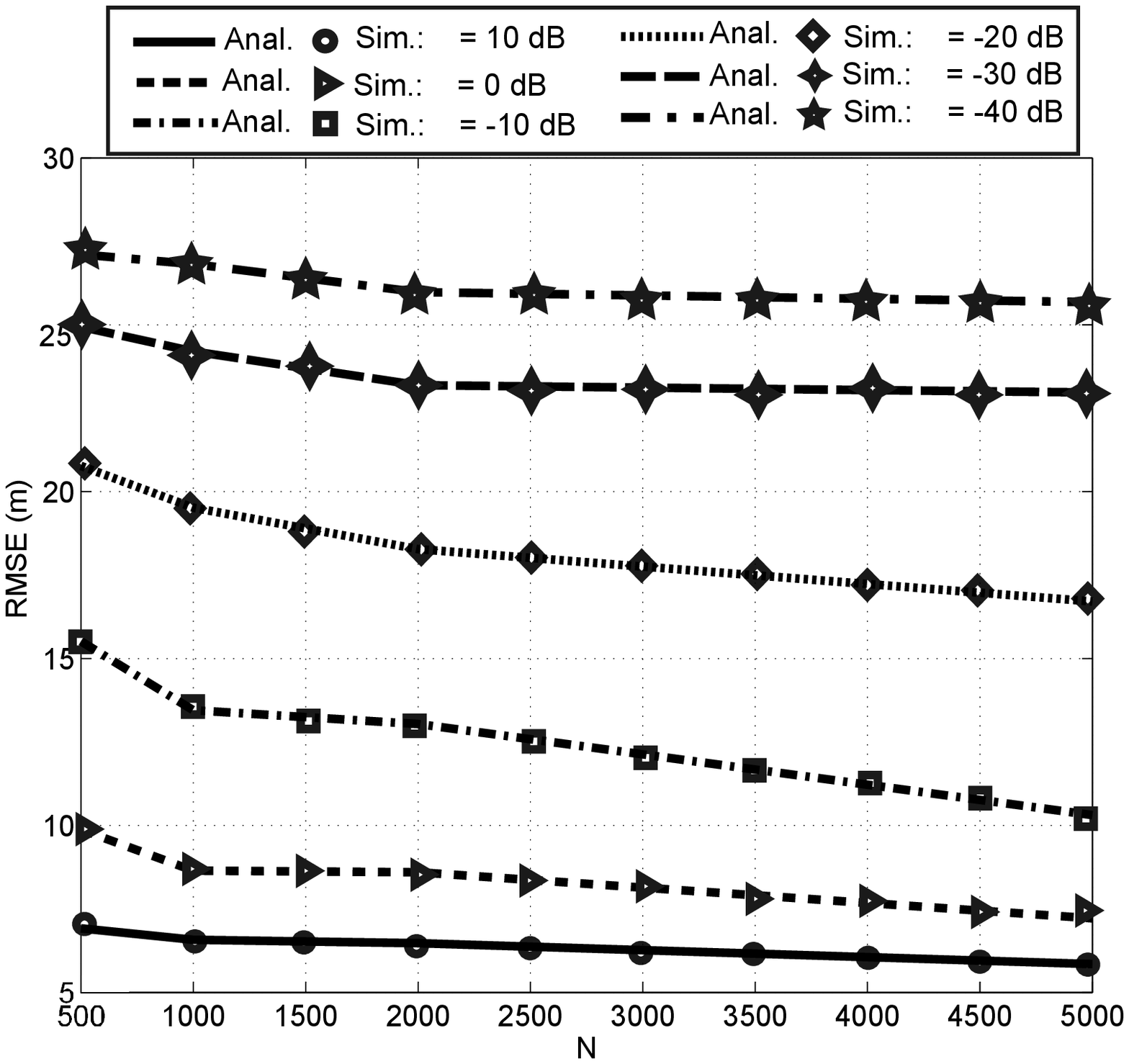}
		\caption{\small Cyclic WCL}
		\label{fig:diff_N_cwcl}
	\end{subfigure}
	\begin{subfigure}[c]{0.6\textwidth}
		\includegraphics[width=\textwidth]{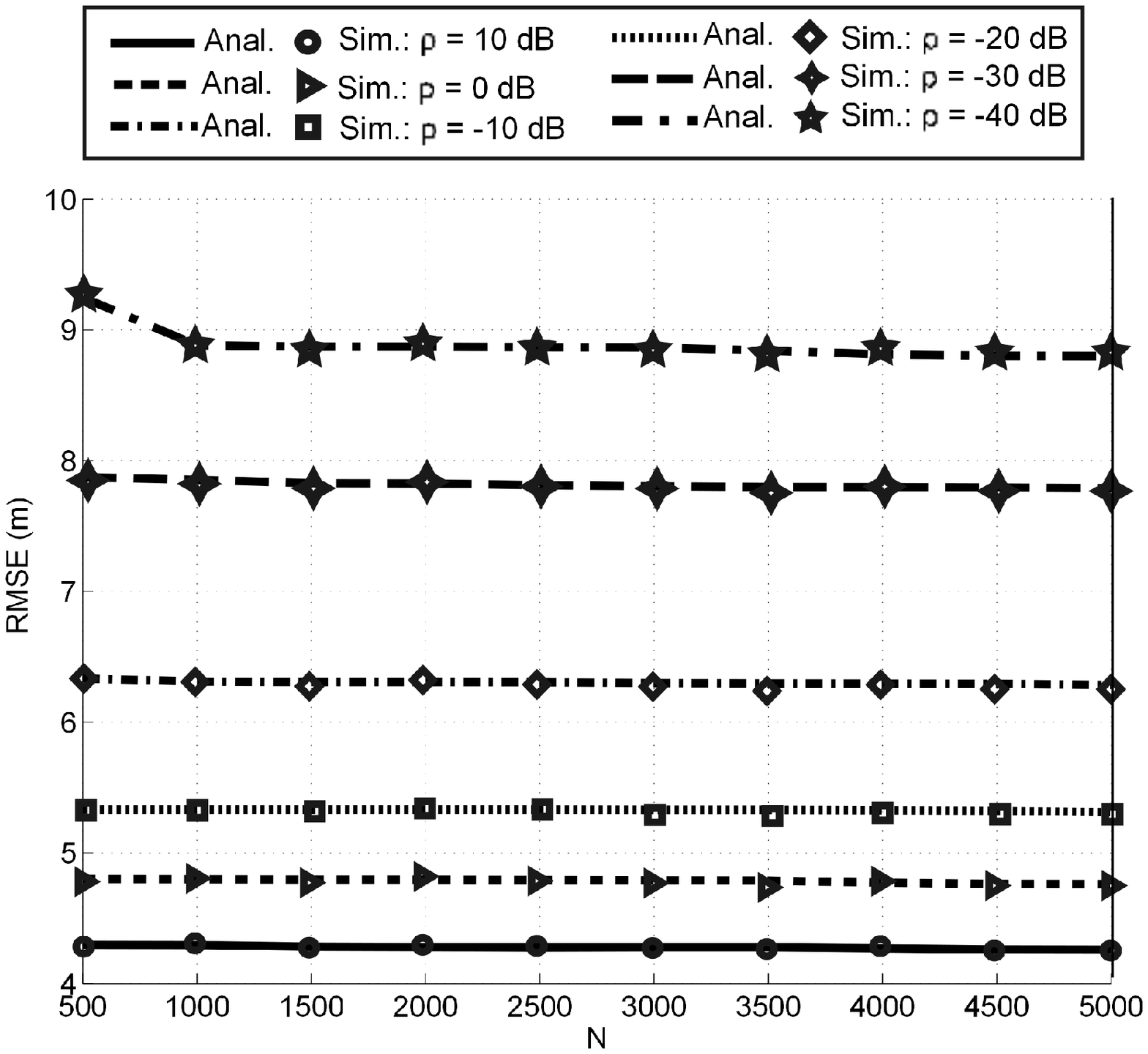}
		\caption{{\small Improved Cyclic WCL}}
		\label{fig:diff_N_icwcl}
	\end{subfigure}
	\caption{\small Impact of number of samples ($N$) on a) Cyclic WCL and b) Improved Cyclic WCL, with $p_t=10$dBm. Number of CRs $K=50$. Target and interferer locations: $[0, 0]$, $[20, 20]$.}
	\label{fig:rmse_vs_rho_diff_N}
\end{figure}
\subsubsection{Impact of CR density}
The impact of increased CR density in the network on the localization error is shown in Fig. \ref{fig:rmse_vs_rho_diff_K}. It has been observed that increasing the number of CRs from $10$ to $200$ at $\rho=10$dB reduces the error by $9$m, as shown in Fig. \ref{fig:diff_K_cwcl}. However, at $\rho=-40$ dB increasing the number of CRs reduces the error by only $0.75$m. This is due to the fact that increasing $K$ increases the number of CRs in the vicinity of the interferer as well which contribute to increased error at higher interferer power ($\rho=-40$dB). Therefore, any gain obtained by increasing the CR density is compensated by increased interferer power which results in essentially a flat curve at $\rho = -40$dB. {Further, the localization error in improved Cyclic WCL, as shown in Fig.} \ref{fig:diff_K_icwcl}, {decreases by approximately $5$m  if $K$ is increased from $10$ to $50$ for all values of $\rho$, while the error is reduced by approximately $1$m if $K$ is increased from $50$ to $200$. Therefore, the performance of the proposed algorithm changes only by a small amount if $K \geq 50$, irrespective of the interferer power.}

\subsubsection{Impact of number of samples N}
In the Cyclic WCL, non-asymptotic estimate of the CAC of the received signal (\ref{eq:cac_rk}) is used to compute weights for each CR location. The non-asymptotic estimates are based on $N$ samples of the received signal $r_k$. Therefore, performance of the Cyclic WCL and the improved Cyclic WCL depends on the value of $N$.
The impact of the value of $N$ on the performance of the  algorithm is shown in Fig. \ref{fig:rmse_vs_rho_diff_N}. It is observed that increasing the number of samples reduces the error in Cyclic WCL. For example, the error is reduced by up to $5$m when $N$ is increased from 500 to 5000. With increased value of $N$, the estimate $\hat{R}_{r_k}^{\alpha_t}$ approaches the true value of $R_{r_k}^{\alpha_t}$ resulting in lower value of the error. {On the other hand, the error in the improved Cyclic WCL does not change significantly with increased $N$ as shown in Fig.} \ref{fig:diff_N_icwcl}. {Therefore, the performance of improved Cyclic WCL is independent of number of samples $N$ as long as $N> \lceil \frac{f_s}{\Delta\hat{\alpha}}\rceil$.}


\begin{figure}
	\centering
	\includegraphics[width=0.7\columnwidth]
	{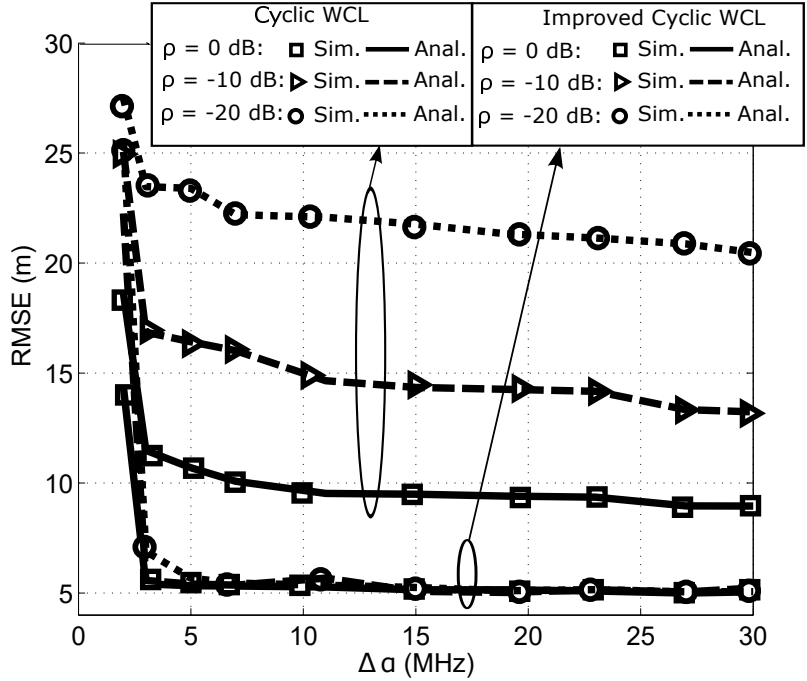}
	\caption{\small RMSE vs $\Delta\alpha = |\alpha_t - \alpha_i|$ with $\alpha_t = 20$ MHz, Estimated cyclic frequency of interferer $\hat{\alpha}_i = 25$ MHz. Estimated difference in cyclic frequencies: $\Delta \hat{\alpha} = |\hat{\alpha}_i - \alpha_t| = 5$ MHz. Number of samples $N=500>10 \lceil \frac{f_s}{\Delta\hat{\alpha}}\rceil$. Target and interferer locations: $[0, 0]$, $[20, 20]$. Suboptimal threshold ${\phi}_0^{sub}$ is used in improved Cyclic WCL.}
	\label{fig:delta_alpha}
\end{figure}

\subsection{Impact of imperfect knowledge of $\alpha_i$}
\label{sec:Results_Delta_Alpha}
In both Cyclic WCL and improved Cyclic WCL, the first step is to select the number of samples $N$ based on $\Delta \alpha = |\alpha_t - \alpha_i|$. In this section, the cyclic frequency of the interferer $\alpha_i$ is varied from $22$MHz to $50$MHz. In order to compute $N$, we assume $\hat{\alpha}_i=25$MHz. The number of samples used are $N=500$, which satisfies $N>10 \lceil \frac{f_s}{\Delta\hat{\alpha}}\rceil =  400$, where $\Delta \hat{\alpha} = |\hat{\alpha}_i - \alpha_t| = 5$MHz. The impact of the imperfect knowledge of $\Delta \alpha$ is shown in Fig. \ref{fig:delta_alpha}. If $\Delta \alpha < \Delta \hat{\alpha} = 5$MHz, the number of samples do not satisfy the condition $N>10 \lceil \frac{f_s}{\Delta\hat{\alpha}}\rceil$, which results in higher interference component in the CAC. This phenomenon results in higher error in both Cyclic WCL and improved Cyclic WCL. On the other hand, if $\Delta \alpha > \Delta \hat{\alpha} = 5$MHz, then the error is reduced and the improvement in the performance of the Cyclic WCL depends on the transmit power ratio ($\rho$). Further, it can be observed that the performance of the improved Cyclic WCL remains the same for different $\rho$ and $\Delta \alpha$ in the regime $\Delta \alpha > \Delta \hat{\alpha}=5$MHz. Therefore, the improved Cyclic WCL is robust to both interference power and error in the estimation of $\alpha_i$.


\subsection{Comparison between traditional WCL, Cyclic WCL and improved Cyclic WCL}
\label{sec:Results_Comparison}
\begin{figure}
	\centering
	\includegraphics[width=0.9\columnwidth]
	{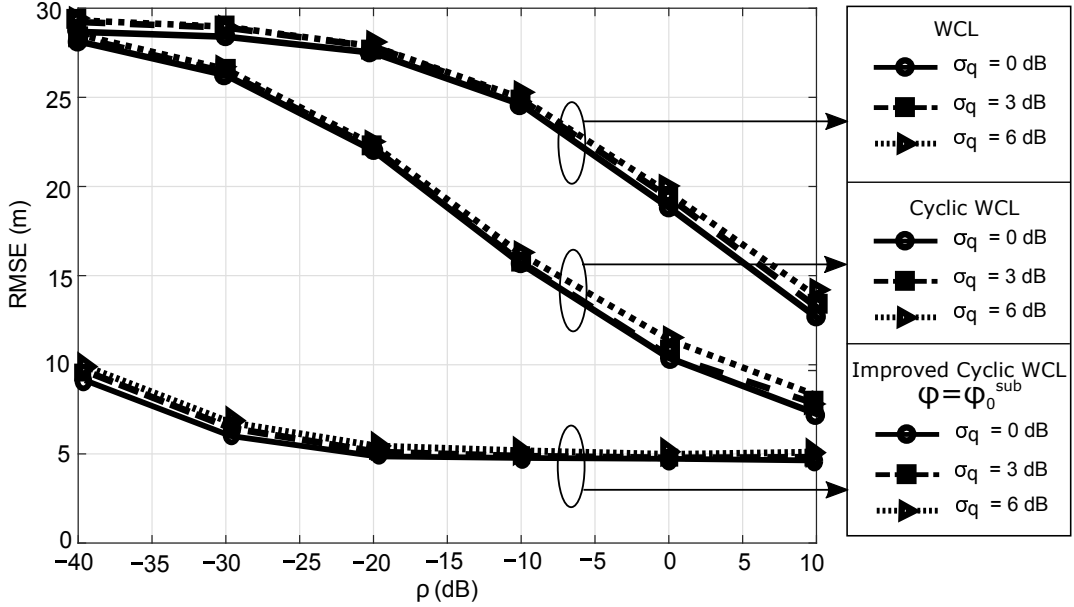}
	\caption{\small Impact of shadowing on WCL, Cyclic WCL, and improved Cyclic WCL with suboptimal FVC threshold. Number of CRs $K = 50$. Number of samples $N=500$. Target and interferer locations: $[0, 0]$, $[20, 20]$.}
	\label{fig:shadowing}
\end{figure}

In this section, we compare the performances of the traditional WCL, Cyclic WCL and the improved Cyclic WCL in shadowing environment. It is observed that, even in shadowing environment ($\sigma_q = 6$ dB), the error in Cyclic WCL is smaller than traditional WCL, as shown in Fig. \ref{fig:shadowing}. In the case of improved Cyclic WCL, the error is reduced by a factor of three as compared to the traditional WCL for $\rho=-40$dB. Further, it has been observed that Cyclic WCL algorithms are robust to shadowing. For example, in improved Cyclic WCL, the error has increased by only $2$m when shadowing variance has increased four fold from $\sigma_q = 0$dB to $\sigma_q=6$dB. This is due to the fact that the shadowing effect over $K$ CRs averages out in the WCL algorithms.

\begin{figure}
	\centering
	\includegraphics[width= 0.8 \columnwidth]{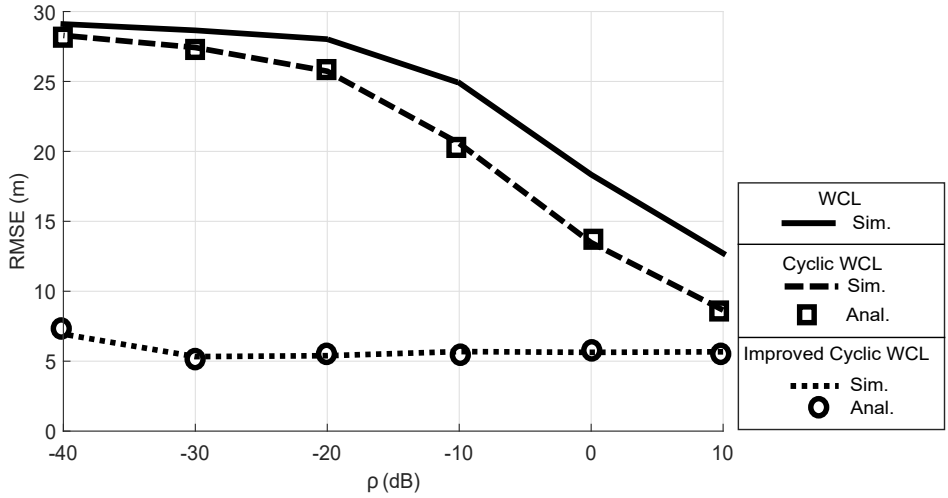}
	\caption{\small RMSE in WCL, Cyclic WCL and improved Cyclic WCL algorithms with OFDM signals. Target and interferer cyclic frequencies $\alpha_t = 250$kHz and $\alpha_i=14$kHz, respectively. Sampling frequency at CRs $f_s=500$kHz. Number of samples $N=100$.  Number of CRs $K=50$. Target and interferer locations: $[0,0]$, $[20,20]$.}
	\label{fig:ofdm_signals}
\end{figure}

Next, we show the performance of proposed algorithms with OFDM signals. In these results, $s_t$ is a WLAN signal with 64 sub-carriers,  and sub-carrier spacing $\Delta f_t =312.5$kHz. The OFDM symbol duration $T_g=4 \mu$s with $T_{cp}=0.8\mu$s. The cyclic frequency of $s_t$ is $\alpha_t=1/T_g=250$kHz.  The interference, $s_i$, is a LTE signal with 1024 sub-carriers and sub-carrier spacing of $\Delta f_i=15$kHz. The OFDM symbol duration $T_h=71.4 \mu$s with $T_{cp}=4.7 \mu$s. The cyclic frequency of the interference is $\alpha_i=1/T_h=14$kHz. Both target and interfering signals carry 4-QAM signal on each sub-carrier, i.e., $c_{l,k}$ and $d_{l,k}$ are 4-QAM symbols. The sampling frequency $f_s = 500$kHz, while the number of samples $N=100>10 \lceil \frac{f_s}{\Delta \alpha} \rceil$. Localization errors in the traditional WCL, Cyclic WCL, and improved Cyclic WCL are shown in Fig. \ref{fig:ofdm_signals}.  The figure also shows analytically computed RMSE with OFDM signals following the derivations in Appendix \ref{app_derivations}. It has been observed that the proposed improved Cyclic WCL continues to provide three times lower error as compared to traditional WCL for $\rho=-40$dB. Further, the comparison of RMSE in improved Cyclic WCL in Fig. \ref{fig:shadowing} and \ref{fig:ofdm_signals} shows that the proposed algorithm performs equally well under OFDM and single carrier signals.

\subsection{Performance under multipath fading channels}
\label{sec:Results_multipath}
The three algorithms are also studied under multipath fading channels suitable for indoor and outdoor settings. For indoor setting, TGn channel model from WLAN standard is used \cite{TGn}. This channel model has delay spread of $40$ns and Doppler spread of $10$Hz. For outdoor setting, extended typical urban (ETU) channel model is considered with $5\mu$s delay spread  and  $300$Hz Doppler spread \cite{LTE_36_101}. This is a commonly used model in LTE cellular system. The channel coefficients $h_{t,k}$ and $h_{i,k}$ are obtained from the aforementioned channel models, where $h_{t,k}$ is the channel between the target and the $k^{th}$ CR, while  $h_{i,k}$ is the channel between interferer and $k^{th}$ CR. The received signal is then obtained as $r_k=\sqrt{p_{t,k}}\left[h_{t,k} \ast s_t \right]+\sqrt{p_{i,k}} \left[h_{i,k} \ast s_i \right]+ w_k$, where $\ast$  denotes the convolution operation. Further implementations of the three algorithms remains the same as in AWGN channel.

\begin{figure}
	\centering
	\includegraphics[width= 0.8 \columnwidth]{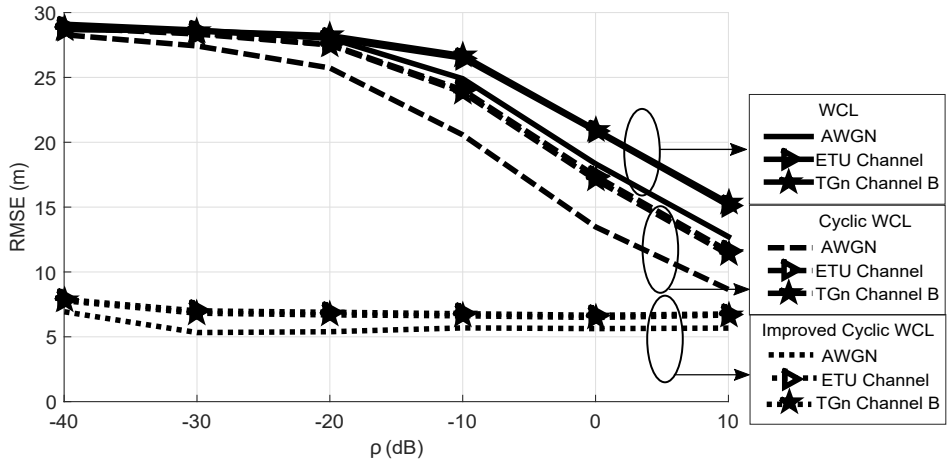}
	\caption{\small RMSE in WCL, Cyclic WCL, and improved Cyclic WCL with different channel models. Signal parameters are same as in Fig. \ref{fig:ofdm_signals}.	}
	\label{fig:nlos_channels}
\end{figure}

The localization errors under the two multipath channels are shown in Fig. \ref{fig:nlos_channels}. As in AWGN case $N=100$ samples are used at sampling rate $500$kHz for localization. It has been observed that the localization errors in all three algorithms increase as compared to AWGN channel. This increase in error can be explained as follows. The signal observation interval at each CR is $100/500$kHz $=0.2$ms, which is smaller than the coherence time of TGn and ETU channels. Therefore, the received power at each CR in this duration is affected by the small scale fading and the impact of fading is not averaged out as the observation duration is smaller than the coherence time. This leads to increased localization errors. However, we can observe that the localization error in WCL and Cyclic WCL is increased by up to $4$m. On the other hand, the error in improved Cyclic WCL increases by up to $2$m only. Therefore, we can conclude that the proposed improved Cyclic WCL more robust against multipath fading channels and continues to provide significant performance gain over the WCL algorithm. 

\section{Summary}
\label{sec:summary_c1}
In this chapter, we have proposed the Cyclic WCL algorithm to mitigate the adverse impact of a spectrally overlapped interference in CR networks on the target localization process. The proposed algorithm uses cyclic autocorrelation of the received signal in order to estimate the target location. Theoretical analysis of the  algorithm is presented in order to compute the error in localization. We have also proposed the improved Cyclic WCL algorithm that identifies CRs in the vicinity of interferer and eliminates them from the localization process to further reduce the error. 

We have studied impacts of interferer's power, its location, and CR density on performance of the proposed algorithms. The improved Cyclic WCL is observed to be robust against interferer's location and its transmit power. The comparison between the traditional WCL and the improved Cyclic WCL shows that the proposed algorithm provides significantly lower error in localization when there is a spectrally overlapped interference in the network. It has been observed that the improved Cyclic WCL is also robust against shadowing and multipath fading environment. The proposed localization algorithm performs equally well with single carrier as well as OFDM signals.

\chapter{Secondary User Selection and Power Allocation in Downlink in Massive MIMO CR Network}
\label{chapter_massive_mimo}


\section{Introduction}
\label{sec:Introduction_c2}
In this chapter, we consider an underlay CR network with a secondary BS (SBS) equipped with a large antenna array that employs massive MIMO technique serve multiple SUs simultaneously.
Massive MIMO and underlay cognitive radio are  being considered for 5G networks in order to accommodate more devices in the available spectrum \cite{andrews2014,boccardi2014a, gupta2015a}. In a massive MIMO system, a base station equipped with a large number of antennas serves multiple users using beamforming techniques in the same time-frequency resource block  \cite{ngo2013a}. On the other hand, in an underlay cognitive radio (CR) network, a secondary base station (SBS) serves its users (secondary users) while keeping the interference to licensed primary users (PUs) below a specified threshold \cite{biglieri2012a}. In underlay CR networks, the SBS transmits the downlink signal in the same time-frequency resource block as the primary transmitter. This is different from traditional interweave cognitive networks where the SBS transmits in an orthogonal time-frequency resource block.

The secondary BS, if equipped with a large number of antennas, can potentially employ beamforming techniques and serve multiple  secondary users (SUs) in the downlink while limiting the interference to primary receivers. However, due to imperfect knowledge of the channels between PUs and the SBS, interference constraints at the PUs, and different rate requirements of SUs, it may not be feasible to serve all the SUs in the network \cite{chaudhari2017a}. Therefore, a judicious selection of SUs and power allocation are required at the SBS in order to simultaneously serve multiple SUs with required rates while limiting the interference to PUs.

\subsection{Related work}
Underlay CR networks with multiple antenna systems have received attention in recent years, since such networks allow concurrent transmissions from  primary and secondary transmitters, thereby increasing the spectrum efficiency \cite{yiu2012, du2012b, du2013b, tsinos2013a,noam2013, noam2014, wang2015, al-ali2016, xiong2016}. The works in \cite{yiu2012, du2012b, tsinos2013a, noam2013, noam2014, al-ali2016} consider small-scale MIMO with approximately ten or fewer antennas at the secondary transmitter. These works consider only one SU in the system, therefore they do not require SU selection. User selection mechanism has been partially considered in \cite{du2013b, xiong2016}. An indirect selection mechanism is implemented in \cite{du2013b} where SUs receiving less than 0dB signal-to-interference-plus-noise-ratio (SINR) are dropped from the downlink transmission. The selection algorithm in \cite{xiong2016} needs to know the number of users to be selected. A massive MIMO system has been employed in the secondary systems in \cite{wang2015, chaudhari2015, chaudhari2017a}. In our previous works \cite{chaudhari2015, chaudhari2017a}, we proposed to use massive MIMO to serve multiple SUs concurrently with primary transmission, while the algorithm in \cite{wang2015} still serves only one SU. A selection algorithm under line-of-sight channels is proposed in \cite{chaudhari2015}, while the feasibility of serving all SUs under Rayleigh fading channels is studied in \cite{chaudhari2017a}. The selection algorithm was not considered in \cite{chaudhari2017a}.

Massive MIMO systems differ from small-scale MIMO systems in \cite{du2012b, du2013b, tsinos2013a, al-ali2016} in the design of beamforming (or precoding) vectors. In small-scale MIMO systems, optimum beamforming vectors are computed using iterative algorithms \cite{du2012b, du2013b, al-ali2016}, and interference alignment \cite{tsinos2013a}. Such approaches become prohibitively expensive in terms of complexity when used with massive MIMO systems. Using linear beamforming techniques such as zero-forcing (ZF), maximal-ratio combining (MRC) or minimum mean-square error (MMSE), the beamforming vectors can be computed using closed-form expressions without requiring any iterative search if the channels and the selected user set is known. In this chapter, we focus on ZF beamforming as it is can also be used to restrict the interference toward PUs.

In an underlay CR network, the interference at primary receivers (PRs) can be eliminated using ZF beamforming if the channels between PRs and the SBS are perfectly known at the SBS. However, due to imperfect CSI in practical networks, there is non-zero leakage interference transmitted towards PRs even when ZF beamforming is used. The magnitude of the interference depends on the power allocated to SUs as well as the set of SUs selected. Therefore, there is a need to design a robust interference control mechanism along with power allocation and user selection in order to limit the interference to PRs below a specified threshold.

\subsection{Summary of contributions and outline}
The main contributions of this work are summarized below.

\begin{enumerate}
	\item
	A new optimization framework is proposed to select the maximum number of SUs in the downlink and obtain power allocation for the selected SUs in order to satisfy their instantaneous rate requirements. The interference to PRs is kept below a specified threshold using margin parameters that compensate for CSI estimation errors. The proposed formulation is different from the formulations in \cite{yoo2006, huang2013, huang2012} which aim to maximize the sum-rate of the selected users and do not have interference constraints.	
	\item A new user selection and power allocation algorithm, called Delete-su-with-Maximum-Power-allocation (DMP), is proposed to obtain a feasible solution for the NP-hard optimization problem. Theoretical analysis of the algorithm is presented in order to compute the average number of SUs achieving the required rate, and average interference to primary receivers.	The proposed algorithm is shown to achieve near-optimal results if the number of antennas at SBS is an order of magnitude larger than the number of users.	
	\item The user selection algorithm in \cite{huang2012} is extended for application in an underlay CR setting. The extended algorithm, called Modified Delete-Minimum-Lambda (MDML), also uses margin parameters and is robust against imperfect CSI. The proposed DMP selection algorithm is shown to serve more users than MDML in an underlay CR network.
\end{enumerate}

\textit{Outline:} This chapter is organized as follows. The system model and the optimization problem are presented in Section \ref{sec:Model_problem_c2}. The DMP and MDML algorithms are presented in Section \ref{sec:solution}. Section \ref{sec:analysis_selection} presents the theoretical analysis and the optimality of the DMP algorithm. Simulation results are presented in Section \ref{sec:results_c2}. Finally, the chapter is summarized in Section \ref{sec:summary_c2}.

\textit{Notations:} We denote vectors by bold, lower-case letters, e.g., $\mathbf{h}$. Matrices are denoted by bold, upper case letters, e.g., $\mathbf{G}$. Scalars are denoted by non-bold letters e.g. $L$. Transpose, conjugate, and Hermitian of vectors and matrices are denoted by $(.)^T$, $(.)^*$, and $(.)^H$, respectively. The norm of a vector $\mathbf{h}$ is denoted by $||\mathbf{h}||$. $\Gamma(k,\theta)$ is the Gamma distribution with shape parameter $k$ and scale parameters $\theta$, whereas $\Gamma(x)$ is the Gamma function. The $i$-th element in the set $\mathcal{S}$ is denoted by $\mathcal{S}(i)$ and the cardinality of the set is denoted by $|\mathcal{S}|$. An empty set is denoted by $\emptyset$.


\section{System model and problem formulation}
\label{sec:Model_problem_c2}
\subsection{System model}
\begin{figure}
	\centering
	\includegraphics[width=0.5\columnwidth]{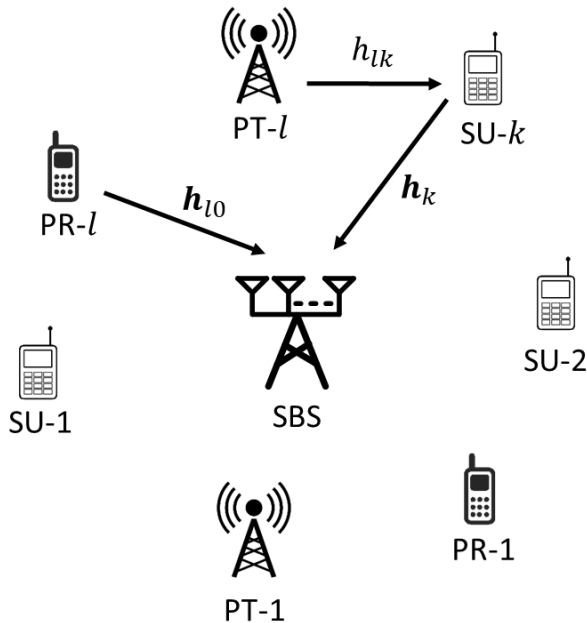}
	\caption{\small Network model showing channels between PT-$l$ and SU-$k$ ($h_{lk}$), PR-$l$ and the SBS ($\mathbf{h}_{l0}$), and SU-$k$ and the SBS ($\mathbf{h}_k$).}
	\label{fig:system_model_c2}
	\vspace{-1mm}
\end{figure}

Consider an underlay CR network with one SBS and $K$ SUs. The SBS is equipped with $M (\gg K)$ antennas. This network coexists with $L$ primary transmitter-receiver pairs. Let $\mathcal{T}=\{1,2,...,L\}$ be the set of primary transmitters (PTs) and  $\mathcal{R}=\{1,2,...,L\}$ be the set of PRs. The SUs, PRs and PTs are assumed to be single antenna terminals. Let $\mathbf{h}_{k} =\sqrt{\beta_k} \mathbf{\tilde{h}}_{k}  \in \mathbb{C}^{M \times 1}$ be the channel between SU-$k$ and the SBS where $\beta_k$ is the slow fading coefficient accounting for attenuation and shadowing and $\mathbf{\tilde{h}}_{k} \sim \mathcal{CN}(0,\mathbf{I})$ \cite{ngo2013a, chien2016, marzetta2010}. The channel between PT-$l$ and SU-$k$ is denoted by ${h_{lk}}=\sqrt{\beta_{lk}} {\tilde{h}_{lk}} \in \mathbb{C}, l\in \mathcal{T}, k=\{1,2,...,K\}$, and ${\tilde{h}_{lk}} \sim \mathcal{CN}(0,1)$. Similarly, the channel between PR-$l$  and the SBS is $\mathbf{h}_{l0}=\sqrt{\beta_{l0}} \mathbf{\tilde{h}}_{l0} \in \mathbb{C}^{M\times 1}, l\in \mathcal{R}, \mathbf{\tilde{h}}_{l0} \sim \mathcal{CN}(0,\mathbf{I})$. We consider a time-division duplex (TDD) systems and the channels are assumed to be reciprocal. { The network is depicted in Fig. \ref{fig:system_model_c2}.}

The SBS has imperfect knowledge of the channels $\mathbf{h}_{l0}, \mathbf{h}_{k}$. The estimates of channels are given by $\mathbf{\hat{h}}_{l0}=\mathbf{{h}}_{l0} + \mathbf{\Delta}_{l0}$ and $\mathbf{\hat{h}}_{k} = \mathbf{h}_{k} + \mathbf{\delta}_k$, respectively, where $\mathbf{\Delta}_{l0} \sim \mathcal{CN}(0,\sigma^2_{\Delta}\mathbf{I})$, and $\mathbf{\delta}_k \sim \mathcal{CN}(0,\sigma^2_{\delta}\mathbf{I})$ are the estimation errors.
We model the quality of CSI between primary and secondary system using $\sigma^2_\Delta = \frac{\sigma^2_w}{P_p}$, while the quality of CSI within the secondary system is modeled using $\sigma^2_\delta = \frac{\sigma^2_w}{P^0}$, where $P_p$ and $P^0$ are transmit powers from PTs and  the SBS, respectively, and $\sigma^2_w$ is the noise power at the SBS and SUs \cite{aquilina2015,razavi2014,maurer2011}. We consider block fading where the channels remain constant for a finite coherence interval.

Let $\mathcal{S}_0=\{1,2,...,K\}$ be the set of all SUs and $\mathcal{S} \subseteq \mathcal{S}_0$ be the set of SUs considered for downlink transmission and $P_k, k\in \mathcal{S}$ be the power allocated to SU-$k$ in the downlink when set ${\mathcal{S}}$ is selected. The ZF transmit vector for SU-$k$ depends on $\mathcal{S}$ and is denoted by {$\mathbf{v}^{\mathcal{S}}_k \in {\mathbb{C}^{M\times 1}}, k \in \mathcal{S}$}. The unit-norm ZF vectors are computed using \cite{huang2012}:
{
\begin{align}
\nonumber \mathbf{v}^{\mathcal{S}}_{\mathcal{S}(i)} &= \frac{\left[\mathbf{G_\mathcal{S}(G_\mathcal{S}^H G_\mathcal{S})^{-1}}\right]_i}{
||\left[\mathbf{G_\mathcal{S}(G_\mathcal{S}^H G_\mathcal{S})^{-1}}\right]_i||},
\\  \mathbf{G}_\mathcal{S} &= [\mathbf{\hat{h}}_{\mathcal{S}(1)}, \mathbf{\hat{h}}_{\mathcal{S}(2)},..., \mathbf{\hat{h}}_{\mathcal{S}(|\mathcal{S}|)}, \mathbf{\hat{h}}_{\mathcal{R}(1)0},...,\mathbf{\hat{h}}_{\mathcal{R}(L)0}],
\label{eq:zf_vectors}
\end{align}}
where $\mathcal{S}(i)$ is the $i$-th entry in $\mathcal{S}$, and $[\mathbf{A}]_i$ is the $i$-th column of matrix $\mathbf{A}$. { In the above equation, the matrix $\mathbf{G}_{\mathcal{S}} \in \mathbb{C}^{M \times (|\mathcal{S}| + L)}$ indicates the channel between SUs in set $\mathcal{S}$, PUs and the SBS.} It should be noted that the ZF vector $\mathbf{v}^{\mathcal{S}}_k$ is in the null-space of the estimated channels to SU-$j, j \in \mathcal{S}, j\neq k$. It is also in the null-space of the estimated channels to PRs. Therefore, $\mathbf{v}^{\mathcal{S}}_k$  satisfies $\mathbf{\hat{h}}^H_j \mathbf{v}^{\mathcal{S}}_k=0, j,k\in \mathcal{S}, j\neq k$ and $\mathbf{\hat{h}}^H_{l0}\mathbf{v}^{\mathcal{S}}_k=0, l \in \mathcal{R}$. 
 
The ZF beamforming vectors are not in the null-space of the true channel $\mathbf{h}^H_{l0}$ when the channel estimation error $\mathbf{\Delta}_{l0}$ is non-zero. Therefore, the interference caused at PR-$l, l\in \mathcal{R}$ is non-zero and can be expressed as:
 \begin{align}
I_l=\sum_{k \in S} I_{kl} = \sum_{k \in S} P_k |\mathbf{h}^H_{l0} \mathbf{v}^{\mathcal{S}}_k|^2, l \in \mathcal{R}, k \in \mathcal{S},
\label{eq:I_l}
 \end{align}
where $I_{kl}$ is the interference contribution of data stream of SU-$k$ towards PR-$l$. The interference $I_l$ depends on power allocation as well as the set of SUs selected. Similarly, the inter-SU interference at SU-$k$ due to the signal transmitted toward SU-$j$ can be expressed as
\begin{align}
I_{jk} = P_j |\mathbf{  h}^H_{k} \mathbf{v}^{\mathcal{S}}_j|^2, k,j \in \mathcal{S}, j \neq k.
\label{eq:I_jk}
\end{align}
Finally, the reverse interference at SU-$k$ is the sum of powers received from PTs: $I_{k} = \sum \limits_{l \in \mathcal{T}}P_p |{h_{lk}}|^2$, where $P_p$ is the power transmitted by PT-$l$.


\subsection{Optimization problem}
\label{sec:problem_c2}
Our goal is to select the maximum number of SUs for downlink transmission in order to satisfy specific instantaneous rate of $R^0_k$ to selected SUs, while keeping the interference towards PRs below $I^0$. The total available power at the SBS is $P^0$. Note that the estimated interference to PR-$l, l \in \mathcal{R}$ based on the estimated channel is $\hat{I}_l= \sum_{k \in \mathcal{S}} P_k |\mathbf{\hat{h}}^H_{l0} \mathbf{v}^{\mathcal{S}}_k|^2 = 0$. Since the true interference $I_l$ is non-zero, we add a margin parameter $\epsilon_1$ to define $\tilde{I}_l = \sum_{k \in \mathcal{S}} P_k (|\mathbf{\hat{h}}^H_{l0} \mathbf{v}^{\mathcal{S}}_k|^2 + \epsilon_1) $ as the new estimated value of the interference with margin.

Further, the instantaneous rate achieved at SU-$k$, when a set $\mathcal{S}$ is selected, is
\begin{align}
R^{\mathcal{S}}_k = \log_2 \left( 1 + \frac{P_k |\mathbf{ h}_{k}^H \mathbf{v}^{\mathcal{S}}_k|^2}{\sigma^2_w + I_k+ \sum \limits_{j \in \mathcal{S}, j \neq k} I_{jk} }\right), k \in \mathcal{S},
\label{eq:R_k_S}
\end{align}
where $\sigma^2_w$ is the noise power at the SU. Due to ZF beamforming, the estimated inter-SU interference will be zero, i.e., $\hat{I}_{jk} =P_j |\mathbf{\hat{h}}^H_k \mathbf{v}^\mathcal{S}_j|^2= 0$ due to $\mathbf{\hat{h}}^H_k \mathbf{v}^\mathcal{S}_j=0, k\neq j$. We use a margin parameter $\epsilon_2$ to compensate for the estimation error in the channels between SBS and SUs. Therefore, the estimated instantaneous rate with margin becomes
\begin{align}
\hat{R}^{\mathcal{S}}_k =  \log_2 \left( 1 + \frac{P_k |\mathbf{ \hat{h}}_{k}^H \mathbf{v}^{\mathcal{S}}_k|^2}{\sigma^2_w + {I}_{k} + \epsilon_2}\right), k \in \mathcal{S}.
\label{eq:R_k_S_est}
\end{align}

{ Note that, unlike $I_l$ and $I_{jk}$, the reverse interference term $I_k$ can be measured at SU-$k$ by observing combined signal received from all PUs during the channel estimation phase.} The optimization problem can then be formulated as follows:
\begin{align}
&\max_{\{\mathcal{S},P_k\}} |\mathcal{S}|
\label{eq:optim2_start}
\\ \text{Subject to}&:
 \tilde{I}_{l}=\sum_{k \in \mathcal{S}} P_k \left(|\mathbf{\hat{h}}^H_{l0} \mathbf{v}^{\mathcal{S}}_k|^2 + \epsilon_1 \right) \leq I^0,~ l \in \mathcal{R}
\label{eq:optim2_const1}
\\ &\hat{R}^{\mathcal{S}}_k  \geq R^0_k,~ k \in \mathcal{S},
\label{eq:optim2_const2}
\\ &\sum_{k\in \mathcal{S}} P_k \leq P^0, P_k \geq 0.
\label{eq:optim2_end}
\end{align}
Selection of parameters $\epsilon_1$ and $\epsilon_2$ is discussed in Section \ref{sec:selection_of_parameters}. By substituting $|\mathbf{\hat{h}}^H_{l0} \mathbf{v}^{\mathcal{S}}_k|^2 = 0$ and rearranging (\ref{eq:optim2_const2}), we obtain the following equivalent optimization problem:
\begin{align}
&\max_{\{\mathcal{S},P_k\}} |\mathcal{S}|
\label{eq:optim3_start}
\\\text{Subject to}&: \sum_{k\in \mathcal{S}} P_k \epsilon_1    \leq I^0,
\label{eq:optim3_const1}
\\ P_k &\geq \frac{(2^{R^0_k} -1)\left(\sigma^2_w + {I}_k + \epsilon_2 \right)}{|\mathbf{  \hat{h}}_k^H \mathbf{v}^{\mathcal{S}}_k|^2}, k \in \mathcal{S},
\label{eq:optim3_middle}
\\& \sum_{k\in \mathcal{S}} P_k \leq P^0.
\label{eq:optim3_end}
\end{align}
Using binary selection variables $s_k \in \{0,1\}$ to indicate whether SU-$k$ is selected ($s_k=1$) or not  ($s_k=0$), we can restate the above problem as:
\begin{align}
(\mathbf{P1}) \max_{\{s_k,P_k\}} &\sum_{k=1}^{K} s_k
\label{eq:optim4_start}
\\\text{Subject to}&:\sum_{k=1}^{K} s_k P_k  \leq \min \left( {I^0/\epsilon_1}, {P^0} \right),
\label{eq:optim4_const1}
\\ P_k &\geq  s_k \frac{(2^{R^0_k} -1)\left(\sigma^2_w +  {I}_k + \epsilon_2\right)}{|\mathbf{\hat{h}}_k^H \mathbf{v}^{\mathcal{S}}_k|^2}, k \in \mathcal{S}_0
\label{eq:optim4_const2}
\\s_k&=1,s_j=0, k \in \mathcal{S},  j \in \mathcal{S}_0 \backslash \mathcal{S}.
\label{eq:optim4_end}
\end{align}
{ The constraint (\ref{eq:optim4_const1}) is obtained by combining (\ref{eq:optim3_const1}) and (\ref{eq:optim3_end}). This constraint indicates that the power allocation is controlled by the interference limit $I^0$ if $I^0/\epsilon_1 \leq P^0$, while it is controlled by the transmit power limit $P^0$ if $I^0/\epsilon_1 > P^0$.}

\section{Selection algorithms and power allocation schemes}
\label{sec:solution}
The optimization problem (\ref{eq:optim4_start})-(\ref{eq:optim4_end}) to compute power allocations and selection variables is a non-convex mixed integer program and an NP-hard problem. Note that the computation of power allocations and selection variables depend on ZF vectors $\mathbf{v}^\mathcal{S}_k$ which in turn depend on the selected users. In order to solve this chicken-and-egg problem, we choose a particular set $\mathcal{S}$ and obtain ZF vectors and power allocation for that set. For a given selected set $\mathcal{S}$, the problem (\ref{eq:optim4_start})-(\ref{eq:optim4_end}) reduces to the following feasibility problem with power allocations as variables:
\begin{align}
	\text{Find}~~& P_k
	\label{eq:optim5_start}
	\\ \text{Subject to}&: \sum_{k\in \mathcal{S}} P_k  \leq \min \left( {I^0/\epsilon_1}, {P^0} \right),
	\label{eq:optim5_const1}
	\\ P_k &\geq  \frac{(2^{R^0_k} -1)\left(\sigma^2_w +  {I}_k + \epsilon_2\right)}{|\mathbf{\hat{h}}_k^H \mathbf{v}^{\mathcal{S}}_k|^2}, k \in \mathcal{S}.
	\label{eq:optim5_end}
\end{align}
If the power allocation
\begin{align}
P^{\mathcal{S}}_k =  \frac{(2^{R^0_k} -1)\left(\sigma^2_w +  {I}_k + \epsilon_2\right)}{|\mathbf{  \hat{h}}^H_k \mathbf{v}^{\mathcal{S}}_k|^2}, k \in \mathcal{S},
\label{eq:P_k_vup}
\end{align}
satisfies the constraint in (\ref{eq:optim5_const1}), then it provides the solution to the power allocation problem for the set $\mathcal{S}$. Note that the above power allocation attempts to satisfy specific instantaneous rate of SU-$k$. Therefore, it is referred to as Qos-Aware-Power-allocation.

Let $K^*$ be the cardinality of optimal sets. The problem (\ref{eq:optim4_start})-(\ref{eq:optim4_end}) can have multiple optimal sets, since multiple sets of the $K^*$ can satisfy the constraints (\ref{eq:optim4_const1})-(\ref{eq:optim4_end}). One approach of obtaining one of the optimal sets is to consider all possible sets of cardinalities $K, K-1,K-2,...,K^*$ one-by-one in decreasing order of cardinality, compute ZF vectors and power allocations by (\ref{eq:zf_vectors}) and (\ref{eq:P_k_vup}), respectively, and check whether the constraints in (\ref{eq:optim5_const1}) are satisfied. Such approach of user selection is prohibitively complex and impractical since the number of sets to be considered increases exponentially with $K$. As an example, for $K=20$ and $K^*=5$, the minimum number of sets to be considered are $\sum_{K'=K^*+1}^{K}$ ${K}\choose{K'}$ $\approx 1$ million. 
Therefore, there is a need to design a low-complexity algorithm to select users and obtain power allocations. 

As our goal is to maximize the cardinality of the set $\mathcal{S}$, we propose an approach which considers only one set of a particular cardinality that is obtained by dropping the SU that requires maximum power in a higher cardinality set. The selection algorithm  is initialized by selecting all the SUs, i.e., $\mathcal{S}= \mathcal{S}_0$. ZF vectors $\mathbf{v}^{\mathcal{S}}_k$ and power allocations $P^{\mathcal{S}}_k$ are computed for the set $\mathcal{S} = \mathcal{S}_0$ using (\ref{eq:zf_vectors}) and (\ref{eq:P_k_vup}), respectively. Then, the condition $\sum_{k \in \mathcal{S}} P^{\mathcal{S}}_k \leq \min(I^0/\epsilon_1, P^0)$ is checked. If the condition is not satisfied, the SU with maximum power allocation is dropped from the set and a set $\mathcal{S} = \mathcal{S} \backslash \{j\}$ of lower cardinality is considered, where $j=\arg \max_{k \in \mathcal{S}} P^{\mathcal{S}}_k$. The ZF vectors and power allocations are re-computed for the new set using (\ref{eq:zf_vectors}) and (\ref{eq:P_k_vup}), respectively. This process is continued until the constraint $\sum_{k \in \mathcal{S}} P^{\mathcal{S}}_k \leq \min(I^0/\epsilon_1, P^0)$ is satisfied. Since the SU with maximum power allocation is dropped in each iteration, the algorithm is called Delete-su-with-Maximum-Power (DMP). Note that dropping of the SU that requires the maximum power causes maximum reduction in $\sum_{k}s_k P_k$ in constraint (\ref{eq:optim4_const1}). This increases the probability that SUs included in set $\mathcal{S}\backslash \{j\}$ will satisfy the constraint (\ref{eq:optim4_const1}). The algorithmic steps are summarized in Algorithm \ref{alg:DMP}. { It should be noted that the SUs which require excess power to satisfy their rate requirements will not be selected by the DMP. For example, if $P^{\mathcal{S}}_k =  \frac{(2^{R^0_k} -1)\left(\sigma^2_w +  {I}_k + \epsilon_2\right)}{|\mathbf{  \hat{h}}^H_k \mathbf{v}^{\mathcal{S}}_k|^2} > \min(I^0/\epsilon_2, P^0)$, then SU-$k$ will not be selected.}

\begin{algorithm}[t]
	\caption{SU Selection Algorithm: DMP}
	\label{alg:DMP}	
	\begin{algorithmic}[1]
		\Statex { Input: channel estimates $\hat{\mathbf{h}}_k, \hat{\mathbf{h}}_{l0}$, reverse interference $I_k$, margins $\epsilon_1, \epsilon_2$, rate requirements $R^0_k$.}
		\State Select all SUs, i.e., $\mathcal{S}=\mathcal{S}_0$ and $s_k=1, \forall k$.
		\State Compute ZF vectors $\mathbf{v}^{\mathcal{S}_0}_k$ and power allocations $P^{\mathcal{S}_0}_k$ using (\ref{eq:zf_vectors}) and (\ref{eq:P_k_vup}), respectively.
		\While{$\mathcal{S}\neq \emptyset $}
		\If{$\sum_{k=1}^{K}s_k P^{\mathcal{S}}_k > \min \left( {I^0/\epsilon_1}, {P^0} \right)$}
		\State Remove SU with max $P^{\mathcal{S}}_k$:
		\label{step:drop}
		\Statex~~~~~~~~~~~~~~a. $j = \arg \max_{k \in \mathcal{S}} P^{\mathcal{S}}_k$.
		\Statex~~~~~~~~~~~~~~b. $\mathcal{S} \leftarrow \mathcal{S} \backslash \{j\}$, $s_{j}=0$.
		\State Update vectors and power allocations:
		\label{step:update}
		\Statex~~~~~~~~~~~~~~a. Compute $\mathbf{v}^{\mathcal{S}}_k$ for set $\mathcal{S}$ using (\ref{eq:zf_vectors}).
		\Statex~~~~~~~~~~~~~~b. Compute $P^{\mathcal{S}}_k$ using (\ref{eq:P_k_vup}).
		\Else~Stop.
		\EndIf
		\EndWhile
		\Statex { Output: set of selected SUs $\mathcal{S}_1^*=\mathcal{S}$, power allocations $P^{\mathcal{S}_1^*}_k = P^{\mathcal{S}}_k$.}
	\end{algorithmic}
\end{algorithm}

The selected set is denoted by $\mathcal{S}_1^*$ and the cardinality of the selected set is $K^*_1 = |\mathcal{S}^*_1|$. Due to imperfect CSI, all the selected SUs may not achieve the rate $R^0_k$. Therefore, we quantify the performance of the algorithm by $K^{**}_1 (<K^*_1)$ which is the number of SUs that achieve rate higher than $R^0_k$:
\begin{align}
K_1^{**} = \sum_{k \in \mathcal{S}^{*}_1} \mathbbm{1} (R^{\mathcal{S}^*_1}_k \geq R^0_k),
\end{align}
where $\mathbbm{1}(.)$ is the indicator function.

We also consider a low-complexity version of this algorithm where step \ref{step:update} of vector and power allocation update in Algorithm \ref{alg:DMP} is omitted. The set selected with this modified version without the vector is denoted by $\mathcal{S}_2^*$ and its cardinality by $K^{*}_2 = |\mathcal{S}_2^*|$. Further,  $K^{**}_2=\sum_{k \in \mathcal{S}^{*}_2} \mathbbm{1} (R^{\mathcal{S}^*_2}_k \geq R^0_k),$ is the number of SUs exceeding the required rate in DMP without vector update.

\begin{algorithm}[t]
	\caption{SU Selection Algorithm: MDML}
	\label{alg:MDML}	
	\begin{algorithmic}[1]
		\Statex { Input: channel estimates $\hat{\mathbf{h}}_k, \hat{\mathbf{h}}_{l0}$, reverse interference $I_k$, margins $\epsilon_1, \epsilon_2$.}
		\State Select all SUs, i.e., $\mathcal{S}=\mathcal{S}_0$ and $s_k=1, \forall k$.
		\State Compute $\mathbf{v}^{\mathcal{S}_0}_k, \lambda^{\mathcal{S}_0}_k, P^{\mathcal{S}_0}_k, \hat{R}(\mathcal{S}_0)$ using (\ref{eq:zf_vectors}), (\ref{eq:lambda_k}), (\ref{eq:P_k_wf}), and (\ref{eq:sum_rate}), respectively.
		\While{$\mathcal{S}\neq \emptyset$}	
		\State Delete SU with minimum lambda:
		\Statex~~~~~~~~~ $j = \arg \min_{k \in \mathcal{S}} \lambda^{\mathcal{S}}_k$.
		\Statex~~~~~~~~~ $\mathcal{S'} \leftarrow \mathcal{S}\backslash \{j\}$.
		\State Compute $\mathbf{v^{\mathcal{S'}}_k}, \lambda^{\mathcal{S'}}_k, P^{\mathcal{S'}}_k, \hat{R}(\mathcal{S'})$ using (\ref{eq:zf_vectors}), (\ref{eq:lambda_k}), (\ref{eq:P_k_wf}), and (\ref{eq:sum_rate}), respectively.
		\If{$\hat{R}(\mathcal{S'}) > \hat{R}(\mathcal{S})$}
		\Statex~~~~~~~~ $\mathcal{S} \leftarrow \mathcal{S'}, s_j =0$.
		\Else~ Stop				
		\EndIf
		\EndWhile
		\Statex { Output: set of selected SUs $\mathcal{S}_M^*=\mathcal{S}$, power allocations $P^{\mathcal{S}_M^*}_k = P^{\mathcal{S}}_k$.}
		
	\end{algorithmic}
\end{algorithm}
\subsubsection{MDML selection algorithm}
\label{sec:DML}
We extend the Delete-Minimum-Lambda (DML) selection scheme presented in \cite{huang2012} to underlay CR network with imperfect CSI. The DML algorithm in \cite{huang2012} selects users while maximizing the sum-rate of selected users. This approach does not take into account the rate constraints $R^{0}_k$ of SUs. It also does not take into account the imperfect CSI. Since DML was proposed for a primary massive MIMO network (CR network was not considered), it also does not include the reverse interference received at SU from primary transmitters. We modify the algorithm to include reverse interference and margin parameters for robustness against imperfect CSI. The Modified-DML (MDML) is described below.

In MDML, the equivalent channel gain between SU-$k, k \in \mathcal{S}$ and SBS is defined as:
\begin{align}
\lambda^{\mathcal{S}}_k = \frac{|\mathbf{  \hat{h}}_k^H \mathbf{v}^{\mathcal{S}}_k|^2}{\sigma^2_w +  {I}_k + \epsilon_2}, k \in \mathcal{S}.
\label{eq:lambda_k}
\end{align}
The power allocation for the SU included in the set $\mathcal{S}$ is obtained by water-filling. In order to satisfy the condition (\ref{eq:optim4_const1}), the maximum power level of $\min \left( {I^0/\epsilon_1}, {P^0} \right)$ is used to compute the power allocation by water-filling as shown below
\begin{align}
P^{\mathcal{S}}_k =  p^{\mathcal{S}}_k/ \lambda^{\mathcal{S}}_k,~
 p^{\mathcal{S}}_k = (\mu \lambda^{\mathcal{S}}_k -1)^{+},  k \in \mathcal{S},
\label{eq:P_k_wf}
\end{align}
where $(x)^+ = \max \{x,0\}$, and $\mu$ is the water level satisfying
\begin{align}
\sum_{k \in \mathcal{S}} \left(\mu - \frac{1}{\lambda^{\mathcal{S}}_k}\right) = \min \left( \frac{I^0}{\epsilon_1}, {P^0} \right).
\label{eq:water_level}
\end{align}
Further, the estimated sum rate for SUs included in $\mathcal{S}$ can be written as
\begin{align}
\hat{R}(\mathcal{S}) = \sum_{k \in \mathcal{S}} \log_2 \left(1+ P^{\mathcal{S}}_k \lambda^{\mathcal{S}}_k\right)
\label{eq:sum_rate}.
\end{align}
The MDML algorithm drops the SU with minimum $\lambda^{\mathcal{S}}_k$, if dropping the SU results in increased sum-rate. The algorithmic steps are summarized in Algorithm \ref{alg:MDML}. The selected set of SUs under this algorithm is denoted by $\mathcal{S}^*_M$.

\section{Analysis of DMP algorithm}
\label{sec:analysis_selection}
In this section, we provide analysis to compute $\mathbb{E}[K^{*}_1]$ and $\mathbb{E}[K^{**}_1]$ under fading channels $\mathbf{h}_k, {h_{lk}}$ and $\mathbf{h}_{l0}$. We also analyze the average interference to primary receivers. Since the coefficients $\beta_k, \beta_{lk}, \beta_{l0}$ change slowly over time, they are assumed to be constant in the analysis \cite{marzetta2010}.

\subsubsection{Average number of SUs served}
\label{sec:E_K_star}
The average number of SUs served using DMP is computed as follows:
\begin{align}
\mathbb{E}[K^{*}_1] = \sum \limits_{k=1}^{K} k \sum \limits_{\mathcal{S}\in \mathcal{S}_k } f(\mathcal{S})g(\mathcal{S}) ,
\end{align}
where $\mathcal{S}_k$ is the set of all sets of cardinality $k$, $f(\mathcal{S})$ is the probability of that the condition $\sum_{k\in S}P^{\mathcal{S}}_k \leq \min (I^0/\epsilon_1, P^0)$ is satisfied:
\begin{align}
f(\mathcal{S}) &= \Pr \left(\sum \limits_{k \in S} P^{\mathcal{S}}_k \leq \min \left(\frac{I^0}{\epsilon_1}, P^0\right) \right), 
\label{eq:f_S}
\end{align}
and $g(\mathcal{S})$ is the probability of arriving at set $\mathcal{S}$ during the algorithmic iterations. Since the set $\mathcal{S}_0$ is always considered, we have $g(\mathcal{S}_0)=1$ and $g(\mathcal{S}), \mathcal{S} \subset \mathcal{S}_0$ can be obtained using the following recursive expression:
\begin{align}	
g(\mathcal{S}) &= \sum_{j, j \notin \mathcal{S} } g(\mathcal{S} \cup \{j\}) P'\left(\{\mathcal{S} \cup \{j\}\} \backslash \{j\} \right),
\label{eq:g_S}
\end{align}
where $ P'\left(\{\mathcal{S} \cup \{j\}\} \backslash \{j\} \right) =  P'(\mathcal{S^+} \backslash \{j\})$ is the probability of dropping SU-$j$ from set $\mathcal{S^+} = \mathcal{S} \cup \{j\}$. This probability can be expressed as:
{\small \begin{align}
\nonumber P'(\mathcal{S}^+ \backslash \{j\} ) &=  (1-f(\mathcal{S}^+)) \Pr \left(P^{\mathcal{S}^+}_j > P^{\mathcal{S}^+}_1,..., P^{\mathcal{S}^+}_j > P^{\mathcal{S}^+}_{|\mathcal{S}^+|}\right),
\\&=(1-f(\mathcal{S}^+)) \int \limits_{0}^{\infty} p_{P^{\mathcal{S}^+}_j}(x)\prod\limits_{i \in \mathcal{S}^+, i\neq j}f_{P^{\mathcal{S}^+}_i}(x)dx,
\label{eq:P_dash_S}
\end{align}}
where $p_{P^{\mathcal{S}^+}_j}(x)$ and $f_{P^{\mathcal{S}^+}_i}(x)$ are the probability density function (pdf) of $P^{\mathcal{S}^+}_j$ and the cumulative distribution function (cdf) of $P^{\mathcal{S}^+}_i$, respectively. In order to evaluate (\ref{eq:f_S}), we need distributions of $P^\mathcal{S}_k$ which can be obtained from Theorem \ref{thm:pj_dist}. The distribution of $\sum_{k \in S}P^{\mathcal{S}_k}$ is required to evaluate (\ref{eq:P_dash_S}) which can be obtained from Corollary \ref{cor:sum_pj_dist}.

\begin{theorem}
	The power allocation $P^{\mathcal{S}}_k$ in (\ref{eq:P_k_vup}) is a Gamma random variable with shape and scale parameters $\kappa^p_k$ and $\theta^p_k$: $P^{\mathcal{S}}_k \sim \Gamma(\kappa^p_k, \gamma^{\mathcal{S}}_k \theta^p_k)$, where
	 \begin{align}
		\nonumber \kappa^p_k &=  \frac{\left(\sigma^2_w +\epsilon_2+ \sum_{l\in \mathcal{T}}P_p \beta_{lk}\right)^2}{ \sum_{l\in \mathcal{T}} (P_p\beta_{lk})^2},
		\\\theta^p_k &= \frac{\sum_{l\in \mathcal{T}}(P_p \beta_{lk})^2 }{\sigma^2_w +  \epsilon_2  +  \sum_{l\in \mathcal{T}} P_p\beta_{lk}}.
		\label{eq:thm1_1}
		\end{align}
	Similarly, for DMP without the vector update step, we have: $P^{\mathcal{S}_0}_k \sim \Gamma(\kappa^p_k, \gamma^{\mathcal{S}_0}_k \theta^p_k)$, where
	\begin{align}
		\nonumber \gamma^{\mathcal{S}}_k = \frac{2^{R^0_k}-1}{(\beta_k + \sigma^2_{\delta}) (M -|\mathcal{S}|-L+1)},
		\\ \gamma^{\mathcal{S}_0}_k = \frac{2^{R^0_k}-1}{(\beta_k + \sigma^2_{\delta}) (M -K-L+1)}.
		\label{eq:thm1_2}
		\end{align}
	\label{thm:pj_dist}
\end{theorem}
\begin{proof}		
	Appendix \ref{app_pj_dist}.	
\end{proof}


\begin{corollary}
	Sum of powers $\sum_{k \in \mathcal{S}} P^{\mathcal{S}}_k$ follows the Gamma distribution: $\sum_{k \in S} P^{\mathcal{S}}_k \sim \Gamma(\kappa_p, \theta_p)$, where
	\begin{align}
		 &\kappa_p = \frac{\left(\sum_{j \in \mathcal{S}} \kappa^p_j \gamma^{\mathcal{S}}_j \theta^p_j\right)^2}{\sum_{j \in \mathcal{S}} \kappa^p_j (\gamma^{\mathcal{S}}_j\theta^p_j)^2 },
		~\theta_p = \frac{\sum_{j \in \mathcal{S}} \kappa^p_j (\gamma^{\mathcal{S}}_j \theta^p_j)^2 }{\sum_{j \in \mathcal{S}} \kappa^p_j \gamma^{\mathcal{S}}_j \theta^p_j},
	\label{eq:k_P_theta_P}
		\end{align}
	\label{cor:sum_pj_dist}
\end{corollary}
\begin{proof}	
	Using Lemma 3 in \cite{hosseini2014}, the sum $\sum_{k \in \mathcal{S}} P^{\mathcal{S}}_k$ is modeled as a Gamma random variable with shape and scale parameters $\kappa_{p}$, and $\theta_{p}$, respectively, as defined in (\ref{eq:k_P_theta_P}). 
\end{proof}


\begin{corollary}
	Consider selection of two sets $\mathcal{S}_1$ and $\mathcal{S}_2$ containing SU-$k$. The power required to achieve rate $R^0_k$ at SU-$k$ with selection of $\mathcal{S}_1$ stochastically dominates the power required to achieve the same rate with the selection of $\mathcal{S}_2$, if $\mathcal{S}_2$ is subset of $\mathcal{S}_1$, i.e., $\Pr(P^{\mathcal{S}_1}_k \geq x)> \Pr(P^{\mathcal{S}_2}_k \geq x)$ for any $x$, if $\mathcal{S}_2 \subset \mathcal{S}_1$.
	\label{cor:dmpvup_P_k}
\end{corollary}
\begin{proof}
	Appendix \ref{app_cor_dmpvup_P_k}.
\end{proof}

\noindent\textit{Remark}: In DMP, if set $\mathcal{S}_1$ does not satisfy constraints in (\ref{eq:optim5_const1}), a subset of $\mathcal{S}_1$, say $\mathcal{S}_2$, is considered by dropping SU-$j$ that consumes the maximum power. Corollary \ref{cor:dmpvup_P_k} implies that the individual power requirements for all SUs still included in $\mathcal{S}_2$ reduce due to dropping of SU-$j$.

The expression for $\mathbb{E}[K^*_2]$ under DMP without vector update is obtained by evaluating $f(\mathcal{S})$ and $P'({S^+\backslash\{j\}})$ using distributions of $P^{\mathcal{S}_0}_k$ instead of $P^{\mathcal{S}}_k$.


\subsubsection{Average number of SUs exceeding the required rate}
\label{sec:E_K_starstar}
The average number of SUs achieving the minimum rate of $R^0_k$ using DMP can be expressed as follows:
\begin{align}
 \mathbb{E}[K^{**}_1] = \sum \limits_{k=1}^{K} \sum \limits_{\mathcal{S}: k \in \mathcal{S}} f(\mathcal{S}) g(\mathcal{S})  \Pr(R^{\mathcal{S}}_k \geq R^0_k),
\label{eq:E_K_starstar_1}
\end{align}
In order to compute the above expression, we need to compute the complementary cdf of $R^\mathcal{S}_k$ which is obtained by Theorem \ref{thm:R_k}.

\begin{theorem}
	The complementary cdf of the achieved rate at SU-k, if the set $\mathcal{S}$ is selected and the power is allocated by (\ref{eq:P_k_vup}), is given by:	
	\begin{align}
		\nonumber &\Pr(R^{\mathcal{S}}_k \geq y) =\\
		&\frac{1}{2\pi}\int \limits_{-\infty}^{\zeta_y} \int \limits_{-\infty}^{\infty}\left[(1-\theta^z_k jt)^{-\kappa^z_k} \prod \limits_{l \in \mathcal{T}} (1-\theta^z_{lk}jt)\right]~e^{-j2\pi wt}dt~dw,
		\label{eq:Pr_R_k}
		\end{align}
	where 
		\begin{align}
		\nonumber \zeta_y= C_y (\sigma^2_w + \epsilon_2) -\sigma^2_w , ~~
		C_y = \frac{\beta_k}{\beta_k + \sigma^2_\delta} \left(\frac{2^{R^0_k}-1}{2^y -1}\right),
		\end{align}
{		\small
	\begin{align}
	\nonumber \kappa^z_k&= \frac{\left(\sum \limits_{j \in \mathcal{S} \backslash \{k\}} {\gamma^{\mathcal{S}}_j \theta^p_j} {\Gamma(\kappa^p_j+1)}/{ \Gamma(\kappa^p_j)}\right)^2}{\sum \limits_{j \in \mathcal{S} \backslash \{k\}} \left({\gamma^{\mathcal{S}}_j \theta^p_j}/ {\Gamma(\kappa^p_j)}\right)^2 \left({2\Gamma(\kappa^p_j+2)\Gamma(\kappa^p_j) - \Gamma^2(\kappa^p_j+1)}\right)}, 
	\\\nonumber \theta^z_k&= \frac{\sigma^2_{\delta}}{\kappa^z_k} \sum \limits_{j \in \mathcal{S} \backslash \{k\}} \gamma^{\mathcal{S}}_j \theta^p_j \frac{{\Gamma(\kappa^p_j+1)} }{ { \Gamma(\kappa^p_j)}}, \text{and}
	\\\theta^z_{lk}&= (1-C_y)P_p\beta_{lk}
	\label{eq:kz_thetaz}
	\end{align}}
	\label{thm:R_k}
\end{theorem}
\begin{proof}	
	Appendix \ref{app_R_k}.
\end{proof}	
\noindent\textit{Remark}: If $\sigma^2_\delta= 0$, then we get $\Pr(R^\mathcal{S}_k \geq R^0_k)=1$. The proof is provided in Appendix \ref{app_R_k}.

The expression $\mathbb{E}[K^{**}_2]$ under DMP without vector update is obtained using the same expression as in the RHS of (\ref{eq:E_K_starstar_1}) by replacing  $\gamma^{\mathcal{S}}_j$ with $\gamma^{\mathcal{S}_0}_j$ in Theorems \ref{thm:pj_dist} and \ref{thm:R_k}.

\subsubsection{Average interference at PR-$l$}
\label{sec:I_l}
The expected value of the interference is computed as follows:
\begin{align}
\mathbb{E}[I_l] = \sum_{\mathcal{S}} g(\mathcal{S}) f(\mathcal{S})\sum_{k \in \mathcal{S}} \mathbb{E}\left[I_{kl}\right],~~ l \in \mathcal{R},
\label{eq:E_I_l}
\end{align}
where $\mathbb{E}\left[I_{kl}\right] = \mathbb{E}\left[P^{\mathcal{S}}_k |\mathbf{h}^H_{l0}\mathbf{v}^{\mathcal{S}}_k|^2\right] = \mathbb{E}\left[P^{\mathcal{S}}_k |\mathbf{\Delta}^H_{l0}\mathbf{v}^{\mathcal{S}}_k|^2\right]$. The second equality follows from $\mathbf{\hat{h}}^H_{l0} \mathbf{v}^{\mathcal{S}}_k =0$ due to zero forcing beamforming. The expression for $\mathbb{E}\left[I_{kl}\right]$ can be written using the distributions of $P^{\mathcal{S}}_k$ and $\mathbf{\Delta}_{l0}$ as follows:
\begin{align}
\mathbb{E}\left[I_{kl}\right] =\mathbb{E}\left[P^{\mathcal{S}}_k |\mathbf{\Delta}^H_{l0}\mathbf{v}^{\mathcal{S}}_k|^2\right] = \gamma^{\mathcal{S}}_k \theta^p_k \sigma^2_{\Delta} \frac{\Gamma(\kappa^p_k+1)}{\Gamma(\kappa^p_k)}.
\label{eq:I_kl_1}
\end{align}
The proof is shown in Appendix \ref{app_I_kl}. Similarly, the expression for $\mathbb{E}[I_l]$ in DMP without vector update is obtained by substituting the following in (\ref{eq:E_I_l}):
\begin{align}
\mathbb{E}\left[I_{kl}\right] = \gamma^{\mathcal{S}_0}_k \theta^p_k \sigma^2_{\Delta} \frac{\Gamma(\kappa^p_k+1)}{\Gamma(\kappa^p_k)}.
\label{eq:I_kl_2}
\end{align}
From (\ref{eq:E_I_l}), (\ref{eq:I_kl_1}), and (\ref{eq:I_kl_2}), we can see that the average interference to PRs is 0 for $\sigma^2_\Delta = 0$.

\subsection{{Optimality of DMP}}
\label{sec:optimality}
As described in Section \ref{sec:solution}, the power allocated to SU-$k$, $P^{\mathcal{S}_0}_k$, remains constant during the algorithmic iterations of DMP without vector update. Therefore, DMP without vector update effectively obtains the solution for the following problem:
\begin{align}
(\mathbf{P2})~~\max_{\{s_k\}} &\sum_{k=1}^{K} s_k
	\label{eq:optim6_start}
	\\\text{Subject to}&:\sum_{k=1}^{K} s_k P^{\mathcal{S}_0}_k  \leq \min \left( {I^0/\epsilon_1}, {P^0} \right),
	\label{eq:optim6_const1}
	\\s_k&=1,s_j=0, k \in \mathcal{S},  j \in \mathcal{S}_0 \backslash \mathcal{S}.
	\label{eq:optim6_end}
\end{align}
The solution obtained by DMP without vector update can be written as follows:
\begin{align}
	\mathcal{S}^*_2 = \arg \max_{\mathcal{S}: \sum_{k\in \mathcal{S}} P^{\mathcal{S}_0}_k  \leq \min \left( {I^0/\epsilon_1}, {P^0} \right)} |\mathcal{S}|.
\end{align}
The solution $\mathcal{S}^*_2$ is the optimal solution for $\mathbf{P2}$, since no set of higher cardinality can satisfy the constraint (\ref{eq:optim6_const1}) for fixed $P^{\mathcal{S}_0}_k$. This is because the proposed algorithm drops the SU with the highest power allocation in each iteration until the constraint (\ref{eq:optim6_const1}) is satisfied and addition of any SU to the set $\mathcal{S}^*_2$ will violate the constraint. 

Further, the optimal solutions of problems $\mathbf{P1}$ and $\mathbf{P2}$ differ due to the difference in power allocations $P^{\mathcal{S}^*}_k$ and $P^{\mathcal{S}_0}_k$, where $\mathcal{S}^*$ indicates the optimal solution set for $\mathbf{P1}$. The power allocations differ due to the difference in the number of nulls in the ZF vectors $\mathbf{v}^{\mathcal{S}^*}_k$ and $\mathbf{v}^{\mathcal{S}_0}_k$, which are denoted by $M-|\mathcal{S}^*|-L+1$ and $M-K-L+1$, respectively. The difference in the power allocations $P^{\mathcal{S}^*}_k$ and $P^{\mathcal{S}_0}_k$ becomes negligible if $M \gg K+L$. From Theorem \ref{thm:pj_dist}, we can also observe that the distribution of $P^{\mathcal{S}^*}_k$ approaches that of $P^{\mathcal{S}_0}_k$ as  $\gamma^{\mathcal{S}^*}_k \rightarrow \gamma^{\mathcal{S}_0}_k$, which occurs if $M \gg K+L$. Therefore, we can conclude that the problem $\mathbf{P1}$ becomes equivalent to $\mathbf{P2}$ for $M \gg K+L$ and  $|\mathcal{S}^*_2|$ approaches the optimal solution $|\mathcal{S}^*|$.

Finally, the number of SUs selected by the DMP algorithm is no less than the number of SUs selected by DMP without vector update, i.e, $|\mathcal{S}_2^*|\leq |\mathcal{S}_1^*|\leq |\mathcal{S}^*|$. This is because the power allocated to each SU in DMP is less that or equal to that in DMP without vector update, i.e., $P^{\mathcal{S}}_k \leq P^{\mathcal{S}_0}_k$ when a set $\mathcal{S}$ is selected. This can also be observed from Corollary \ref{cor:dmpvup_P_k}. The condition for selection of a set $\mathcal{S}$ under DMP {\small$\left(\sum \limits_{k \in \mathcal{S}} P^{\mathcal{S}}_k \leq \min \left(\frac{I^0}{\epsilon_1}, P^0\right)\right)$}  is always satisfied if the condition under DMP without vector update is satisfied {\small$\left(\sum \limits_{k \in \mathcal{S}} P^{\mathcal{S}_0}_k \leq \min \left(\frac{I^0}{\epsilon_1}, P^0\right)\right)$}, while the converse is not true. Therefore, we get $|\mathcal{S}_2^*|\leq |\mathcal{S}_1^*|\leq |\mathcal{S}^*|$. This phenomenon can be intuitively explained as follows. When the set $\mathcal{S}_0$ is selected initially, the number of degrees of freedom in the beamforming, after adding $L+K-1$ nulls, is $M-K-L+1$. The number of degrees of freedom increments if the ZF vectors are updated after dropping an SU. Therefore, the power requirements of SUs which are not dropped reduce as the ZF vectors $\mathbf{v}_k$ are better aligned with channels $\mathbf{\hat{h}}_k$. The reduction in power requirements implies that more SUs can be kept in the downlink, while satisfying the constraints of the problem $\mathbf{P1}$. Therefore, the number of SUs selected by the DMP is no less than the number of SUs selected by DMP without vector update.

\subsection{Selection of algorithm parameters}
\label{sec:selection_of_parameters}
The optimization framework $\mathbf{P1}$ in (\ref{eq:optim4_start})-(\ref{eq:optim4_end}) involves various parameters. In this section, we provide discussion on the selection of  these parameters. The parameters can be broadly classified into two categories: 1) network dependent fixed parameters: $P^0, I^0, R^0_k$, and 2) proposed margin parameters: $\epsilon_1, \epsilon_2$. 

\subsubsection{Network dependent parameters}
\label{sec: network_dependent_parameters}
The network dependent parameters are decided by the operators of secondary and primary networks. Consider, for example, that primary and secondary networks coexist in 3.5GHz band as CBRS users where primary system is Priority Access License (PAL) user and secondary system is General Access Authorization (GAA) user \cite{fcc_15_47A1, ye2016b}. The value of $P^0$ will be determined using the power amplifier used at the SBS. Typical value of $P^0=40$dBm is used for BS under sub-6GHz bands. The rate constraints $R^0_k$ are determined by the operator of the secondary network depending on the QoS requirements for the SUs. The interference constraint $I^0$ is determined by the operator of primary network. The value of $I^0$ determines the SINR degradation of PUs due to the coexisting SUs. For example, if SINR degradation of $<1$dB is desired then $I^0$ should be set such that $I^0/\sigma^2_w< -6$dB, where $\sigma^2_w$ is the noise power at the PU.

\subsubsection{Proposed margin parameters}
\label{sec: proposed_margin_parameters}
Once the values of $P^0, I^0$ and $R^0_k$ are fixed, the algorithm specific margin parameters $\epsilon_1 $ and $\epsilon_2$ are set as described next. Margin parameters $\epsilon_1$ and $\epsilon_2$ are used in the proposed optimization framework in order to compensate for interference to PRs and inter-SU interference, respectively, resulting form imperfect CSI estimates. 

In order to select appropriate value of $\epsilon_1$, let us consider average value of the true interference $I_l$ at PR-$l$ for given channel estimates $\mathbf{\hat{h}}_{l0}$ and a selected set $\mathcal{S}$:
\begin{align}
\nonumber \mathbb{E}[I_l | \mathbf{\hat{h}}_{l0}, \mathcal{S}] &= \sum_{k \in \mathcal{S}} P^\mathcal{S}_k \mathbb{E}[|\mathbf{h}^H_{l0} \mathbf{v}^{\mathcal{S}}_k|^2  | \mathbf{\hat{h}}_{l0}, \mathcal{S}], l \in \mathcal{R}, k \in \mathcal{S},\\
 &= \sum_{k \in \mathcal{S}} P^\mathcal{S}_k \sigma^2_\Delta.
\end{align}
The last equality in the above equation follows from the fact that beamforming vectors $\mathbf{v}^\mathcal{S}_k$ are unit vectors that are in nullspace of estimated channels $\mathbf{\hat{h}}^H_{l0}$. Further, the proposed optimization problem ensures that $\sum_{k \in S} P^\mathcal{S}_k \epsilon_1 \leq I^0$ due to constraint in (\ref{eq:optim3_const1}). Therefore, the average interference $\mathbb{E}[I_l | \mathbf{\hat{h}}_{l0}, \mathcal{S}]$ is below the threshold $I^0$ if $\epsilon_1 \geq \sigma^2_\Delta$.

Now, let us consider the selection of $\epsilon_2$. From (\ref{eq:R_k_S}) and (\ref{eq:R_k_S_est}), we can see that the variable $\epsilon_2$ serves as a placeholder for inter-SU interference $\sum_{j \in \mathcal{S}, j\neq k}I_{jk}$. The variable $\epsilon_2$ ensures that higher power is allocated to SU-$k$ to compensate for the inter-SU interference as seen from (\ref{eq:P_k_vup}). Therefore, the value of $\epsilon_2$ should be selected such that $\epsilon_2 \geq \sum_{j, j\neq k}I_{jk}$. However, the instantaneous value of inter-SU interference is unknown. We propose to set the parameter value such that it exceeds the expected value of inter-SU interference for given channel estimates $\mathbf{\hat{h}}_{j}, \mathbf{\hat{h}}_{k}$ and a selected set $\mathcal{S}$:
\begin{align}
\nonumber \epsilon_2 &\geq \mathbb{E}\left[\sum_{j \in \mathcal{S}, j\neq k}I_{jk} \bigg | \mathbf{\hat{h}}_{j},\mathbf{\hat{h}}_{k}, \mathcal{S}\right],\\
\nonumber & = \mathbb{E}\left[\sum_{j \in {\mathcal{S}}, j\neq k}P^{\mathcal{S}}_j |\mathbf{{h}}^H_k \mathbf{v}^{\mathcal{S}}_j|^2  \bigg | \mathbf{\hat{h}}_{j},\mathbf{\hat{h}}_{k}, \mathcal{S}\right], k,j \in{\mathcal{S}}, k\neq j,\\
& \stackrel{(a)}{=} \sum_{j \in {\mathcal{S}}, j\neq k}P^{\mathcal{S}}_j \mathbb{E}[|\delta_k^H \mathbf{v}^{\mathcal{S}}_j |^2],\\
& \stackrel{(b)}{=} \sum_{j \in {\mathcal{S}}, j\neq k}P^{\mathcal{S}}_j \sigma^2_\delta.
\label{eq:epsilon_2_final}
\end{align}
The equality $(a)$ results due to the fact that $\mathbf{\hat{h}}_k \mathbf{v}^{\mathcal{S}}_j = 0$. The equality $(b)$ results from $\delta_k \sim CN(0,\sigma^2_\delta \mathbf{I})$ and $||\mathbf{v}^{\mathcal{S}}_k||=1$. Thus, $\epsilon_2$ should be greater than $\sum_{j \in {\mathcal{S}}, j\neq k}P^{\mathcal{S}}_j \sigma^2_\delta$. However, the power allocations $P^{\mathcal{S}}_j$ are not known in advance. Therefore, we set $\epsilon_2\geq P^0 \sigma^2_\delta$, which ensures that the condition in (\ref{eq:epsilon_2_final}) is satisfied since $P^0 \sigma^2_\delta \geq \sum_{j \in {\mathcal{S}}, j\neq k}P^{\mathcal{S}}_j \sigma^2_\delta$.

\subsubsection{Optimum margins $\epsilon_1, \epsilon_2$}
\label{sec: optimum_margins}
We observe that larger $\epsilon_1$ in (\ref{eq:optim5_const1}) results in admitting fewer SUs in the downlink. Similarly, larger $\epsilon_2$ results in larger power allocation according to (\ref{eq:P_k_vup}), further resulting in dropping of SUs due to the constraint $\sum_{k\in \mathcal{S}} P_k \leq \min(I^0/\epsilon_1, P^0)$ in (\ref{eq:optim5_const1}). In order to satisfy the rate and interference constraints while admitting maximum number of SUs in the downlink, it is necessary to set $\epsilon_1$ and $\epsilon_2$ to the smallest possible values. Therefore, the setting $\epsilon_1=\sigma^2_\Delta$ and $\epsilon_2 = P^0 \sigma^2_\delta$ results in serving maximum number of SUs with given interference and rate constraints.

\subsection{{Complexity analysis}}
\label{sec:complexity_c2}
Computational complexity of DMP as well as MDML is dominated by the computation of ZF vectors. For a set $\mathcal{S}$, the complexity of obtaining the ZF vectors is $\mathcal{O}\left(M (|\mathcal{S}| +L)^3\right)$ \cite{huang2012}. Since ZF vectors are updated in each iteration of DMP and MDML until a feasible set is reached, the worst case complexity of both algorithms is $\mathcal{O}\left(M K (K+L)^3 \right)$, while the worst case complexity of DMP without vector update is $\mathcal{O}\left(M (K+L)^3\right)$. 

As shown in the previous section, the solution $|\mathcal{S}^*_2|$ obtained by DMP without vector update approaches the optimal value with large $M$. Therefore, we can conclude that near-optimal number of SUs can be selected by the proposed algorithm while reducing the computational complexity by a factor of $K$ as compared to MDML.

\section{Simulation results}
\label{sec:results_c2}
In the results shown below, the noise power $\sigma^2_w$ is assumed to be $-100$dBm, the transmitted power from primary transmitters is $P_p=20$dBm, the transmit power limit is $P^0=40$dBm. The variance of error is modeled assuming reciprocal channels in a time-division duplex system as $\sigma^2_\Delta = \sigma^2_w/P_p$ and $\sigma^2_\delta = \sigma^2_w/P^0$ \cite{aquilina2015,razavi2014,maurer2011}. 
We consider uniformly distributed SUs and primary transmitters and receivers in a circular cell of radius $2$km with the SBS at the center. The minimum distance between the SBS and SUs is $100$m \cite{ngo2013, ngo2013a}. For each realization of locations, the slow fading coefficients between two nodes are computed as $\beta = \rho d^{-3.8}$, where $d$ is the distance between the two nodes and $\rho$ is a log-normal shadowing variable with standard deviation $\sigma_{s} = 8$dB. The margin parameters are set as $\epsilon_1 =  \sigma^2_\Delta$ and $\epsilon_2 = P^0 \sigma^2_\delta$. We simulate the algorithms for 1000 realizations of the channel for each realization of locations. Analytical and simulation results are averaged over 1000 realizations of the locations.

\begin{figure}
	\centering
	\includegraphics[width=0.7\columnwidth]{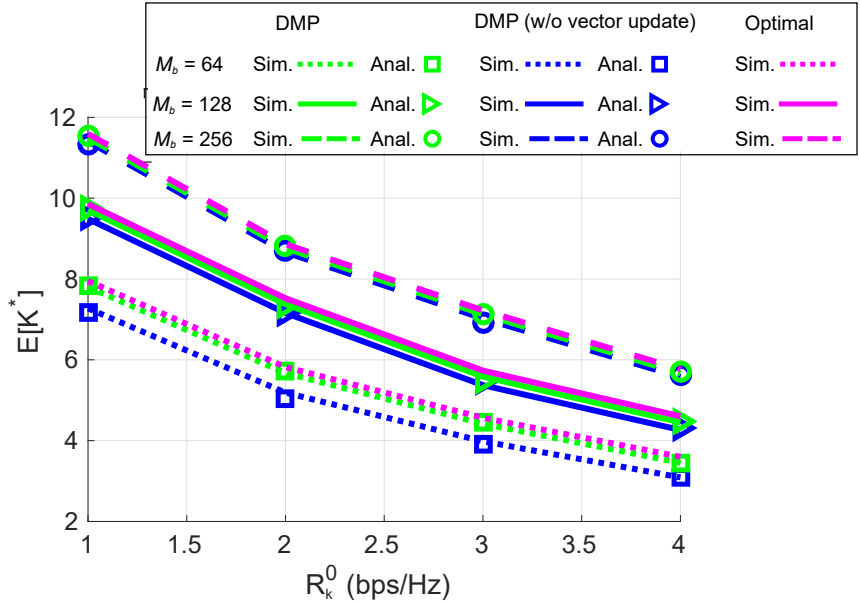}
	\caption{\small Comparison of number of SUs selected by DMP and optimal selection. $L=4$, $K=20$, $I^0 = -106$ dBm.}
	\label{fig:opt}
\end{figure}

\textit{{Comparison with optimal solution:}}
The comparison of the average number of SUs selected by DMP and optimal selection is shown in Fig. \ref{fig:opt}. The optimal solution is obtained by considering all possible sets of cardinalities $K, K-1,K-2,...,K^*$ one-by-one in decreasing order of cardinality, computing ZF vectors and power allocations by (\ref{eq:P_k_vup}), until the constraints in (\ref{eq:optim5_const1}) are satisfied. We observe that the number of SUs selected by DMP is very similar to that by optimal selection and the difference between $E[|\mathcal{S}^*_1|] \approx E[|\mathcal{S}^*|]$. As the number of antennas increased from 64 to 256, the difference between the performance of DMP without vector update and optimal selection becomes negligible as explained in Section \ref{sec:optimality}.

\begin{figure}[t!]
	\centering	
	\begin{subfigure}[b]{\columnwidth}
		\centering
		\includegraphics[width=0.7\columnwidth]{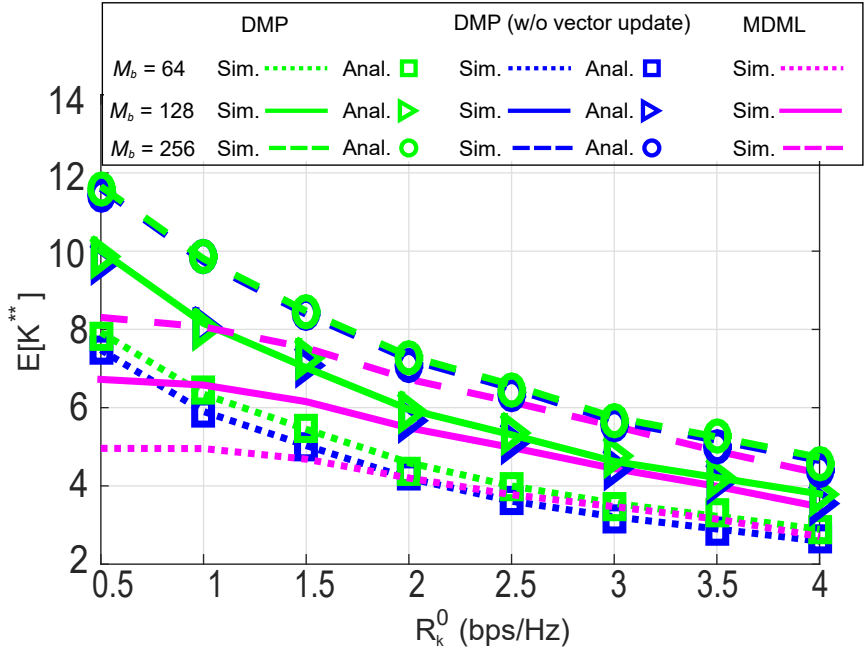}
	\end{subfigure}		
	\caption{\small Impact of rate constraints. $L=4$, $K=20$, $I^0 = -106$ dBm.}
	\label{fig:diff_R0}
\end{figure}

\textit{Impact of $R^0_k$:} The proposed DMP algorithm is designed to satisfy the minimum rate required by the SUs unlike MDML which does not take into account the rate requirements. Therefore, it can be observed that the DMP serves more SUs exceeding the minimum required rate than MDML when $R^0_k$ is uniformly distributed in the range $(0,4]$ as seen in Fig. \ref{fig:diff_R0}. The performance of the DMP and MDML becomes similar as the rate requirements increase to $4$bps/Hz. Further, it can be observed that the performance curves of the DMP and the MDML converge at a higher rate for large number of antennas. This indicates that the performance gain obtained by the DMP over MDML increases for a given rate requirement as the number of antennas increase.


\begin{figure}[t]
	\centering	
	\begin{subfigure}[b]{\columnwidth}
		\centering
		\includegraphics[width=0.7 \columnwidth]{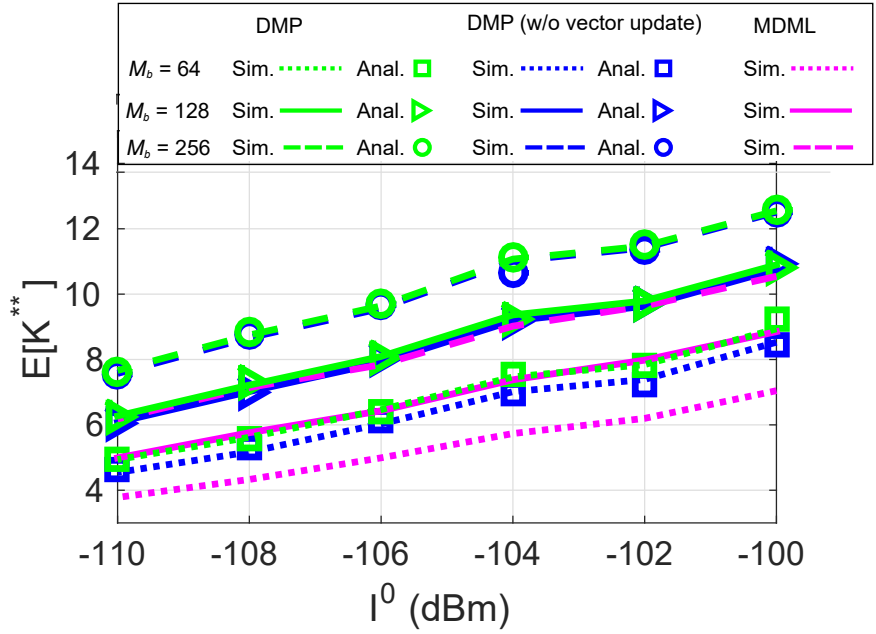}
	\end{subfigure}		
	\caption{\small Impact of interference constraints. $L=4$, $K=20$. $R^0_k=1$ bps/Hz.}
	\label{fig:diff_I0}
		\vspace{-4mm}
\end{figure}

\textit{Impact of $I^0$:} The interference threshold $I^0$ limits the total transmitted power below $I^0/\epsilon_1$, thereby limiting the number of SUs served in both DMP and MDML. It should be noted that the interference of $-100$, $-106$ and $-110$ dBm results in SINR loss of $3, 0.97$ and $0.41$dB, respectively at primary receivers. As shown in  Fig. \ref{fig:diff_I0}, the number of SUs served by the three algorithms increases by 1.5 times with increased interference threshold from $-110$dBm to $-100$dBm at the cost of reduced { signal-to-interference-plus-noise ratio (SINR)} at primary receivers. It can be observed that the performance gain obtained by the DMP over MDML is consistent for different interference thresholds. 

\begin{figure}[t!]
	\centering	
	\begin{subfigure}[b]{\columnwidth}
		\centering
		\includegraphics[width=0.7 \columnwidth]{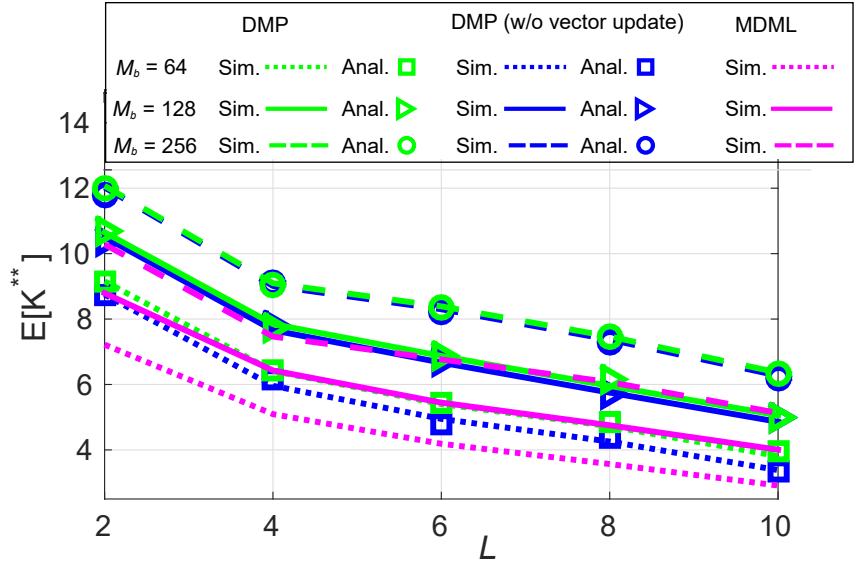}
		\label{fig:diff_Lp_E_k_starstar}
	\end{subfigure}		
	\caption{\small Impact number of primary pairs. $K=20$,$I^0 = -106$ dBm. $R^0_k=1$ bps/Hz.}
	\label{fig:diff_Lp}
\end{figure}

\textit{Impact of number of primary tx-rx pairs $L$:} Increased number of PTs in the network increases the reverse interference $I_k$ to SUs. This results in increased power requirement $P_k$ for the SU-$k$ to satisfy the rate according to (\ref{eq:P_k_vup}). This increased power requirement in turn results in dropping of more SUs in step 6 of the DMP algorithm. Therefore, the number of SUs served by the proposed algorithm reduces with higher $L$ as shown in Fig. \ref{fig:diff_Lp}.

\begin{figure}[t!]
	\centering	
	\begin{subfigure}[b]{\columnwidth}
		\centering
		\includegraphics[width=0.7\columnwidth]{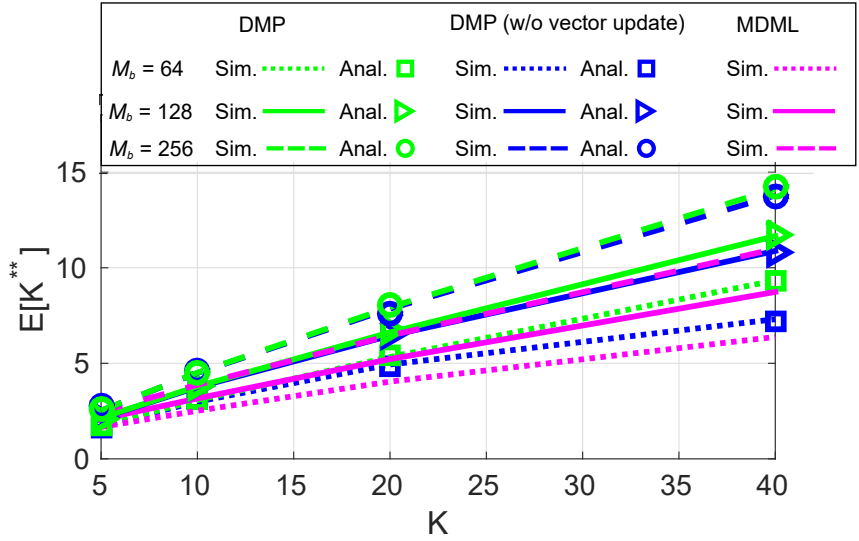}
		\label{fig:diff_K_unif_R0_E_k_starstar}
	\end{subfigure}		
	\caption{\small Impact of total number of SUs $K$ in the network. $L=4$, $I^0 = -106$ dBm.  $R^0_k$ is uniformly distributed in (0,4] bps/Hz.}
	\label{fig:diff_K_unif_R0}
\end{figure}

\textit{Impact of total number of SUs:}
The impact of increasing number SUs is shown Fig. \ref{fig:diff_K_unif_R0}. The number of SUs exceeding the required rate increases almost linearly for $M=128$ and $M=256$ under DMP when the rate requirements $R^0_k$ are uniformly distributed in the range $ (0,4]$ bps/Hz. 
We also observe that the difference in the performance of the DMP with and without vector update reduces with increased number of antennas. This is due to the fact that the ratio $\gamma^{\mathcal{S}_0}/\gamma^{\mathcal{S}^*_1}$ is close to one which results in similar power allocations for SUs with and without vector update, thereby resulting in similar number of SUs being dropped in the step \ref{step:drop} of the DMP algorithm.

\begin{figure}
	\centering	
	\begin{subfigure}[b]{\columnwidth}
		\centering
		\includegraphics[width=0.6\columnwidth]{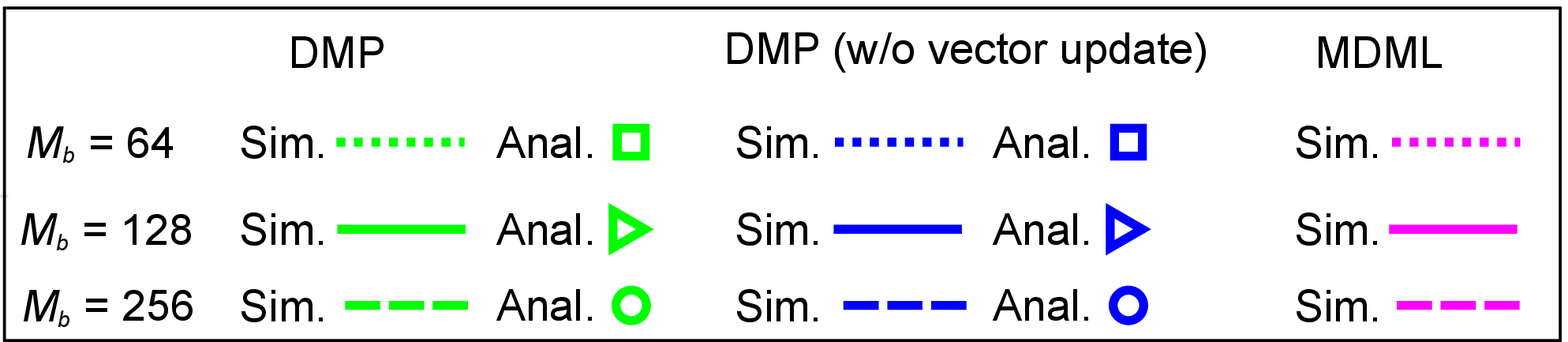}
		\label{fig:legend_diff_eta_ep_unif_R0}
	\end{subfigure}			
	\\
	\begin{subfigure}[b]{\columnwidth}
		\centering
		\includegraphics[width=0.6\columnwidth]{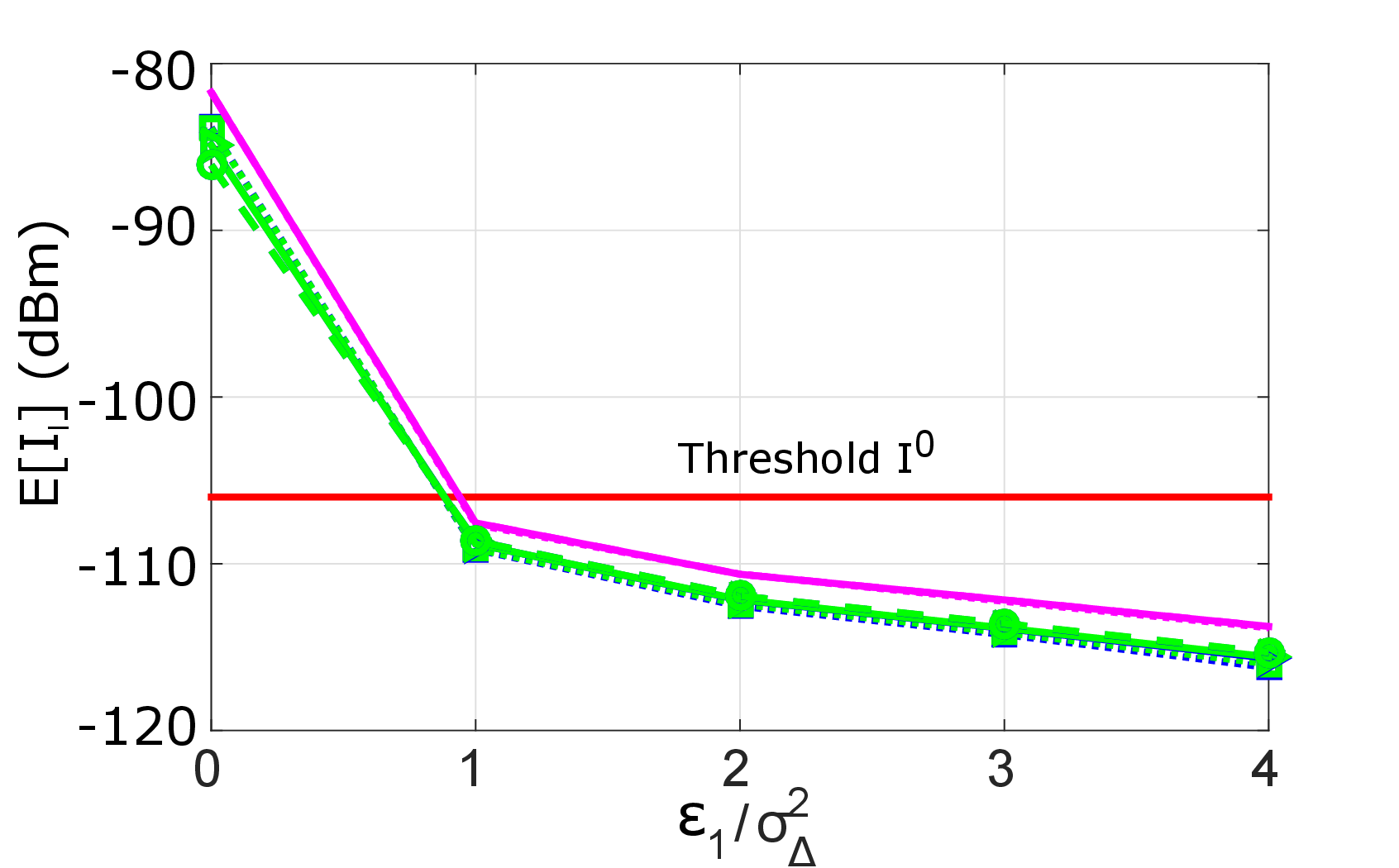}
		\caption{\small Avg. interference to primary receivers. Results hold for $\frac{\epsilon_2}{P^0 \sigma^2_\delta} \in  [0,4]$.}
		\label{fig:diff_eta_ep_unif_R0_avg_int}
	\end{subfigure}	
	\begin{subfigure}[b]{\columnwidth}
		\centering
		\includegraphics[width=0.6\columnwidth]{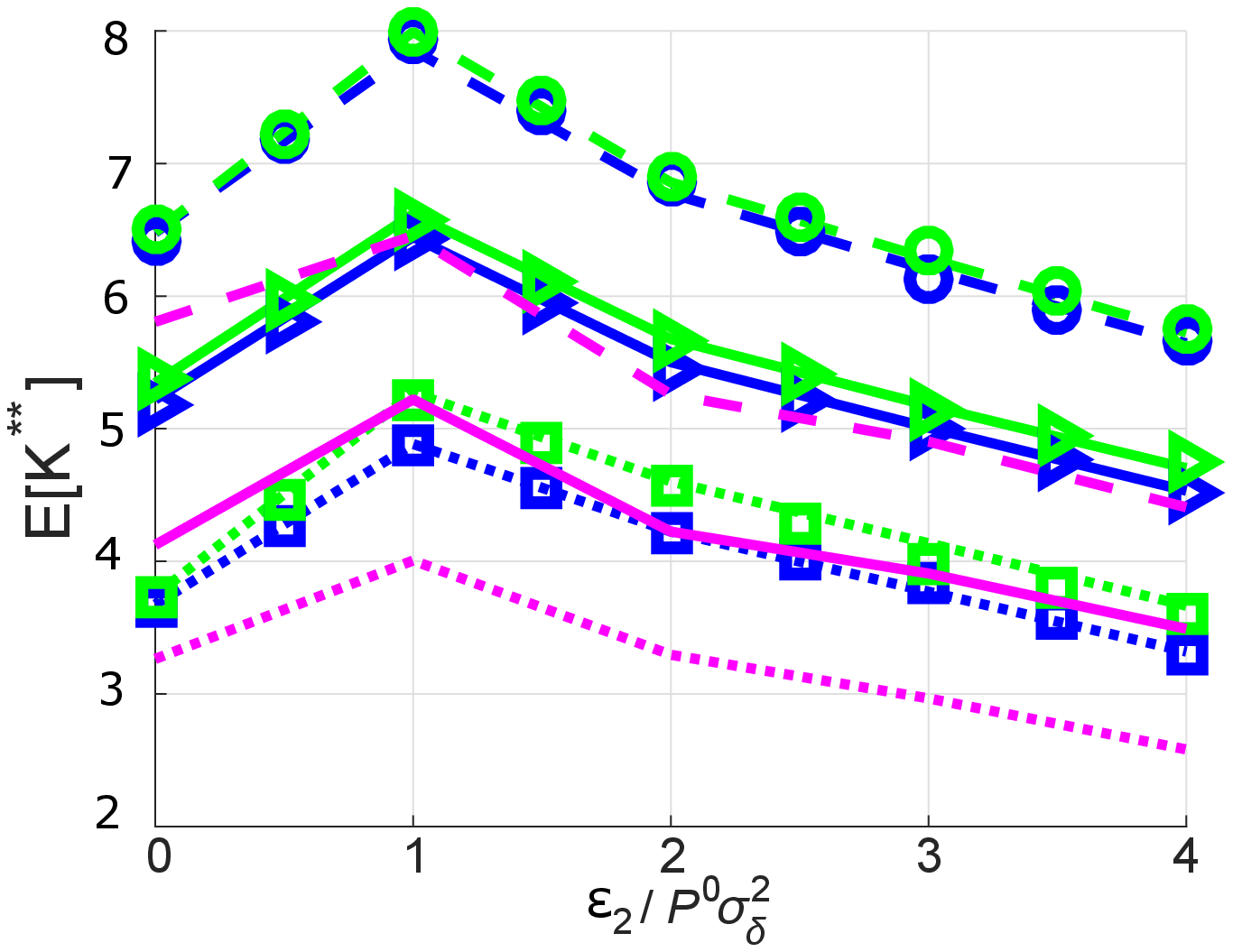}
		\caption{\small Average \# SUs served with minimum rate $R^0_k$ for $\epsilon_1 = \sigma^2_\Delta = \sigma^2_w / P_p$.}
		\label{fig:diff_eta_ep_unif_R0_E_k_starstar}
	\end{subfigure}			
	\caption{\small Impact of margin parameters. $K=20$, $I^0 = -106$ dBm. $R^0_k$ is uniformly distributed in (0,4] bps/Hz.}
	\label{fig:diff_eta_ep_unif_R0}
\end{figure}

\textit{Optimality of margins:} The margin parameters $\epsilon_1$ and $\epsilon_2$ are used to protect the PRs from the interference under imperfect CSI. In order to study the impact of margins on the performance, we plot the average interference and the number of SUs served for different values of $\epsilon_1$ and $\epsilon_2$. As shown in Fig. \ref{fig:diff_eta_ep_unif_R0_avg_int}, the average interference remains below the threshold for $\epsilon_1>\sigma^2_\Delta$. This result holds for values of $\epsilon_2/ P^0 \sigma^2_\delta$ in range [0,4], because the variable $\epsilon_2$ does not significantly affect the average interference. As mentioned in Section \ref{sec:selection_of_parameters}, larger values of $\epsilon_1$ result in smaller the number of SUs served. Therefore, we keep $\epsilon_1=\sigma^2_\Delta = \sigma^2_w/ P_p$ and plot $E[K^{**}]$ for range of values of $\epsilon_2$, as shown in Fig. \ref{fig:diff_eta_ep_unif_R0_E_k_starstar}. For $\epsilon_2< P^0 \sigma^2_\delta$, fewer SUs receive the required rate due to inter-SU interference, while for $\epsilon_2 > P^0 \sigma^2_\delta$ fewer SUs are admitted in the downlink due to large power allocation. Therefore, we see that the maximum number of SUs are served for $\epsilon_2=P^0 \sigma^2_\delta$ as described in Section \ref{sec: optimum_margins}.

\section{Summary}
\label{sec:summary_c2}
In this chapter, we proposed an optimization framework in order to serve the maximum number of SUs in an underlay CR network consisting of a secondary BS equipped with a large number of antennas. The proposed framework uses margin parameters to limit the interference to PUs below a specified threshold under imperfect knowledge of CSI. A new user selection and power allocation algorithm, referred to as DMP, is proposed that is based on ZF beamforming and power allocation that satisfies specific rate requirements of selected SUs. Theoretical analysis is presented to compute the number of SUs selected and the interference caused at PUs by the proposed algorithm. Results show that the proposed DMP algorithm serves more SUs than modified DML algorithm for lower rate requirements. As the rate requirements for the SUs increase, the performance of the modified DML algorithm approaches that of DMP. A low complexity version of DMP without ZF vector update is also studied. This algorithm reduces the complexity by a factor of the number of SUs and provides similar performance as DMP with vector update when the number of SBS antennas is an order of magnitude larger than the number of SUs. The analysis and simulation results show that the number of SUs selected by the proposed algorithm approaches the optimal solution if the number of SBS antennas is an order of magnitude larger than the number of SUs and PUs in the network. 

\chapter{Power Control and Frequency Band Selection Policies for Underlay MIMO Cognitive Radio}
\label{chapter_policy}


\section{Introduction}
\label{sec:Introduction_c3}
As described in the previous chapters, the spectral efficiency can be improved by allowing co-existence of unlicensed secondary users (SUs) with licensed primary users (PUs) in the same frequency band. Cognitive radio (CR) networks allow such co-existence under two paradigms: {interweave} and underlay \cite{biglieri2012a, tanab2017}. In an {interweave} CR network, the SU can transmit only in \textit{empty} time-slots when PUs are inactive in order to avoid interfering with the them. The achievable rate of the SU is further improved if there are multiple frequency bands available for transmission. {In a multi-band {interweave} CR network}, the SU can maximize its achievable rate by predicting which frequency band will have an empty time slot and then tunning to that band for transmission. Thus, the SU can dynamically hop to a different frequency band in each time-slot in search of an empty time slot to maximize its rate. This prediction-based band hopping is achieved by the multi-armed bandit (MAB) framework \cite{zhao2008a, tekin2011, liu2013a, dai2014, ouyang2014, oksanen2015, wang2016a, maghsudi2016, raj2018}. In the MAB framework, the SU learns the on-off activity of PUs in different bands in order to predict which band (arm) will be empty in the next time slot. However, the achievable rate of the SU in the {interweave} network is limited by the PU activity since the probability finding an empty slot is low when the PU activity is high. Further, a costly RF front-end is required at the SU to hop to a different band in each slot.

The achievable rate of SU can be improved if it is allowed to transmit even when the PU is active. The underlay CR paradigm allows the SU to transmit concurrently with PUs as long as the the interference towards primary receiver is below a specified limit \cite{biglieri2012a}. 
The SU can transmit concurrently with the PU, if the SU is equipped with multiple antennas and employs beamforming techniques to steer its signal in the null space of channels to primary receiver in order to contain the interference \cite{tsinos2013, noam2013}. The null space to primary receiver is estimated using the received auto-covariance matrix at the SU during a previous slot when receiver was the transmitter \cite{tsinos2013, gao2010, yi2009, yi2010}. Since the channel between SU and PU evolves due to fading during these time slots, the SU cannot not perfectly eliminate the interference towards the primary receiver using only null steering. Therefore, transmit power control is required along with null steering to limit the interference. The transmit power from the SU depends on the time between transmit and receive modes of PUs, i.e., the link reversal time of the PU link. In other words, the power transmitted from SU depends on the traffic pattern of the PU transmitter-receiver link as well as the temporal correlation that determines the rate of channel fading. Such transmit power control has not been considered in underlay MIMO CR literature and is addressed in this work. In an underlay CR network, the rate of the SU link depends on the transmit power as well as beamforming gain achieved after null steering. Therefore, {for a multi-band underlay CR network}, the band selection policy needs to take into account transmit power, beamforming gain and PU traffic statistics in each frequency band.

\subsection{Related work}
Frequency band selection using MAB based prediction has been considered for interweave cognitive radio in \cite{zhao2008a, tekin2011, liu2013a, dai2014, ouyang2014, oksanen2015, wang2016a}. In these works, the problem is cast as a restless MAB where each frequency band is modeled as an independent arm of the bandit problem. The term \textit{restless} implies that the physical channels in each band keep evolving even when that band is not selected by the SU, which holds for wireless channels. The goal of the band selection policies using restless MAB is to maximize the expected rate at the SU. Since the optimal solution to a general restless MAB problem is intractable \cite{zhao2008a, ouyang2014}, most of the works consider special cases. The special cases include policies based on a binary channel model as well as myopic policies where the goal is to maximize immediate rate in the next time slot. In the works \cite{zhao2008a, tekin2011, dai2014, oksanen2015}, a binary channel model was considered, where the SU receives reward (rate) 0 if the selected band is occupied by the PU, otherwise it receives rate 1. This model is suitable in the interweave CR network where SU transmits only when PU is inactive. For the binary channel model, the optimality of myopic band selection policy was shown in \cite{zhao2008a} for two frequency bands under the condition that the channel state evolves independently from one slot to the next. An online learning based band selection was proposed in \cite{dai2014} that implements the myopic policy in \cite{zhao2008a} without prior knowledge of PU activity. The work in \cite{tekin2011} considered a more general case where the state of the binary channel is modeled as a Markov chain (Gilbert-Elliot model). The authors proposed regenerative cycle algorithm (RCA) that outperforms the selection scheme in \cite{zhao2008a}. A recency based band selection policy was introduced in \cite{oksanen2015}, where the SU selects a suboptimal band less frequently and thus provides better performance as compared to earlier policies in a binary channel model with independent or Markovian evolution.

Binary channel models are not suitable for band selection in underlay CR network where the SU can receive a non-zero rate even when PU is active in the selected band. The rate received in this case depends on the beamforming gain between secondary transmitter and receiver as well as the transmit power. In order to model the beamforming gain, a multi-state channel model is required. Multi-state channels are considered for restless MAB problems in \cite{ouyang2014, wang2016a, liu2013a}. The optimality of myopic policy is established in \cite{ouyang2014} for a multi-state channel under the condition that the rate received by the SU in different channel states is sufficiently separated. This condition, however, may not hold in a real world channel with continuous state space. The work in \cite{wang2016a} established the optimality of the myopic policy when $F-1$ out of $F$ channels are selected by SU in each time slot. A policy, called deterministic sequencing of exploration and exploitation (DSEE) was constructed in \cite{liu2013a}. Under this policy, the SU stays on one band for multiple consecutive slots, called epochs, and the epoch length grows geometrically. It has been shown that the DSEE outperforms RCA for multi-state channels.

In an underlay CR network, if the transmit power is known, then the frequency band selection policy can be constructed by aforementioned restless MAB approaches such as DSEE. However, the existing works do not consider transmit power control for such a band selection problem in underlay CR networks. 

\subsection{Summary of contributions and outline}
In this chapter, we first propose fixed and dynamic power control schemes for a SU with multiple antennas. In the fixed power scheme, the SU transmits fixed power when the PU is active in that time slot, while in the dynamic power control scheme, the transmit power from the SU changes in each time slot. We show that the transmit power in a given frequency band depends on the traffic statistics of the PU transmitter-receiver links and the temporal correlation of the channels. 

Next, we analyze the following categories of band selection policies that use the above power control schemes: fixed band fixed power (FBFP), fixed band dynamic power (FBDP), dynamic band fixed power (DBFP) and clairvoyant policy. In the FBFP and FBDP policies, the SU stays on one frequency band and uses fixed or dynamic power control. The band selection policies based on restless MAB, such as DSEE, fall under DBFP category along with round robin and random band selection policies. The SU may hop to different frequency band under the DBFP policies. We also analyze the performance of a genie-aided clairvoyant policy that selects the frequency band providing the maximum gain in each slot. We compare the performance of these policies in terms of rate received at SU and interference towards PU.

The main contributions of this chapter are summarized below.
\begin{enumerate}
	\item Expressions for transmit power are derived for fixed and dynamic power control schemes as functions of link reversal time of the PU transmitter-receiver link and temporal correlation of channels. It is observed that the transmit power and thus the rate of the SU increase as the PU link reversal time decreases. 
	\item We show that the dynamic power control policy provides higher rate to SU than the fixed power control, i.e., FBDP provides higher rate than FBFP. Both polices keep the interference leakage towards PU below the specified limit. It is also shown that the DBFP polices, such as round robin, random, and DSEE, cause higher interference to PUs as compared to the fixed band policies.
	\item The expression is derived for the gap between the rate achieved by an optimal genie-aided clairvoyant policy and the FBFP policy. It is shown that this gap reduces under slow-varying channels and as the number of SU antennas is increased. This implies that the SU does not loose significant amount of rate by staying on one frequency band.
\end{enumerate}

\textit{Outline:} This chapter is organized as follows. The system model and problem statement are described in Section \ref{sec:Model_problem}. Power control and band selection policies are discussed in Section \ref{sec:policies}. Analytical comparison of the policies is presented in Section \ref{sec:analysis} while numerical results are shown in Section \ref{sec:results}. Finally, concluding remarks and future extensions are discussed in Section \ref{sec:Conclusion}.

\textit{Notations:} We denote vectors by bold, lower-case letters, e.g., $\mathbf{h}$. Matrices are denoted by bold, upper case letters, e.g., $\mathbf{G}$. Scalars are denoted by non-bold letters e.g. $L$. Transpose, conjugate, and Hermitian of vectors and matrices are denoted by $(.)^T$, $(.)^*$, and $(.)^H$, respectively. The norm of a vector $\mathbf{h}$ is denoted by $||\mathbf{h}||$. $\Gamma(x)$ is the Gamma function, while $\gamma(M,x)$ is the incomplete Gamma function defined as $\int_{0}^{x}t^{M-1}e^{-t} dt$. $\mathbb{E}[.]$ denotes the expectation operator, while $\mathbb{E}_x[.]$ is the expectation with respect to random variable $x$.


\begin{figure*}
	\centering
	\includegraphics[width=0.8\columnwidth]{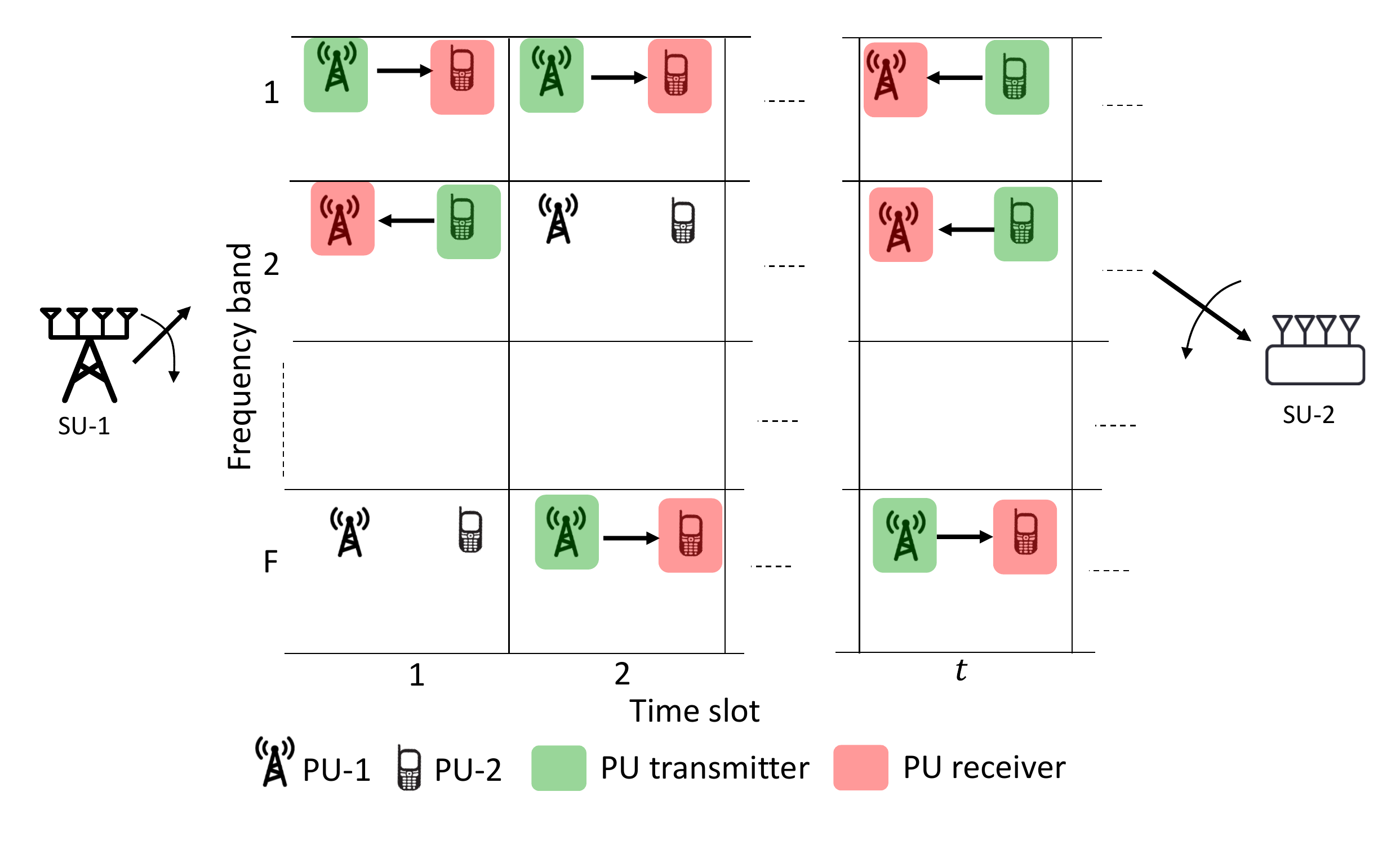}
	\vspace{-5mm}
	\caption{\small System model: SUs select any one out of $F$ available bands at time slot-$t$. Each band is occupied by a PU transmitter-receiver link.}
	\label{fig:system_model}
	\vspace{-1mm}
\end{figure*}

\section{System model and problem formulation}
\label{sec:Model_problem}
\subsection{System model}
\begin{figure}
	\centering
	\includegraphics[width=0.5\columnwidth]{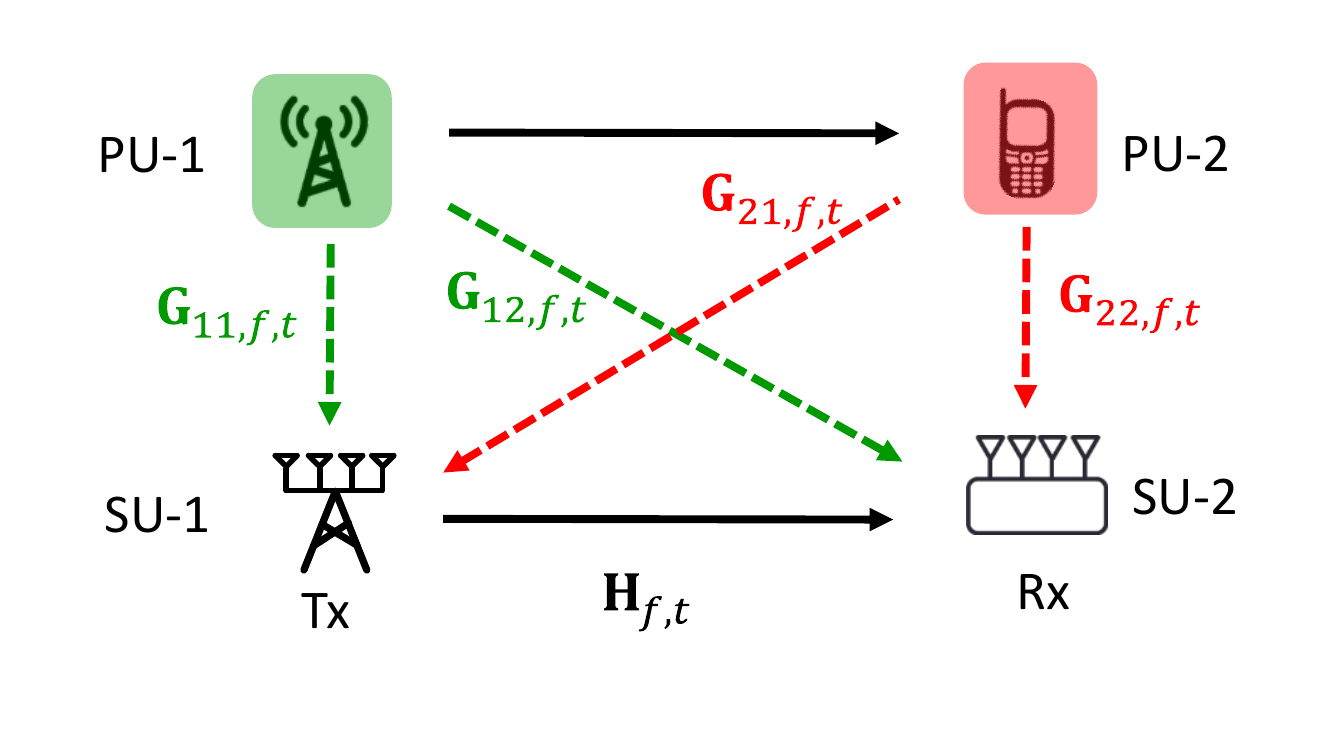}
	\vspace{-5mm}
	\caption{\small MIMO channels in frequency band-$f$ in slot $t$ with $s_{f,t}=1$. The channels in green are between PU transmitter and SUs, while the ones in red are between PU receiver and SUs.}
	\label{fig:network_model}
	\vspace{-1mm}
\end{figure}

Consider an underlay CR network in which SU transmit-receive pair, SU-1 and SU-2, selects one out of $F$ available frequency bands as shown in Fig. \ref{fig:system_model}. Let $M_s$ be the number of antennas at SU-1 and SU-2. Each band is occupied by a pair of PUs as shown in Fig. \ref{fig:network_model}. Let $M_p (<M_s)$ be the number of antennas at PUs. The PU-1 and PU-2 in each pair change role from transmitter to receiver according to a Markov chain.

The PU link in frequency band $f$ is in one of the three states at time slot $t$: 1) state-0: both PUs are silent, 2) state-1: PU-1 is the transmitter and PU-2 is the receiver,  and 3) state-2: PU-1 is the receiver and PU-2 is the transmitter. The state of the PU link is denoted by $s_{f,t} \in \{0,1,2\}$. The transition between the states is determined by transition probability matrix $\mathbf{T}_{f}$ as shown below:
\begin{align}
\mathbf{T}_{f} = 
\begin{bmatrix}
p_{00,f} & p_{01,f} &  p_{02,f} \\
p_{10,f}& p_{11,f}  & p_{12,f} \\
p_{20,f} & p_{21,f} &  p_{22,f} \\
\end{bmatrix},
\label{eq:TPF}
\end{align}
where $p_{kl,f} = \Pr(s_{f,t+1}=l|s_{f,t}=k), k,l \in \{0,1,2\}$ is the probability that PU link goes from state-$k$ in slot $t$ to state-$l$ in slot $t+1$. The steady state probability of PU link being in state $k$ in any slot $t$ is denoted by $\pi_{k,f} = \Pr(s_{f,t}=k), k \in \{0,1,2\}$ such that $\sum_k \pi_{k,f} = 1, \forall f$. The matrix $\mathbf{T}_f$ depends on the traffic configuration of the PU link. In order evaluate the policies, we will consider TDD LTE traffic models specified 3GPP 36.211 \cite{3gpp2017} to construct $\mathbf{T}_f$. Without the loss of generality, we consider the SU-1 is the transmitter and SU-2 is receiver in the secondary network.

Consider that the SU selects band $f$ in slot $t$ and the PU link is in state $s_{f,t}=1$. Then, the channel between SUs and PUs are shown in Fig. \ref{fig:network_model}: $\mathbf{H}_{f,t} \in \mathbb{C}^{M_s\times M_s}$ is the channel between SU-1 and SU-2, while $\mathbf{G}_{ij,f,t} \in \mathbb{C}^{M_s\times M_p}, i,j \in \{1,2\}$ denote channel between PU-$i$ and SU-$j$ in time slot $t$. We assume that the channels remain unchanged for the duration of time slot and evolve from slot $t$ to slot $t+1$ according to the Gauss-Markov model as follows \cite{sadeghi2008, so2015a}:

\begin{align}
\mathbf{H}_{f,t+1} =\alpha_f \mathbf{H}_{f,t} +\sqrt{1-\alpha_f^2} \Delta\mathbf{H}_{f,t},
\\\mathbf{G}_{ij, f,t+1} =\alpha_f \mathbf{G}_{ij,f,t} +\sqrt{1-\alpha_f^2} \Delta\mathbf{G}_{ij,f,t}, 
\label{eq:gauss_markov}
\end{align}
where $\alpha_f = J_0(2\pi f_d T_{slot})$ is the temporal correlation coefficient, $J_0(.)$ is the 0th order Bessel function, $f_d$ is the Doppler rate, and $T_{slot}$ is the duration of slot.  The matrices $\Delta\mathbf{H}_{f,t}, \Delta\mathbf{G}_{ij,f,t} \sim \mathcal{CN}(0, \mathbf{I})$ are i.i.d. channel update matrices in slot $t$. We assume that the channels are reciprocal. Initial distributions of the MIMO channels are $\mathbf{H}_{f,0} \sim \mathcal{CN}(0, \mathbf{I}) $ and $\mathbf{G}_{f,ij,0} \sim \mathcal{CN}(0, \mathbf{I})$. We consider a normalized channel model with identity covariance matrix for each flat fading MIMO channel as also used in \cite{dey2014, ho2017}. Distance based path-loss is not modeled since it does not affect the null space of channels $\mathbf{G}_{ij,f,t}$.

\subsubsection{Null space computation}
\label{sec:null_space_computation}
\begin{figure}
	\centering
	\includegraphics[width=0.5\columnwidth]{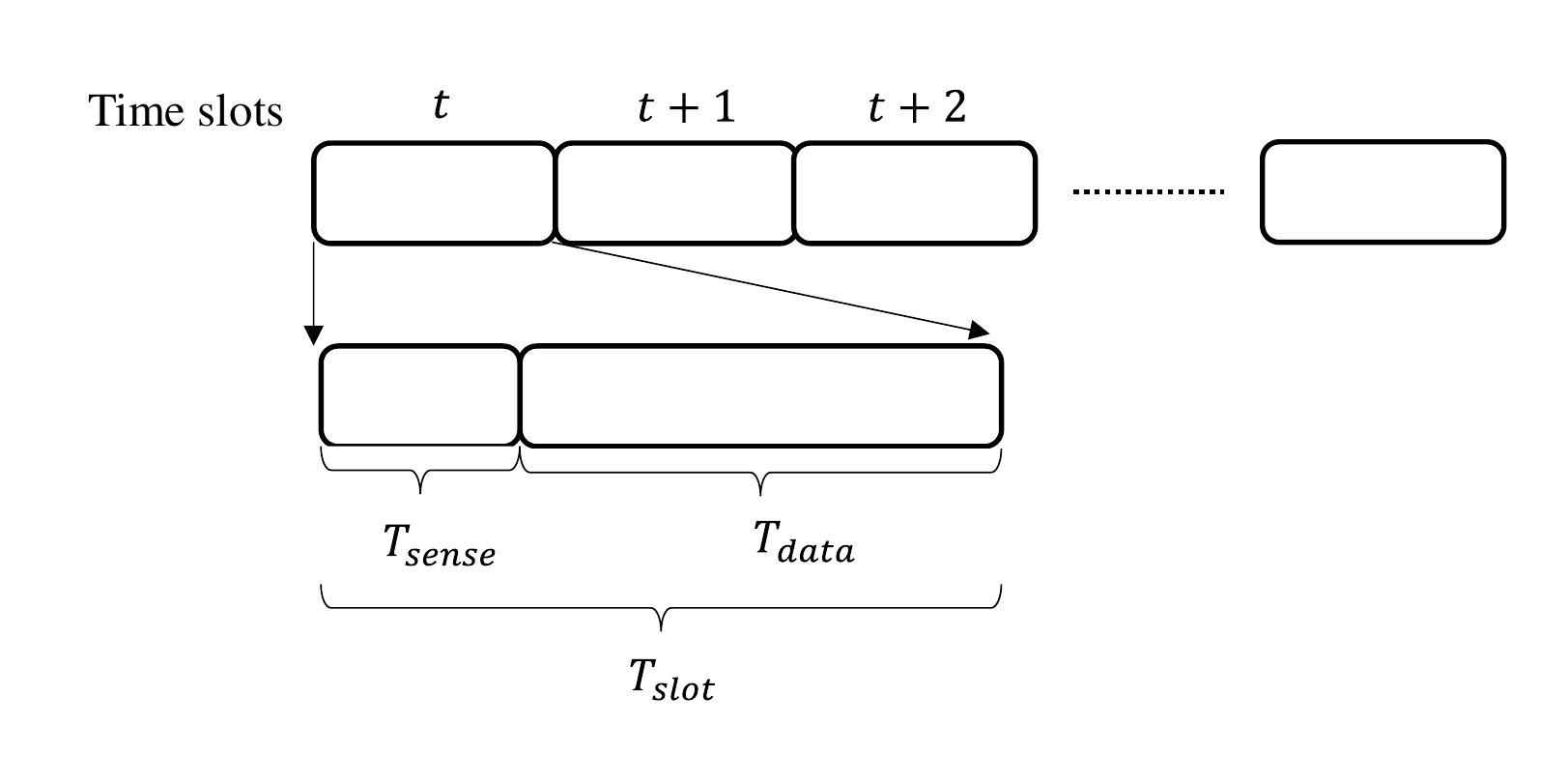}
	\vspace{-8mm}
	\caption{\small Time slot structure. SU selects frequency band $f$ at the beginning of slot and stays on that band for the duration $T_{slot}$.}
	\label{fig:slot_structure}
	\vspace{-1mm}
\end{figure}

The SU pair employs transceiver beamforming to transmit its signal in the null space  of channels to PUs. This ensures that the interference from SU transmitter to PU receiver and PU transmitter to SU receiver is minimized. The null space of the channels to PUs is obtained during the sensing duration $T_{sense}$ of each time slot. As shown in Fig. \ref{fig:slot_structure}, one time-slot consists of sensing duration $T_{sense}$ to obtain null spaces and SU data transmission $T_{data}$ \cite{kaushik2016}. In the sensing duration, SU-1 and SU-2 receive the signal from the PU transmitter and compute the null space of the channel.

\begin{figure}[t!]
	\centering	
	\begin{subfigure}[b]{0.45\columnwidth}
		\centering	
		\includegraphics[width=0.8\columnwidth]{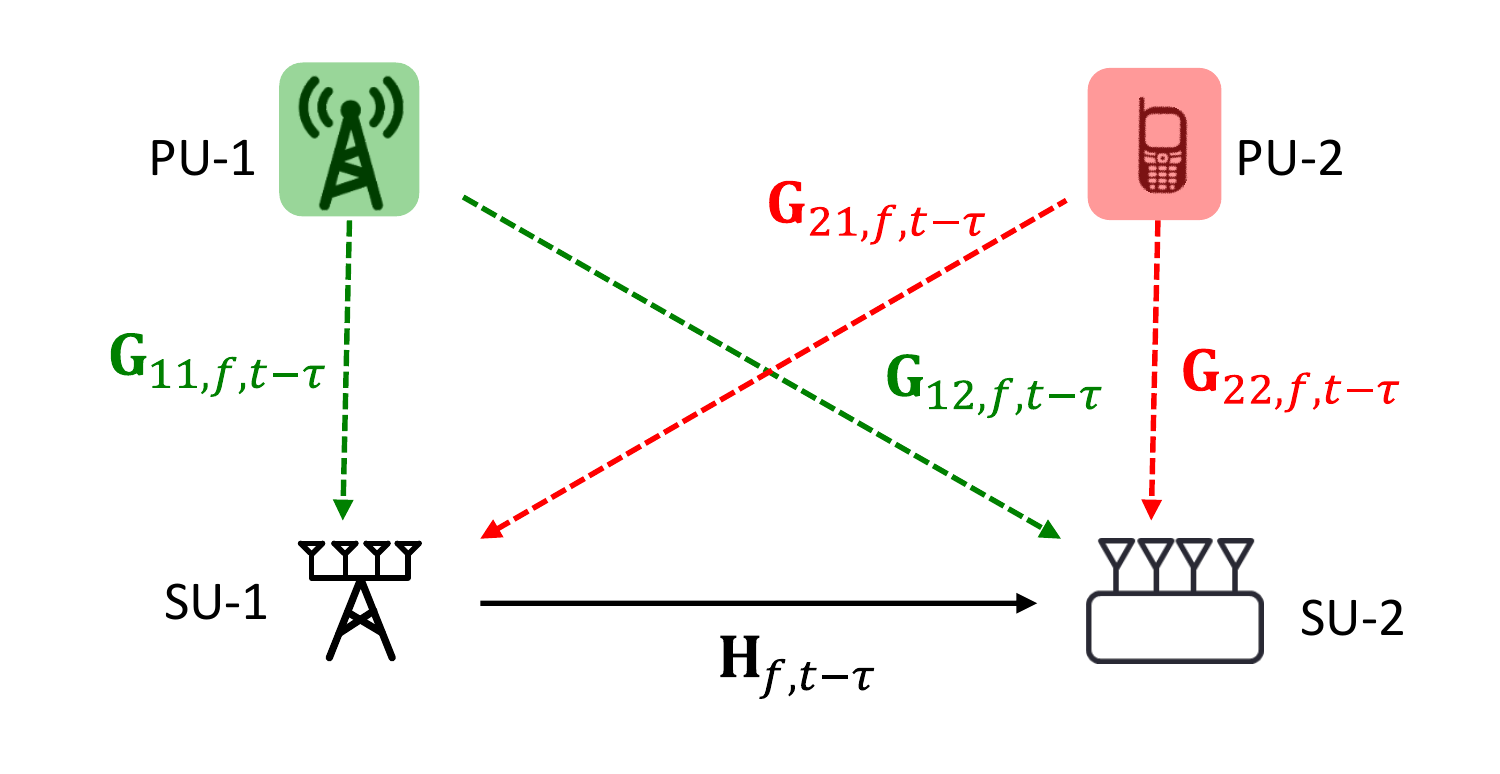}
		\caption{\small Slot $t-\tau$: ($s_{f,t-\tau}=1$) PU-1 is tx.}
		\label{fig:snt_t_tau}
	\end{subfigure}			
	\begin{subfigure}[b]{0.45\columnwidth}
		\centering	
		\includegraphics[width=0.8\columnwidth]{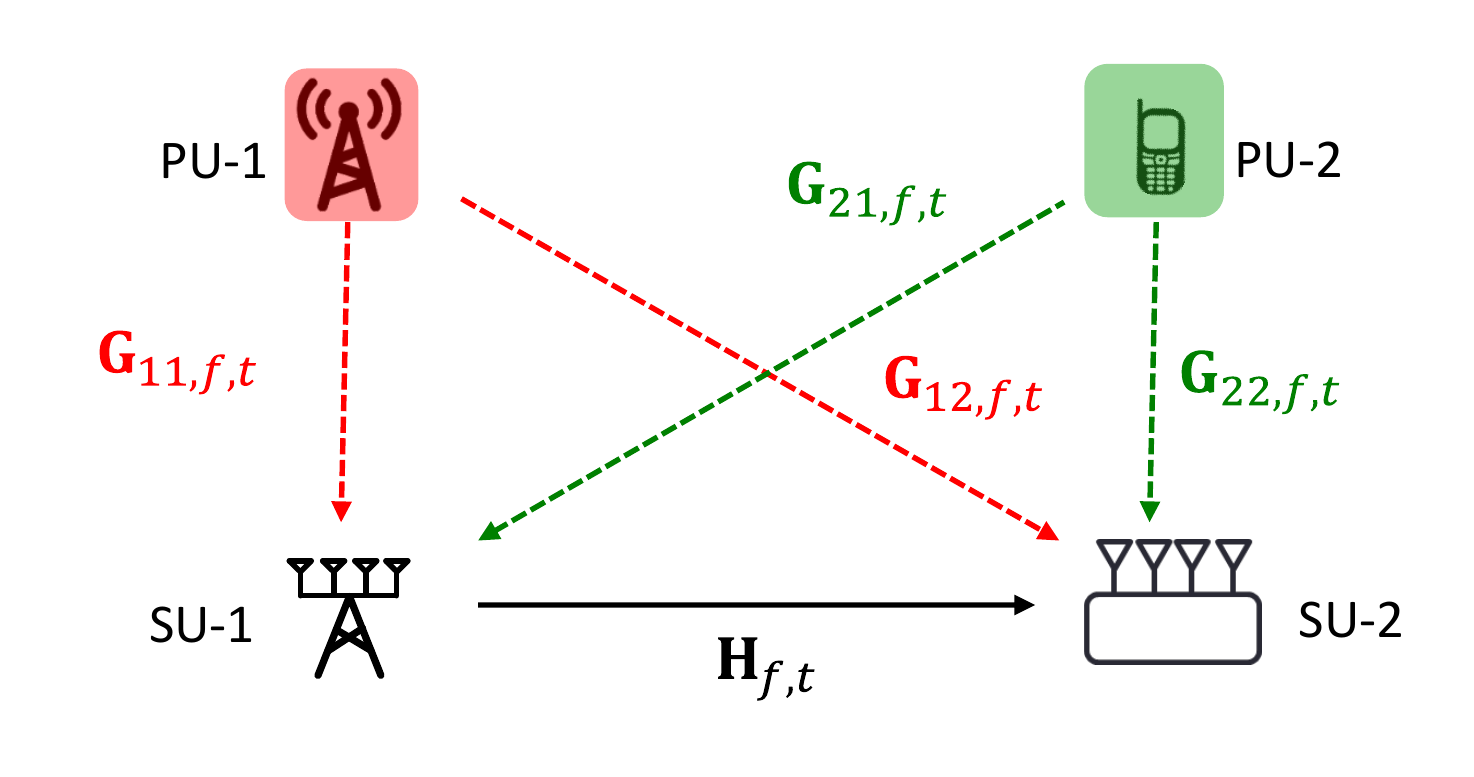}
		\caption{\small Slot $t$:  ($s_{f,t}=2$) PU-2 is tx.}
		\label{fig:snt_t}
	\end{subfigure}		
	\caption{\small Null space computation during sensing duration. SUs compute null space of channels shown in green by sensing the signal received from PU transmitter.}
	\label{fig:snt}	
	\vspace{-1mm}
\end{figure}

Consider that SU selects band $f$ in slot $t$ when PU link state is $s_{f,t}=2$, i.e., PU-2 is the transmitter and PU-1 is the receiver. Let us assume that PU-1 was the transmitter $\tau_{f,t}$ slots ago, i.e., $s_{f,t-{\tau}_{f,t}}=1$. For simplicity of notations, we drop subscripts from $\tau_{f,t}$. As shown in Fig. \ref{fig:snt}, in slot $t-\tau$, SUs obtain the null space of the channels $\mathbf{G}_{11,f,t-\tau}$ and $\mathbf{G}_{12,f,t-\tau}$ using the received autocovariance matrices. Let $\mathbf{y}_{1,f,t-\tau}(n)$ and $\mathbf{y}_{2,f,t-\tau}(n)$ be the received signal vectors at SU-1 and SU-2 during sensing duration of slot $t-\tau$ as expressed below:
\begin{align}
\mathbf{y}_{1,f,t-\tau}(n) = \mathbf{G}_{11,f,t-\tau} \mathbf{x}_1(n) + \mathbf{w}(n), n = {0,2,\cdots, N-1},
\\ \mathbf{y}_{2,f,t-\tau}(n) = \mathbf{G}_{12,f,t-\tau} \mathbf{x}_1(n) + \mathbf{w}(n), n = {0,2,\cdots,N-1},
\end{align}
where $\mathbf{x}_1(n)\in \mathbb{C}^{M_p\times 1}$ is the transmitted signal vector from PU-1, $\mathbf{w}(n)\sim \mathcal{CN}(0, \mathbf{I})$ is the noise vector, and { $N=\frac{T_{sense}}{T_s}$ is the number of samples collected and $T_s$ is the sampling duration}. Let $\mathbf{A}_{1,f,t-\tau} \in \mathbb{C}^{M_s \times (M_s - M_p)}$ be the null space matrix of channel $\mathbf{G}_{11, f, t-\tau}$ and $\mathbf{B}_{1,f,t-\tau} \in \mathbb{C}^{M_s \times (M_s - M_p)}$ be the null space matrix of channel $\mathbf{G}_{12, f, t-\tau}$. Matrix $\mathbf{A}_{1,f,t-\tau}$ contains columns in the null space of the received covariance matrix $\mathbf{\hat{Q}}_{1,f,t-\tau} = \frac{1}{N}\sum\limits_{n=0}^{N-1} \mathbf{y}_{1,f,t-\tau}(n)\mathbf{y}^H_{1,f,t-\tau}(n)$ and are obtained by eigenvalue decomposition (EVD) of $\mathbf{\hat{Q}}_{1,f,t-\tau}$ at SU-1. Similarly,  matrix $\mathbf{B}_{1,f,t-\tau}$ contains columns in the null space of the covariance matrix $\mathbf{\hat{Q}}_{2,f,t-\tau} =$

\noindent $\frac{1}{N} \sum\limits_{n=0}^{N-1} \mathbf{y}_{2,f,t-\tau}(n) \mathbf{y}^H_{2,f,t-\tau}(n)$ and are obtained by eigenvalue decomposition (EVD) of $\mathbf{\hat{Q}}_{2,f,t-\tau}$ at SU-2. Similarly, in slot $t$, SU-1 obtains null space $\mathbf{A}_{2,f,t}$ of channel matrix $\mathbf{G}_{21,f,t}$, while SU-2 obtains null space $\mathbf{B}_{2,f,t}$ of channel matrix $\mathbf{G}_{22,f,t}$. {SUs can estimate the state of PU link, $s_{f,t}$, based on the received signal from PUs in the sensing duration as described in Appendix \ref{app:st_est}.}

\subsubsection{Transceiver beamforming at SUs}
\label{sec:transceiver_beamforming}
Let us consider SU signal transmission from SU-1 to SU-2 in slot $t$. As shown in Fig. \ref{fig:snt_t}, in this slot, PU-2 is the transmitter and PU-1 is the receiver. In order to mitigate the interference towards PU receiver (PU-1), SU-1 needs to transmit its signal in the null space $\mathbf{A}_{1,f,t}$ of $\mathbf{G}_{11,f,t}$. However, this null space is not available at SU-1 in slot $t$, since PU-1 is not transmitting. Therefore, SU-1 utilizes the null space  $\mathbf{A}_{1,f,t-\tau}$ for precoding which was obtained in slot $t-\tau$ when PU-1 was the transmitter. On the other hand, on the receiver side, SU-2 utilizes the null space $\mathbf{B}_{2,f,t}$ for receiver combining to mitigate the interference from PU-2. In order to maximize the beamforming gain of SU link, the SU utilizes the maximum eigenmode of the equivalent channel $\mathbf{H}_{eq,f,t} = \mathbf{B}_{2,f,t}^H\mathbf{H}_{f,t}\mathbf{A}_{1,f,t-\tau}$. If the transmitted power  in slot $t$ is denoted by $P_t$, the achievable rate of the SU link is given by

\begin{align}
R_{f,t} = \frac{T_{data}}{T_{slot}}\log_2\left(1 + \frac{ P_t \Gamma_{f,t}}{\sigma^2_w}\right) = \frac{T_{data}}{T_{slot}}\log_2\left(1 + { P_t \Gamma_{f,t}}\right),
\label{eq:rate}
\end{align}
where $\Gamma_{f,t} $ is the maximum eigenvalue of $\mathbf{H}_{eq,f,t}$ and $\sigma^2_w=1$ is the noise power, and $\frac{T_{data}}{T_{slot}}$ is the fraction of time slot used for signal transmission \cite{raj2018}. Note that if both the PUs are silent in slot $t$, i.e., $s_{f,t}=0$, we get $\mathbf{H}_{eq,f,t}=\mathbf{H}_{f,t}$. The rank of $\mathbf{H}_{eq,f,t}$ is $M_s$ when $s_{f,t}=0$, while the rank is $M_s-M_p$ when  $s_{f,t}=\{1,2\}$. 

\subsubsection{Interference leakage towards PU receiver}
\label{sec:interference_to_pu}
As mentioned above, the SU transmitter cannot access the null space of channel to PU receiver in slot $t$. It uses the null space obtained $\tau$ slots ago when the PU receiver was the transmitter. This results in non-zero interference towards the PU receiver. {The expected interference leakage towards PU receiver in slot $t$ under the Gauss-Markov evolution can be written as follows:
	\begin{align}
	\mathbb{E}[I_{f,t}]= \mathbb{E}\left[ P_t ||\mathbf{G}^H_{11,f,t}\mathbf{v}_t||^2 \right] = P_t M_p (1-\alpha_f^{2\tau}),
	\label{eq:avg_int}
	\end{align}
	where the transmit beamforming vector, $\mathbf{v}_t$, is the principle right singular vector of $\textbf{H}_{eq,f,t}$ and $M_p$ is the rank of the channel $\mathbf{G}_{11,f,t}$}. The proof of the second equality is shown in Appendix \ref{app:avg_int}. From the above expression, we observe that the interference to PU in frequency bands with smaller correlation $\alpha_f$ will be higher. Further, higher value of $\tau$ implies larger interference to PU, i.e., older the null space to PU receiver, higher will be the interference.  

Note that the expectation in (\ref{eq:avg_int}) is with respect to the random variations $\Delta \mathbf{G}_{11}$ in the Gauss-Markov model. In this expression, $\tau$ is assumed to be constant. The variable $\tau$ indicates \text{how old} the null space is at any given slot $t$. The expected value of the interference with respect to the random variable $\tau$ can be written as below:

\begin{align}
\mathbb{E}_{\tau}\left[ \mathbb{E}[I_{f,t} | \tau] \right]=  \mathbb{E}_{\tau} \left[P_t M_p (1-\alpha_f^{2\tau})\right].
\label{eq:avg_int_tau}
\end{align}


\subsection{Power control and band selection problem}
\label{sec:problem}
Let $a_t \in \{1,2,...,F\}$ denote the frequency band selected by the SU link and $P_t$ be the transmit power in time slot $t$ . Then, the power control and frequency band selection problem for $t=1,2,...$ can be written as follows:
\begin{align}
\nonumber \{a^*_t, P^*_t\} &= \arg\max_{a_t,P_t} \mathbb{E}\left[R_{a_t,t} \right],
\\ \nonumber  \text{Subject to:~~}& \mathbb{E}_{\tau}\left[ \mathbb{E}[I_{a_t,t} | \tau] \right] \leq I^0,
\\ &P_t \leq P^0,
\end{align}
where $I^0$ is the threshold on the interference towards PU receiver and $P^0$ is the maximum transmit power. By substituting for $R_{f,t}$ and $\mathbb{E}[I_{f,t}|\tau]$ from (\ref{eq:rate}) and (\ref{eq:avg_int_tau}), respectively, we get the following problem statement:
\begin{align}
\nonumber \textbf{(P1)}~~ \nonumber \{a^*_t, P^*_t\} &= \arg\max_{a_t,P_t}\mathbb{E} \left[\log_2\left(1 + { P_t \Gamma_{a_t,t}}\right)\right],
\\ \text{Subject to:~~}& \mathbb{E}_{\tau} \left[P_t M_p (1-\alpha_{a_t}^{2\tau})\right] \leq I^0,
\label{eq:int_cont}
\\ &P_t \leq P^0.
\label{eq:power_cont}
\end{align}

\section{Power control and band selection policies}
\label{sec:policies}
In order to compute the transmit power $P_t$, we assume that $\mathbf{T}_f$ and $\alpha_f$ are known at the SU. The transition probability matrix $\mathbf{T}_f$ can be obtained if the knowledge of the PU traffic model is shared by the PU operator with the SU. For example, if PUs in band $f$ are LTE base station and user following TDD traffic models specified in 3GPP 36.211 \cite{3gpp2017}, then SU can construct $\mathbf{T}_f$ from frame structure of the traffic model. Further, the parameter $\alpha_f$ of the Guass-Markov process can be estimated using parameter estimation methods for the first-order auto-regressive process \cite{verbout1998, urteaga2017}. Another option is to assume the maximum Doppler rate $f_{d, max}$ to obtain $\alpha_f = J_0(2\pi f_{d,max}T_{slot})$.

{In this section, we describe two power control schemes: fixed and dynamic. Under the fixed power control scheme, the SU transmits power $P_t = P^{fix}_f$ if it is on frequency band $f$  in slot $t$ and the band is occupied by PUs, i.e., $s_{f,t}\in \{1,2\}$. Under the dynamic power control scheme, the SU transmits power $P_t = P^{dyn}_{f,t}$ if it is on frequency band $f$  in slot $t$ and the band is occupied by PUs, i.e., $s_{f,t}\in \{1,2\}$. The expressions for $P^{fix}_f$ and $P^{dyn}_{f,t}$ are derived in this section.	Note that the SU transmits maximum power $P_t = P^0$ if the PUs are inactive, i.e., $s_{f,t}\in \{0\}$.} In section \ref{sec:band_selection_policies}, we study the performance of frequency band selection policies using the transmit power determined in this section.

\subsection{Power control policies}
\label{sec:power_control_policies}

We first determine the maximum fixed transmit power in band $f$ from SU in order to satisfy constraints (\ref{eq:int_cont}) and (\ref{eq:power_cont}). Let $P_t = P^{fix}_f$ be the the maximum fixed transmit power in band $f$ from SU. The power $P^{fix}_f$ needs to satisfy the following constraints:
\begin{align}
P^{fix}_f M_p \mathbb{E}_\tau \left[1-\alpha_f^{2\tau}\right] \leq I^0,
\\ P^{fix}_f \leq P^0.
\end{align}

Therefore, the maximum fixed transmit power in band $f$ is:
\begin{align}
P^{fix}_f = \min \left( \frac{I^0}{M_p \mathbb{E}_\tau \left[1-\alpha_f^{2\tau}\right]},  P^0  \right)= \left( \frac{I^0}{M_p g(\alpha_f, \mathbf{T}_f)}, P^0 \right) 
\label{eq:p_n_fix_def}
\end{align}
where
\begin{align}
g(\alpha_f, \mathbf{T}_f) =
\frac{\sum \limits_{i} (1-\alpha_f^{2i}) \left(\pi_{1,f} \sum \limits_{s\in \{0,2\}} p_{1s}p_{s2\backslash 1}^{(i-1)} + \pi_{2,f} \sum \limits_{s\in \{0,1\}} p_{2s}p_{s1\backslash 2}^{(i-1)} \right)}{(\pi_{1,f} + \pi_{2,f})},
\label{eq:p_n_fix}
\end{align}
and $p_{ss' \backslash s''}^{(i)}$ is the probability of PU link going from state $s$ to state $s'$ in $i$ slots without hitting state $s''$, where $s,s',s'' \in \{0,1,2\}$. $p_{ss'' \backslash s''}^{(i)}$ is obtained from the transition probabilities in $\mathbf{T}_f$. The proof of the last equality is shown in Appendix \ref{app:p_n_fix}.

The SU can also dynamically control the transmit power in each slot to satisfy the constraints (\ref{eq:int_cont}) and (\ref{eq:power_cont}). Let $P^{dyn}_{f,t}$ be the transmit power if SU stays on band $f$ in slot $t$. The interference constraint is satisfied if we have $P^{dyn}_{f,t} M_p (1-\alpha_f^{2\tau}) \leq I^0$. Therefore, the maximum  dynamic power transmitted from the SU is 
\begin{align}
P^{dyn}_{f,t} = \min \left(\frac{I^0}{M_p (1-\alpha_f^{2 \tau})} , P^0\right).
\label{eq:p_n_dyn}
\end{align}

From expressions (\ref{eq:p_n_fix_def}) and (\ref{eq:p_n_dyn}), we can observe that the transmit power increases if $\alpha_f$ is increased. The power also increases of the PU link reversal time $\tau$ is decreased. Thus, the transmit power depends on the temporal correlation as well as the traffic statistic of the PU link. Therefore, the SU can transmit maximum power in the frequency band that has high correlation or small PU link reversal time $\tau$.

Note that the power control is required only when the PU is active in the band selected by SU in slot $t$, i.e., $s_{a_t,t}=\{1,2\}$. The SU transmits maximum power $P_t = P^0$ if $s_{a_t,t}=0$, since there will be no interference to PU receiver in such a slot. 

\subsection{Band selection policies}
\label{sec:band_selection_policies}

Next, we describe four types of polices for band selection using aforementioned power control schemes, namely fixed band fixed power (FBFP), fixed band dynamic power (FBDP), dynamic band fixed power (DBFP), and clairvoyant policy. In the FBFP and FBDP policies, the SU determines which frequency band allows maximum transmit power according to (\ref{eq:p_n_fix}) and (\ref{eq:p_n_dyn}) and stays on that frequency band. On the other hand, in DBFP policies, SU can hop to different frequency bands. The clairvoyant policy assumes a genie SU that can observe all $F$ frequency bands simultaneously in all the slots to maximize its rate.

\subsubsection{Fixed band fixed power (FBFP)}
\label{sec:fbfp}
The simplest band selection and power control policy for the SU is to stay on one frequency band, say $f^*$, and use a fixed transmit power $P^{fix}_{f^*}$. This policy is called fixed band fixed power (FBFP). The band $f^*$ is the frequency band that allows SU to transmit maximum power while satisfying the interference and power constraints. Therefore, we have
\begin{align}
a_t = f^{*} = \arg\max_f P^{fix}_f.
\label{eq:FBFP}
\end{align}

\subsubsection{Fixed band dynamic power (FBDP)}
\label{sec:fbdp}
In this policy as well, the SU stays on one frequency band that allows maximum transmit power. However, the transmit power $P^{dyn}_{f^*,t}$ is dynamically controlled in each slot when the PU is active. The frequency band selected by the SU is same as in the FBFP policy:
\begin{align}
a_t = f^{*} = \arg\max_f P^{fix}_f = \arg\max_f P^{dyn}_{f,t}.
\label{eq:FBDP}
\end{align}

\subsubsection{Dynamic band fixed power (DBFP)}
\label{sec:dbfp}
Under DBFP category, we consider three policies: random, round robin, and DSEE in \cite{liu2013a}. In all three policies, we assume that SU utilizes fixed transmit power $P^{fix}_f$ from (\ref{eq:p_n_fix}) in band $f$. 

In the random policy, the SU selects the frequency band $f$ randomly. Each band has equal probability of getting selected in slot $t$. In the round robin policy, as the name suggests, the SU selects frequency band in round robin fashion. In the DSEE policy presented in \cite{liu2013a}, the band selection problem is treated as a restless multi-armed bandit (MAB) problem. For a fixed transmit power, the problem (\textbf{P1}) can be cast as a MAB problem. Therefore, we apply the DSEE policy in order to dynamically select the frequency band. In the DSEE policy, we consider that if SU is on band $f$ in slot $t$, it uses transmit power $P^{fix}_f$ from (\ref{eq:p_n_fix}) and receives rate (reward) $R_{f,t} = \log_2 \left(1 + P^{fix}_f \Gamma_{f,t} \right)$. The DSEE band selection policy is implemented as described in \cite[Section II.B]{liu2013a}.

\subsubsection{Clairvoyant policy}
\label{sec:clair}
We compare the rate achieved in the aforementioned policies with an ideal,  clairvoyant policy, where a genie-aided SU can observe all $F$ bands simultaneously in each slot $t$, compute null spaces and beamforming vectors in each band and then select the one which provides maximum rate. {In this policy, dynamic power control $P^{dyn}_{f,t}$ is used since it provides higher rate than fixed power, as shown in Theorem \ref{thm:fbdp_v_fbfp} in Section \ref{sec:analysis}}. Therefore, the frequency band selected by the clairvoyant policy in slot $t$, $f_{c,t}$, is given by
\begin{align}
a_t = f_{c,t} &= \arg \max_{f} \log_2 \left(1 + {P^{dyn}_{f,t} \Gamma_{f,t}} \right).
\end{align}
The clairvoyant policy provides an upper bound on achievable rate in the given setting.

\section{Analysis of the policies}
\label{sec:analysis}
In this section, we analyze the achievable rate and interference leakage towards PU under the policies described in the previous section.
\subsubsection{Analysis of FBFP policy}
\label{sec:fbfp_rate}
The expected achievable rate under FBFP policy, if the SU stays on band $f^*$, is denoted by $\mathbb{E}[{R^{(1)}_{f^*,t}}]$. We drop the asterisk in the subscript to simplify the notation. This expression also holds for any frequency band $f \in [1,F]$. The expected rate can be expressed as follows:
{\footnotesize
	\begin{align}
	\mathbb{E}[{R^{(1)}_{f,t}}] = \frac{T_{data}}{T_{slot}} \left[\pi_{0,f} \mathbb{E}\left[\log_2\left(1 + {P^0 \Gamma_{f, t}} \right) \vert s_{f,t}=0\right] 
	+ (1-\pi_{0,f})  \mathbb{E}\left[\log_2\left(1 + {P^{fix}_{f} \Gamma_{f, t}} \right) \vert s_{f,t}=\{1,2\}\right]\right].
	\label{eq:fbfp_rate}
	\end{align}}
The first term in above expression is the expected rate when PUs are silent, while the second term is the rate when either PU is active.  As mentioned in Section \ref{sec:Model_problem}, the equivalent channel matrix $\mathbf{H}_{eq,f, t}$ has rank $M_s$ if PU is silent and rank $M_s-M_p$ if one PU is transmitting. Therefore, the expectations in (\ref{eq:fbfp_rate}) can be expressed using the distribution of maximum eigenvalue of rank $M_s$ and $M_s-M_p$ matrices as follows:
\begin{align}
\small
\mathbb{E}\left[\log_2\left(1 + {P^0 \Gamma_{f, t}} \right) \vert s_{f,t}=0\right] 
&= \int_{0}^{\infty}\log_2 \left(1 + {P^{fix}_{f} x} \right) f_{M_s}(x)dx
\\ \mathbb{E}\left[\log_2\left(1 + {P^{fix}_{f} \Gamma_{f, t}} \right) \vert s_{f,t}=\{1,2\}\right]
&= \int_{0}^{\infty}\log_2 \left(1 + {P^{fix}_{f} x} \right) f_{M_s-M_p}(x)dx,
\label{eq:fbfp_rate2}
\end{align}
where $f_{M_s}(x)$ is the probability density function (pdf) of the largest eigenvalue of Hermitian matrix $\mathbf{H}^H_{eq,f,t}\mathbf{H}_{eq,f,t}$ of rank $M_s$. The pdf of the largest eigenvalue is computed using the cumulative distribution function (cdf) $F_{M_s}(x)$ as follows:
\begin{align}
f_{M_s}(x) = \frac{d}{dx}F_{M_s}(x) = \frac{x^{M_s-1}e^{-x}}{\Gamma(M_s)},
\label{eq:pdf_max_eig}
\end{align}
where $F_{M_s}(x) = \frac{\gamma(M_s,x)}{\Gamma(M_s)}$ is the cdf as given in \cite[Eq. 9]{kang2003}, $\gamma(.,.)$ is the incomplete Gamma function and $\Gamma(.)$ is the Gamma function. The expression (\ref{eq:fbfp_rate2}) is computed using the (\ref{eq:pdf_max_eig}) and the expected rate $\mathbb{E}[R^{(1)}_{f,t}]$ is evaluated by substituting (\ref{eq:fbfp_rate2}) in (\ref{eq:fbfp_rate}). Note that the expected interference towards PU remains under FBFP is $I^0$ since the power control scheme ensures that the constraint (\ref{eq:int_cont}) is satisfied with equality.

\subsubsection{Analysis of FBDP policy}
\label{sec:fbdp_rate}
In FBDP, the power is dynamically changed per slot in order to control the interference to the PU. Note that the power $P^{dyn}_{f,t}$ is a function of $\tau$ as shown in (\ref{eq:p_n_dyn}). Therefore, while computing the expected achievable rate, maximum eigenvalue as well as $\tau$ are treated as random variables. The expected rate can be expressed as below:
{\small
	\begin{align}
	\mathbb{E}[R^{(2)}_{f,t}] =  \frac{T_{data}}{T_{slot}} \left[\pi_{0,f} \mathbb{E}\left[\log_2\left(1 + {P^0 \Gamma_{f, t}} \right) \vert s_{f,t}=0\right] 
	+ (1-\pi_{0,f})  \mathbb{E}\left[\log_2\left(1 + {P^{dyn}_{f,t} \Gamma_{f, t}} \right) \vert s_{f,t}=\{1,2\}\right]\right].
	\label{eq:fbdp_rate_v1}
	\end{align}}
The first term is same as the first term in (\ref{eq:fbfp_rate}). The expectation in the second term is computed as follows:
\begin{align}
\mathbb{E}\left[\log_2\left(1 + {P^{dyn}_{f,t} \Gamma_{f, t}} \right) \right] 
= \mathbb{E}_\tau \left[ \mathbb{E}_{\Gamma}\left[\log_2\left(1+ {P^{dyn}_{f,t} \Gamma_{f,t}} \right) \vert \tau\right] \right],
\end{align}
where $\mathbb{E}_\Gamma$ is the expectation with respect the the maximum eigenvalue assuming $\tau$ is a constant. The condition $s_{f,t}=\{1,2\}$  is dropped from the above expression to simplify the notation. The outer expectation $\mathbb{E}_\tau$ is with respect to $\tau$. The inner expectation is computed by substituting $P^{fix}_f$ with $P^{dyn}_{f,t}$ in (\ref{eq:fbfp_rate2}). Let $z(i) = \mathbb{E}_{\Gamma}\left[\log_2\left(1+ {P^{dyn}_{f,t} \Gamma_{f,t}} \right) \vert \tau=i \right]$ be the inner expectation for $\tau=i$. Then, the outer expectation is given by
\begin{align}
\mathbb{E}_{\tau} [z(i)] = \sum_{i} z(i) \times \Pr(\tau=i),
\label{eq:fbdp_exp_outer}
\end{align}
where $\Pr(\tau=i)$ is computed using 

\begin{align}
\Pr (\tau=i) = \pi_{2,f} \sum_{s\in \{0,1\}} p_{2s}p_{s1\backslash 2}^{(i-1)} + \pi_{1,f} \sum_{s\in \{0,2\}} p_{1s}p_{s2\backslash 1}^{(i-1)}.
\label{eq:pr_tau}
\end{align}
The derivation for the above expression is provided in Appendix \ref{app:p_n_fix}. The expected rate is computed by substituting (\ref{eq:fbdp_exp_outer}) in (\ref{eq:fbdp_rate_v1}). At this point, we state the following theorem comparing the expected rates in FBDP and FBFP:
\begin{theorem}
	The expected rate under fixed band dynamic power (FBDP) exceeds the expected rate under fixed band fixed power (FBFP) policy, i.e., $\mathbb{E}[R^{(2)}_{f,t}]\geq \mathbb{E}[R^{(1)}_{f,t}]$.
	\label{thm:fbdp_v_fbfp}
\end{theorem}
\begin{proof}
	Appendix \ref{app:fbdp_v_fbfp}.
\end{proof}

\subsubsection{Analysis of DBFP policies}
\label{sec:dbdp_rate}
In DBFP policies, the SU uses fixed transmit power $P^{fix}_{f}$ when it is on band $f$. In round robin and random band selection policies, each band is selected for same number of times on an average. Therefore, the expected rate under these two policies will be equal. Let $\mathbb{E}[{R^{(3)}_{f,t}}]$ be the expected rate under random and round robin policies. Since each band is visited with equal probability, the expected rate can be written as:
\begin{align}
\mathbb{E}[{R^{(3)}_{f,t}}] =\frac{T_{data}}{T_{slot}} \frac{1}{F}\sum_{f'=1}^{F} \mathbb{E}[R^{(1)}_{f',t}],
\end{align}
where $\mathbb{E}[R^{(1)}_{f',t}]$ is the expected rate under FBFP policy if the SU stays on band $f'$. The expression for $\mathbb{E}[R^{(1)}_{f',t}]$ is obtained from (\ref{eq:fbfp_rate}) by substituting $f=f'$. It can be observed that expected rate under round robin or random band selection is lower than the rate in FBFP, i.e., $\mathbb{E}[{R^{(3)}_{f,t}}]\leq \mathbb{E}[{R^{(1)}_{f,t}}]$. This is because the FBFP policy selects the band that maximizes expected rate, therefore hopping to a different frequency bands does not improve the achievable rate of the SU link. 

For the DSEE policy proposed in \cite{liu2013a}, the performance of DSEE is measured in terms of regret, i.e., difference between the rate received in fixed band policy and the rate received in DSEE. Since the regret is shown to be positive in \cite[Theorem 1]{liu2013a}, we can conclude that DSEE provides lower rate as compared to the fixed band policy (FBFP). Therefore, these DBFP policies using fixed transmit power $P^{fix}_{f}$ do not provide higher rate as compared to the FBFP policy.

The interference towards PU under these policies is higher than the threshold as stated in the following theorem.

\begin{theorem}
	In dynamic band fixed power polices (DBFP), the expected interference leakage towards PU exceeds the threshold $I^0$.
	\label{thm:int_dbfp}
\end{theorem}
\begin{proof}
	{Consider that SU follows a DBFP policy, for example, round robin band selection policy and the SU is on band $f$ in slot $t$. Let $s_{f,t}=1$, i.e. PU-1 is the transmitter and PU-2 is the receiver. Under the robin robin policy, SU was on the same band during previous slots $t-F, t-2F, t-3F,\cdots$. The null space to PU receiver (PU-2) was obtained in slot $t-kF$ when PU-2 was the transmitter where $k = \arg \min_{k} \left( s_{f,t-kF}=2 \right)$. Therefore $\tau'=kF$ is the PU link reversal time perceived by the SU under this policy. We can see that $\tau'$ is larger than the PU link reversal time $\tau = \arg \min_{k} (s_{f,t-k}=2)$ in FBFP and FBDP policies wherein the SU stays on the same band and we have $\tau'>\tau$. This holds true for other DBFP policies as well. The expected interference towards PU in band $f$ under DBFP policies is given by $\mathbb{E}_{\tau'}[P^{fix}_f M_p (1-\alpha_f^{2\tau'})]=P^{fix}_fM_p \mathbb{E}_{\tau'}[ (1-\alpha_f^{2\tau'})]$. Since $\tau' > \tau$, we have $\mathbb{E}_{\tau'}[ (1-\alpha_f^{2\tau'})] > \mathbb{E}_{\tau}[ (1-\alpha_f^{2\tau})]$. Further, since the fixed power is given as $P^{fix}_f = I^0/M_p \mathbb{E}_{\tau}[(1-\alpha_f^{2\tau})]$, the average interference in band $f$ under DBFP policies is $P^{fix}_f M_p \mathbb{E}_{\tau'}[ (1-\alpha_f^{2\tau'})] = I^0 \frac{\mathbb{E}_{\tau'}[ (1-\alpha_f^{2\tau'})]}{\mathbb{E}_{\tau}[ (1-\alpha_f^{2\tau})]} > I^0$.}
\end{proof}
\begin{corollary}
	There exists no fixed transmit power for DBFP policies that provides higher rate than the single band policy while satisfying the interference constraint (\ref{eq:int_cont}).
	\label{cor:dbfp}
\end{corollary}
\begin{proof}
	Since the fixed power $P^{fix}_{f}$ incurs interference above the threshold $I^0$, one way of satisfying the interference constraint is to transmit lower power $P^{f} < P^{fix}_{f}$ that will satisfy the interference constraint in band $f$. However, this approach  reduces the rate below $\mathbb{E}[R^{(3)}_{f,t}]$ which was already lower than single band policy. Therefore, there is no transmit power that will increase the rate of dynamic band polices while satisfying the interference constraint.
\end{proof}
\subsubsection{Analysis of clairvoyant policy}
\label{sec:clair_rate}
In the clairvoyant policy, we assume that a genie-aided SU observes all $F$ frequency band in each slot, computes the null space to PUs in all bands and then selects the band offering the maximum rate. In this section, we analyze the expected gain of clairvoyant policy over FBFP. By doing so, we can find an upper bound on the achievable rate.

Let us define the expected gain of clairvoyant policy over FBFP as follows:
\begin{align}
\mathbb{E}[g^{(c)}_t] = \mathbb{E} [R^{(c)}_{f,t} - R^{(1)}_{f,t}],
\label{eq:clair_gain}
\end{align}

For simplicity, we consider that clairvoyant policy provides gain in slot $t$ if a) the PU is active in band selected by FBFP, i.e., $s_{f^*,t}=\{1,2\}$ and b) there exists a band $f'$ with no active PU. Under this condition the beamforming gain and transmitted power in band $f'$ will be higher than in band $f^*$. The probability of satisfying this condition in a time slot is $(1-\pi_{0,f^*}) \left(1- \prod_{f' \neq f^*} (1-\pi_{0,f'}\right)$ and the expected gain is given as:

\begin{align}
\mathbb{E}[g^{(c)}_t] = \frac{T_{data}}{T_{slot}} (1-\pi_{0,f^*}) \left(1- \prod_{f' \neq f^*} (1-\pi_{0,f'})\right)  \mathbb{E}\left[\log_2 \left( \frac{1+ P^0 x}{1 + P^{fix}_{f^*} y } \right)\right],
\label{eq:clair_gain2}
\end{align}
where $x$ is a random variable with pdf $f_{M_s}(x)$ as mentioned in (\ref{eq:pdf_max_eig}) and  $y$ is a random variable with pdf $f_{M_s-M_p}(y)$. The expectation on the RHS can be written as follows:
\begin{align}
\mathbb{E}\left[\log_2 \left( \frac{1+ P^0 x}{1 + P^{fix}_{f^*} y } \right)\right] =\mathbb{E}_y\left[ \mathbb{E}_x \left[\log_2 \left( \frac{1+ P^0 x}{1 + P^{fix}_{f^*} y } \right) \vert y\right]\right]
\label{eq:clair_gain3}
\end{align}
For a given value of $y$, the inner expectation is a concave function of $x$, it has an upper bound as follows:
\begin{align}
\mathbb{E}_x \left[\log_2 \left( \frac{1+ P^0 x}{1 + P^{fix}_{f^*} y } \right) \vert y \right] \leq \log_2  \left( \frac{1+ P^0 \mathbb{E} [x]}{1 + P^{fix}_{f^*} y } \right) 
= \log_2  \left( \frac{1+ P^0 M_s}{1 + P^{fix}_{f^*} y } \right) 
\label{eq:clair_gain4}
\end{align}
Substituting the above inequality in (\ref{eq:clair_gain3}), we get 
\begin{align}
\mathbb{E}\left[\log_2 \left( \frac{1+ P^0 x}{1 + P^{fix}_{f^*} y } \right)\right] &\leq \mathbb{E}_y\left[ \log_2 \left( \frac{1+ P^0 M_s}{1 + P^{fix}_{f^*} y } \right) \right]
\\   &= \frac{1}{\Gamma(M_s-M_p)}\int\limits_{0}^{\infty} \log_2 \left( \frac{1+ P^0 M_s}{1 + P^{fix}_{f^*} y } \right) y^{M_s - M_p -1} e^{-y}dy.
\label{eq:clair_gain5}
\end{align}
Therefore, the expected gain $ \mathbb{E}[g^{(c)}_t]$ in (\ref{eq:clair_gain2}) us upper bounded as follows:
{\footnotesize\begin{align}
\mathbb{E}[g^{(c)}_t] \leq \frac{T_{data}}{T_{slot}} \frac{(1-\pi_{0,f^*}) \left(1- \prod \limits_{f' \neq f^*} (1-\pi_{0,f'})\right)}{\Gamma(M_s-M_p)}  
\int\limits_{0}^{\infty} \log_2 \left( \frac{1+ P^0 M_s}{1 + P^{fix}_{f^*} y } \right) y^{M_s - M_p -1} e^{-y}dy
=  g^{c}_{max}
\label{eq:clair_gain6}
\end{align}}

Note that, in the above equation, $g^{c}_{max}$ depends on the temporal correlation $\alpha_{f}$ through $P^{fix}_{f}$. The transition probabilities of PU links also affect the gain through $P^{fix}_{f^*} =$

\noindent $ \min \left(\frac{I^0}{M_p g(\alpha_{f^*}, \mathbf{T}_{f^*})} , P^0 \right)$. We can see that as $\alpha_{f^*} \rightarrow 1$, the power $P^{fix}_{f^*} \rightarrow P^0$ and the difference between $P^0 M_s$ and $P^{fix}_{f^*}y$ reduces for any given $y$. Therefore, higher temporal correlation decreases the gain of clairvoyant policy. It should be noted that the clairvoyant policy provides the maximum rate amongst all possible policies. If the gain of this policy over FBFP is small, then it means that there exists no policy that achieves significantly higher rate than the fixed band policy.


\begin{table}
	\centering
	\caption{\small PU traffic configurations}
	\begin{tabular}{|c|c|c|}
		\hline
		Traffic config. & Transition probability matrix & $\mathbb{E}[\tau]$\\
		\hline \hline
		0 & $\mathbf{T}_f = \begin{bmatrix}
		0 & 0 & 1	\\
		1 & 0 & 0	\\
		0 & 0.2 & 0.8\\
		\end{bmatrix}$ 	 & 4.43\\ \hline
		1 & $\mathbf{T}_f = \begin{bmatrix}
		0 & 0 & 1 	\\
		0.67 & 0.33 & 0	\\
		0 & 0.5  &0.5 \\
		\end{bmatrix}$ 	 & 1.83\\ \hline
		2 & $\mathbf{T}_f =  \begin{bmatrix}
		0 & 0 & 1\\
		0.4 & 0.6 & 0\\
		0 & 1 & 0\\
		\end{bmatrix}$ 	 & 1.83\\ \hline
		3 & $\mathbf{T}_f = \begin{bmatrix}
		0 & 0 & 1\\
		0.2 &0.8 &0\\
		0 & 0.33 & 0.67\\
		\end{bmatrix}$ 	 & 4.11\\ \hline	
		4 & $\mathbf{T}_f = \begin{bmatrix}
		0 & 0 & 1\\
		0.17 & 0.83 & 0\\
		0 & 0.5 & 0.5\\
		\end{bmatrix}$ 	 & 4.67\\ \hline
		5 & $\mathbf{T}_f = \begin{bmatrix}
		0 & 0 & 1\\
		0.14 & 0.86 &	0 \\
		0 & 1 & 0\\
		\end{bmatrix}$ 	 & 5.67 \\ \hline
		6 & $\mathbf{T}_f = \begin{bmatrix}
		0 & 0 & 1\\
		1 & 0 & 0\\
		0 & 0.4 & 0.6 \\
		\end{bmatrix}$ 	 & 2.17 \\ \hline
	\end{tabular}
	\label{tab:traffic_config}
	\vspace{-5mm}
	
\end{table}

\section{Simulation results}
\label{sec:results}
In this section, we compare the performance of the policies in terms of achievable rate at SU and interference towards PU. For the simulations, the matrix $\mathbf{T}_{f}$ for PU link is constructed using traffic models of TDD LTE in {3GPP 36.211} \cite{3gpp2017} considering PU-1 is the base station and PU-2 is UE. Therefore, downlink is treated as $s_{f,t}=1$ and uplink is $s_{f,t}=2$. The transition probabilities $p_{kl,f} = \Pr(s_{f,t+1}=l|s_{f,t}=k), k,l \in \{0,1,2\}$ under these configurations are computed and are shown in Table \ref{tab:traffic_config} along with the average time for link reversal $\mathbb{E}[\tau]$ under the model. It is computed using the following expression:
\begin{align}
\mathbb{E}[\tau] = \sum_{i} i  \times \Pr (\tau=i),
\end{align}
where $\Pr (\tau=i)$ is obtained from (\ref{eq:pr_tau}). The temporal fading coefficient is modeled as $\alpha_f = J_0(2\pi f_d T_{slot})$, where $J_0(.)$ is the 0-th order Bessel function, $f_d$ is the Doppler frequency and $T_{slot}= 1$ ms. The fraction of time slot used for SU data transmission is $T_{data}/T_{slot} = 0.8$, while that for sensing is  $T_{sense}/T_{slot} = 0.2$ \cite{gabran2011}. The number of antennas at the SUs and PUs are $M_s=4$ and $M_p=1$, respectively, unless specified otherwise. Total transmit power and interference thresholds are $P^0/\sigma^2_w = 20$dB and $I^0/\sigma^2_w = -10$dB. Analytical and simulation results are shown for the power control and band selection policies.

\begin{figure}[t!]
	\centering		
	\begin{subfigure}[b]{\columnwidth}
		\centering
		\includegraphics[width=0.55\columnwidth]{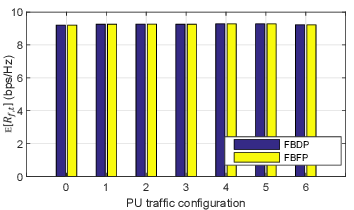}
		\vspace{-3mm}
		\caption{\small $\alpha=0.9998$ ($f_d = 5$Hz)}
		\label{fig:fbfp_vs_fbdp_large_alpha}
	\end{subfigure}
	\begin{subfigure}[b]{\columnwidth}
		\centering
		\includegraphics[width=0.55\columnwidth]{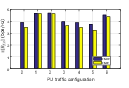}
		\vspace{-3mm}
		\caption{\small $\alpha=0.9938$ ($f_d = 25$Hz)}
		\label{fig:fbfp_vs_fbdp_small_alpha}
	\end{subfigure}		
	\vspace{-5mm}
	\caption{\small Comparison between the achievable rate of SU under FBFP and FBDP with different PU traffic configurations. $F=1$. $I^0/\sigma^2_w=-10$dB. $P^0/\sigma^2_w=20$dB.}
	\label{fig:fbfp_vs_fbdp_traffic_model}
	\vspace{-1mm}
\end{figure}

\begin{figure}[t!]
	\centering		
	\includegraphics[width=0.5\columnwidth]{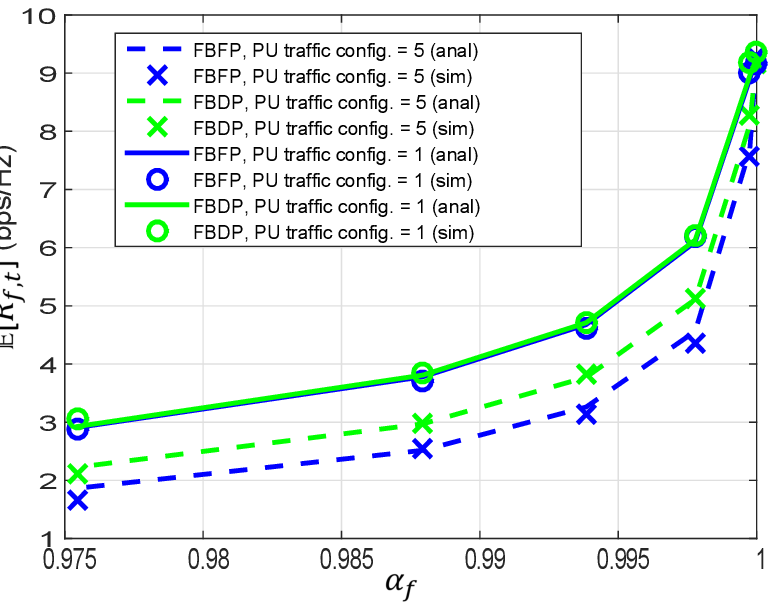}
	\caption{\small Rate at SU under FBFP and FBDP for different temporal correlations $\alpha_f \in [0.9755, 0.9998]$ or Doppler rate $f_d \in [5, 50]$. $F=1$. $I^0/\sigma^2_w=-10$dB. $P^0/\sigma^2_w=20$dB.}	
	\label{fig:fbfp_vd_fbdp_alpha}
	\vspace{-1mm}
\end{figure}

\subsubsection{Comparison between FBFP and FBDP for $F=1$}
\label{sec:results_single_band}
First, we compare the performance of fixed band policies: FBFP and FBDP for $F=1$ under the PU traffic models described above. The average rate under the two polices is shown in Fig. \ref{fig:fbfp_vs_fbdp_traffic_model} for $\alpha_f = 0.9998$ and $\alpha_f = 0.9938$. We observe that both the policies provide same rate when  $\alpha_f = 0.9998$. For $\alpha_f \rightarrow 1$, the dynamic power $P^{dyn}_{f,t} \rightarrow P^0$ and it does not change significantly with $\tau$, hence it becomes approximately constant as in FBFP. Therefore, the two policies provide same rate as shown in Fig.\ref{fig:fbfp_vs_fbdp_large_alpha}. On the other hand, as the temporal correlation decreases to $\alpha_f = 0.9938$ as shown in Fig. \ref{fig:fbfp_vs_fbdp_small_alpha}, the rate achievable rate differs under the two policies. The rate is maximum under PU traffic models 1 and 2, while its smallest under PU traffic model 5. It can be observed that the achievable rate in this case varies inversely with the average link reversal time $\mathbb{E}[\tau]$ shown in Table \ref{tab:traffic_config}. For smaller link reversal time, the PU switches its role from transmitter to receiver in short duration. Therefore, the SU transmitter has more accurate null space to the PU receiver in a given time slot and it can transmit higher power while still keeping the interference below the threshold. This in turn results in higher rate for the SU. Further, it can be seen that the rate under the two polices reduces with smaller temporal correlation as shown in Fig. \ref{fig:fbfp_vd_fbdp_alpha}. This is due to the fact for larger $\mathbb{E}[\tau]$ or smaller $\alpha_f$ the SU transmitter needs to transmit lower power $P^{fix}_{f}$ and $P^{dyn}_{f,t}$ according to (\ref{eq:p_n_fix}) and (\ref{eq:p_n_dyn}), respectively, which in turn reduces the achievable rate. These results also confirm Theorem \ref{thm:fbdp_v_fbfp} since rate under FBDP is no smaller than in FBFP.

It can also be observed from Fig. \ref{fig:fbfp_vs_fbdp_small_alpha} and \ref{fig:fbfp_vd_fbdp_alpha} that the difference in the rate of SU link under FBFP and FBDP is negligible if PU traffic configuration is 1 and 2, i.e., when $T_{f,rev}$ is small. As the link reversal time approaches 1 as in the case of traffic configurations 1 and 2, the transmitted power  under the two polices become similar: $P^{dyn}_{f,t} \approx P^{fix}_{f} \rightarrow \frac{I^0}{M_p(1-\alpha_f^2)}$, resulting in similar achievable rates.

\begin{figure}[t!]
	\centering		
	\includegraphics[width=0.6\columnwidth]{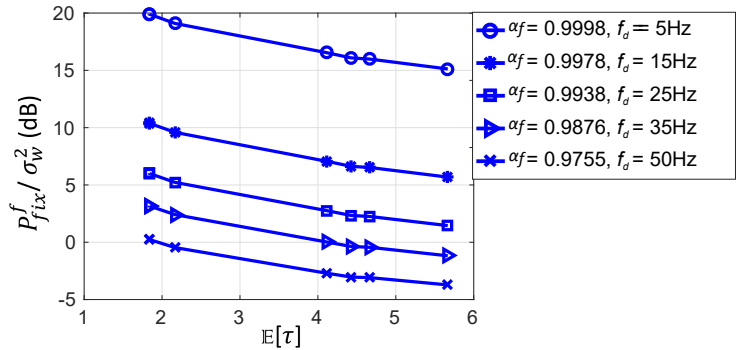}
	\caption{\small Fixed transmit power as a function of $\alpha_f$ and PU link reversal time $\mathbb{E}[\tau]$.}
	\label{fig:P_fix_v_T_rev}
	\vspace{-1mm}
\end{figure}

The fixed band polices FBFP and FBDP select the band that maximizes the power $P^{fix}_{f}$, as this band maximizes the expected rate. The transmitted power is inversely proportional to the average link reversal time of the traffic models as shown in Fig. \ref{fig:P_fix_v_T_rev}. We can observe that if there are multiple bands available with same  temporal correlations $\alpha_f$, then the fixed band policies select the band with lowest $\mathbb{E}[\tau]$. Similarly, if the link reversal time is same in different bands, the policies would select the band with maximum temporal correlation.

\begin{figure}[t!]
	\centering		
	\begin{subfigure}[b]{\columnwidth}
		\centering
		\includegraphics[width=0.6\columnwidth]{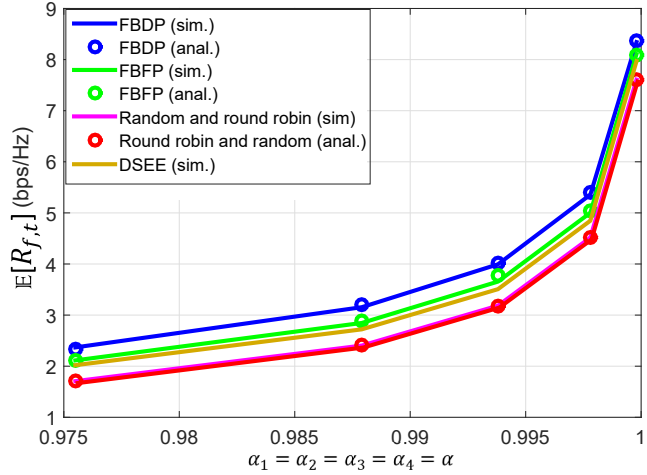}
		\caption{\small Rate under different policies.}
		\label{fig:all_policies_rate}
	\end{subfigure}
	\begin{subfigure}[b]{\columnwidth}
		\centering
		\includegraphics[width=0.6\columnwidth]{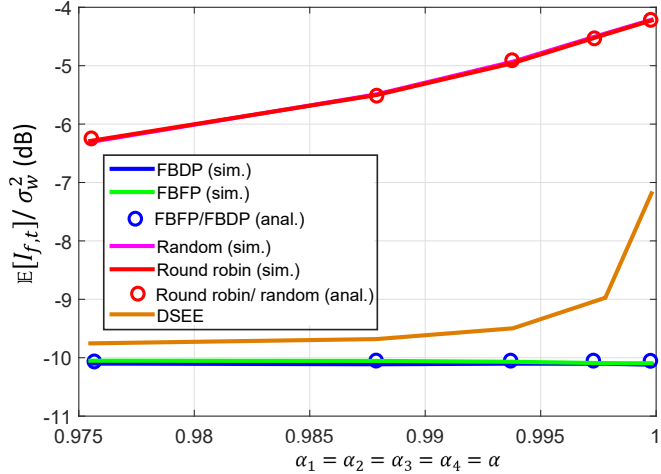}
		\caption{\small Average interference towards PU under different policies.}
		\label{fig:all_policies_int}
	\end{subfigure}		
	\caption{\small Comparison between policies with $F=4$ bands. PUs in band 1, 2, 3, and 4 follow traffic configurations 0, 3, 4, and 5, respectively. $I^0/\sigma^2_w=-10$dB. $P^0/\sigma^2_w=20$dB.}
	\label{fig:all_policies}
	\vspace{-1mm}
\end{figure}

\subsubsection{Comparison of fixed and dynamic band policies for $F=4$}
\label{sec:results_multiband}
In this section, we compare FBFP and FBDP with DBFP policies when $F=4$ and bands 1, 2, 3, and 4 have PU traffic configurations $0,3,4$ and $5$, respectively. We selected the 4 traffic configurations with smallest $\mathbb{E}[\tau]$ so that powers  $P^{dyn}_{f,t}$ and $P^{fix}_f$ are not similar. The achievable rate and interference towards PU is under different policies is shown in Fig. \ref{fig:all_policies}. We can see that the FBDP policy provides higher rate than other policies as shown in Fig. \ref{fig:all_policies_rate}. Further, the interference under dynamic band policies is higher than the required threshold as mentioned in Theorem \ref{thm:int_dbfp}. It is interesting to note that the interference under round robin, random and DSEE is not only higher than FBDP and FBFP, but it also increases with increased temporal correlation. This counter-intuitive observation can be explained as follows.

For a given traffic configuration, the function $g(\alpha_f, \mathbf{T}_f)$ in (\ref{eq:p_n_fix}) depends only on the temporal correlation $\alpha_f$. As the temporal correlation increases $g(\alpha_f, \mathbf{T}_f)$ reduces and higher power $P^{fix}_f$ is transmitted by the SU. While computing the transmit power $P^{fix}_f$, the underlying assumption is that the SU-1 has the latest null space to PU receiver. However, since SU is hopping to different bands in multi-band policies, it has older null space than what is assumed in the computation. This in turn increases the interference towards PU as shown in Fig. \ref{fig:all_policies_int}. The interference under DSEE is lower than in random and round robin due to the fact that the SU stays on the same band for a longer time under DSEE before exploring other bands \cite{liu2013a}. On the other hand, SU hops to different bands more frequently under round robin and random policies. This results in older null spaces at SU-1 causing significantly high interference towards PU receiver.

Note that to limit the interference below $I^0$, DBFP policies need to reduce the transmit power, which would further reduce the rate of the SU link. Therefore, the SU cannot simultaneously contain the interference and provide higher rate in DBFP policies as compared to fixed band policies as mentioned in Corollary \ref{cor:dbfp}.

\begin{figure}[t!]
	\centering		
	\includegraphics[width=0.6\columnwidth]{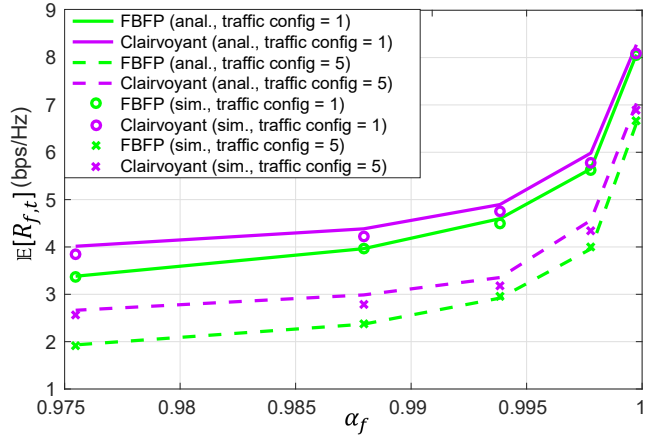}
	\caption{\small Impact of temporal correlation on the rate under FBFP and clairvoyant policies with $F=4$ bands each following same traffic configuration. $I^0/\sigma^2_w=-10$dB. $P^0/\sigma^2_w=20$dB. $M_s=4$.}
	\label{fig:clair_fbfp_alpha}
	\vspace{-5mm}
\end{figure}

\begin{figure}[t!]
	\centering		
	\includegraphics[width=0.6\columnwidth]{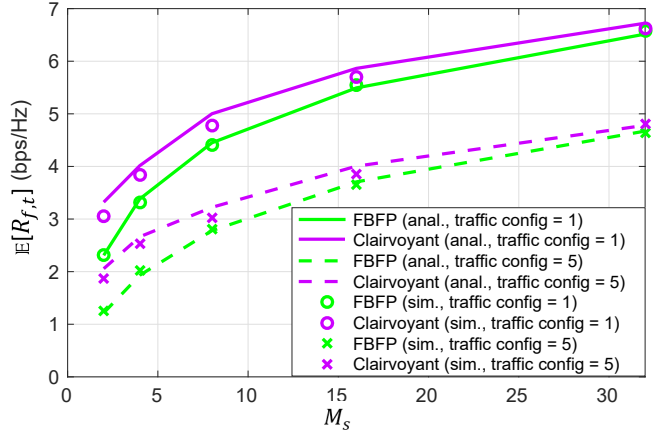}
	\caption{\small Impact of number of SU antennas on the rate under FBFP and clairvoyant policies with $F=4$ bands each following same traffic configuration. $I^0/\sigma^2_w=-10$dB. $P^0/\sigma^2_w=20$dB. $\alpha_f = 0.9755$ in each band ($f_d=50$Hz).}
	\label{fig:clair_fbfp_Ms}
	\vspace{-1mm}
\end{figure}

\subsubsection{Comparison with clairvoyant policy}
\label{sec:results_fbdp_clairvoyant}
In this section, we compare the achievable rate of clairvoyant policy and FBFP assuming all $F=4$ channels follow the same traffic configuration and temporal correlations. We consider two traffic configurations $1$ and $5$ with maximum and minimum link reversal time, respectively. It can be observed in Fig. \ref{fig:clair_fbfp_alpha} that the gain of clairvoyant policy reduces as temporal correlation increases as explained in Section \ref{sec:clair_rate}. The impact of increasing the number antennas at SUs is shown in Fig. \ref{fig:clair_fbfp_Ms}. The gain of clairvoyant policy reduces with increased number of antennas. In other words, the SU does not loose significant amount of rate even if it stays in one band. Therefore, finding empty time slots in other bands in clairvoyant policy is less beneficial in terms of increasing the rate of the SU. From these observations, we conclude that the gap between clairvoyant policy and the fixed band policy can be reduced by increasing the number of antennas at SUs.


\section{Conclusion and future work}
\label{sec:Conclusion}
\begin{figure}
	\centering
	\includegraphics[width=0.6\columnwidth]{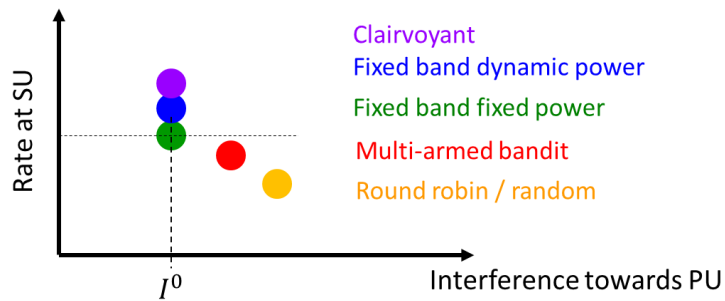}
	\caption{\small {Relative performance of policies studied in this chapter.}}
	\label{fig:policy_fig_last}
	\vspace{-1mm}
\end{figure}

We studied power control and frequency band selection policies for multi-band underlay MIMO cognitive radio with the objective of maximizing the rate of the SU while keeping the interference leakage towards PUs below specified level. First, we derived expressions for transmit power from SU for fixed and dynamic power control schemes. Then, the we studied the performance of band selection policies that use the proposed transmit power schemes. In the fixed band policies, we proved that the FBDP policy provides higher rate than the FBFP policy, while both policies keep the interference towards PU to the specified threshold. We showed that the DBFP policies, such as round robin, random and the DSEE policy based on multi-armed bandit framework, result in higher interference to PUs as compared to fixed band policies. In conclusion, the relative performance of the policies can be represented as shown in Fig. \ref{fig:policy_fig_last} which shows rate at SU versus interference to PU under the policies studied in this chapter. As observed in the figure, the genie-aided clairvoyant policy provides maximum rate at the SU. We have provided an expression for the gap between the rate achieved in optimal clairvoyant policy and the FBFP policy. We show that the gap is reduced under slow-varying channels or as the number of SU antennas is increased.

\section*{Future extension}
The fixed band policies considered in this chapter for a network with one SU pair can be extended to a more general CR network where $K$ SU pairs are coordinated by one central node that assigns frequency bands to SU pairs. Let us consider that SU pairs are indexed by $n$. In order to implement the fixed band policy, the central node computes the transmit power and corresponding rate $R_{k,f}$ for each SU pair $n$ and frequency band $f$ using (\ref{eq:fbfp_rate}). Then, it assigns frequency bands to the SU pairs by solving a weighted bipartite matching problem with the objective of maximizing the sum rate $\sum_{k,f}R_{k,f}$ for the CR network. Once the assignment is decided, the SU pairs can stay on the assigned band and utilize the proposed power control scheme to transmit their signals while limiting the interference to PUs.

\chapter{Conclusion}
\label{chapter_conclusion}
\section{Summary of contributions}
In this dissertation, we developed interference mitigation and resource allocation techniques for underlay CR networks in order to enable spectrum sharing between PUs and SUs. 

In Chapter \ref{chapter_localization}, we proposed a location estimation algorithm to estimate the location coordinates of a PU in a CR network in the presence of a spectrally overlapped interference. The proposed localization algorithm, called Cyclic WCL, estimates the location as a weighted average of location coordinates of the SUs in the network. The weights are computed using the cyclic autocorrelation (CAC) of the received signal at the cyclic frequency of the PU signal, which is assumed to have distinct cyclostationary properties as compared to the interferer. It is shown that the Cylic WCL provides lower estimation error as compared to previous WCL algorithm that uses only the received signal strength to estimate the location. Thus, the proposed algorithm mitigates the impact of a spectrally overlapped interference by separating the PU signal from the interference in the cyclostationary domain. 

The location estimation error is further reduced by eliminating the SUs in the vicinity of the interferer. Such SUs have corrupted received signal due to higher power interference component and inclusion of such SUs in the localization causes high estimation error. This elimination of SUs is implemented in the improved Cyclic WCL algorithm. In order to identify and eliminate SUs in the presence of the interferer, we proposed a statistic called feature variation coefficient (FVC) which is the ratio of variance and the mean of the square of absolute value of the CAC of the received signal at the SU. It has been observed that the proposed improved Cyclic WCL is robust to interference power level as well as shadowing effects. The estimated PU location can be used in smart routing in a CR network to steer the SU signal away from the PU.

In Chapter \ref{chapter_massive_mimo}, we considered a dual problem of interference control and SU selection in a CR network with one massive MIMO secondary base station (SBS) and multiple SUs. An optimization framework is developed in order to serve maximum number of SUs in the downlink while keeping the interference to PUs below a specified limit. In this proposed optimization framework, we take into account rate requirements for each SU. We also consider that only imperfect CSI is available at the SBS and the SBS utilizes zero-forcing beamforming in order to minimize the interference to PUs. However, due to imperfect CSI between the SBS and PUs, the interference to PU is not zero and depends on the number of SUs selected and the power allocated to them. Similarly, due to imperfect CSI between the SBS and SUs, the SUs may not receive the required rate as predicted by the SBS during power allocation. In order to resolve these challenges caused by imperfect CSI, we introduce two margin variables in the optimization problem. These variables ensure that each SU is allocated sufficiently high power to meet its rate requirement under the imperfect CSI and the number of SUs served are limited in order to keep the average interference to PUs below the specified limit.

Since the optimization framework involves integer selection variable along with continuous power variables, it is a mixed integer program and a non-convex problem. We propose a low-complexity algorithm, called DMP, for SU selection and power allocation. The analysis shows that the proposed algorithm achieves optimal performance, i.e. it serves maximum number of SUs in a massive MIMO regime when number SBS antennas is an order of magnitude large than the number of SUs and PUs in the network.

In Chapter \ref{chapter_policy}, frequency band selection and power control policies are studied for underlay MIMO CR which can choose any one frequency band at each time slot. As in the previous chapter, we also consider interference constraints in order to determine the transmit power for the SU. We show that the transmit power in each band depends on the traffic statistics of the PU transmitter-receiver link in that band as well as the temporal correlation of the channels that determined how fast the channels change from one slot to the next. The transmit power in each band is shown to be inversely proportional to the PU link reversal time which is defined as the average time taken by the PU transmitter to switch its role to receiver. The transmit power is also shown to be inversely proportional to the temporal correlation of the channels.

We analyzed four classes of band selection and power control policies: 1) fixed band fixed power (FBFP), 2) fixed band dynamic power (FBDP), 3) dynamic band fixed power (DBFP) and 4) a genie-aided clairvoyant policy. We show that the FBDP policy that dynamically updates transmit power of the SU in each slot provides higher rate that FBFP policy which uses fixed transmit power. It has been observed that the DBFP policies, such as round robin, random band selection, where SU dynamically hops to a different frequency band in each time slot incur higher interference to PUs than the FBFP and FBDP policies where SU stays on one frequency band. Finally, we provide an expression for the gap between the rate achieved by a genie-aided clairvoyant policy and the FBFP policy. It has been observed that the gap reduces if the temporal correlation is high, i.e., channels are slow-varying, and as the number of antennas at the SU is increased. This implies that if the SU is equipped with a large number of antennas or if the channels fading rate is slow, then staying on one frequency band with smallest PU link reversal time maximizes the rate of the SU link.

\section{Future works}
In this dissertation, we developed localization and resource allocation methods to enable efficient spectrum sharing in underlay CR network. The techniques presented here can be extended in the following directions.

In the Cyclic WCL localization algorithm, we estimate location coordinates of one PU in the presence of one spectrally overlapped interferer. This model can be further generalized to include localization of PU with multiple interfering signals. In the current system model, we consider single antenna SUs. In order to isolate the PU signal in the presence of multiple interfering signals, SUs equipped with multiple antennas can be incorporated in this algorithm.  

In Chapter \ref{chapter_massive_mimo}, a user selection and power allocation algorithm was developed in order to serve maximum number of SUs in one time-slot. In this mechanism, there are some SUs which would not be selected for the downlink transmission in the current time-slot and thus receive zero rate. The proposed algorithm can be extended to schedule such SUs in future time-slots so that the time-averaged rate at those SUs is non-zero. Such extension would require a scheduling mechanism that runs on the top of the proposed algorithm and decides which SUs should be considered by the proposed DMP selection algorithm in each time-slot for the downlink. Further, in the system model, we considered only one SBS equipped with a large antenna array.  This model can be extended to include multiple SBSs in the underlay CR network. Such a model would enable the operator of the secondary network to assign SUs to SBSs and ensure that fewer SUs are dropped in the downlink. A coordinated association algorithm can be developed where SBSs coordinate to determine the association between SUs and SBSs. This, however, comes with additional complexity of obtaining the CSI between all SUs and SBSs. Low-complexity algorithms will be required to establish such association.

In Chapter \ref{chapter_policy}, we considered frequency and power control policies for a CR network with one SU transmitter-receiver pair. The analysis presented in this work tells us that for a SU equipped with multiple antennas, the policy of staying on one frequency band becomes approximately optimal as the number of antennas increase. Such a fixed band policy makes it easy to assign frequency bands to SUs in a more general CR network with multiple SU pairs coordinated by one central node. If the underlay CR network has multiple frequency bands each with a given PU traffic pattern and channel fading rate, then the central coordinating node can compute the average rate achieved by a SU pair in each band. Then, it can assign one frequency band for each SU pair using bipartite matching in order to maximize the sum rate of the SUs. Once the assignment is decided, the SU transmitter-receiver pairs can stay on the assigned fixed band and utilize fixed power to transmit their signals while limiting the interference to PUs.

In summary, the methods developed in this work will enable efficient spectrum sharing by maximizing the capacity of unlicensed secondary users while protecting the licensed primary users from harmful interference.

\appendix
\chapter{Appendix for Chapter \ref{chapter_localization}}

\section{Proof: location estimate is a ratio of quadratic form in $\boldsymbol{\hat \theta}$}
\label{app_rq}
From (\ref{eq:xt}), we have the location estimate as
\begin{align}
\nonumber
{{\hat x}_t} &= \frac{{\sum\limits_{k = 1}^K {|{{\hat R}_{{r_k}}^{\alpha_t}}{|^2}} {x_k}}}
{{\sum\limits_{k = 1}^K {|{{\hat R}_{{r_k}}^{\alpha_t}}{|^2}} }} = \frac{{{{\sum\limits_{k = 1}^K {\left\| {\left[ \begin{gathered}
  {\boldsymbol{\hat \theta_r }^T} \hfill \\
  {\boldsymbol{\hat \theta_i }^T} \hfill \\ 
\end{gathered}  \right]\mathbf{p_k}} \right\|} }^2}{x_k}}}
{{\sum\limits_{k = 1}^K {{{\left\| {\left[ \begin{gathered}
  {\boldsymbol{\hat \theta_r }^T} \hfill \\
  {\boldsymbol{\hat \theta_i }^T} \hfill \\ 
\end{gathered}  \right]\mathbf{p_k}} \right\|}^2}} }} \text{ and}
\end{align}
\begin{align}
\hat{x}_t= \frac{{\sum\limits_{k = 1}^K {\mathbf{p_k}^T\left[ {{\boldsymbol{\hat \theta_r }}{\text{ }}{\boldsymbol{\hat \theta_i }}} \right]} {\text{ }}{x_k}\left[ \begin{gathered}
  {\boldsymbol{\hat \theta_r }^T} \hfill \\
  {\boldsymbol{\hat \theta_i }^T} \hfill \\ 
\end{gathered}  \right]\mathbf{p_k}}}
{{\sum\limits_{k = 1}^K {\mathbf{p_k}^T\left[ {{\boldsymbol{\hat \theta_r }}{\text{ }}{\boldsymbol{\hat \theta_i }}} \right]} {\text{ }}\left[ \begin{gathered}
  {\boldsymbol{\hat \theta_r }^T} \hfill \\
  {\boldsymbol{\hat \theta_i }^T} \hfill \\ 
\end{gathered}  \right]\mathbf{p_k}}}.
\end{align}
After rearranging the terms and taking the summation operator inside, we have
\begin{align}
\nonumber
{{\hat x}_t} = \frac{{Tr\left( {\left[ \begin{gathered}
  {\boldsymbol{\hat \theta_r }^T} \hfill \\
  {\boldsymbol{\hat \theta_i }^T} \hfill \\ 
\end{gathered}  \right]\mathop {\sum {\text{ }}}\limits_{k = 1}^K \mathbf{p_k}{\text{ }}{x_k}\mathbf{p_k}^T\left[ {{\boldsymbol{\hat \theta_r }}{\text{ }}{\boldsymbol{\hat \theta_i }}} \right]} \right)}}
{{Tr\left( {\left[ \begin{gathered}
  {\boldsymbol{\hat \theta_r }^T} \hfill \\
  {\boldsymbol{\hat \theta_i }^T} \hfill \\ 
\end{gathered}  \right]\mathop \sum \limits_{k = 1}^K  \mathbf{p_k}\mathbf{p_k}^T\left[  {{\boldsymbol{\hat \theta_r }}{\text{ }}{\boldsymbol{\hat \theta_i }}} \right]} \right)}}.
\end{align}
Further, using the definitions of $\boldsymbol{P}$ and $\boldsymbol{X}$ from (\ref{eq:matrix_def}), we get
\begin{align}
\nonumber
{{\hat x}_t} &= \frac{{Tr\left( {\left[ \begin{gathered}
  {\boldsymbol{\hat \theta_r }^T} \hfill \\
  {\boldsymbol{\hat \theta_i }^T} \hfill \\ 
\end{gathered}  \right]\mathbf{PXP}^T\left[  {{\boldsymbol{\hat \theta_r }}{\text{ }}{\boldsymbol{\hat \theta_i }}} \right]} \right)}}
{{Tr\left( {\left[ \begin{gathered}
  {\boldsymbol{\hat \theta_r }^T} \hfill \\
  {\boldsymbol{\hat \theta_i }^T} \hfill \\ 
\end{gathered}  \right]\mathbf{PP}^T\left[  {{\boldsymbol{\hat \theta_r }}{\text{ }}{\boldsymbol{\hat \theta_i }}} \right]} \right)}} \\
\nonumber &= \frac{{\left[ {{\boldsymbol{\hat \theta_r }^T}{\text{ }}{\boldsymbol{\hat \theta_i }^T}} \right]\left[ {\begin{array}{*{20}{c}}
   \mathbf{PXP}^T & \boldsymbol{0}  \\
   \boldsymbol{0} & \mathbf{PXP}^T  \\
 \end{array} } \right]\left[ \begin{gathered}
  {\boldsymbol{\hat \theta_r }} \hfill \\
  {\boldsymbol{\hat \theta_i }} \hfill \\ 
\end{gathered}  \right]}}
{{\left[ {{\boldsymbol{\hat \theta_r }^T}{\text{ }}{\boldsymbol{\hat \theta_i }^T}} \right]\left[ {\begin{array}{*{20}{c}}
   \mathbf{PP}^T & \boldsymbol{0}  \\
   \boldsymbol{0} & \mathbf{PP}^T  \\
 \end{array} } \right]\left[ \begin{gathered}
  {\boldsymbol{\hat \theta_r }} \hfill \\
  {\boldsymbol{\hat \theta_i }} \hfill \\ 
\end{gathered}  \right]}}= \frac{{{\boldsymbol{\hat \theta }^T}\mathbf{A_x}\boldsymbol{\hat \theta} }}
{{{\boldsymbol{\hat \theta }^T}\mathbf{B}\boldsymbol{\hat \theta} }},
\end{align}
where $\mathbf{A_x} = \diag(\mathbf{PXP^T, PXP^T})$ and $\mathbf{B} =\diag(\mathbf{PP^T, PP^T}) $.


\section{Derivations of $\mathbb{E}[\boldsymbol{\hat \theta}]$ and $\boldsymbol{\Sigma_{\hat \theta}}$ in terms of transmitted symbols and pulse shape.}
\label{app_derivations}
The mean and covariance matrix of $\boldsymbol{\hat{\theta}} = [\boldsymbol{\hat{\theta_r}}^T  \boldsymbol{\hat{\theta_i}}^T ]^T$ are defined
as follows:
\begin{align}
\nonumber
\mathbb{E}[\boldsymbol{\hat{\theta}}]=[\mathbb{E}[\boldsymbol{\hat{\theta}_{r}}]^{T},\mathbb{E}[\boldsymbol{\hat{\theta}_{i}}]^{T}]^{T} \text{, }
\boldsymbol{\Sigma_{\hat{\theta}}}=\mathbb{E}[\boldsymbol{\hat{\theta}\hat{\theta}}^{T}]-\mathbb{E}[\boldsymbol{\hat{\theta}}]\mathbb{E}[\boldsymbol{\hat{\theta}}]^{T},
\end{align}
where $\mathbb{E}[\boldsymbol{\hat{\theta}\hat{\theta}}^{T}]=\left[\begin{array}{cc}
\mathbb{E}[\boldsymbol{\hat{\theta}_{r}\hat{\theta}_{r}}^{T}] & \mathbb{E}[\boldsymbol{\hat{\theta}_{r}\hat{\theta}_{i}}^{T}]\\
\mathbb{E}[\boldsymbol{\hat{\theta}_{i}\hat{\theta}_{r}}^{T}] & \mathbb{E}[\boldsymbol{\hat{\theta}_{i}\hat{\theta}_{i}}^{T}]
\end{array}\right]$

In order to compute $\mathbb{E}[\boldsymbol{\hat{\theta}}]$ and $\boldsymbol{\Sigma_{\hat{\theta}}}$, the terms of the form \\$\mathbb{E}[\operatorname {Re}\{\hat{R}_{u}^{\alpha_t}\}\operatorname {Im}\{\hat{R}_{v}^{\alpha_t}\}]$, $\mathbb{E}[\operatorname {Re}\{\hat{R}_{u}^{\alpha_t}\}\operatorname {Im}\{\hat{R}_{uv}^{\alpha_t}\}]$ and $\mathbb{E}[\operatorname {Re}\{\hat{R}_{uv}^{\alpha_t}\}\operatorname {Im}\{\hat{R}_{uv}^{\alpha_t}\}]$ are required, where $u,v \in \{s_t, s_i, w_k\}$. Further, it should be noted that the four terms of the form \\$\mathbb{E}[\operatorname {Re}\{\hat{R}_{u}^{\alpha_t}\}\operatorname {Re}\{\hat{R}_{v}^{\alpha_t}\}]$, $\mathbb{E}[\operatorname {Im}\{\hat{R}_{u}^{\alpha_t}\}\operatorname {Re}\{\hat{R}_{v}^{\alpha_t}\}]$, $\mathbb{E}[\operatorname {Re}\{\hat{R}_{u}^{\alpha_t}\}\operatorname {Im}\{\hat{R}_{v}^{\alpha_t}\}]$ and $\mathbb{E}[\operatorname {Im}\{\hat{R}_{u}^{\alpha_t}\}\operatorname {Im}\{\hat{R}_{v}^{\alpha_t}\}]$ can be obtained from a single term $\mathbb{E}[\hat{R}_{u}\hat{R}_{v}^*]$, by replacing $e^{-j2\pi\alpha_t(n-m)T_s}$ with \\$\operatorname{cos}(2\pi\alpha_tnT_s)\operatorname{cos}(2\pi\alpha_tmT_s)$, $\operatorname{sin}(2\pi\alpha_tnT_s)\operatorname{cos}(2\pi\alpha_tmT_s)$, $\operatorname{cos}(2\pi\alpha_tnT_s)\operatorname{sin}(2\pi\alpha_tmT_s)$ and \\ $\operatorname{sin}(2\pi\alpha_tnT_s)\operatorname{sin}(2\pi\alpha_tmT_s)$, respectively. Therefore, only the computations of the form $\mathbb{E}[\hat{R}_{u}^{\alpha_t}\hat{R}_{v}^{\alpha_t*}]$ are shown.


\subsection{Moments of CACs and CCCs of signals $s_t$ and $s_i$}
\label{sec:moments_cac_ccc}
The moments of CACs and CCCs are computed using the moments of $s_t$ and $s_i$, as shown later in this section. Therefore, the moments of $s_t$ and $s_i$ are computed first. 


Since $a_l's$ and $c_{\kappa,l}$ zero mean, we have $\mathbb{E}[s_{t}(n)]=0$ under both single- and multi-carrier models. In single carrier signal $s_t$, the second moments of $s_t$ are
\begin{align}
\small
\nonumber
\mathbb{E}[s_{t}(n)s_{t}(m)]&=\sum_{k=-\infty}^{\infty}\sum_{l=-\infty}^{\infty}\mathbb{E}[a_{k}a_{l}]g_{n,k}g_{m,l}e^{j2\pi f_t(n+m)T_s}\\\nonumber
&=\mathbb{E}[a^{2}]e^{j2\pi f_t(n+m)T_s}\sum_{k=-\infty}^{\infty}g_{n,k}g_{m,k} \text{, and}
\\
\mathbb{E}[s_{t}(n)s_{t}^{*}(m)]&=\mathbb{E}[|a|^{2}]e^{j2\pi f_t(n-m)T_s}\sum_{k=-\infty}^{\infty}g_{n,k}g_{m,k}.
\label{eq:app1}
\end{align}

Further, the subscript $l$ is dropped after taking the expectation in the derivations, since $a_l's$ and $c_{\kappa,l}$ are i.i.d. for all $l$.
In order to compute $\mathbb{E}[|s_{t}(n)|^{2}|s_{t}(m)|^{2}]$ for single carrier signals, note that $|s_{t}(n)|^{2}= \sum_{k=-\infty}^{\infty}\sum_{l=-\infty}^{\infty}a_{k}a_{l}^{*}g_{n,k}g_{n,l}$ and $|s_{t}(m)|^{2}=\sum_{p=-\infty}^{\infty}\sum_{q=-\infty}^{\infty}a_{p}a_{q}^{*}g_{m,p}g_{m,q}$. Then, we have
\begin{align}
\nonumber \mathbb{E}[|s_{t}(n)|^{2}|s_{t}(m)|^{2}]=\sum_{k=-\infty}^{\infty}\sum_{l=-\infty}^{\infty}\sum_{p=-\infty}^{\infty}\sum_{q=-\infty}^{\infty}\mathbb{E}[a_{k}a_{l}^{*}a_{p}a_{q}^{*}]g_{n,k}g_{n,l}g_{m,p}g_{m,q}.
\end{align}
Since $a_k's$ are i.i.d. and zero mean, we have:
\begin{multline}
\mathbb{E}[|s_{t}(n)|^{2}|s_{t}(m)|^{2}] = \mathbb{E}[|a|^{4}]\sum_{k=-\infty}^{\infty}g^{2}_{n,k}g^{2}_{m,k} 
 + \mathbb{E}[|a|^{2}]^{2}\sum_{k=-\infty}^{\infty}\sum_{p=-\infty, p \neq k}^{\infty}g^{2}_{n,k}g^{2}_{m,p} 
\\+|\mathbb{E}[a^{2}]|^{2}\sum_{k=-\infty}^{\infty}\sum_{l=-\infty, l \neq k}^{\infty}g_{n,k}g_{n,l}g_{m,k}g_{m,l}
+\mathbb{E}[|a|^{2}]^{2}\sum_{k=-\infty}^{\infty}\sum_{l=-\infty, l \neq k}^{\infty}g_{n,k}g_{n,l}g_{m,l}g_{m,k}.
\label{eq:app2}
\end{multline}

 For the OFDM signal, $\mathbb{E}[s_{t}(n)s_{t}(m)]$, $\mathbb{E}[s_{t}(n)s_{t}^{*}(m)]$, and $\mathbb{E}[|s_{t}(n)|^{2}|s_{t}(m)|^{2}]$ are obtained by making the following substitutions in (\ref{eq:app1}) and (\ref{eq:app2}). First, $\mathbb{E}[a^2]$ is substituted with  $\sum_{\kappa \in \mathbb{K}}\mathbb{E}[c_{\kappa}^2]e^{j2\pi \kappa \Delta f_t (n+m)T_s}$, $\mathbb{E}[|a|^2]$ is substituted with $\sum_{\kappa \in \mathbb{K}}\mathbb{E}[|c_{\kappa}|^2]e^{j2\pi \kappa \Delta f_t (n-m)T_s}$, where $\mathbb{K} = \left \{- \frac{N_{c,t}}{2},..., \frac{N_{c,t}}{2}-1 \right\}$. Finally, $\mathbb{E}[|a|^4]$ is substituted with 
\begin{align}
\small
\nonumber \sum_{\kappa \in \mathbb{K}} \mathbb{E}[|c_{\kappa}|^4] + \sum_{\kappa_1 \in \mathbb{K}} \sum\limits_{\substack{\kappa_2 \in \mathbb{K}\\ \kappa_1 \neq \kappa_2}}\mathbb{E}[c_{\kappa_1}^2]\mathbb{E}[(c^{*}_{\kappa_2})^2]e^{j 2 \pi (\kappa_1 - \kappa_2)\Delta f_t(n+m)T_s } 
\\+\sum_{\kappa_1 \in \mathbb{K}} \sum\limits_{\substack{\kappa_2 \in \mathbb{K}\\ \kappa_1 \neq \kappa_2}}\mathbb{E}[|c_{\kappa_1}|^2]\mathbb{E}[|c_{\kappa_2}|^2]\left(1+ e^{j 2 \pi (\kappa_1 - \kappa_2)\Delta f_t(n-m)T_s }\right).
\label{eq:app2_2}
\end{align}



Similarly, the moments are computed for $s_{i}(n)$ using single carrier and OFDM signal models.
From (\ref{eq:app1}) and (\ref{eq:app2_2}), first and second order moments of the estimates of the CAC are obtained as follows:
\begin{align}
\nonumber
\mathbb{E}[ \hat{R}_{s_{t}}^{\alpha_t}]&=\frac{1}{N}\sum_{n=0}^{N-1}\mathbb{E}[|s_{t}(n)|^{2}]e^{-j2\pi \alpha_t n T_s} \text{, and} \\
\mathbb{E}[|\hat{R}_{s_{t}}^{\alpha_t}|^2]&=\frac{1}{N^{2}}\sum_{n=0}^{N-1}\sum_{m=0}^{N-1}\mathbb{E}[|s_{t}(n)|^{2}|s_{t}(m)|^{2}]e^{-j2\pi \alpha_t (n-m) T_s}.
\label{eq:app3}
\end{align}
 Further, the cross correlation between estimates of CACs of $s_t$ and $s_i$ is
\begin{align}
\small \mathbb{E}[\hat{R}_{s_{t}}^{\alpha_t}\hat{R}_{s_{i}}^{\alpha_t*}]=\frac{1}{N^{2}}\sum_{n=0}^{N-1}\sum_{m=0}^{N-1}\mathbb{E}[|s_{t}(n)|^{2}]\mathbb{E}[|s_{i}(m)|^{2}]e^{-j2\pi \alpha_t (n-m) T_s}.
\end{align}
Similarly, the mean of the CCCs between $s_t$ and $s_i$ is
\begin{align}
\nonumber
\small \mathbb{E}[\hat{R}_{s_{t}s_{i}}^{\alpha_t}]=&E \left[\frac{1}{N}\sum_{n=0}^{N-1}2\operatorname {Re}\{s_{t}(n)s_{i}^{*}(n\}e^{-j2\pi\alpha nT_{s}}\right]\\=&\frac{1}{N}\sum_{n=0}^{N-1}(\mathbb{E}[s_{t}(n)]\mathbb{E}[s_{i}(n)]^{*}+\mathbb{E}[s_{t}(n)]^{*}\mathbb{E}[s_{i}(n)])e^{-j2\pi \alpha_t n T_s}=0.
\label{eq:app5}
\end{align}
The above equation follows from the fact that $s_t$ and $s_i$ are independent and
zero mean signals. From the same argument, we can show that $\mathbb{E}[\hat{R}_{s_{t}s_{i}}^{\alpha_t}\hat{R}_{s_{i}}^{\alpha_t*}]=0$ and $\mathbb{E}[\hat{R}_{s_{t}s_{i}}\hat{R}_{s_{t}}^*]=0$.

Finally, in order to compute the second moment of CCC $\mathbb{E}[\hat{R}_{s_{t}s_{i}}\hat{R}_{s_{t}s_{i}}^*]$, let is define $Z$ as follows:
\begin{align}
\nonumber
Z=&\mathbb{E}[2\operatorname {Re}\{s_{t}(n)s_{i}^{*}(n)\}2\operatorname {Re}\{s_{t}(m)s_{i}^{*}(m)\}] \\= &2\operatorname {Re}\{\mathbb{E}[s_{t}(n)s_{t}(m)]\mathbb{E}[s_{i}(n)s_{i}(m)]^{*}\}+2\operatorname {Re}\{\mathbb{E}[s_{t}(n)s_{t}^{*}(m)]\mathbb{E}[s_{i}^{*}(n)s_{i}(m)]\}.
\label{eq:app7}
\end{align}
Then, it can be shown that
\begin{align}
\mathbb{E}[\hat{R}_{s_{t}s_{i}}^{\alpha_t}\hat{R}_{s_{t}s_{i}}^{\alpha_t*}]=\frac{1}{N^{2}}\sum_{n=0}^{N-1}\sum_{m=0}^{N-1}Ze^{-j2\pi \alpha_t (n-m)T_s}.
\label{eq:app8}
\end{align}
In the equations (\ref{eq:app3})-(\ref{eq:app8}), the required moments of $s_t$ and $s_i$ are obtained using (\ref{eq:app1}) and (\ref{eq:app2}).
The moments related to the CAC of the noise are obtained using (\ref{eq:app3})-(\ref{eq:app7}) by substituting $s_t$ and $s_i$ by $w$ and are also derived in \cite{Rebeiz2013}. The moments are as follows:
\begin{align}
\nonumber
\mathbb{E}[\hat{R}_{w}^{\alpha_t}] = \frac{\sigma_{w}^{2}}{N}e^{-j\pi\alpha_t(N-1)T_{s}}\frac{sin(\pi\alpha_t NT_{s})}{sin(\pi\alpha_t T_{s})}, \\
\mathbb{E}[\hat{R}_{w}^{\alpha_t}\hat{R}_{w}^{\alpha_t*}]=\frac{\sigma_{w}^{4}}{N^{2}}+\frac{\sigma_{w}^{4}}{N^{2}}\left(\frac{sin^{2}(\pi\alpha_t NT_{s})}{sin^{2}(\pi\alpha_t T_{s})}\right).
\label{eq:app10}
\end{align}

In order to compute $\mathbb{E}[\hat{R}_{s_{t}w}\hat{R}_{s_{t}w}^*]$, we substitute $s_i$ by $w$ in (\ref{eq:app8}) and use 
\begin{align}
\nonumber
Y=&2\operatorname {Re}\{\mathbb{E}[s_{t}(n)s_{t}(m)]\mathbb{E}[w(n)w(m)]^{*}\}\\&+2\operatorname {Re}\{\mathbb{E}[s_{t}(n)s_{t}^{*}(m)]\mathbb{E}[w^{*}(n)w(m)]\}.
\end{align}
Then we have, 
\begin{align}
\small \mathbb{E}[\hat{R}_{s_{t}w}^{\alpha_t}\hat{R}_{s_{t}w}^{\alpha_t*}]=\frac{2\sigma_{w}^{2}}{N^{2}}\sum_{n=0}^{N-1}\mathbb{E}[|s_{t}(n)|^{2}] \approx  \frac{2\sigma_{w}^{2}}{N}.
\label{eq:app11}
\end{align}
The above equation follows from the fact that, for $m\neq n$, $Y=0$, and for $m=n$, $Y=2\operatorname {Re}\{\mathbb{E}[|s_{t}(n)|^{2}]\mathbb{E}[|w(n)|^{2}]\}=2\sigma_{w}^{2}\mathbb{E}[|s_{t}(n)|^{2}]$, since $\mathbb{E}[w^{2}(n)]=0$. The approximation holds since $s_t$ is a signal with unit average energy. Similarly we can show that  $\mathbb{E}[\hat{R}_{s_{i}w}^{\alpha_t}\hat{R}_{s_{i}w}^{\alpha_t*}]\approx  \frac{2\sigma_{w}^{2}}{N}$. Using (\ref{eq:app1}) - (\ref{eq:app11}), the mean $\mathbb{E}[\boldsymbol{\hat{\theta}}]$ and covariance matrix $\boldsymbol{\Sigma_{\hat{\theta}}}$ are computed.

\section{Derivation of lower bound on $N$}
\label{sec:lower_bound_N}
In this section, we further analyze the behavior of the CAC of the interferer signal at cyclic frequency $\alpha_t$ as a function of $N$. From derivations in Appendix B, we have
\begin{align}
\nonumber
\mathbb{E}[\hat{R}_{s_{i}}^{\alpha_t}]&=\frac{\mathbb{E}[|b|^{2}]}{N}\sum_{n=0}^{N-1}\sum_{k=-\infty}^{\infty}h^{2}_{n,k}e^{-j2\pi\alpha_{t}nT_{s}}\\&=\frac{\mathbb{E}[|b|^{2}]}{N}\sum_{n=0}^{N-1}\sum_{k=-\infty}^{\infty}h^{2}(nT_{s}-kT_{h})e^{-j2\pi\alpha_{t}nT_{s}}.
\label{eq:exp_R_si}
\end{align}

Let us define $h'(t) = \sum_{k=-\infty}^{\infty}h^2(t-kT_h)$ and let $H'(\alpha)$ be the Fourier transform of $h'(t)$ at frequency $\alpha$. The period of $h'(t)$ is $T_h$ and its Fourier spectrum consists of spectral lines at frequencies $l/T_h, l=1,2,3,...$. $h'(t)$ can be expressed in terms of Fourier series coefficients $H'(l/T_h)$ as $h'(t)=\sum_{l=-\infty}^{\infty}H'(l/T_h)e^{j2\pi \left(\frac{l}{T_h}\right)t}$. The magnitude of the first spectral line at $1/T_h$ is greater than other spectral lines, since $T_h$ is the fundamental period of $h'(t)$. The first spectral line is located at at the symbol rate of the interference signal $\alpha_i=1/T_h$. Further, $h'(nT_s) =  \sum_{k=-\infty}^{\infty}h^2(nT_s-kT_h)$ is obtained by sampling $h'(t)$ at sampling rate $1/T_s > \text{max}(\alpha_t,\alpha_i)$. Let us define windowed discrete time Fourier transform of $h'(nT_s)$ at frequency $\alpha$ as $H'(\alpha, N, T_s) = \frac{1}{N} \sum_{n=0}^{N-1}h'(nT_s)e^{-j2\pi\alpha nT_s}$, where $N$ is the length of the rectangular window. It should be noted that $f_s = 1/T_s >\text{max}(\alpha_t,\alpha_i)$ to avoid aliased components in $H'(\alpha, N, T_s)$. Thus, (\ref{eq:exp_R_si}) can be expressed as $\mathbb{E}[\hat{R}_{s_{i}}^{\alpha_t}]=\mathbb{E}[|b|^{2}] H'(\alpha_t, N, T_s)$.

In order to find a condition on $N$ such that the interference caused by $h'(nT_s)$ at $\alpha_t$ is negligible, we consider only the first spectral line located at $\alpha_i=1/T_h$. If the power of this spectral component is reduced to a negligible amount at $\alpha_t$, the remaining spectral components (at $l/T_h, l>1$) will have even less power $\alpha_t$ and will not cause significant interference. Then, (\ref{eq:exp_R_si}) can be written as:
\begin{align}
\nonumber
\mathbb{E}[\hat{R}_{s_{i}}^{\alpha_t}]&={\mathbb{E}[|b|^{2}]}H'(\alpha_t, N, T_s) \\\nonumber&= \frac{\mathbb{E}[|b|^{2}]}{N} \sum_{n=0}^{N-1}h'(nT_s)e^{-j2\pi\alpha nT_s}
\\\nonumber
&\approx \frac{\mathbb{E}[|b|^{2}]}{N}\sum_{n=0}^{N-1}H'(\alpha_i)e^{-j2\pi\Delta\alpha nT_s} \\&= \frac{\mathbb{E}[|b|^{2}] H'(\alpha_i)}{N} \frac{sin(\pi\Delta\alpha NT_s)}{sin(\pi\Delta\alpha T_s)}e^{-j\pi\Delta\alpha(N-1)T_s},
\label{eq:exp_R_si_2}
\end{align}
where $\Delta\alpha = |\alpha_t - \alpha_i|$ and $H'(\alpha_i)$ is Fourier series coefficient at $\alpha_i=1/T_h$ defined as $H(\alpha_i)=\frac{1}{T_h}\int_{0}^{T_h}h'(t)e^{j2\pi\alpha_i t}$. In (\ref{eq:exp_R_si_2}), the power of interference caused $s_i$ at $\alpha_t$ is negligible if   $\left(\frac{sin(\pi\Delta\alpha NT_s)}{sin(\pi\Delta\alpha T_s)}\right)^2 \approx 0$. In order to find a lower bound on number of samples $N$, we consider $\left(\frac{sin(\pi\Delta\alpha NT_s)}{sin(\pi\Delta\alpha T_s)}\right)^2 \approx 0$ if $\left(\frac{sin(\pi\Delta\alpha NT_s)}{sin(\pi\Delta\alpha T_s)}\right)^2 < 0.001$, i.e., the power of side-lobe is $< -30$ dB. This condition is satisfied if there are at least ten side-lobes of the sinc function $\left(\frac{sin(\pi\Delta\alpha NT_s)}{sin(\pi\Delta\alpha T_s)}\right)^2$ in the interval $[\alpha_t, \alpha_i]$. Therefore, the required lower bound on $N$ is $N>10 \lceil\frac{f_s}{\Delta \alpha}\rceil =10 \lceil \frac{f_s}{|\alpha_i-\alpha_t|} \rceil$.


\section{Proof: $\phi_k$ is a strictly monotonically decreasing function of $\rho_k$ and its value lies between 0 and 1.}
\label{sec:phi_k}
From the definition, $\phi_k = \frac{{\operatorname{var} ({{\hat R}_{{r_k}}^{\alpha_t}})}}
{{\mathbb{E}[|{{\hat R}_{{r_k}}^{\alpha_t}}{|^2}]}}$, where ${\operatorname{var} ({{\hat R}_{{r_k}}^{\alpha_t}})}$ and ${\mathbb{E}[|{{\hat R}_{{r_k}}^{\alpha_t}}{|^2}]}$ are given as
\begin{align}
\nonumber
{\operatorname{var} ({{\hat R}_{{r_k}}^{\alpha_t}})} =& p_{t,k}^2{\operatorname{var} ({{\hat R}_{{s_t}}^{\alpha_t}})} + p_{i,k}^2{\operatorname{var} ({{\hat R}_{{s_i}}^{\alpha_t}})} + {\operatorname{var} ({{\hat R}_{{w}}^{\alpha_t}})} 
\\\nonumber&+ p_{t,k}p_{i,k}\mathbb{E}[|{{\hat R}_{{s_ts_i}}^{\alpha_t}}|^2] + p_{t,k}\mathbb{E}[|{{\hat R}_{{s_tw}}^{\alpha_t}}|^2] + p_{i,k}\mathbb{E}[|{{\hat R}_{{s_iw}}^{\alpha_t}}|^2],
\end{align}
\begin{align}
\nonumber
{\mathbb{E}[|{{\hat R}_{{r_k}}^{\alpha_t}}{|^2}]} =& p_{t,k}^2 \mathbb{E}[|{{\hat R}_{{s_t}}^{\alpha_t}}|^2]+ p_{i,k}^2\mathbb{E}[|{{\hat R}_{{s_i}}^{\alpha_t}}|^2] + \mathbb{E}[|{{\hat R}_{{w}}^{\alpha_t}}|^2] 
\\\nonumber &+ p_{t,k}p_{i,k}\mathbb{E}[|{{\hat R}_{{s_ts_i}}^{\alpha_t}}|^2] + p_{t,k}\mathbb{E}[|{{\hat R}_{{s_tw}}^{\alpha_t}}|^2] + p_{i,k}\mathbb{E}[|{{\hat R}_{{s_iw}}^{\alpha_t}}|^2].
\end{align}
The above equations follows from $\mathbb{E}[{{\hat R}_{{s_t}}^{\alpha_t}}{{\hat R}_{{s_ts_i}}^{\alpha_t*}}] = \mathbb{E}[{{\hat R}_{{s_t}}^{\alpha_t}}{{\hat R}_{{s_tw}}^{\alpha_t*}}]=\mathbb{E}[{{\hat R}_{{s_i}}^{\alpha_t}}{{\hat R}_{{s_ts_i}}^{\alpha_t*}}] = \mathbb{E}[{{\hat R}_{{s_i}}^{\alpha_t}}{{\hat R}_{{s_tw}}^{\alpha_t*}}]=\mathbb{E}[{{\hat R}_{{s_ts_i}}^{\alpha_t}}]=0$, which has been proved in the Appendix B. 
From Appendix B, $\mathbb{E}[|{{\hat R}_{{w}}^{\alpha_t}}|^2]$  and ${\operatorname{var} ({{\hat R}_{{w}}^{\alpha_t}})}$ are proportional to $\frac{\sigma_w^4}{N^2}$  and $\mathbb{E}[|{{\hat R}_{{s_tw}}^{\alpha_t}}|^2] = \mathbb{E}[|{{\hat R}_{{s_iw}}^{\alpha_t}}|^2]  \approx \frac{2\sigma_w^2}{N}$. We assume that the terms $\mathbb{E}[|{{\hat R}_{{w}}^{\alpha_t}}|^2]$, ${\operatorname{var} ({{\hat R}_{{w}}^{\alpha_t}})}$, $\mathbb{E}[|{{\hat R}_{{s_tw}}^{\alpha_t}}|^2]$ and $ \mathbb{E}[|{{\hat R}_{{s_iw}}^{\alpha_t}}|^2]$ are negligibly small for noise PSD = $-174$dBm/Hz, observed bandwidth at each CR = $f_s/2$ and $N > 10 \lceil f_s/\Delta \alpha \rceil$. For $N > 10 \lceil f_s/\Delta \alpha \rceil$, we have ${\operatorname{var} ({{\hat R}_{{s_i}}^{\alpha_t}})} =\mathbb{E}[|{{\hat R}_{{s_i}}^{\alpha_t}}|^2] $, since $\mathbb{E}[{{\hat R}_{{s_i}}^{\alpha_t}}] = 0$ as shown in Appendix C. Then, $\phi_k$ reduces to the following
\begin{align}
{\phi _k} &= \frac{{\operatorname{var} ({{\hat R}_{{r_k}}^{\alpha_t}})}}
{{\mathbb{E}[|{{\hat R}_{{r_k}}^{\alpha_t}}{|^2}]}} = \frac{\rho_k^2\operatorname{var} ({{\hat R}_{{s_t}}^{\alpha_t}}) + \mathbb{E}[|{{\hat R}_{{s_i}}^{\alpha_t}}{|^2}] + \rho_k \mathbb{E}[|{{\hat R}_{{s_t}{s_i}}^{\alpha_t}}{|^2}]}
{\rho_k^2\mathbb{E}[|{{\hat R}_{{s_t}}^{\alpha_t}}{|^2}] + \mathbb{E}[|{{\hat R}_{{s_i}}^{\alpha_t}}{|^2}] + \rho_k \mathbb{E}[|{{\hat R}_{{s_t}{s_i}}^{\alpha_t}}{|^2}]}
\nonumber
\\&= \frac{v_k}
{e_k} = \frac{\rho_k^2 v_{t} + e_{i} + \rho_k e_{ti}}
{\rho_k^2 e_t + e_i + \rho_k e_{ti}}.
\label{eq:app_phi2}
\end{align}
For simplicity of notations, operators ${\operatorname{var}(.)}$ and $\mathbb{E}[|.|^2]$ are replaced by variables $v$ and $e$, respectively. The theoretical value of $\phi_k$ in (\ref{eq:app_phi2}) is computed using moments derived in Appendix B. We note that all the variables in above equation are non-negative. 

Let $\rho_{k1} > \rho_{k2}$, then we have ${\phi _{k1}} = \frac{\rho_{k1}^2 v_{t} + e_{i} + \rho_{k1} e_{ti}}
{\rho_{k1}^2 e_t + e_i + \rho_{k1} e_{ti}}
$ and ${\phi _{k2}} = \frac{\rho_{k2}^2 v_{t} + e_{i} + \rho_{k2} e_{ti}}
{\rho_{k2}^2 e_t + e_i + \rho_{k2} e_{ti}}$.
To prove that $\phi_{k1}<\phi_{k2}$, consider the following steps:
\begin{align}       
{\phi _{k1}} &\lesseqgtr {\phi _{k2}} 
\nonumber\\
\frac{{\rho _{k1}^2{v_t} + {v_i} + {\rho _{k1}}{v_{ti}}}}
{{\rho _{k1}^2{e_t} + {v_i} + {\rho _{k1}}{v_{ti}}}} &\lesseqgtr \frac{{\rho _{k2}^2{v_t} + {v_i} + {\rho _{k2}}{v_{ti}}}}
{{\rho _{k2}^2{e_t} + {v_i} + {\rho _{k2}}{v_{ti}}}}.
\end{align}
After cross-multiplying with denominators and removing the common terms on both sides, we have:
\begin{align}
 [{e_i}(\rho _{k1}^2 - \rho _{k2}^2) + {e_{ti}}{\rho _{k1}}{\rho _{k2}}({\rho _{k1}} - {\rho _{k2}})]({v_t} - {e_t}) &\lesseqgtr 0.
\label{eq:inequality}
\end{align}
Since $\rho_{k1}>\rho_{k2}$ and $v_t - e_t = \operatorname{var} ({{\hat R}_{{s_t}}}) - \mathbb{E}[|{{\hat R}_{{s_t}}}{|^2}] = - |\mathbb{E}[{\hat R}_{{s_t}}]|^2 < 0$, the LHS in the above equation is less than zero. Therefore, ${\phi _{k1}} < {\phi _{k2}}$. Hence, $\phi_k$ is a strictly monotonically decreasing function of $\rho_k$. Further, $\phi_k$ is a ratio of two non-negative quantities, $\phi_k \geq 0$. And,
\begin{align}
{\phi _k} &= \frac{{\operatorname{var} ({{\hat R}_{{r_k}}^{\alpha_t}})}}
{{\mathbb{E}[|{{\hat R}_{{r_k}}^{\alpha_t}}{|^2}]}} = 1 - \frac{{|\mathbb{E}[{{\hat R}_{{r_k}}^{\alpha_t}}{}]|^2}}{{\mathbb{E}[|{{\hat R}_{{r_k}}^{\alpha_t}}{|^2}]}}.
\end{align}
Since $\frac{{|\mathbb{E}[{{\hat R}_{{r_k}}^{\alpha_t}}{}]|^2}}{{\mathbb{E}[|{{\hat R}_{{r_k}}^{\alpha_t}}{|^2}]}} \geq 0$, $\phi_k$ can not exceed 1. Therefore, $0 \leq \phi_k \leq 1$.


\section{Derivations of $\mu_{v_s} $, $\sigma_{v_s}^2$, $\mu_{e_s} $, $\sigma_{e_s}^2$ and $\sigma_{v_se_s}$}
From Section \ref{sec:phi_k_M}, we have
$\mu_{v_{s}} = \operatorname{var}(\hat{R}_{r_k}^{\alpha_t})  \text{ and }
\sigma_{v_{s}}^2 = \frac{1}{M}\left[{\mu_4 - \frac{M-3}{M-1}\mu_{v_{s}}^2}\right]$,
where $\mu_4 = E\{|\hat{R}_{r_k}^{\alpha_t} - \mathbb{E}[\hat{R}_{r_k}^{\alpha_t}]|^4\} = 2[\operatorname{var}({\hat{R}_{r_k}^{\alpha_t}})]^2 $ from \cite{Miller1969}. Further we have, $\mu_{e_{s}} =  \mathbb{E}[|\hat{R}_{r_k}^{\alpha_t}|^2] \text{ and }
\sigma_{e_{s}}^2 = \operatorname{var}(|\hat{R}_{r_k}^{\alpha_t}|^2)/M
$.
Note that $|\hat{R}_{r_k}^{\alpha_t}|^2$ can be expressed in terms of $\boldsymbol{\hat{\theta}}$, $\mathbf{p_k}$ using (\ref{eq:vector_def}) and (\ref{eq:cac_rk_expanded}) as 
\begin{align}
|\hat{R}_{r_k}^{\alpha_t}|^2 = {{\left[ {{\boldsymbol{\hat \theta_r }^T}{\text{ }}{\boldsymbol{\hat \theta_i }^T}} \right]\left[ {\begin{array}{*{20}{c}}
   \mathbf{p_kp_k}^T & \boldsymbol{0}  \\
   \boldsymbol{0} & \mathbf{p_kp_k}^T  \\
 \end{array} } \right]\left[ \begin{gathered}
  {\boldsymbol{\hat \theta_r }} \hfill \\
  {\boldsymbol{\hat \theta_i }} \hfill \\ 
\end{gathered}  \right]}}
= {{{\boldsymbol{\hat \theta }^T}\mathbf{A_k}\boldsymbol{\hat \theta} }},
\end{align}
where $\mathbf{A_k} = \diag(\mathbf{p_kp_k}^T, \mathbf{p_kp_k}^T)$. Therefore, $|\hat{R}_{r_k}|^2$ is a quadratic form in Gaussian vector ${\boldsymbol{\hat \theta }}$. Hence, the mean and variance of $|\hat{R}_{r_k}|^2$ can be written using \cite[Eqn. 6]{Magnus1986} as follows:
\begin{align}
\nonumber
\mathbb{E}[|\hat{R}_{r_k}^{\alpha_t}|^2] &= Tr(\mathbf{A_k}\boldsymbol{\Sigma_{\hat \theta}}) + \mathbb{E}[\boldsymbol{\hat \theta}]^T \mathbf{A_k}\mathbb{E}[\boldsymbol{\hat \theta}], \text{ }
\\
\operatorname{var}(|\hat{R}_{r_k}^{\alpha_t}|^2) &=  2Tr[(\mathbf{A_k} \boldsymbol{\Sigma_{\hat \theta}})^2] + 4\mathbb{E}[\boldsymbol{\hat \theta}]^T \mathbf{A_k}\boldsymbol{\Sigma_{\hat \theta}} \mathbf{A_k}\mathbb{E}[\boldsymbol{\hat \theta}]
\label{eq:app_e_3}
\end{align}
Computations of $\mathbb{E}[\boldsymbol{\hat \theta}]$ and $\Sigma_{\hat \theta}$ are shown in Appendix A. $\mathbb{E}[\hat{R}_{r_k}^{\alpha_t}]$ is also obtained from the moments derived in Appendix A. Using (\ref{eq:app_e_3}), $\mu_{v_s} $, $\sigma_{v_s}^2$, $\mu_{e_s} $, $\sigma_{e_s}^2$ are computed.

Next, we have to compute $\mathbb{E}[{v_s}{e_s}] $ in order to obtain $\sigma_{v_se_s} = \mathbb{E}[{v_s}{e_s}] -  \mu_{v_s} \mu_{e_s}$, where $v_s$ and $e_s$ are sample variance and sample second order moment of $\hat{R}_{r_k}^{\alpha_t}$ using $M$ samples. The sample mean ($m_s$), the sample second order moment and sample variance of $\hat{R}_{r_k}^{\alpha_t}$ are defined in (\ref{eq:sample_fvc}). Therefore, $\mathbb{E}[{v_s}{e_s}] $ becomes
\begin{align}
\mathbb{E}[{v_s}{e_s}] =\frac{M-1}{M}\left[\mathbb{E}[v_s^2] + \mathbb{E}[v_s |m_s|^2]\right],
\label{eq:corr_v_e}
\end{align}
 where $\mathbb{E}[v_s^2] = \sigma_{v_s}^2 + \mu_{v_s}^2$, and  $\mathbb{E}[v_s|m_s|^2]$ is obtained as follows. Although counter-intuitive, it has been shown in \cite{Shanmugam2008}, that the correlation coefficient between the sample mean $m_s$ and the sample variance $v_s$ are is zero if the samples are taken from a symmetric distribution such as Gaussian distribution. Therefore, the correlation coefficient between $v_s$ and $m_s$ is zero, i.e., $v_s$ and $m_s$ are uncorrelated.
Since $v_s$ and $m_s$ are Gaussian variables, they are also independent due to uncorrelatedness. Therefore, $v_s$ and $|m_s|^2$ are also independent which gives us $\mathbb{E}[v_s|m_s|^2] = \mathbb{E}[v_s]\mathbb{E}[|m_s|^2] = \mu_{v_s}\mathbb{E}[|m_s|^2]$, where $\mathbb{E}[|m_s|^2] = var(\hat{R}_{r_k})/M + |\mathbb{E}[\hat{R}_{r_k}]|^2 $. Substituting in (\ref{eq:corr_v_e}) we get:
$\mathbb{E}[{v_s}{e_s}] = \frac{M-1}{M} \left[\sigma_{v_s}^2 + \mu_{v_s}^2 +\mu_{v_s}\left(\frac{\operatorname{var}(\hat{R}_{r_k}^{\alpha_t})}{M} + |\mathbb{E}[\hat{R}_{r_k}^{\alpha_t}]|^2 \right) \right].$

\chapter{Appendix for Chapter \ref{chapter_massive_mimo}}

\section{Derivation of distributions of $P^{\mathcal{S}}_k$ and $P^{\mathcal{S}_0}_k$}
\label{app_pj_dist}
From (\ref{eq:P_k_vup}), $P^{\mathcal{S}}_j$ can be expressed as
\begin{align}
P^{\mathcal{S}}_k =  \frac{(2^{R^0_k} -1)\left(\sigma^2_w + {I}_k+\epsilon_2\right)}{|\mathbf{  \hat{h}}^H_k \mathbf{v}^{\mathcal{S}}_k|^2} = (2^{R^0_k}-1)\frac{X}{Y},
\end{align}
where $X = \sigma^2_w + {I}_k + \epsilon_2$ and $Y = |\mathbf{\hat{h}}^H_k \mathbf{v}^{\mathcal{S}}_k|^2 = |({\mathbf{h}^H_k + \mathbf{\delta}^H_k}) \mathbf{v}^{\mathcal{S}}_k|^2$. The vector $\mathbf{h}^H_k + \mathbf{\delta}^H_k \sim CN(0, \beta_k +\sigma^2_\delta)$ is an isotropic vector, while the vector $\mathbf{v}^{\mathcal{S}}_k$ spans $M-|\mathcal{S}|-L+1$ dimensional space due to $|\mathcal{S}|-1+L$ nulls. Therefore, $Y$ is modeled as a Gamma random variable with shape and scale parameters $M - |\mathcal{S}|-L+1$ and $\beta_k + \sigma^2_\delta$, respectively \cite[lemma 1]{hosseini2014} \cite{chaudhari2017a}, i.e., $Y \sim \Gamma(M - |\mathcal{S}|-L+1, \beta_k + \sigma^2_\delta)$. For simplicity of the analysis, we approximate $Y$ with its average value: $Y=(M - |\mathcal{S}|-L+1)(\beta_k + \sigma^2_\delta)$. This approximation is valid because the variance of $Y$, $(M - |S|-L+1)(\beta_k + \sigma^2_\delta)^2$, is negligible as compared to its mean  $(M - |\mathcal{S}|-L+1)(\beta_k + \sigma^2_\delta)$.

Further, consider the variable $X = \sigma^2_w + {I}_{k} +\epsilon_2$. We define a constant $C=\sigma^2_w + \epsilon_2$. Each term in the summation ${I}_k = \sum_{l \in \mathcal{T}}P_p |{{h}_{lk}}|^2 $ is modeled as a Gamma random variable with distribution $\Gamma(1,P_p\beta_{lk})$. Therefore, the mean of $X$ is $C+ \sum_{l\in \mathcal{T}}P_p \beta_{lk}$, while its variance is $\sum_{l\in \mathcal{T}}{P_p}^2\beta_{lk}^2$. We model $X$ as a Gamma random variable with size and shape parameters $\kappa^p_k$ and $\theta^p_k$, respectively, i.e., $X \sim \Gamma(\kappa^p_k, \theta^p_k)$ Therefore, we have
\begin{align}
\nonumber \mathbb{E}[X] = \kappa^p_k\theta^p_k= C+\sum_{l\in \mathcal{T}} P_p \beta_{lk} ,
\end{align}
\begin{align}
\mathbf{var}[X] = \kappa^p_k(\theta^p_k)^2=  \sum_{l\in \mathcal{T}}{P_p}^2\beta_{lk} ^2.
\label{eq: app_kPk_theta_Pk}
\end{align}
By solving for $\kappa^p_k$ and $\theta^p_k$, we obtain (\ref{eq:thm1_1}). 

Finally, $P^{\mathcal{S}}_k = \frac{2^{R^0_k}-1}{(M - |S|-L+1)(\beta_k + \sigma^2_\delta)}  X = \gamma^{\mathcal{S}}_k X$ is a Gamma random variable with size and shape parameters $\kappa^p_k$ and $\gamma^{\mathcal{S}}_k \theta^p_k$, respectively, i.e., $P^{\mathcal{S}}_k \sim \Gamma(\kappa^p_k, \gamma^{\mathcal{S}}_k \theta^p_k)$. Note that $P^{\mathcal{S}}_k, k=1,2,...$ are independent variables since they are functions of independent random variables ${{h}_{lk}}$.

The distribution of $P^{\mathcal{S}_0}_k$ is obtained by following the above derivation with $Y=(M-K-L+1)(\beta_k +\sigma^2_\delta)$. This is due to the fact that the vector $\mathbf{v}^{\mathcal{S}_0}_k$, in this case, spans $M-K-L+1$ dimensional space due to $K-1+L$ nulls.


\section{Proof of corollary \ref{cor:dmpvup_P_k}}
\label{app_cor_dmpvup_P_k}
Consider two sets $\mathcal{S}_1$ and $\mathcal{S}_2$ containing SU-$k$ such that $\mathcal{S}_2 \subset \mathcal{S}_1$ and $|\mathcal{S}_1|> |\mathcal{S}_2|$. Since $P^{\mathcal{S}_1}_k$ and $P^{\mathcal{S}_2}_k$ are Gamma random variables, their CDFs can be written as follows:
\begin{align}
\nonumber \Pr(P^{\mathcal{S}_1}_k \leq x) = \frac{1}{\Gamma(\kappa^p_k)} \int_{0}^{x/(\gamma^{\mathcal{S}_1}_k\theta^p_k)} t^{\kappa^p_k-1}e^{-t}dt
\\\Pr(P^{\mathcal{S}_2}_k \leq x) =\frac{1}{\Gamma(\kappa^p_k)} \int_{0}^{x/(\gamma^{\mathcal{S}_2}_k\theta^p_k)} t^{\kappa^p_k-1}e^{-t}dt.
\end{align}
Since $\gamma^{\mathcal{S}_1}_k > \gamma^{\mathcal{S}_2}_k$, we get $\Pr(P^{\mathcal{S}_1}_k \leq x) < \Pr(P^{\mathcal{S}_2}_k \leq x)$ or $\Pr(P^{\mathcal{S}_1}_k \geq x) > \Pr(P^{\mathcal{S}_2}_k \geq x)$. 


\section{Derivation of $\Pr(R^{\mathcal{S}}_k \geq y)$}
\label{app_R_k}
The rate achieved at SU-$k$ is given as
\begin{align}
R^{\mathcal{S}}_{k} = \log_2 \left( 1 + \frac{P^{\mathcal{S}}_k |\mathbf{ h}_{k}^H \mathbf{v}^{\mathcal{S}}_k|^2}{\sigma^2_w + I_k+ \sum \limits_{j \in \mathcal{S}, j \neq k} I_{jk} }\right), k \in \mathcal{S},
\end{align}
where $\mathcal{S}$ is the selected set. Substituting for $P_k^{\mathcal{S}}$ from (\ref{eq:P_k_vup}) in the above equation, we get
\begin{align}
 \Pr(R^{\mathcal{S}}_k \geq y)   =
\Pr \left(\frac{|\mathbf{ h}_{k}^H \mathbf{v}^{\mathcal{S}}_k|^2}{|\mathbf{ \hat{h}}_{k}^H \mathbf{v}^{\mathcal{S}}_k|^2} \frac{\sigma^2_w + {I}_k + \epsilon_2}{\sigma^2_w + I_k+ \sum \limits_{j \in \mathcal{S}, j \neq k} I_{jk} } \geq \frac{2^y-1}{2^{R^0_k}-1}\right).
\label{eq:app_Pr_R_K}
\end{align}
Similar to the variable $Y$ in the previous appendix, variables $|\mathbf{ h}_{k}^H \mathbf{v}^{\mathcal{S}}_k|^2$ and $|\mathbf{\hat{ h}}_{k}^H \mathbf{v}^{\mathcal{S}}_k|^2$ are approximated with their average values $(M-|\mathcal{S}|-L+1)\beta_k$ and $(M-|\mathcal{S}|-L+1)(\beta_k + \sigma^2_\delta)$, respectively. Let us define $C_y$ as follows:
\begin{align}
C_y = \frac{|\mathbf{ h}_{k}^H \mathbf{v}^{\mathcal{S}}_k|^2}{|\mathbf{ \hat{h}}_{k}^H \mathbf{v}^{\mathcal{S}}_k|^2}\left(\frac{2^{R^0_k}-1}{2^y-1}\right) =  \frac{\beta_k}{\beta_k + \sigma^2_\delta} \left(\frac{2^{R^0_k}-1}{2^y -1}\right).
\end{align}
Substituting the above equation, ${I}_{k}=   \sum_l P_p |{{h}_{lk}}|^2 $, and $I_{jk}=P_j |\mathbf{  h}_{k}^H \mathbf{v}_j|^2$ in (\ref{eq:app_Pr_R_K}), we can rewrite the equation as follows:
\begin{align}
 \Pr(R^{\mathcal{S}}_k \geq y) = &\Pr\left( (1-C_y)\sum_{l\in \mathcal{T}}P_p|{h_{lk}}|^2+ \sum_{j \in \mathcal{S}, j \neq k}P^{\mathcal{S}}_j |\mathbf{  h}_{k}^H \mathbf{v}_j|^2\leq \zeta_y\right)
\\= &\Pr\left(\sum_{l\in \mathcal{T}}Z_{lk} + Z_k \leq \zeta_y \right),
\label{eq:app_P_R_K_2}
\end{align}
where $Z_{lk}=(1-C_y)P_p|{h_{lk}}|^2, Z_k= \sum_{j \in \mathcal{S}, j \neq k}P^{\mathcal{S}}_j |\mathbf{  h}_{k}^H \mathbf{v}^{\mathcal{S}}_j|^2 $, and $\zeta_y$ as defined in (\ref{eq:kz_thetaz}). Since $Z_{lk}$ and $Z_k$ are independent random variables, the cdf in the RHS of (\ref{eq:app_P_R_K_2}) can be expressed in terms of Fourier transforms of the characteristic functions these variables. Therefore, we derive the characteristic functions of $Z_{lk}$ and $Z_k$. The variable $Z_{lk}$ is a Gamma random variable $\sim \Gamma(1,(1-C_y)P_p\beta_{lk})$ with characteristic function:
\begin{align}
\phi_{lk}(jt) = (1-\theta^z_{lk}jt),
\label{eq:app_phi_lk}
\end{align}
where $\theta^z_{lk} = (1-C_y)P_p\beta_{lk}$.
Further, the variable $Z_k= \sum_{j \in \mathcal{S}, j \neq k}P^{\mathcal{S}}_j |\mathbf{h}_{k}^H \mathbf{v}^{\mathcal{S}}_j|^2$ can be written as $Z_k= \sum \limits_{j \in \mathcal{S}, j \neq k}P^{\mathcal{S}} |\mathbf{\hat{h}}_{k}^H \mathbf{v}^{\mathcal{S}}_j  + \mathbf{\delta}_k^H \mathbf{v}^{\mathcal{S}}_j|^2$ $=  \sum_{j \in \mathcal{S}, j \neq k}P^{\mathcal{S}}_j |\mathbf{\delta}_k^H \mathbf{v}^{\mathcal{S}}_j|^2$. The second equality follows from $\mathbf{\hat{h}_k v^{\mathcal{S}}_j} =0$ due to zero forcing beamforming. The term $|\mathbf{\delta}_k^H \mathbf{v}^{\mathcal{S}}_j|^2$ is the projection of isotropic vector $\delta_k \sim CN(0,\sigma^2_\delta)$ on uncorrelated space spanned by $\mathbf{v}^{\mathcal{S}}_k$, which gives $|\mathbf{\delta}_k^H \mathbf{v}^{\mathcal{S}}_j|^2 \sim \Gamma(1,\sigma^2_\delta)$ \cite[lemma 3]{hosseini2014}. Therefore, $P^{\mathcal{S}}_j |\mathbf{\delta}_k^H \mathbf{v}^{\mathcal{S}}_j|^2$ is a product of two Gamma random variables and is approximated as a Gamma random variable \cite{coelho2014}. The mean and the variance of $P^{\mathcal{S}}_j |\mathbf{\delta}_k^H \mathbf{v}^{\mathcal{S}}_j|^2$ are given below:
\begin{align}
\nonumber\mathbb{E}[P^{\mathcal{S}}_j |\mathbf{\delta}_k^H \mathbf{v}^{\mathcal{S}}_j|^2]&= \sigma^2_\delta \gamma^{\mathcal{S}}_j \theta^p_j \frac{\Gamma(\kappa^p_j+1)}{\Gamma(\kappa^p_j)}
\\ \mathbf{var}[P^{\mathcal{S}}_j |\mathbf{\delta}_k^H \mathbf{v}^{\mathcal{S}}_j|^2]&= (\sigma^2_\delta \gamma^{\mathcal{S}}_j \theta^p_j)^2 \frac{2\Gamma(\kappa^p_j+2)\Gamma(\kappa^p_j)- \Gamma^2(\kappa^p_j+1)}{\Gamma^2(\kappa^p_j)}.
\label{eq:app_mean_var_zk}
\end{align}
The variable $Z_k$ is modeled as a Gamma random variable with shape parameter $\kappa^z_k$ and shape scale parameter $\theta^z_k$. The parameters are computed using moment matching method \cite[lemma 3]{hosseini2014} by solving the following two equations for $\kappa^z_k$ and $\theta^z_k$:
\begin{align}
\nonumber \kappa^z_k{\theta^z_k}=\sum_{j \in \mathcal{S}, j \neq k} \mathbb{E}[P^{\mathcal{S}}_j |\mathbf{\delta}_k^H \mathbf{v}^{\mathcal{S}}_j|^2],
\\\kappa^z_k({\theta^z_k})^2 =\sum_{j \in \mathcal{S}, j \neq k} \mathbf{var}[P^{\mathcal{S}}_j |\mathbf{\delta}_k^H \mathbf{v}^{\mathcal{S}}_j|^2].
\label{eq:app_kz_thetaz}
\end{align}
Expressions in (\ref{eq:kz_thetaz}) follow from (\ref{eq:app_mean_var_zk}) and (\ref{eq:app_kz_thetaz}). The characteristic function of the Gamma random variable $Z_k$ is as follows \cite{Garcia2008}:
\begin{align}
\phi_k(jt) = (1-\theta^z_k jt )^{-\kappa^z_k}.
\label{eq:app_phi_k}
\end{align}
Since $Z_{lk}$ and $Z_k$ are independent random variables, the cdf in the RHS of (\ref{eq:app_P_R_K_2}) can be written in terms of the Fourier transform of the product of characteristic functions of these random variables as follows:
\begin{align}
\Pr(R^{\mathcal{S}}_k \geq y) = \frac{1}{2\pi}\int \limits_{-\infty}^{\zeta_y} \int\limits_{-\infty}^{\infty}\left(\prod_{l\in \mathcal{T}}\phi_{lk}(jt)\right) \phi_k(jt) e^{-j2\pi wt} dt dw.
\end{align}
Substituting for $\phi_{lk}(jt)$ and $\phi_k(jt)$ from (\ref{eq:app_phi_lk}) and (\ref{eq:app_phi_k}), respectively, we obtain the expression (\ref{eq:Pr_R_k}). The expression for $\Pr(R^{\mathcal{S}}_k \geq y)$ under DMP without vector update is computed by following the above derivation and replacing $\gamma^{\mathcal{S}}_j$ by $\gamma^{\mathcal{S}_0}_j$ in (\ref{eq:app_mean_var_zk}).

If we have $\sigma^2_\delta = 0$, $\mathbf{\hat{h}}_{k}=\mathbf{{h}}_{k}$, and $I_{jk}=0$. Then, substituting $y=R^0_k$ in
(\ref{eq:app_Pr_R_K}), we get $\Pr(R^\mathcal{S}_k \geq R^0_k)=1$.

%
%
%


\section{Derivation of $\mathbb{E}[I_{lk}]$}
\label{app_I_kl}
In DMP, we have $\mathbb{E}\left[I_{kl}\right] =\mathbb{E}[P^{\mathcal{S}}_k |\mathbf{\Delta}^H_{l0}\mathbf{v}^{\mathcal{S}}_k|^2]$. The expression $\mathbb{E}[P^{\mathcal{S}}_k |\mathbf{\Delta}^H_{l0}\mathbf{v}^{\mathcal{S}}_k|^2]$ is obtained by following the derivation of $\mathbb{E}[P^{\mathcal{S}}_j |\mathbf{\delta}_k^H \mathbf{v}^{\mathcal{S}}_j|^2]$ in Appendix \ref{app_R_k} and replacing $\mathbf{\delta}_k$ and $\sigma^2_{\delta}$ with $\mathbf{\Delta}_{l0}$ and $\sigma^2_{\Delta}$, respectively. In DMP without vector update, the expression for $\mathbb{E}\left[I_{kl}\right]$ is obtained by replacing $\gamma^{\mathcal{S}}_k$ with $\gamma^{\mathcal{S}_0}_k$ in (\ref{eq:I_kl_1}).

\chapter{Appendix for Chapter \ref{chapter_policy}}

\section{Estimation of PU link state}
\label{app:st_est}
{SU-1 and SU-2 can independently estimate $s_{f,t}$ based on the received signals during the sensing duration. Here, we describe estimation at SU-1. Whether the PU link is active, i.e. $s_{f,t}\in \{1,2\}$ or inactive, i.e. $s_{f,t}=0$, is identified by energy detection \cite{soltanmohammadi2013}. In order to identify $s_{f,t}=1$ from $s_{f,t}=2$, the signal received at SU-1 are classified using a hypothesis test. Consider that SU-1 computes the received covariance $\mathbf{\hat{Q}}_{1,f,t-\tau}$ and null space $\mathbf{A}_{1,f,t-\tau}$ in slot $t-\tau$ and labels the PU state as $s_{f,t-\tau}=1$. Then in slot $t$, SU-1 computes the covariance matrix $\mathbf{\hat{Q}}_{1,f,t}$. In order to determine whether $s_{f,t}=1$ or $s_{f,t}=2$, SU-1 runs a binary hypothesis test: $\mathcal{H}_1$ indicates $s_{f,t}=1$ and $\mathcal{H}_2$ indicates  $s_{f,t}=2$. For the test, SU-1 uses the signal power received in the previously computed null space $P_{null} = \text{Tr}(\mathbf{A}_{1,f,t-\tau}^H \mathbf{\hat{Q}}_{1,f,t}\mathbf{A}_{1,f,t-\tau})$ \cite{chaudhari2017}. The hypothesis test is described as:
	{\small \begin{align}
		P_{null} =  \text{Tr}(\mathbf{A}_{1,f,t-\tau}^H \mathbf{\hat{Q}}_{1,f,t}\mathbf{A}_{1,f,t-\tau}) \mathop{\gtreqless}_{\mathcal{H}_1}^{\mathcal{H}_2} P_{th},
		\label{eq:hypothesis}
		\end{align}}
	
	\vspace{-4mm}
	\noindent where $P_{null}$ is the component of estimated received power $\text{Tr}(\mathbf{\hat{Q}}_{1,f,t})$ in the subspace spanned by columns of $\mathbf{A}_{1,f,t-\tau}$. Under $\mathcal{H}_1$, the asymptotic estimate of the received power is given as 
	{\small \begin{align}
		\text{Tr}(\mathbf{{Q}}_{1,f,t}) = \text{Tr}\left(\mathbf{G}_{11,f,t} \mathbb{E}\left[\mathbf{x}_1(n)\mathbf{x}^H_1(n)\right]\mathbf{G}^H_{11,f,t} +\sigma^2_w \mathbf{I}\right) = P_{1,x} \text{Tr}\left(\mathbf{G}_{11,f,t} \mathbf{G}^H_{11,f,t}\right) +M_s \sigma^2_w,
		\end{align}}
	
	\vspace{-6mm}
	\noindent
	where $P_{1,x}$ is the transmit power from PU-1 and $\mathbb{E}\left[\mathbf{x}_1(n)\mathbf{x}^H_1(n)\right] = P_{1,x} \mathbf{I}$. However, since SU has only non-asymptotic estimate, the estimated received power $\text{Tr}(\mathbf{\hat{Q}}_{1,f,t})$ is modeled as a Gaussian random variable with mean $\mu = \text{Tr}(\mathbf{{Q}}_{1,f,t}) = M_s\sigma^2_w (\text{SNR}  + 1)$ and variance $\sigma^2 = \frac{1}{N}\left(\mu^2 + \sigma^4_w\right)$, where $\text{SNR} = \frac{ P_{1,x} \text{Tr}\left(\mathbf{G}_{11,f,t} \mathbf{G}^H_{11,f,t}\right)}{M_s \sigma^2_w}$ \cite[Eq. (12)]{laghate2017}. Further, using the Gauss-Markov model
	{\small \begin{align}
		\mathbf{{G}}_{11,f,t} = \alpha_f^\tau \mathbf{{G}}_{11,f,t-\tau} + \sqrt{1-\alpha_f^2} \sum_{\tau' = 0}^{\tau-1} \alpha_f^{\tau - \tau' - 1} \Delta \mathbf{{G}}_{11,f,t-\tau'}
		\end{align}}
	
	\vspace{-6mm}
	\noindent and the definition of correlation in channel vectors in \cite[Eq. (8)]{choi2005}, we can see that the columns of $\mathbf{{G}}_{11,f,t}$ and $\mathbf{{G}}_{11,f,t-\tau}$ are correlated with correlation $\alpha_f^{2\tau}$. As shown in \cite[Fig. 2]{choi2005}, the correlation in channel vectors results in the same correlation in eigenvectors if $\alpha_f^{2\tau}\geq 0.7$. Therefore, we get
	{\small\begin{align}
		\frac{\mathbb{E}[|\mathbf{{G}}_{11,f,t}(:,k)\mathbf{{G}}^H_{11,f,t-\tau}(:,k)|^2]}{\mathbb{E}\left[||\mathbf{{G}}_{11,f,t}(:,k)||^2\right] \mathbb{E}\left[||\mathbf{{G}}_{11,f,t-\tau}(:,k)||^2\right]} = \mathbb{E}[|\mathbf{{A}}_{1,f,t}(:,l)\mathbf{{A}}^H_{1,f,t-\tau}(:,l)|^2] =\alpha_f^{2\tau}, 
		\end{align}}
	
	\vspace{-2mm}
	\noindent where $k \in \{1,2\cdots,M_p\}, l\in\{1,2,\cdots,M_s-M_p\}$, and $\mathbf{A}(:,l)$ is the $l$th column of matrix $\mathbf{A}$. Further, using the orthogonality between $\mathbf{{A}}_{1,f,t-\tau}$ and $\mathbf{{G}}_{1,f,t-\tau}$ and the Guass-Markov model, we can write the correlation between $\mathbf{{A}}_{1,f,t-\tau}$ and $\mathbf{{G}}_{1,f,t}$ as follows:
	{\small\begin{align}
		\mathbb{E}[|\mathbf{{G}}_{11,f,t}(:,k)\mathbf{{A}}^H_{1,f,t-\tau}(:,l)|^2] =  (1-\alpha_f^{2\tau})\mathbb{E}\left[||\mathbf{{G}}_{11,f,t}(:,k)||^2\right] =(1-\alpha_f^{2\tau})\text{Tr}(\mathbf{G}_{11,f,t}(:,k)\mathbf{G}^H_{11,f,t}(:,k)),
		\end{align}}
	
	\vspace{-9mm}
	\noindent Therefore, the mean of $P_{null}$ under $\mathcal{H}_1$ is as follows:
	{\begin{align}
		\nonumber \mu_{P }=\mathbb{E}[P_{null}\vert \mathcal{H}_1] &= \text{Tr}(\mathbf{A}_{1,f,t-\tau}^H \mathbf{Q}_{1,f,t}\mathbf{A}_{1,f,t-\tau}) = (1-\alpha_f^{2\tau})P_{1,x}\text{Tr}\left(\mathbf{G}_{11,f,t} \mathbf{G}^H_{11,f,t}\right) + M_s \sigma^2_w \\
		&= (1-\alpha_f^{2\tau})\mu + \alpha_f^{2\tau}M_s \sigma^2_w.
		\label{eq:mean_P_null_H1}
		\end{align}}
	
	\vspace{-7mm}
	\noindent  The variance due to non-asymptotic estimation is expressed as in \cite[Eq.8]{chaudhari2017}:
	{ \begin{align}
		\sigma^2_{P} = \text{Var}({P_{null}}\vert \mathcal{H}_1) = (1-\alpha_f^{2\tau})^2 \sigma^2 = \frac{(1-\alpha_f^{2\tau})^2}{N}(\mu^2 + \sigma^4_w).
		\label{eq:var_P_null_H1}
		\end{align}}
	
	\vspace{-6mm}
	The probability of miss-classifying $s_{f,t}=2$ when $s_{f,t}=1$ is $p_{m}=Q\left(\frac{P_{th}- \mu_P}{\sigma_P}\right)$, where $Q(.)$ is the Q-function. For a fixed $p_m$, the threshold can be set as $P_{th} = Q^{-1}(p_m)\sigma_P + \mu_P$ and the estimate of $\mu_P$ and $\sigma_P$ are computed by replacing $\mu= \text{Tr}(\mathbf{Q}_{1,f,t})$ in (\ref{eq:mean_P_null_H1}) with $\text{Tr}(\mathbf{\hat{Q}}_{1,f,t})$. 
	
	Under $\mathcal{H}_2$, we have $\mathbf{Q}_{1,f,t} = P_{2,x} \mathbf{G}_{21,f,t}\mathbf{G}^H_{21,f,t} + \sigma^2_w  \mathbf{I}$, where $P_{2,x}$ is the transmit power from PU-2 and SNR is $\frac{P_{2,x} \text{Tr}\left(\mathbf{G}_{21,f,t} \mathbf{G}^H_{21,f,t}\right)}{{M_s \sigma^2_w}}$.	Since the columns of $\mathbf{A}_{1,f,t-\tau}$ and $\mathbf{G}_{21,f,t}$ are uncorrelated, $P_{null}$ is a gamma random variable with shape parameter $\kappa$ and scale parameter $\theta$ given as follows:
		{ \begin{align}
			\kappa = \frac{M_s M_p \left(P_{2,x} + \sigma^2_w +\sigma^2_w/N \right)^2}{P^2_{2,x} + \sigma^4_w +\sigma^4_w/N^2},~
			\theta = \frac{P^2_{2,x} + \sigma^4_w +\sigma^4_w/N^2}{P_{2,x} + \sigma^2_w +\sigma^2_w/N},
			\label{eq:P_null_H2}
			\end{align}}
		
		\vspace{-5mm}
		\noindent The above follows from \cite[Lemma 2 and 3]{hosseini2014}. Thus, for a fixed threshold $P_{th}$, the probability of error in state estimation is given as 	
		{ \begin{align}
			p_e &=  \pi_{1,f} \Pr\left(P_{null} > P_{th} \vert \mathcal{H}_1\right) + \pi_{2,f} \Pr\left(P_{null} \leq P_{th} \vert \mathcal{H}_2\right )\\
			&= \pi_{1,f} Q\left(\frac{P_{th} - \mu_{P}}{\sigma_{P }}\right) + \pi_{2,f} \mathcal{F}(\kappa,\theta, P_{th}),
			\label{eq:state_est_error}
			\end{align}}
		where $\mathcal{F}(\kappa,\theta, P_{th}) = \gamma\left(\kappa, \frac{P_{th}}{\theta}\right) / \Gamma(\kappa)$ is the CDF of $P_{null}$  under $\mathcal{H}_2$.

		If there is an error in the state estimation, SU-1 utilizes incorrect null space for precoding, which results in higher interference leakage to PU receiver. For example, if $s_{f,t}=1$ and estimated state is $s_{f,t}=2$, then the expected interference to PU receiver is $\mathbb{E}[P_t ||\mathbf{G}_{12,f,t}^H \mathbf{v}_t||^2]=P_t M_p$. Therefore, the expected interference to PU receiver considering the state estimation error is
		\begin{align}
		\mathbb{E}[I_{f,t}] = (1-p_e) P_tM_p (1-\alpha_f^{2\tau}) + p_e P_t M_p,
		\label{eq:avg_int_v2}
		\end{align}
		where the first term follows from the discussion in Section \ref{sec:interference_to_pu}. For $p_e \ll 	\frac{1-\alpha_f^{2\tau}}{2-\alpha_f^{2\tau}}$, we can ignore the impact of error in state estimation and approximate (\ref{eq:avg_int_v2}) with (\ref{eq:avg_int}). The estimation error is plotted as a function of SNR in Fig. \ref{fig:fig12} and \ref{fig:fig34} under PU traffic configurations 4 and 5 which have largest average link reversal time. Fig. \ref{fig:fig12} shows the plot for $\alpha_f=0.9755$, while Fig. \ref{fig:fig34} shows the plot for $\alpha_f = 0.9999$. The threshold is set as $P_{th} = Q^{-1}(10^{-4})\mu_P + \sigma_P$ for $p_m=10^{-4}$.	In these figures, the red line shows the bound $\Lambda(\alpha_f, \tau) = 0.01\left(\frac{1-\alpha_f^{2\tau}}{2-\alpha_f^{2\tau}}\right)$. From Fig. \ref{fig:fig1}, \ref{fig:fig2}, \ref{fig:fig3}, and \ref{fig:fig4}, we can observe that the condition  $p_e \leq \Lambda(\alpha_f, \tau)$ is satisfied for $\text{SNR}\geq 3$dB if $\tau\leq 10$ when the number of antennas is $M_s=4$ and $M_p=1$. 
		
		The impact of increasing the number of SU and PU antennas is shown in Fig. \ref{fig:fig56}. Increasing $M_s$ for fixed $M_p$ reduces the estimation error as shown in Fig. \ref{fig:fig5}. Similarly, we can see in  Fig. \ref{fig:fig6} that increasing $M_p$ for fixed $M_s$ reduces the estimation error. This is due to the fact that the shape parameter $\kappa$ of $P_{null}$ under $\mathcal{H}_2$ is proportional to $M_s M_p$ as shown in (\ref{eq:P_null_H2}). Increasing $M_s$ or $M_p$ increases the mean of $P_{null}$ under $\mathcal{H}_2$ which results in lower probability of error.

\begin{figure}[t!]
	\centering	
	\begin{subfigure}[b]{0.49\columnwidth}
		\centering	
		\includegraphics[width=1.1\columnwidth]{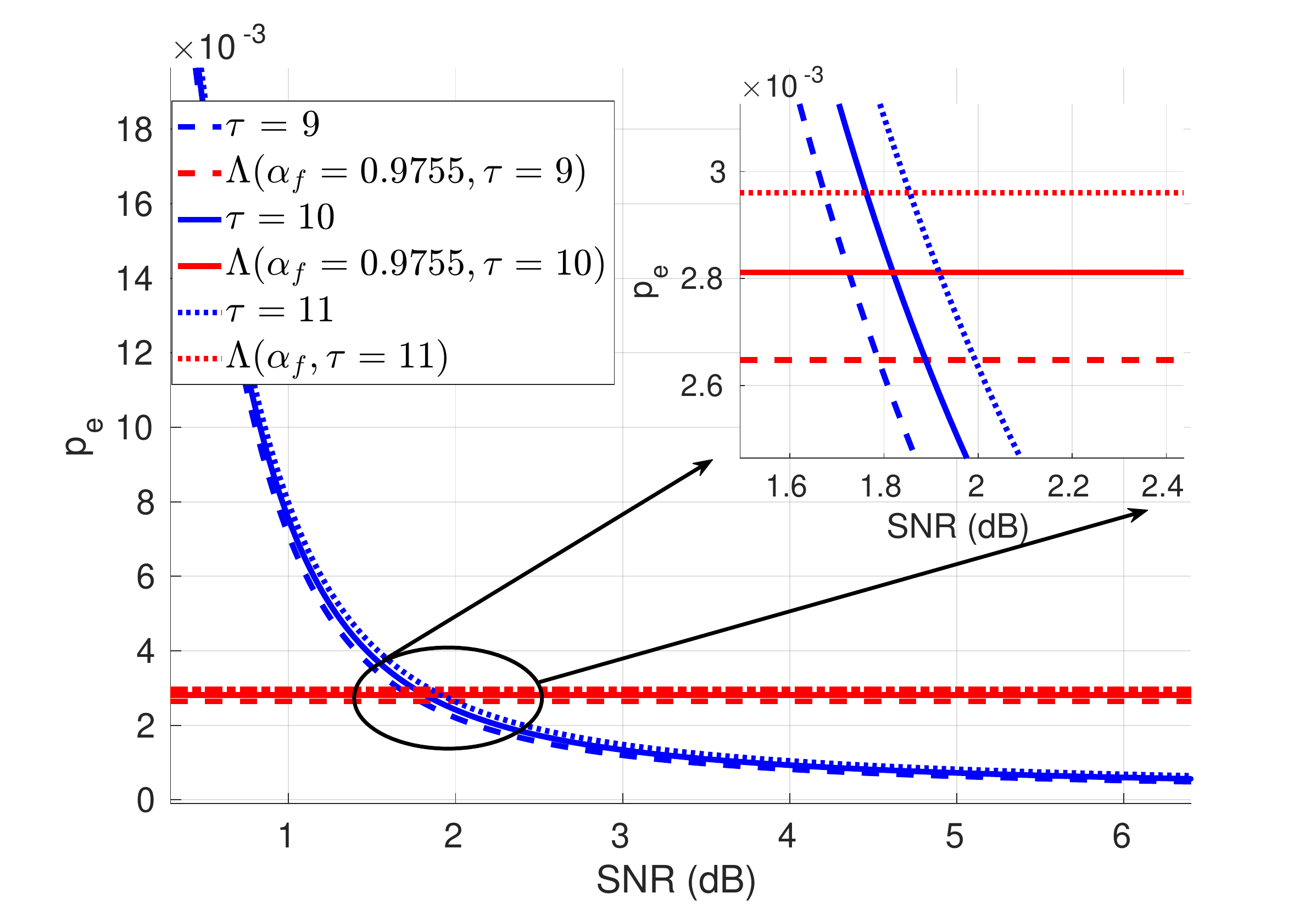}
		\caption{\small Traffic config. = 4.}
		\label{fig:fig1}
	\end{subfigure}			
	\begin{subfigure}[b]{0.49\columnwidth}
		\centering	
		\includegraphics[width=1.1\columnwidth]{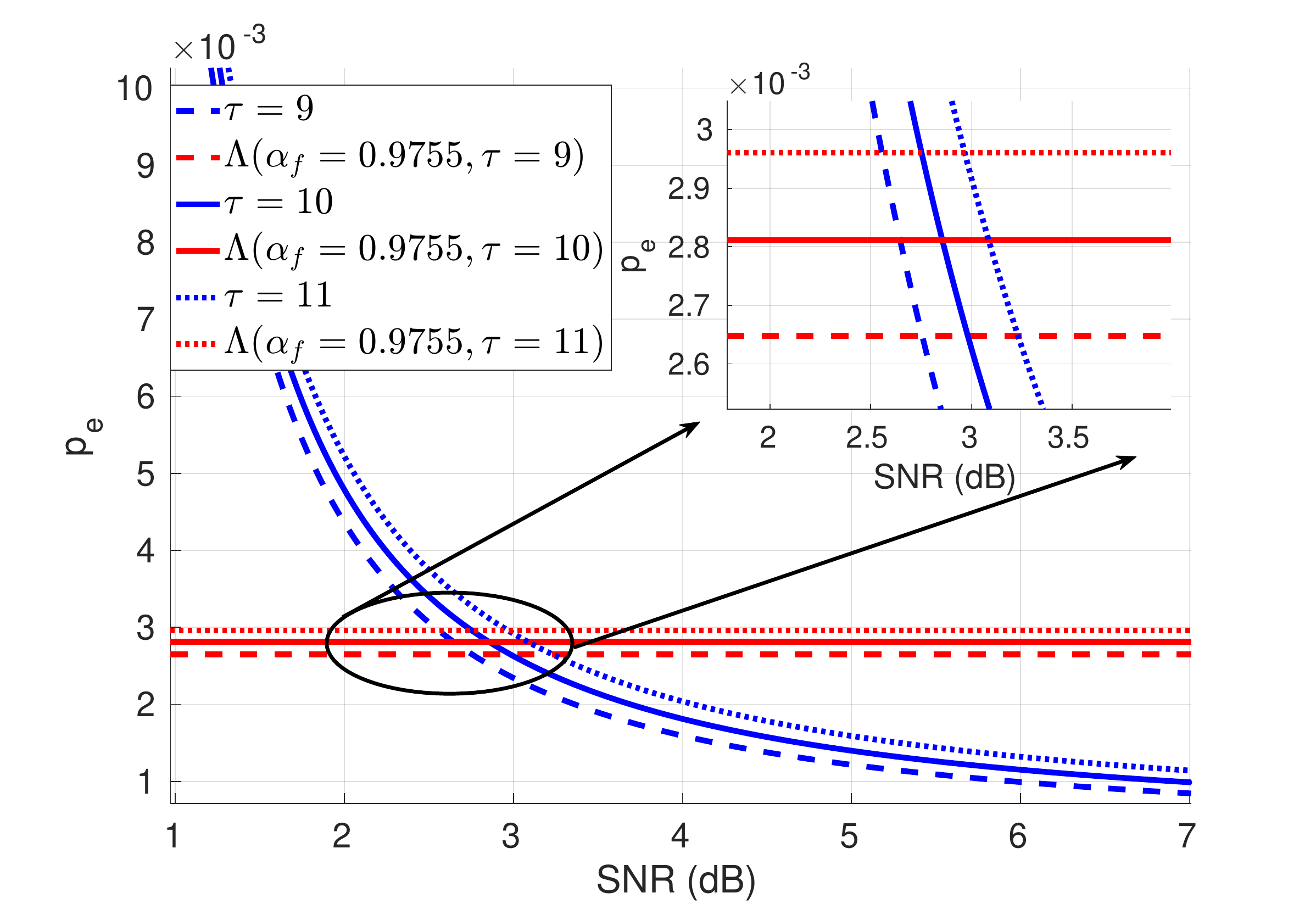}
		\caption{\small Traffic config. = 5.}
		\label{fig:fig2}
	\end{subfigure}		
	\caption{\small $p_e$ vs SNR at different $\tau$ under traffic configuration 4 and 5, $\alpha_f = 0.9755, M_s = 4, M_p=1, N=200, \Lambda(\alpha_f, \tau) = 0.01\left(\frac{1-\alpha_f^{2\tau}}{2-\alpha_f^{2\tau}}\right)$.}
	\label{fig:fig12}	
	\vspace{-1mm}
\end{figure}

\begin{figure}[t!]
	\centering	
	\begin{subfigure}[b]{0.49\columnwidth}
		\centering	
		\includegraphics[width=1.1\columnwidth]{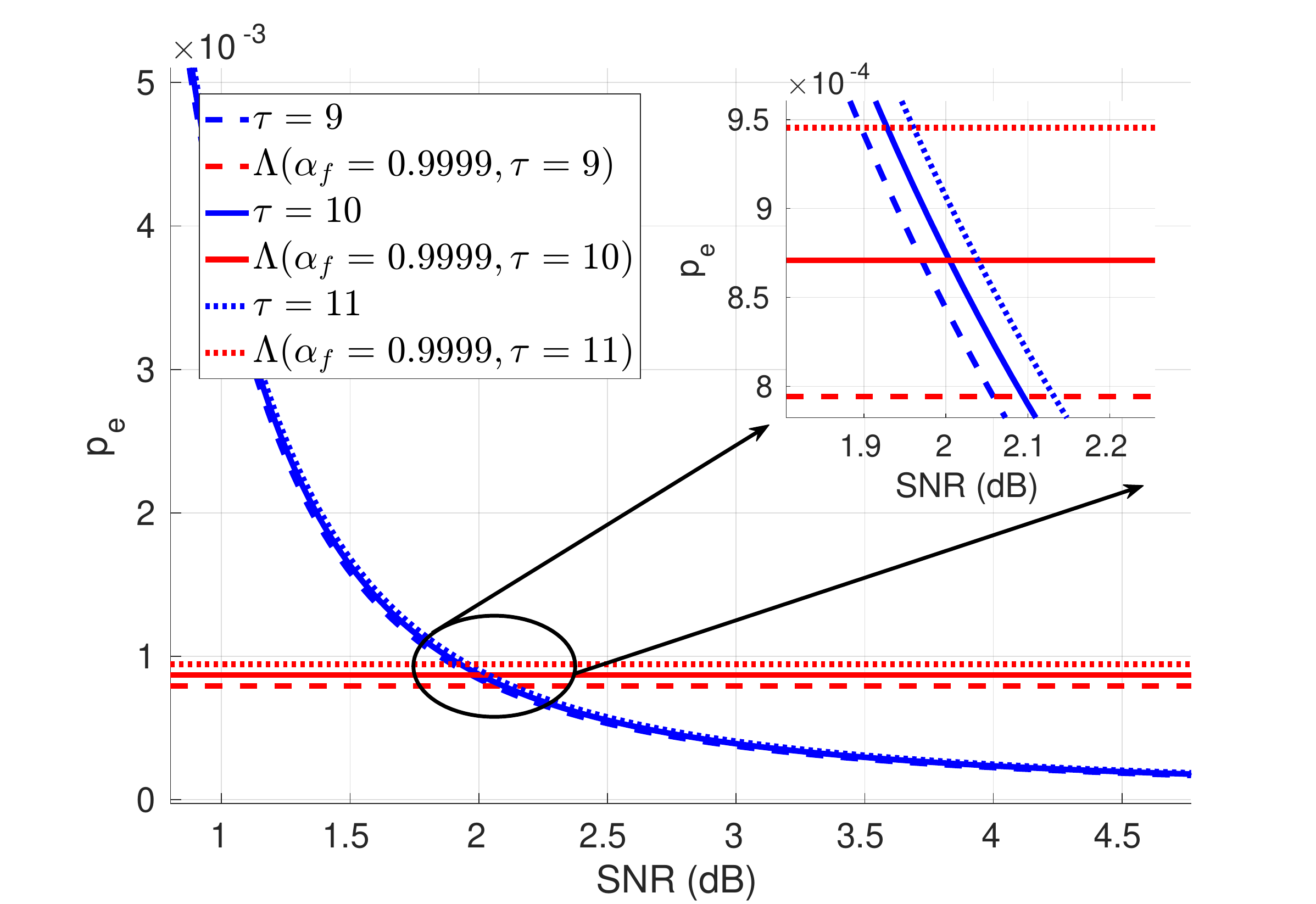}
		\caption{\small Traffic config = 4.}
		\label{fig:fig3}
	\end{subfigure}			
	\begin{subfigure}[b]{0.49\columnwidth}
		\centering	
		\includegraphics[width=1.1\columnwidth]{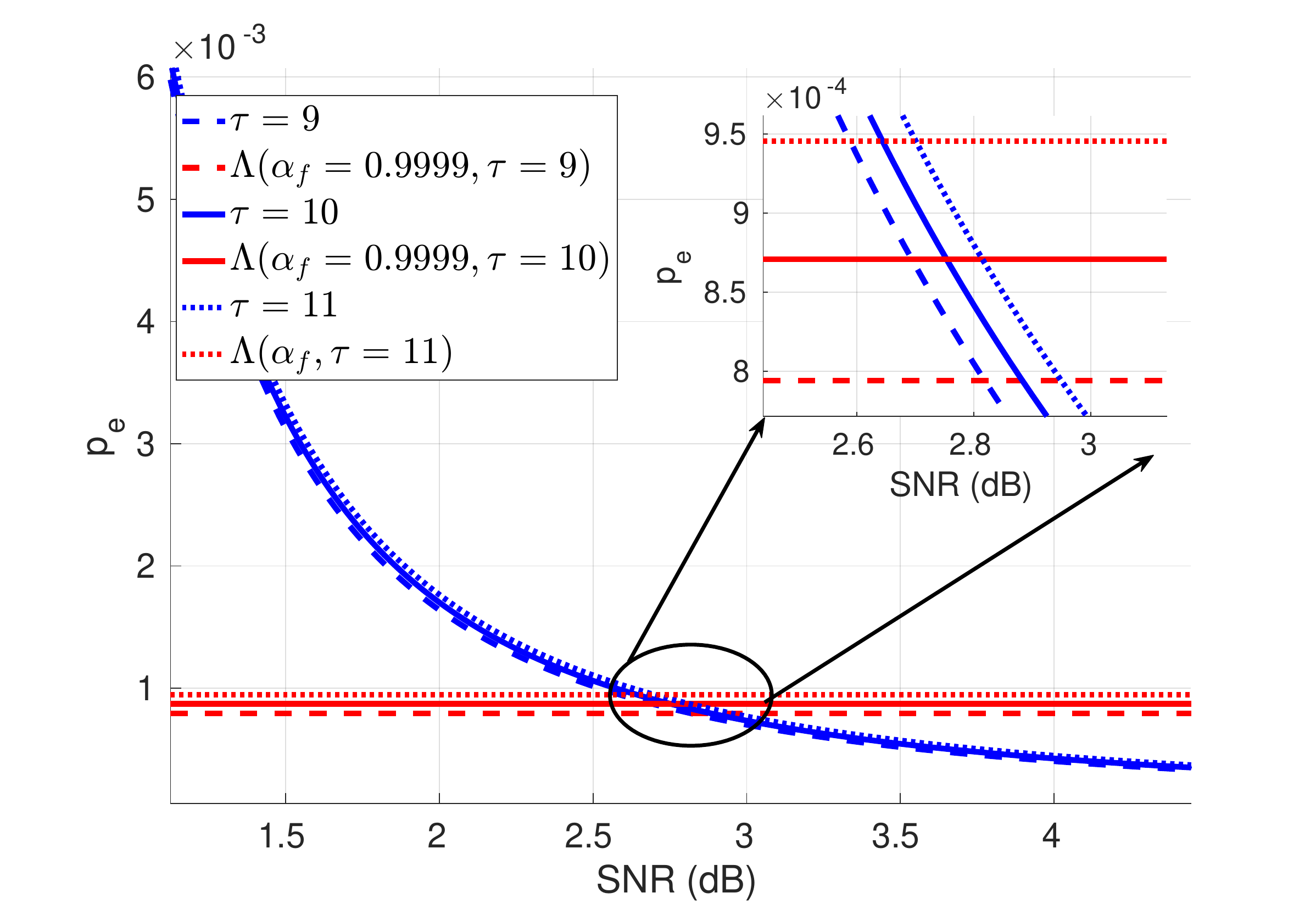}
		\caption{\small Traffic config = 5.}
		\label{fig:fig4}
	\end{subfigure}		
	\caption{\small $p_e$ vs SNR at different $\tau$ under traffic configuration 4 and 5, $\alpha_f = 0.9999, M_s = 4, M_p=1, N=200,  \Lambda(\alpha_f, \tau) = 0.01\left(\frac{1-\alpha_f^{2\tau}}{2-\alpha_f^{2\tau}}\right)$.}
	\label{fig:fig34}	
	\vspace{-1mm}
\end{figure}

\begin{figure}[t!]
	\centering	
	\begin{subfigure}[b]{0.49\columnwidth}
		\centering	
		\includegraphics[width=1.1\columnwidth]{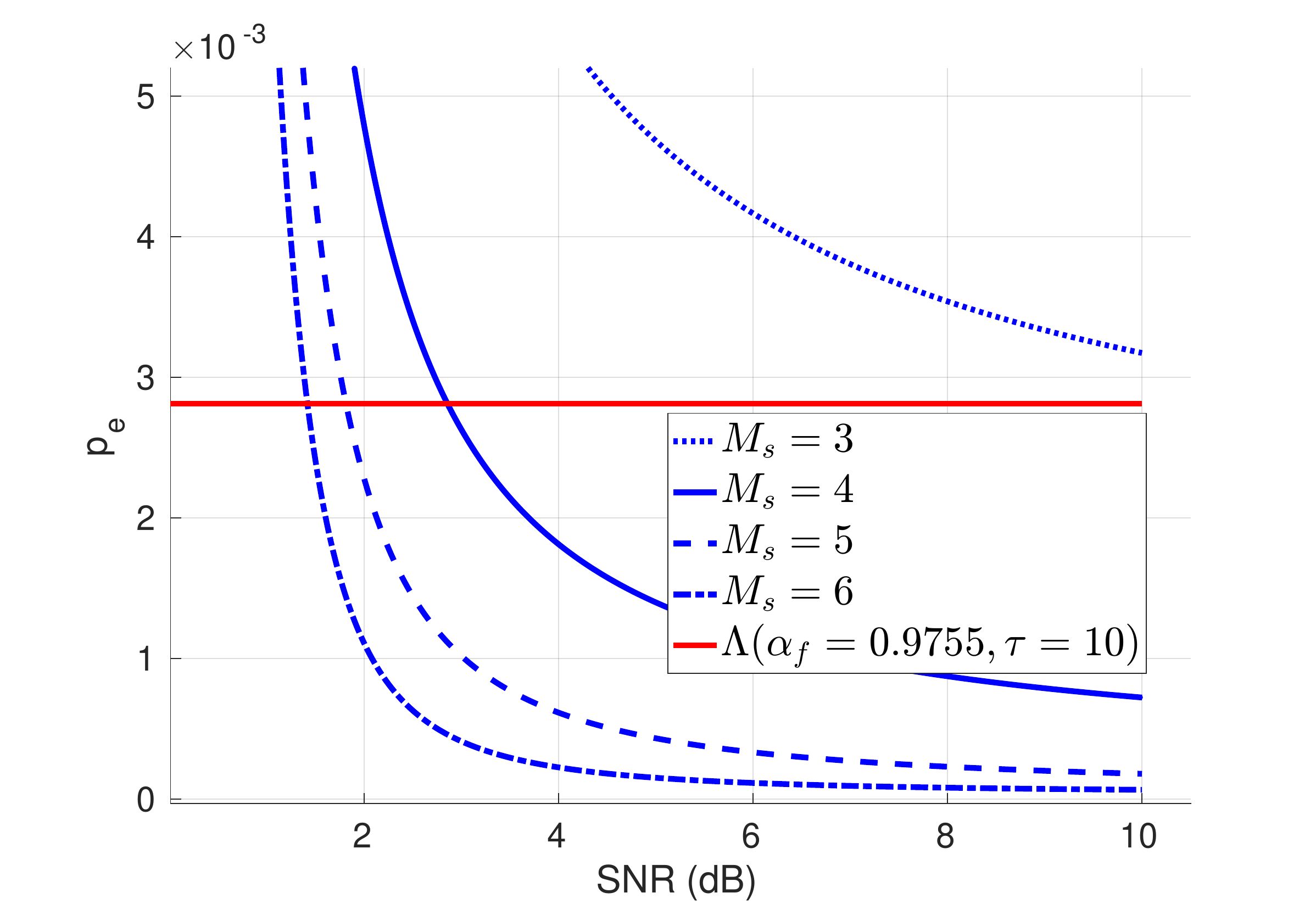}
		\caption{\small $M_p=1, M_s \in \{3,4,5,6\}$.}
		\label{fig:fig5}
	\end{subfigure}			
	\begin{subfigure}[b]{0.49\columnwidth}
		\centering	
		\includegraphics[width=1.1\columnwidth]{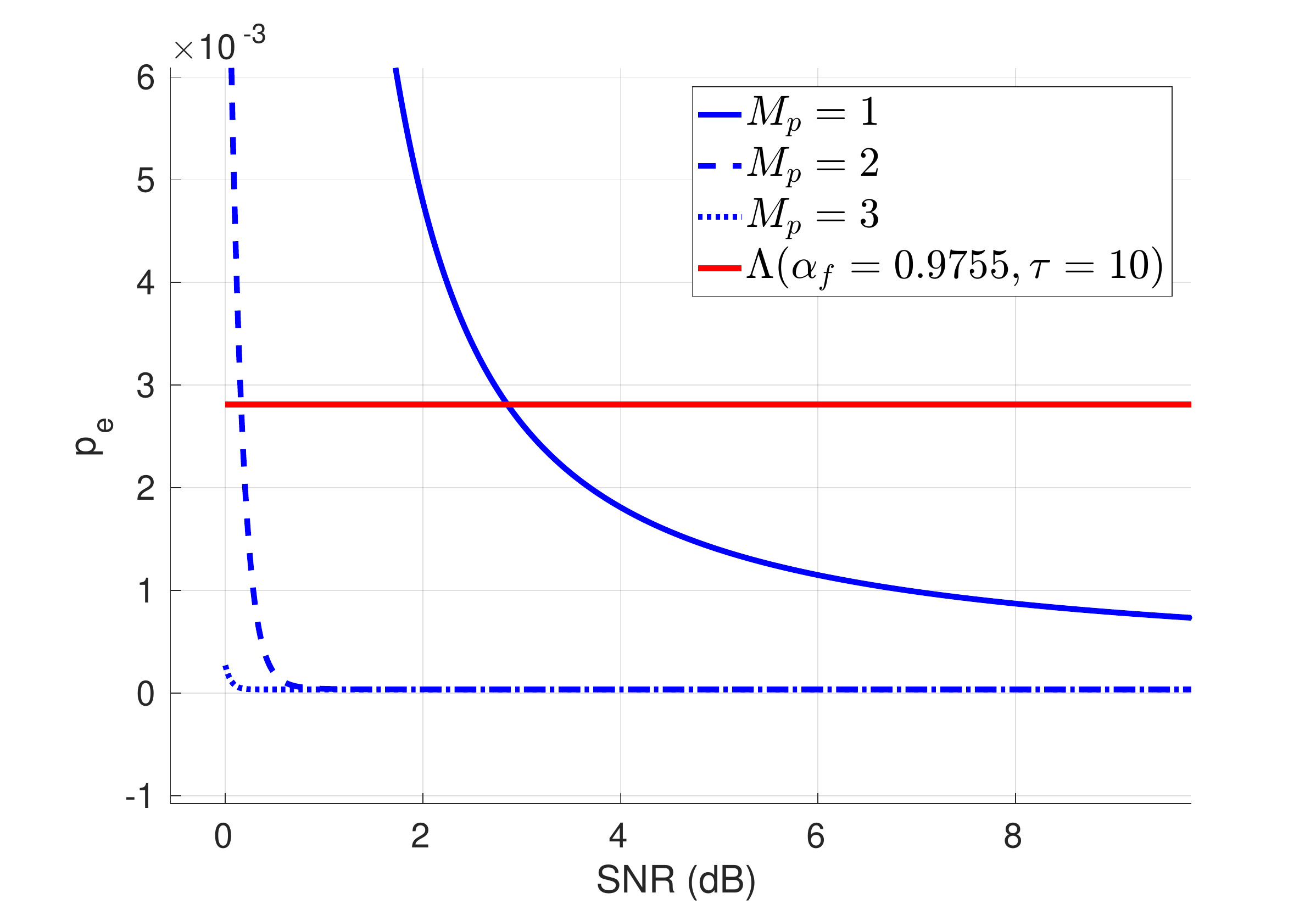}
		\caption{\small $M_s=4, M_p \in \{1,2,3\}$.}
		\label{fig:fig6}
	\end{subfigure}		
	\caption{\small $p_e$ vs SNR with different number of antennas. Traffic configuration 5, $\alpha_f = 0.9755,  N=200, \tau=10,\Lambda(\alpha_f, \tau) = 0.01\left(\frac{1-\alpha_f^{2\tau}}{2-\alpha_f^{2\tau}}\right)$.}
	\label{fig:fig56}	
	\vspace{-1mm}
\end{figure}


\section{Proof: $\mathbb{E}[P_t ||\mathbf{G}^H_{11,f,t}\mathbf{v}_t||^2] = P_t M_p (1-\alpha_f^{2\tau})$}
\label{app:avg_int}
Using the Gauss-Markov model, the relationship between $\mathbf{G}^H_{11,f,t}$ and $\mathbf{G}^H_{11,f,t-\tau}$ can be expressed as follows:
{\small\begin{align}
\mathbf{G}^H_{11,f,t} = \alpha_f^\tau \mathbf{G}^H_{11,f,t-\tau} + \sqrt{1-\alpha_f^2} \sum_{\tau' = 0}^{\tau-1} \alpha_f^{\tau - \tau' - 1} \Delta \mathbf{G}^H_{11,f,t-\tau'}.
\end{align}}
The beamforming vector $\mathbf{v}_t$ is in the null space of $\mathbf{G}^H_{11,f,t-\tau}$, i.e., $\mathbf{G}^H_{11,f,t-\tau} \mathbf{v}_t =0$. Therefore, 
{\begin{align}
 \mathbb{E}& \left[ P_t||\mathbf{G}^H_{11,f,t} \mathbf{v}_t||^2 \right] =
P_t (1-\alpha_f^2) \mathbb{E}\left[||\sum_{\tau' = 0}^{\tau-1} \alpha_f^{\tau - \tau' - 1} \Delta \mathbf{G}^H_{11,f,t-\tau'} \mathbf{v}_t ||^2\right].
\end{align}}

\vspace{-7mm}
\noindent Using the fact that channel evolutions $\Delta \mathbf{G}^H_{11,f,t-\tau'}\sim \mathcal{CN}(0,\mathbf{I})$ are i.i.d., and $\mathbf{v}_t$ is a unit-norm vector, the above equation reduces to
{\small \begin{align}
\mathbb{E} \left[ P_t||\mathbf{G}^H_{11,f,t} \mathbf{v}_t||^2 \right] = P_t M_p(1-\alpha_f^2)  \sum_{\tau' = 0}^{\tau-1} \alpha_f^{2(\tau - \tau' - 1)} = P_t M_p (1-\alpha_f^{2\tau}),
\end{align}}

\vspace{-7mm}
\noindent where $M_p$ is the rank of $\mathbf{G}_{11,f,t}$.

\section{Expression for $P^{fix}_f = \frac{I^0}{M_p\mathbb{E}_\tau [1-\alpha_f^{2\tau}]}$ }
\label{app:p_n_fix}
Let us define $g(\alpha_f, \mathbf{T}_f) = \mathbb{E}_\tau [1-\alpha_f^{2\tau}]$. The expectation can be written as
{\footnotesize \begin{align}
\nonumber g(\alpha_f, \mathbf{T}_f)  &= \sum_{i \in \mathbb{N}} (1 -\alpha_f^{2i }) \times \Pr(\tau=i \vert s_{f,t}=\{1,2\})
\\& = \sum_{i \in \mathbb{N}} (1 -\alpha_f^{2i }) \left[ \frac{\pi_{1,f}}{\pi_{1,f} + \pi_{2,f}} \Pr(\tau=i \vert s_{f,t}=1)\right]
+ \sum_{i \in \mathbb{N}} (1 -\alpha_f^{2i }) \left[ \frac{\pi_{2,f}}{\pi_{1,f} + \pi_{2,f}}  \Pr(\tau=i \vert s_{f,t}=2)\right]
\label{eq:f1}
\end{align}}

\vspace{-7mm}
\noindent where $ \mathbb{N}$ is a set of natural numbers and $\Pr(\tau=i | s_{f,t}=1)$ is the probability that the null space of $i$ slots old when $s_{f,t}=1$. This probability can be written as follows:
{\small \begin{align}
\Pr(\tau=i\vert s_{f,t}=1) 
= \frac{\pi_{2,f}}{\pi_{1,f}} \Pr(s_t=1, s_{t-1}\neq 2,...,s_{t-(i-1)}\neq 2 | s_{t-i}=2)
=\frac{\pi_{2,f}}{\pi_{1,f}} \sum_{s\in \{0,1\}} p_{2s}p_{s1\backslash 2}^{(i-1)}.
\label{eq:tau_st_1}
\end{align}}

\vspace{-7mm}
\noindent where $p_{ss' \backslash s''}^{(i)}$ is the probability of PU link going from state $s$ to state $s'$ in $i$ slots without hitting state $s''$. Similarly, 
\begin{align}
\Pr(\tau=i\vert s_{f,t}=2) = \frac{\pi_{1,f}}{\pi_{2,f}}  \sum_{s\in \{0,2\}} p_{1s}p_{s2\backslash 1}^{(i-1)},
\label{eq:tau_st_2}
\end{align}
Substituting (\ref{eq:tau_st_2}) and (\ref{eq:tau_st_1}) in  (\ref{eq:f1}) and then in (\ref{eq:p_n_fix_def}), we get the required expression in (\ref{eq:p_n_fix}).

\section{Proof of theorem \ref{thm:fbdp_v_fbfp}}
\label{app:fbdp_v_fbfp}
Let us consider that SU stays on frequency band $f$ under fixed and dynamic power policies. The expected rates under the two policies can be written as follows:
{\small
\begin{align}
\nonumber \mathbb{E}[R^{(1)}_{f,t}] =  (1- \pi_{0,f}) \mathbb{E}[R^{(1)}_{f,t} | s_{f,t}=\{1,2\}] + \pi_{0,f} \mathbb{E}[R^{(1)}_{f,t} | s_{f,t}=\{0\}],
\\ \mathbb{E}[R^{(2)}_{f,t}] =  (1- \pi_{0,f}) \mathbb{E}[R^{(2)}_{f,t} | s_{f,t}=\{1,2\}] + \pi_{0,f} \mathbb{E}[R^{(2)}_{f,t} | s_{f,t}=\{0\}].
\end{align}}
Since both SU are on the same band and it transmits same transmit power $P^0$ under the two policies when PU is silent, i.e., $s_{f,t}=0$, we have $\mathbb{E}[R^{(1)}_{f,t} | s_{f,t}=\{0\}]= \mathbb{E}[R^{(2)}_{f,t} | s_{f,t}=\{0\}]$. Therefore, to prove  $\mathbb{E}[R^{(2)}_{f,t}]\geq \mathbb{E}[R^{(1)}_{f,t}]$, it is sufficient to prove that $\mathbb{E}[R^{(2)}_{f,t} \vert s_{f,t}=\{1,2\}]  \geq \mathbb{E}[R^{(1)}_{f,t} \vert s_{f,t}=\{1,2\}] $. Equivalently, we need to prove that
{\footnotesize \begin{align}
 \mathbb{E}\left[\log_2\left(1 + \frac{I^0}{M_p(1-\alpha_f^{2\tau})} {\Gamma_{f,t}}\right) \vert s_{f,t}=\{1,2\} \right] 
\geq \mathbb{E}_{\Gamma}\left[\log_2\left(1 + \frac{I^0}{M_p\mathbb{E}_\tau [1-\alpha_f^{2\tau}]} {\Gamma_{f,t}}\right) \vert s_{f,t}=\{1,2\} \right].
\label{eq:r2_v_r1}
\end{align}}

\noindent The expectation in the LHS can be split in terms of expectation with respect to $\tau$ and $\Gamma$ as:
{\footnotesize \begin{align}
\mathbb{E}\left[\log_2\left(1 + \frac{I^0}{M_p(1-\alpha_f^{2\tau})} {\Gamma}\right) \vert s_{f,t}=\{1,2\} \right] =
 \mathbb{E}_{\Gamma}\left[ \mathbb{E}_{\tau} \left[\log_2\left(1 + \frac{I^0}{M_p(1-\alpha_f^{2\tau})} {\Gamma}\right) \vert s_{f,t}=\{1,2\} \right] \right].
\end{align}}

\noindent For simplicity of notations, we have dropped suffixes from $\Gamma$. In order to prove the inequality in (\ref{eq:r2_v_r1}), it is sufficient to prove that for any given value of $\Gamma_{f,t}$ the following holds:
{\footnotesize\begin{align}
 \mathbb{E}_{\tau} \left[\log_2\left(1 + \frac{I^0}{M_p(1-\alpha_f^{2\tau})} {\Gamma}\right) \vert s_{f,t}=\{1,2\} \right] \geq \log_2\left(1 + \frac{I^0}{M_p\mathbb{E}_\tau [1-\alpha_f^{2\tau}]}{\Gamma}\right).
\label{eq:eqvt}
\end{align}}

\noindent Using Bayes' rule, the term in the LHS can be written as:
{\footnotesize \begin{align}
\nonumber \mathbb{E}_{\tau}\left[\log_2\left(1 + \frac{I^0}{M_p(1-\alpha_f^{2\tau})} {\Gamma}\right) \vert s_{f,t}=\{1,2\} \right] 
 =& \frac{\pi_{1,f}}{\pi_{1,f} + \pi_{2,f}} \mathbb{E}\left[\log_2\left(1 + \frac{I^0}{M_p(1-\alpha_f^{2\tau})} {\Gamma}\right) \vert s_{f,t}=\{1\} \right]  
\\ &+ \frac{\pi_{2,f}}{\pi_{1,f} + \pi_{2,f}} \mathbb{E}\left[\log_2\left(1 + \frac{I^0}{M_p(1-\alpha_f^{2\tau})} {\Gamma}\right) \vert s_{f,t}=\{2\} \right].
\label{eq:eqvt2}
\end{align}}

Let us define $a = \frac{I^0}{M_p(1-\alpha_f^{2\tau})} | s_{f,t}=1$, i.e., $a$ is a random variable with value $\frac{I^0}{M_p(1-\alpha_f^{2\tau})}$ when $s_{f,t}=1$. Similarly, let 
$b = \frac{I^0}{M_p(1-\alpha_f^{2\tau})} | s_{f,t}=2$. Note that $\Gamma$ follows the same distribution under $s_{f,t}=1$ and $s_{f,t}=2$. Therefore, we can write (\ref{eq:eqvt2}) as follows

{\footnotesize \begin{align}
 \mathbb{E}_{\tau} &\left[\log_2\left(1 + \frac{I^0}{M_p(1-\alpha_f^{2\tau})} {\Gamma}\right) \vert s_{f,t}=\{1,2\} \right] = \frac{\pi_{1,f}}{\pi_{1,f} + \pi_{2,f}} \mathbb{E}_{\tau}\left[\log_2(1+a \Gamma)\right] + \frac{\pi_{2,f}}{\pi_{1,f} + \pi_{2,f}} \mathbb{E}_{\tau} \left[\log_2(1+b \Gamma)\right].
\label{eq:eqvt3}
\end{align}}

\vspace{-5mm}
\noindent Using the property of log function, we have {\small$\frac{\pi_{1,f}}{\pi_{1,f} + \pi_{2,f}} \mathbb{E}_{\tau}\left[\log_2(1+a \Gamma)\right] \geq -\frac{\pi_{1,f}}{\pi_{1,f} + \pi_{2,f}} \log_2 \mathbb{E}_\tau(\frac{1}{1+a\Gamma})$} and $\frac{\pi_{2,f}}{\pi_{1,f} + \pi_{2,f}} \mathbb{E}_{\tau}\left[\log_2(1+b \Gamma)\right] \geq -\frac{\pi_{2,f}}{\pi_{1,f} + \pi_{2,f}} \log_2 \mathbb{E}_\tau(\frac{1}{1+b\Gamma})$.

Now let us define $c = \frac{I^0}{M_p\mathbb{E}_{\tau}[1-\alpha_f^{2\tau}]}$. The term in the RHS of (\ref{eq:eqvt}) is then 
\begin{align}
\log_2\left(1 + \frac{I^0}{M_p\mathbb{E}_\tau [1-\alpha_f^{2\tau}]} {\Gamma}\right) = \log_2(1+ c\Gamma).
\label{eq:eqvt4}
\end{align}
Therefore, using (\ref{eq:eqvt3}), (\ref{eq:eqvt4}), we can say that to prove (\ref{eq:eqvt}) hold for any value of $\Gamma$, it is sufficient to prove the following:
\begin{align}
-\frac{\pi_{1,f}}{\pi_{1,f} + \pi_{2,f}} \log_2 \left(\mathbb{E}_\tau \left[\frac{1}{1+a\Gamma}\right] \right) -\frac{\pi_{2,f}}{\pi_{1,f} + \pi_{2,f}} \log_2 \mathbb{E}_\tau \left( \left[\frac{1}{1+b\Gamma} \right]\right) \geq \log_2 (1 + c\Gamma),
\end{align}
or equivalently
{\small \begin{align}
 \log_2 \left[ \left(\mathbb{E}_\tau \left[\frac{1}{1+a\Gamma}\right] \right)^{-\pi_{1,f}} \left(\mathbb{E}_\tau \left[\frac{1}{1+b\Gamma}\right] \right)^{-\pi_{2,f}}  \right] \geq  \log_2\left[ (1 + c\Gamma)^{{\pi_{1,f} + \pi_{2,f} }}\right].
\end{align}}

\vspace{-5mm} \noindent Since, logarithm is a monotonically increasing function, it is sufficient to prove that
{\small \begin{align}
\left(\mathbb{E}_\tau \left[\frac{1}{1+a\Gamma}\right] \right)^{-\pi_{1,f}} & \left(\mathbb{E}_\tau \left[\frac{1}{1+b\Gamma}\right] \right)^{-\pi_{2,f}}  \geq   (1 + c\Gamma)^{\pi_{1,f} } (1 + c\Gamma)^{\pi_{2,f} },
\end{align}}

\vspace{-5mm}
\noindent or equivalently
{\small \begin{align}
1 \geq   \left[(1 + c\Gamma)\left(\mathbb{E}\frac{1}{1+a\Gamma}\right)\right]^{\pi_{1,f} }   \left[(1 + c\Gamma)\left(\mathbb{E}\frac{1}{1+b\Gamma}\right)\right] ^{\pi_{2,f} }.
\label{eq:eqvt5}
\end{align}}

\vspace{-5mm}
\noindent Since $0\leq \pi_{1,f}, \pi_{2,f} \leq 1$, the above inequality holds if each term in the square bracket is $\leq 1$. To prove that is the case, we first express the relationship between random variables $a,b$ and $c$ is as $c =  \frac{\pi_{1,f} + \pi_{2,f} }{\mathbb{E}[1/a] + \mathbb{E}[1/b]}$. Therefore, we have
{\begin{align}
(1 + c\Gamma)\mathbb{E}\left[\frac{1}{1+a\Gamma}\right] = \left(1 + \frac{(\pi_{1,f} + \pi_{2,f})\Gamma }{\mathbb{E}[1/a] + \mathbb{E}[1/b]} \right)\mathbb{E}\left[\frac{1}{1+a\Gamma}\right]  
\end{align}}

\vspace{-5mm}
\noindent Note that $a,b,\Gamma \geq 0$ and $0 \leq \pi_{1,f} +\pi_{2,f} \leq 1$. Therefore, to prove that the above term is $\leq 1$, it is sufficient to show that
{\small\begin{align}
\Gamma \mathbb{E} \left[ \frac{1}{1+a\Gamma}\right] \leq \mathbb{E}\left[\frac{1}{a}\right] \left(1- \mathbb{E}\left[\frac{1}{1+a\Gamma}\right] \right)
\end{align}}

\vspace{-5mm}
\noindent Since $\Gamma$ is a constant in the above equation, we can re-arrange the LHS to get the following requirement for the proof:
{\small \begin{align}
\mathbb{E} \left[ \frac{1}{a} \left(\frac{a \Gamma}{1+a\Gamma } \right)\right] \leq  \mathbb{E}\left[\frac{1}{a}\right]\mathbb{E}\left[\frac{a\Gamma}{1+a\Gamma}\right]
\label{eq:eqvt6}
\end{align}}

\vspace{-5mm}
\noindent Note that for any given $\Gamma\geq 0$, the random variables $1/a$ and $a\Gamma /(1+a\Gamma) $ are negatively correlated. Therefore, (\ref{eq:eqvt6}) always holds for any value of $\Gamma$ and (\ref{eq:r2_v_r1}) is always true, which completes the required proof of $\mathbb{E}[R^{(2)}_{f,t}] \geq \mathbb{E}[R^{(1)}_{f,t}]$ (Theorem \ref{thm:fbdp_v_fbfp}).

\bibliographystyle{uclathes}
\bibliography{IEEEabrv,dissertation_refs}    

\end {document}